\documentclass[prd,preprint,tightenlines,floatfix,showpacs,preprintnumbers,nofootinbib,eqsecnum,superscriptaddress]{revtex4-1}

\usepackage{graphicx}
 \usepackage{amssymb}
  \usepackage{amsmath}
   \usepackage{amsfonts}
    \usepackage{epsfig}
     \usepackage{bm}
      \usepackage{multirow}
\usepackage{dcolumn}
\usepackage{braket}

\usepackage{color}

\newcommand{\p}{\partial}
\newcommand{\twosidep}[1]{\stackrel{\leftrightarrow}{\p_{#1}}}
\newcommand{\leftsidep}[1]{\stackrel{\leftarrow}{\p_{#1}}}
\newcommand{\rightsidep}[1]{\stackrel{\rightarrow}{\p_{#1}}}
\begin{document}

\title{
Exclusive central diffractive production of \\scalar and pseudoscalar mesons;
tensorial vs. vectorial pomeron}

\author{Piotr Lebiedowicz}
 \email{Piotr.Lebiedowicz@ifj.edu.pl}
\affiliation{Institute of Nuclear Physics PAN, PL-31-342 Cracow, Poland}

\author{Otto Nachtmann}
 \email{O.Nachtmann@thphys.uni-heidelberg.de}
\affiliation{Institut f\"ur Theoretische Physik, Universit\"at Heidelberg Philosophenweg 16, D-69120 Heidelberg, Germany}

\author{Antoni Szczurek}
 \email{Antoni.Szczurek@ifj.edu.pl}
\affiliation{Institute of Nuclear Physics PAN, PL-31-342 Cracow, Poland}
\affiliation{University of Rzesz\'ow, PL-35-959 Rzesz\'ow, Poland}

\begin{abstract}
We discuss consequences of the models of ``tensorial pomeron'' and ``vectorial pomeron''
for exclusive diffractive production of scalar and pseudoscalar mesons 
in proton-proton collisions.
Diffractive production of $f_{0}(980)$, $f_{0}(1370)$, $f_{0}(1500)$,
$\eta$, and $\eta'(958)$ mesons is discussed.
Different pomeron-pomeron-meson tensorial coupling structures are possible in general.
In most cases two lowest orbital angular momentum - spin couplings 
are necessary to describe experimental differential distributions.
For $f_{0}(980)$ and $\eta$ production reggeon-pomeron,
pomeron-reggeon, and reggeon-reggeon exchanges are included in addition, which
seems to be necessary at relatively low energies.
The theoretical results are compared with the WA102 experimental data.
Correlations in azimuthal angle between outgoing protons,
distributions in rapidities and transverse momenta of outgoing protons and mesons,
in a special ``glueball filter variable'',
as well as some two-dimensional distributions are presented.
We discuss differences between results of the vectorial and tensorial pomeron models.
We show that high-energy central production, in particular of pseudoscalar mesons,
could provide crucial information on the spin structure of the soft pomeron.
\end{abstract}

\pacs{13.87.Ce, 13.60.Le, 13.85.Lg}

\maketitle

\section{Introduction}
\label{section:Introduction}
Double pomeron exchange mechanism is known to be responsible for 
high-energy central production of mesons with $I^{G} = 0^{+}$.
While it is clear that the effective pomeron must be a colour singlet
the spin structure of the pomeron and its coupling to hadrons is, however,
not finally established. 
It is commonly assumed that the pomeron has effectively a vectorial nature; 
see for instance \cite{DDLN,FR,Close} for the history and many references.
This model of the pomeron is being questioned in \cite{talkN, EMN13}.
Recent activity in the field concentrated rather on perturbative aspects of the pomeron. 
For instance, the production of heavy objects 
($\chi_{c}$~mesons \cite{chic, LPS11}, Higgs bosons \cite{MPS11}, 
dijets \cite{MPS11}, $W^{+}W^{-}$~pairs \cite{LPS13}, etc.) 
has been considered in the language of unintegrated gluon distributions.
Exclusive $\pi^{+}\pi^{-}$ \cite{LS10, LPS11, HKRS} and $K^{+} K^{-}$ \cite{LS12} pairs 
production mediated by pomeron-pomeron fusion
has been a subject of both theoretical and experimental studies.
Particularly interesting is the transition 
between the nonperturbative (small meson transverse momenta)
and perturbative (large meson transverse momenta) regimes.
Here we wish to concentrate rather on 
central exclusive meson production in the nonperturbative region 
using the notion of effective pomeron.
In general, such an object may have a nontrivial spin structure.

In the present analysis we explore the hypothesis
of ``tensorial pomeron'' in the central meson production.
The theoretical arguments for considering 
an effective tensorial ansatz for the nonperturbative pomeron are sketched in \cite{talkN}
and are discussed in detail \cite{EMN13}.
Hadronic correlation observables could be particularly sensitive to the spin aspects of the pomeron.

Indeed, tests for the helicity structure of the pomeron have been devised in \cite{ANDL97}
for diffractive contributions to electron-proton scattering, that is,
for virtual-photon--proton reactions.
For central meson production in proton-proton collisions such tests
were discussed in \cite{Close} and in the following we shall
compare our results with those of Ref.~\cite{Close} whenever suitable.

There are some attempts to obtain the pomeron-pomeron-meson vertex in special models of the pomeron. 
In~\cite{Close} results were obtained from the assumption that
the pomeron acts as a $J^{PC} = 1^{++}$ conserved and non-conserved current.
The general structure of helicity amplitudes of the simple Regge behaviour
was also considered in Ref.~\cite{KKMR03,PRSG05}.
On the other hand, the detailed structure of the amplitudes
depends on dynamics and cannot be predicted from the general principles of Regge theory.
The mechanism for central production of scalar glueball
based on the ``instanton'' structure of QCD vacuum
was considered in \cite{EK98, K99, KL01, SZ03}.

In the present paper we shall consider some examples 
of central meson production and compare results of our calculations for the ``tensorial pomeron''
with those for the ``vectorial pomeron'' as well as with experimental data whenever possible.
Pragmatic consequences will be drawn.
Predictions for experiments at RHIC, Tevatron, and LHC are rather straightforward
and will be presented elsewhere.

The aim of the present study is to explore the potential of exclusive processes
in order to better pin down the nature of the pomeron exchange.
Therefore, we shall limit ourselves to Born level calculations leaving other,
more complicated, effects for further studies.
Nevertheless, we hope that our studies will be useful for planned 
or just being carried out experiments.

Our paper is organised as follows.
In Section \ref{section:Formalism} we discuss the formalism.
We present amplitudes for the exclusive production of scalar and pseudoscalar mesons
and we also briefly report some experimental activity in this field.
In Section \ref{section:Results} we compare results of our calculations with existing data,
mostly those from the WA102 experiment 
\cite{WA102_PLB397,WA102_PLB427,WA102_PLB462,WA102_PLB467,WA102_PLB474,kirk00}.
In Appendices \ref{section:Tensorial_Pomeron} and \ref{section:Vectorial_Pomeron}
we discuss properties and useful relations for the tensorial and vectorial pomeron, respectively.
In Appendices \ref{section:Covariant_Couplings} and \ref{section:Kinematic_Relations}
we have collected some useful formulae concerning details of the calculations.
Central production of mesons with spin greater than zero will be discussed in a separate paper.

\section{Formalism}
\label{section:Formalism}

\subsection{Basic elements}
\label{subsection:Basic_elements}
We shall study exclusive central meson production
in proton-proton collisions at high energies
\begin{eqnarray}
p(p_{a},\lambda_{a}) + p(p_{b},\lambda_{b}) \to
p(p_{1},\lambda_{1}) + M(k) + p(p_{2},\lambda_{2}) \,.
\label{2to3}
\end{eqnarray}
Here $p_{a,b}$, $p_{1,2}$ and $\lambda_{a,b}$, $\lambda_{1,2}$
denote, respectively, the four-momenta and helicities of the protons and
$M (k)$ denotes a meson with $I^{G} = 0^{+}$ and four-momentum $k$.
Our kinematic variables are defined as follows
\begin{eqnarray}
&&q_{1} = p_{a} - p_{1}, \quad
q_{2} = p_{b} - p_{2}, \quad
k = q_{1} + q_{2}, \nonumber \\ 
&&s = (p_{a} + p_{b})^{2} = (p_{1} + p_{2} + k)^{2}, \quad
s_{13} = (p_{1} + k)^{2}, \quad
s_{23} = (p_{2} + k)^{2}, \nonumber \\
&&t_{1} = q_{1}^{2}, \quad
t_{2} = q_{2}^{2}, \quad
m_{M}^{2} = k^{2}\,.
\label{2to3_kinematics}
\end{eqnarray}
For the totally antisymmetric symbol $\varepsilon_{\mu \nu \rho \sigma}$
we use the convention $\varepsilon_{0123} = 1$.
Further kinematic relations, in particular those valid in the high-energy small-angle limit,
are discussed in Appendix~\ref{section:Kinematic_Relations}.

\begin{figure}[!ht]
\includegraphics[width=0.3\textwidth]{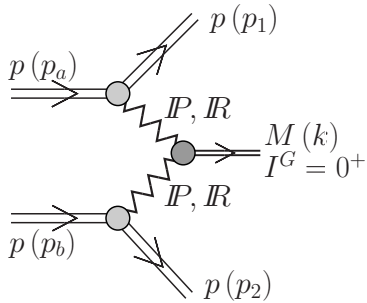}
  \caption{\label{fig:pp_pompom_ppmes}
  \small
The exchange mechanisms for central exclusive meson production in proton-proton collisions.}
\end{figure}

At high c.m.~energies $\sqrt{s}$ the dominant contribution to (\ref{2to3})
comes from pomeron-pomeron ($I\!\!P-I\!\!P$) fusion;
see Fig.~\ref{fig:pp_pompom_ppmes}. Non-leading terms arise from
reggeon-pomeron ($I\!\!R-I\!\!P$) and reggeon-reggeon ($I\!\!R-I\!\!R$) exchanges.
We shall be mainly interested in the $I\!\!P I\!\!P$-fusion giving the meson $M$.
It is clear from Fig.~\ref{fig:pp_pompom_ppmes} that in order
to calculate this contribution we must know the $I\!\!P p p$ vertex,
the effective $I\!\!P$ propagator and the $I\!\!P I\!\!P M$ vertex.
This propagator and these vertices will now be discussed, both, 
for the tensorial and vectorial ansatz for the pomeron $I\!\!P$.

\subsection{Scalar and pseudoscalar meson production}
\label{subsection:Scalar_and_pseudoscalar_meson_production}

In this section we study central production of scalar and pseudoscalar mesons,
that is, the reaction (\ref{2to3})
with $J^{PC} = 0^{++}$ and $0^{-+}$ mesons $M$.
We shall consider pomeron-pomeron fusion, see Fig.~\ref{fig:pp_pompom_ppmes},
for both, the tensorial- and the vectorial-pomeron approaches.
In Table~\ref{tab:table2_A1} of Appendix~\ref{section:Tensorial_Pomeron}
we list mesons $M$ in which we are interested.
There we also give the values of the lowest orbital angular momentum $l$
and of the corresponding total spin $S$
which can lead to the production of $M$ in the fictitious fusion of two
tensorial and vectorial ``pomeron particles''.
The lower the values of $l$ is,
the lower is the angular momentum barrier in the reaction.

We discuss first the tensor-pomeron case.
For scalar mesons, $J^{PC} = 0^{++}$, the effective Lagrangians
and the vertices for $I\!\!P I\!\!P \to M$ are discussed in
Appendix~\ref{section:Tensorial_Pomeron}.
For the tensorial pomeron the vertex corresponding to the lowest values of $(l,S)$,
that is $(l,S) = (0,0)$ plus $(2,2)$,
is given in (\ref{vertex_pompomS}).
For pseudoscalar mesons, $J^{PC} = 0^{-+}$,
the tensorial pomeron-pomeron-meson $(I\!\!P I\!\!P \tilde{M})$ coupling
corresponding to $(l,S) = (1,1)$, see Table~\ref{tab:table1_A1} of Appendix~\ref{section:Tensorial_Pomeron},
has the form
\begin{eqnarray}
{\cal L}_{I\!\!P I\!\!P \tilde{M}}'(x) = 
-\dfrac{2}{M_{0}} \, g_{I\!\!P I\!\!P \tilde{M}}' \, 
\left[ \partial_{\rho} I\!\!P_{\mu \nu}(x) \right] \, 
\left[ \partial_{\sigma} I\!\!P_{\kappa \lambda}(x) \right] \, 
g^{\mu \kappa} \, \varepsilon^{\nu \lambda \rho \sigma} \,
\tilde\chi(x) \,.
\label{Lagrangian_pseudoscalar}
\end{eqnarray}
Here $\tilde\chi(x)$ and $I\!\!P_{\mu \nu}(x)$ are the pseudoscalar meson 
and effective tensor-pomeron field operators, respectively; 
$M_{0} \equiv 1$~GeV, and $g_{I\!\!P I\!\!P \tilde{M}}'$ is a dimensionless coupling constant.
%
\begin{figure}[!ht]
(a)\includegraphics[width=0.3\textwidth]{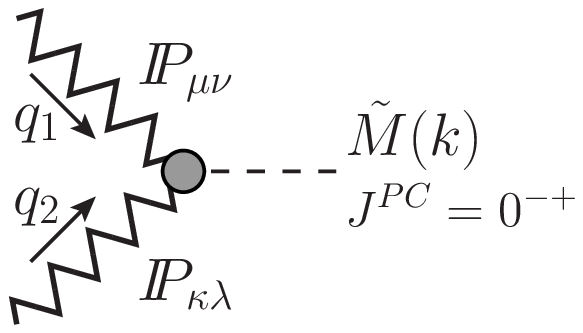}
(b)\includegraphics[width=0.3\textwidth]{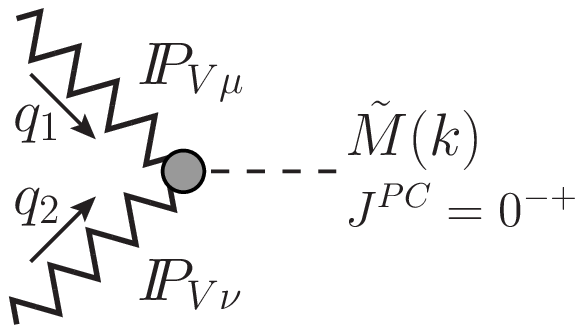}
  \caption{\label{fig:ptMp_pom}
  \small
A sketch of the pomeron-pomeron-pseudoscalar meson vertex 
for the tensorial (a) and vectorial (b) pomeron fusion.}
\end{figure}
%
The $I\!\!P I\!\!P \tilde{M}$ vertex corresponding to $(l,S) = (1,1)$ obtained from
(\ref{Lagrangian_pseudoscalar}), see Fig.~\ref{fig:ptMp_pom}~(a), including a form factor,
reads as follows:
\begin{eqnarray}
i\Gamma_{\mu \nu,\kappa \lambda}'^{(I\!\!P I\!\!P \to \tilde{M})}(q_{1}, q_{2}) &=&
i \, \frac{g_{I\!\!P I\!\!P \tilde{M}}'}{2 M_{0}} \,
\left( g_{\mu \kappa} \varepsilon_{\nu \lambda \rho \sigma}
      +g_{\nu \kappa} \varepsilon_{\mu \lambda \rho \sigma}
      +g_{\mu \lambda}\varepsilon_{\nu \kappa \rho \sigma}
      +g_{\nu \lambda}\varepsilon_{\mu \kappa \rho \sigma} \right)
(q_{1}-q_{2})^{\rho} k^{\sigma} \nonumber \\
&&\times F_{I\!\!P I\!\!P \tilde{M}}(q_{1}^{2}, q_{2}^{2})\,,
\label{vertex_pompomPS}
\end{eqnarray}
where the meson four-momentum $k = q_{1} + q_{2}$.
Another form for the $I\!\!P I\!\!P \tilde{M}$ coupling
corresponding to $(l,S) = (3,3)$ is
\begin{eqnarray}
{\cal L}_{I\!\!P I\!\!P \tilde{M}}''(x) &=&
-\dfrac{g_{I\!\!P I\!\!P \tilde{M}}''}{M_{0}^{3}} \,
\varepsilon^{\mu_{1} \mu_{2} \nu_{1} \nu_{2}} \, ( \partial_{\mu_{1}} \tilde\chi(x) )\nonumber \\
&& \times  [ \left( \partial_{\mu_{3}} I\!\!P_{\mu_{4} \nu_{1}}(x) -
                    \partial_{\mu_{4}} I\!\!P_{\mu_{3} \nu_{1}}(x) \right) 
\twosidep{\mu_{2}} 
            \left( \partial^{\mu_{3}} I\!\!P^{\mu_{4}}_{\quad \nu_{2}}(x) -
                   \partial^{\mu_{4}} I\!\!P^{\mu_{3}}_{\quad \nu_{2}}(x) \right) ]\, ,
\label{Lagrangian_pseudoscalar_bis}
\end{eqnarray}
where the asymmetric derivative has the form
$\twosidep{\mu} = \rightsidep{\mu} - \leftsidep{\mu}$.
From (\ref{Lagrangian_pseudoscalar_bis}) we get the vertex, including a form factor, as follows
\begin{eqnarray}
i\Gamma_{\mu \nu,\kappa \lambda}''^{(I\!\!P I\!\!P \to \tilde{M})}(q_{1}, q_{2}) &=&
i \, \frac{g_{I\!\!P I\!\!P \tilde{M}}''}{M_{0}^{3}} \, 
\lbrace \varepsilon_{\nu \lambda \rho \sigma} \left[ q_{1 \kappa} q_{2 \mu} - (q_{1} q_{2}) g_{\mu \kappa} \right] +
\varepsilon_{\mu \lambda \rho \sigma} \left[ q_{1 \kappa} q_{2 \nu} - (q_{1} q_{2}) g_{\nu \kappa} \right]  \nonumber \\
&& +\, 
\varepsilon_{\nu \kappa \rho \sigma}  \left[ q_{1 \lambda} q_{2 \mu} - (q_{1} q_{2}) g_{\mu \lambda} \right] +
\varepsilon_{\mu \kappa \rho \sigma}  \left[ q_{1 \lambda} q_{2 \nu} - (q_{1} q_{2}) g_{\nu \lambda} \right] 
\rbrace 
(q_{1}-q_{2})^{\rho} k^{\sigma} \nonumber \\
&& \times 
F_{I\!\!P I\!\!P \tilde{M}}(q_{1}^{2}, q_{2}^{2}) \,
\label{vertex_pompomPS_bis}
\end{eqnarray}
with $g_{I\!\!P I\!\!P \tilde{M}}''$ a dimensionless coupling constant.
As complete vertex we take the sum of 
(\ref{vertex_pompomPS}) and (\ref{vertex_pompomPS_bis})
\begin{eqnarray}
i\Gamma_{\mu \nu,\kappa \lambda}^{(I\!\!P I\!\!P \to \tilde{M})}(q_{1}, q_{2}) =
i\Gamma_{\mu \nu,\kappa \lambda}'^{(I\!\!P I\!\!P \to \tilde{M})}(q_{1}, q_{2}) +
i\Gamma_{\mu \nu,\kappa \lambda}''^{(I\!\!P I\!\!P \to \tilde{M})}(q_{1}, q_{2}) \,.
\label{vertex_pompomPS_sum}
\end{eqnarray}
It can be checked that this vertex satisfies the identities
\begin{eqnarray}
&& \Gamma_{\mu \nu, \kappa \lambda}^{(I\!\!P I\!\!P \to \tilde{M})}(q_{1}, q_{2}) =
\Gamma_{\kappa \lambda, \mu \nu}^{(I\!\!P I\!\!P \to \tilde{M})}(q_{2}, q_{1}) \,,\nonumber \\
&& g^{\mu \nu} \Gamma_{\mu \nu, \kappa \lambda}^{(I\!\!P I\!\!P \to \tilde{M})}(q_{1}, q_{2}) = 0\,, \quad 
g^{\kappa \lambda} \Gamma_{\mu \nu, \kappa \lambda}^{(I\!\!P I\!\!P \to \tilde{M})}(q_{1}, q_{2}) = 0 \,.
\label{eqn2_gamma}
\end{eqnarray}

Now we can write down the $I\!\!P I\!\!P$-fusion contributions to
the Born amplitudes for the scalar and pseudoscalar meson exclusive production.
We find for a $0^{++}$ meson $M$
\begin{eqnarray}
&&\Braket{p(p_{1},\lambda_{1}),p(p_{2},\lambda_{2}),M(k)|{\cal T}|p(p_{a},\lambda_{a}),p(p_{b},\lambda_{b})}
\mid_{I\!\!P I\!\!P} \;\equiv \nonumber\\
&&{\cal M}^{2 \to 3}_{\lambda_{a}\lambda_{b} \to \lambda_{1}\lambda_{2} M} \mid_{I\!\!P I\!\!P} \;=
(-i) 
\bar{u}(p_{1},\lambda_{1}) i \Gamma_{\mu_{1} \nu_{1}}^{(I\!\!P pp)}(p_{1},p_{a}) u(p_{a},\lambda_{a})
\nonumber \\  
&&\qquad \qquad \qquad \qquad \quad \times 
i \Delta^{(I\!\!P) \, \mu_{1} \nu_{1}, \kappa_{1} \lambda_{1}}(s_{13},t_{1})  \;\;
i \Gamma_{\kappa_{1} \lambda_{1}, \kappa_{2} \lambda_{2}}^{(I\!\!P I\!\!P \to M)}(q_{1},q_{2}) \;\;
i \Delta^{(I\!\!P) \, \kappa_{2} \lambda_{2}, \mu_{2} \nu_{2}}(s_{23},t_{2}) 
\nonumber \\ 
&&\qquad \qquad \qquad \qquad \quad \times 
\bar{u}(p_{2},\lambda_{2}) i \Gamma_{\mu_{2} \nu_{2}}^{(I\!\!P pp)}(p_{2},p_{b}) u(p_{b},\lambda_{b})\,.
\label{2to3_amp}
\end{eqnarray}
Here $\Delta^{(I\!\!P)}$ and $\Gamma^{(I\!\!P pp)}$ 
denote the effective propagator and proton vertex function, respectively, for the tensorial pomeron.
For the explicit expressions, see Appendix~\ref{section:Tensorial_Pomeron},
(\ref{vertex_pomNN}) to (\ref{eqn2_delta}) and for the $I\!\!P I\!\!P M$ vertex (\ref{vertex_pompomS}).
For a pseudoscalar meson $\tilde{M}$ the amplitude is similar with
$\Gamma_{\kappa_{1} \lambda_{1}, \kappa_{2} \lambda_{2}}^{(I\!\!P I\!\!P \to M)}$
replaced by 
$\Gamma_{\kappa_{1} \lambda_{1}, \kappa_{2} \lambda_{2}}^{(I\!\!P I\!\!P \to \tilde{M})}$
in (\ref{2to3_amp}).

Explicitly we obtain from (\ref{2to3_amp}),
using the expressions from Appendix~\ref{section:Tensorial_Pomeron},
the amplitude for exclusive production of a scalar meson $M$ as
\begin{eqnarray}
&&{\cal M}^{2 \to 3}_{\lambda_{a}\lambda_{b} \to \lambda_{1}\lambda_{2} M} 
\mid_{I\!\!P I\!\!P} \;=
- \left(3 \beta_{I\!\!P NN} \right)^{2} \,  F_{1}(t_{1}) \, F_{1}(t_{2}) \, F_{I\!\!P I\!\!P M}(t_{1},t_{2}) 
\nonumber \\ &&
\times\bar{u}(p_{1},\lambda_{1}) \gamma^{\mu_{1}} (p_{1}+p_{a})^{\nu_{1}} u(p_{a},\lambda_{a})\;
\frac{1}{2 s_{13}} \left( -i s_{13} \alpha'_{I\!\!P} \right)^{\alpha_{I\!\!P}(t_{1})-1}
\nonumber \\ &&
\times 
\left.\Bigl[
g_{I\!\!P I\!\!P M}' M_{0} 
\Bigl( g_{\mu_{1} \mu_{2}} g_{\nu_{1} \nu_{2}} + 
  g_{\mu_{1} \nu_{2}} g_{\nu_{1} \mu_{2}} -
  \frac{1}{2} g_{\mu_{1} \nu_{1}} g_{\mu_{2} \nu_{2}} \Bigr) 
+ \frac{g_{I\!\!P I\!\!P M}''}{2 M_{0}} \right.
 \nonumber \\
&& \left. \times 
\Bigr( 
q_{1 \mu_{2}} q_{2 \mu_{1}} g_{\nu_{1} \nu_{2}} + 
q_{1 \mu_{2}} q_{2 \nu_{1}} g_{\mu_{1} \nu_{2}} +
q_{1 \nu_{2}} q_{2 \mu_{1}} g_{\nu_{1} \mu_{2}} + 
q_{1 \nu_{2}} q_{2 \nu_{1}} g_{\mu_{1} \mu_{2}} -
2 (q_{1}q_{2}) (g_{\mu_{1} \mu_{2}} g_{\nu_{1} \nu_{2}} 
              + g_{\nu_{1} \mu_{2}} g_{\mu_{1} \nu_{2}} ) \Bigr)
\right.\Bigr] \nonumber \\ 
&& \times 
\frac{1}{2 s_{23}} \left(-i s_{23} \alpha'_{I\!\!P}\right)^{\alpha_{I\!\!P}(t_{2})-1}\;
\bar{u}(p_{2},\lambda_{2}) \gamma^{\mu_{2}} (p_{2}+p_{b})^{\nu_{2}} u(p_{b},\lambda_{b})\,.
\label{2to3_scalar}
\end{eqnarray}
The coupling constants $\beta_{I\!\!P NN}$, $g_{I\!\!P I\!\!P M}'$, and $g_{I\!\!P I\!\!P M}''$ 
are defined in (\ref{vertex_pomNN}), (\ref{formula_A1_17}), and (\ref{formula_A1_17_bis}),
and the form factors $F_{1}$ and $F_{I\!\!P I\!\!P M}$
in (\ref{Fpomproton}) and (\ref{Fpompommeson_pion}), respectively.
Similarly, we obtain the amplitude for production of a pseudoscalar meson $\tilde{M}$ as
\begin{eqnarray}
&&{\cal M}^{2 \to 3}_{\lambda_{a}\lambda_{b} \to \lambda_{1}\lambda_{2} \tilde{M}} 
\mid_{I\!\!P I\!\!P} \;=
-(3 \beta_{I\!\!P NN})^{2} \,  F_{1}(t_{1}) \, F_{1}(t_{2}) \, F_{I\!\!P I\!\!P \tilde{M}}(t_{1},t_{2})
\nonumber \\ &&
\times\bar{u}(p_{1},\lambda_{1}) \gamma^{\mu_{1}} (p_{1}+p_{a})^{\nu_{1}} u(p_{a},\lambda_{a})\;
\frac{1}{2 s_{13}} \left(-i s_{13} \alpha'_{I\!\!P}\right)^{\alpha_{I\!\!P}(t_{1})-1}
\nonumber \\ &&
\times 
\left.\Bigl[
\Bigl( \frac{g_{I\!\!P I\!\!P \tilde{M}}}{2 M_{0}} - 
       \frac{g_{I\!\!P I\!\!P \tilde{M}}''}{M_{0}^{3}} (q_{1} q_{2}) \Bigr) 
\Bigl( g_{\mu_{1} \mu_{2}} \varepsilon_{\nu_{1} \nu_{2} \rho \sigma}
      +g_{\nu_{1} \mu_{2}} \varepsilon_{\mu_{1} \nu_{2} \rho \sigma}
      +g_{\mu_{1} \nu_{2}} \varepsilon_{\nu_{1} \mu_{2} \rho \sigma}
      +g_{\nu_{1} \nu_{2}} \varepsilon_{\mu_{1} \mu_{2} \rho \sigma} \Bigr)
\right.
 \nonumber \\
&& \left.
+ \frac{g_{I\!\!P I\!\!P \tilde{M}}''}{M_{0}^{3}}
\Bigl(  \varepsilon_{\nu_{1} \nu_{2} \rho \sigma} q_{1 \mu_{2}} q_{2 \mu_{1}} 
      + \varepsilon_{\mu_{1} \nu_{2} \rho \sigma} q_{1 \mu_{2}} q_{2 \nu_{1}}
      + \varepsilon_{\nu_{1} \mu_{2} \rho \sigma} q_{1 \nu_{2}} q_{2 \mu_{1}}
      + \varepsilon_{\mu_{1} \mu_{2} \rho \sigma} q_{1 \nu_{2}} q_{2 \nu_{1}} \Bigr)
\right.\Bigr] 
( q_{1}-q_{2} )^{\rho} k^{\sigma}
\nonumber \\ &&
\times 
\frac{1}{2 s_{23}} \left(-i s_{23} \alpha'_{I\!\!P}\right)^{\alpha_{I\!\!P}(t_{2})-1}\;
\bar{u}(p_{2},\lambda_{2}) \gamma^{\mu_{2}} (p_{2}+p_{b})^{\nu_{2}} u(p_{b},\lambda_{b})\,;
\label{2to3_pseudoscalar}
\end{eqnarray}
see (\ref{vertex_pompomPS}), (\ref{vertex_pompomPS_bis}), (\ref{vertex_pomNN}), (\ref{Fpomproton}),
and (\ref{Fpompommeson_pion}).

The same steps can now be repeated in the model of the vector pomeron.
The Born amplitude for the production of a $0^{++}$ meson $M$ via
$I\!\!P_{V} I\!\!P_{V}$-fusion can be written as
\begin{eqnarray}
{\cal M}^{2 \to 3}_{\lambda_{a}\lambda_{b} \to \lambda_{1}\lambda_{2} M} \mid_{I\!\!P_{V} I\!\!P_{V}}
&=&
(-i) 
\bar{u}(p_{1},\lambda_{1}) i \Gamma_{\mu_{1}}^{(I\!\!P_{V} pp)}(p_{1},p_{a}) u(p_{a},\lambda_{a})
\nonumber \\ && \times 
i \Delta^{(I\!\!P_{V})\,\mu_{1} \nu_{1}}(s_{13},t_{1})  \;\;
i \Gamma_{\nu_{1} \nu_{2}}^{(I\!\!P_{V} I\!\!P_{V} \to M)}(q_{1},q_{2}) \;\;
i \Delta^{(I\!\!P_{V})\,\nu_{2} \mu_{2}}(s_{23},t_{2}) 
\nonumber \\ &&\times 
\bar{u}(p_{2},\lambda_{2}) i \Gamma_{\mu_{2}}^{(I\!\!P_{V} pp)}(p_{2},p_{b}) u(p_{b},\lambda_{b})\,.
\label{2to3_amp_pomV}
\end{eqnarray}
The effective Lagrangian
and the vertices for $I\!\!P_{V} I\!\!P_{V} \to M$ are discussed in
Appendix~\ref{section:Vectorial_Pomeron};
see (\ref{vertex_pomVNN}), (\ref{prop_pomV}), and (\ref{vertex_pomVpomVS}).
Explicitly we obtain
\begin{eqnarray}
{\cal M}^{2 \to 3}_{\lambda_{a}\lambda_{b} \to \lambda_{1}\lambda_{2} M} 
\mid_{I\!\!P_{V} I\!\!P_{V}} &=&
- (3 \beta_{I\!\!P NN})^{2} \,  F_{1}(t_{1}) \, F_{1}(t_{2}) \, F_{I\!\!P I\!\!P M}(t_{1},t_{2})
\nonumber \\ &&
\times \bar{u}(p_{1},\lambda_{1}) \gamma^{\mu_{1}} u(p_{a},\lambda_{a})\,
g_{\mu_{1} \nu_{1}}\, \left(-i s_{13} \alpha'_{I\!\!P}\right)^{\alpha_{I\!\!P}(t_{1})-1}\,
\nonumber \\ &&
\times 
\Bigl[
  \frac{2}{M_{0}} \,g_{I\!\!P_{V} I\!\!P_{V} M}' \, g^{\nu_{1} \nu_{2}}\,
+ \frac{2}{M_{0}^{3}} \,g_{I\!\!P_{V} I\!\!P_{V} M}'' \, 
  \Bigl( q_{2}^{\nu_{1}} q_{1}^{\nu_{2}} - (q_{1} q_{2}) g^{\nu_{1} \nu_{2}} \Bigr)
\Bigr]
\nonumber \\ &&
\times g_{\nu_{2} \mu_{2}} \, \left(-i s_{23} \alpha'_{I\!\!P}\right)^{\alpha_{I\!\!P}(t_{2})-1}\,
\bar{u}(p_{2},\lambda_{2}) \gamma^{\mu_{2}} u(p_{b},\lambda_{b})\,.
\label{2to3_scalar_pomV}
\end{eqnarray}
%

%
Now we turn to the production of a pseudoscalar meson $\tilde{M}$
via $I\!\!P_{V} I\!\!P_{V}$-fusion.
The first step is to construct an effective coupling Lagrangian
$I\!\!P_{V} I\!\!P_{V} \tilde{M}$.
Traditionally this is done in analogy to the $\gamma \gamma \pi^{0}$ coupling
which is given by the Adler-Bell-Jackiw anomaly
(for a review see chapter 22 of \cite{Weinberg}).
In this way we get
\begin{eqnarray}
{\cal L}_{I\!\!P_{V} I\!\!P_{V} \tilde{M}}'(x) = 
\frac{g_{I\!\!P_{V} I\!\!P_{V} \tilde{M}}'}{16 M_{0}} \, 
\left[ \partial_{\mu} I\!\!P_{V \nu}(x) - \partial_{\nu} I\!\!P_{V \mu}(x) \right] \, 
\left[ \partial_{\rho} I\!\!P_{V \sigma}(x) - \partial_{\sigma} I\!\!P_{V \rho}(x) \right] \, 
\varepsilon^{\mu \nu \rho \sigma} \,
\tilde\chi(x)\nonumber \\
\label{Lagrangian_pseudoscalar_pomV}
\end{eqnarray}
with $g_{I\!\!P_{V} I\!\!P_{V} \tilde{M}}'$ a dimensionless coupling constant.

The corresponding vertex, including a form factor, reads as follows (see Fig.~\ref{fig:ptMp_pom}~(b)):
\begin{eqnarray}
i\Gamma_{\mu \nu}'^{(I\!\!P_{V} I\!\!P_{V} \to \tilde{M})}(q_{1}, q_{2}) =
i \, \frac{g_{I\!\!P_{V} I\!\!P_{V} \tilde{M}}'}{2 M_{0}} \,
\varepsilon_{\mu \nu \rho \sigma}q_{1}^{\rho} q_{2}^{\sigma} \, 
F_{I\!\!P I\!\!P \tilde{M}}(q_{1}^{2}, q_{2}^{2}) \,.
\label{vertex_pomVpomVPS}
\end{eqnarray}
It is easy to see that in the fictitious reaction (\ref{pomV_to_pomV})
the coupling (\ref{Lagrangian_pseudoscalar_pomV}),
(\ref{vertex_pomVpomVPS}) gives $(l,S) = (1,1)$.
Note that in our framework we have for $I\!\!P_{T} I\!\!P_{T}$-fusion
two values, $(l,S) = (1,1)$ and $(3,3)$, which can lead to a pseudoscalar meson;
see Table~\ref{tab:table1_A1} in Appendix~\ref{section:Tensorial_Pomeron}.
Correspondingly, we have two independent couplings,
(\ref{Lagrangian_pseudoscalar}) and (\ref{Lagrangian_pseudoscalar_bis}).
For $I\!\!P_{V} I\!\!P_{V}$-fusion, on the other hand, we find from 
Table~\ref{tab:table1_A2} in Appendix~\ref{section:Vectorial_Pomeron}
that only $(l,S) = (1,1)$ can lead to a pseudoscalar meson,
thus, only the coupling (\ref{Lagrangian_pseudoscalar_pomV}) is possible there.
This clear difference between the $I\!\!P_{T}$ and $I\!\!P_{V}$ ans\"atze
can be exploited for experimentally distinguishing the two cases.

The amplitude for the production of a $J^{PC} = 0^{-+}$ meson $\tilde{M}$
via $I\!\!P_{V} I\!\!P_{V}$-fusion can now be written down as 
in (\ref{2to3_amp_pomV}) with the $I\!\!P_{V} I\!\!P_{V} \tilde{M}$ vertex from
(\ref{vertex_pomVpomVPS}). Explicitly this gives
\begin{eqnarray}
{\cal M}^{2 \to 3}_{\lambda_{a}\lambda_{b} \to \lambda_{1}\lambda_{2} \tilde{M}} 
\mid_{I\!\!P_{V} I\!\!P_{V}} \,=\,
&& -(3 \beta_{I\!\!P NN})^{2}\, F_{1}(t_{1})\,F_{1}(t_{2})\, F_{I\!\!P I\!\!P \tilde{M}}(t_{1},t_{2}) \,
 \frac{g_{I\!\!P_{V} I\!\!P_{V} \tilde{M}}'}{2 M_{0}^{3}}
\nonumber \\ && \times 
 \bar{u}(p_{1},\lambda_{1}) \gamma^{\mu_{1}} u(p_{a}, \lambda_{a})\,
 g_{\mu_{1} \nu_{1}}\, \left(-i s_{13} \alpha'_{I\!\!P}\right)^{\alpha_{I\!\!P}(t_{1})-1}
\nonumber \\ && \times 
 \epsilon^{\nu_{1} \nu_{2} \rho \sigma} \, q_{1 \rho} q_{2 \sigma}
\nonumber \\ && \times 
 g_{\nu_{2} \mu_{2}}\, \left(-i s_{23} \alpha'_{I\!\!P}\right)^{\alpha_{I\!\!P}(t_{2})-1}\,
 \bar{u}(p_{2},\lambda_{2}) \gamma^{\mu_{2}} u(p_{b}, \lambda_{b}) \,.
\label{2to3_pseudoscalar_pomV}
\end{eqnarray}
%

In \cite{KMV99} also (vector pomeron)-(vector pomeron) fusion 
was considered as the dominant mechanism of the $\eta'$-meson production.
In order to estimate this contribution, 
the Donnachie-Landshoff energy dependence of the pomeron exchange \cite{DL} was used.

We shall now consider the high-energy small-angle limit, 
see Appendix~\ref{section:Kinematic_Relations}, for both
the tensorial and vectorial pomeron fusion reactions giving
the mesons $M$ and $\tilde{M}$.
With (\ref{D_8a}) to (\ref{D_11a}) we get from 
(\ref{2to3_scalar}) and (\ref{2to3_pseudoscalar})
for the tensorial pomeron
\begin{eqnarray}
{\cal M}^{2 \to 3}_{\lambda_{a}\lambda_{b} \to \lambda_{1}\lambda_{2} M} 
\mid_{I\!\!P I\!\!P} &\cong&
- 2s \, (3 \beta_{I\!\!P NN})^{2} \,  F_{1}(t_{1}) \, F_{1}(t_{2}) \, F_{I\!\!P I\!\!P M}(t_{1},t_{2}) 
\nonumber \\ &&
\times \frac{M_{0}}{m_{M}^{2}} 
\Bigl( g_{I\!\!P I\!\!P M}' + 
       g_{I\!\!P I\!\!P M}'' \, \frac{1}{M_{0}^{2}} \, \vec{p}_{1\perp} \cdot \vec{p}_{2\perp} \Bigr)
\nonumber \\ &&
\times \left(-i s_{13} \alpha'_{I\!\!P}\right)^{\alpha_{I\!\!P}(t_{1})-1}\;
       \left(-i s_{23} \alpha'_{I\!\!P}\right)^{\alpha_{I\!\!P}(t_{2})-1}
\nonumber \\ &&
\times \delta_{\lambda_{1}\lambda_{a}} \delta_{\lambda_{2}\lambda_{b}}\,,
\label{2to3_scalar_limit}\\
{\cal M}^{2 \to 3}_{\lambda_{a}\lambda_{b} \to \lambda_{1}\lambda_{2} \tilde{M}} 
\mid_{I\!\!P I\!\!P} &\cong&
- (3 \beta_{I\!\!P NN})^{2} \, F_{1}(t_{1}) \, F_{1}(t_{2}) \, F_{I\!\!P I\!\!P \tilde{M}}(t_{1},t_{2})
\nonumber \\ &&
\times \frac{1}{m_{\tilde{M}}^{2}}
\Bigl[ \frac{g_{I\!\!P I\!\!P \tilde{M}}'}{M_{0}} + 
       \frac{g_{I\!\!P I\!\!P \tilde{M}}''}{s M_{0}^{3}} \, 
       \Bigl( (q_{1},p_{2}+p_{b})(q_{2},p_{1}+p_{a})-
(q_{1},q_{2})(p_{1}+p_{a},p_{2}+p_{b}) \Bigr) 
\Bigr]
\nonumber \\ &&
\times \varepsilon_{\mu \nu \rho \sigma} 
(p_{1} + p_{a})^{\mu} (p_{2} + p_{b})^{\nu} (q_{1} - q_{2})^{\rho} k^{\sigma}
\nonumber \\ &&
\times \left(-i s_{13} \alpha'_{I\!\!P}\right)^{\alpha_{I\!\!P}(t_{1})-1}\;
       \left(-i s_{23} \alpha'_{I\!\!P}\right)^{\alpha_{I\!\!P}(t_{2})-1}
\nonumber \\ &&
\times \delta_{\lambda_{1}\lambda_{a}} \delta_{\lambda_{2}\lambda_{b}} \nonumber\\
&\cong&
- 4s \, (3 \beta_{I\!\!P NN})^{2} \, F_{1}(t_{1}) \, F_{1}(t_{2}) \, F_{I\!\!P I\!\!P \tilde{M}}(t_{1},t_{2})
\nonumber \\ &&
\times \frac{1}{m_{\tilde{M}}^{2} M_{0}} |\vec{p}_{1\perp}| |\vec{p}_{2\perp}| \sin\phi_{pp}
\Bigl( g_{I\!\!P I\!\!P \tilde{M}}' + 
       g_{I\!\!P I\!\!P \tilde{M}}'' \, 
       \frac{2}{M_{0}^{2}} |\vec{p}_{1\perp}| |\vec{p}_{2\perp}| \cos\phi_{pp}
\Bigr)
\nonumber \\ &&
\times \left(-i s_{13} \alpha'_{I\!\!P}\right)^{\alpha_{I\!\!P}(t_{1})-1}\;
       \left(-i s_{23} \alpha'_{I\!\!P}\right)^{\alpha_{I\!\!P}(t_{2})-1}
\nonumber \\ &&
\times \delta_{\lambda_{1}\lambda_{a}} \delta_{\lambda_{2}\lambda_{b}}\,.
\label{2to3_pseudoscalar_limit2}
\end{eqnarray}
For the vectorial pomeron we get in this limit 
from (\ref{2to3_scalar_pomV}) and (\ref{2to3_pseudoscalar_pomV})
the expressions 
(\ref{2to3_scalar_limit}) and (\ref{2to3_pseudoscalar_limit2}),
respectively, but with the replacements:
\begin{eqnarray}
&& g_{I\!\!P I\!\!P M}' \to \frac{2 m_{M}^{2}}{M_{0}^{2}} \, g_{I\!\!P_{V} I\!\!P_{V} M}' \,,
\quad 
g_{I\!\!P I\!\!P M}'' \to \frac{2 m_{M}^{2}}{M_{0}^{2}} \, g_{I\!\!P_{V} I\!\!P_{V} M}''\,,
\label{couplings_scalar}\\
&& g_{I\!\!P I\!\!P \tilde{M}}' \to \frac{m_{\tilde{M}}^{2}}{4 M_{0}^{2}} \, 
g_{I\!\!P_{V} I\!\!P_{V} \tilde{M}}' \,,
\quad 
g_{I\!\!P I\!\!P \tilde{M}}'' \to 0 \,.
\label{couplings_pseudoscalar}
\end{eqnarray}
We see that for the vectorial pomeron the term $\propto \cos\phi_{pp} \sin\phi_{pp}$
in (\ref{2to3_pseudoscalar_limit2}) is absent.

Going now from high to intermediate collision energies we must expect
besides pomeron-pomeron fusion
also reggeon-pomeron (pomeron-reggeon) and reggeon-reggeon fusion
to become important; see Fig.~\ref{fig:pp_pompom_ppmes}.
The relevant scales for these non-leading terms 
should be given by the subenergies squared 
$s_{13}$ and $s_{23}$ in (\ref{2to3_kinematics}).
We have to consider for the first non-leading contributions those from
the Regge trajectories with intercept $\alpha_{I\!\!R}(0) \approx 0.5$,
that is, the $f_{2}$, $a_{2}$, $\omega$ and $\rho$ trajectories
which we shall denote by 
$f_{2 I\!\!R}$, $a_{2 I\!\!R}$, $\omega_{I\!\!R}$ and $\rho_{I\!\!R}$, respectively.
In Ref.~\cite{EMN13} effective propagators for these reggeons and
reggeon-proton-proton vertices are given.
The $C = +1$ reggeons $f_{2 I\!\!R}$ and $a_{2 I\!\!R}$
are treated as effective tensor exchanges,
the $C = -1$ reggeons $\omega_{I\!\!R}$ and $\rho_{I\!\!R}$
as effective vector exchanges.
We shall make use of the results of \cite{EMN13} in the following.

To give an example we discuss the contribution of $\omega_{I\!\!R} \omega_{I\!\!R}$-fusion
to the production of a pseudoscalar meson $\tilde{M}$;
see Fig.~\ref{fig:pp_pompom_ppmes} with $I\!\!R = \omega_{I\!\!R}$
and $M \to \tilde{M}$.
The effective $\omega_{I\!\!R}$ propagator and the $\omega_{I\!\!R} pp$ vertex
are given in \cite{EMN13} as follows:
\begin{itemize}
\item $\omega_{I\!\!R}$ propagator
\begin{eqnarray}
i\Delta_{\mu \nu}^{(\omega_{I\!\!R})}(s,t) =
i \, g_{\mu \nu} \, \frac{1}{M_{-}^{2}} \,
\left(-i s \alpha'_{I\!\!R}\right)^{\alpha_{I\!\!R}(t)-1}
\label{prop_omega}
\end{eqnarray}
with the parameters (see \cite{DDLN}) of the Regge trajectory
\begin{eqnarray}
&&\alpha_{I\!\!R}(t) = \alpha_{I\!\!R}(0)+\alpha'_{I\!\!R}\,t \,,\nonumber\\
&&\alpha_{I\!\!R}(0) = 0.5475,\quad \alpha'_{I\!\!R} = 0.9 \; \mathrm{GeV}^{-2}\,,
\label{reggeon_trajectory}
\end{eqnarray}
and the mass scale $M_{-} = 1.41$~GeV.

\item $\omega_{I\!\!R} pp$ vertex
\begin{eqnarray}
i\Gamma_{\mu}^{(\omega_{I\!\!R} pp)}(p',p)=
-i \,g_{\omega_{I\!\!R} pp} \,  F_{1}\bigl((p'-p)^2\bigr) \, \gamma_{\mu}\, ,
\label{vertex_omeRNN}
\end{eqnarray}
where $g_{\omega_{I\!\!R} pp} = 8.65$.
\end{itemize}

For the $\omega_{I\!\!R} \omega_{I\!\!R} \tilde{M}$ vertex we shall make an ansatz
in complete analogy to 
(\ref{Lagrangian_pseudoscalar_pomV}), (\ref{vertex_pomVpomVPS})
for the vectorial pomeron.
We get then
\begin{eqnarray}
i\Gamma_{\mu \nu}^{(\omega_{I\!\!R} \omega_{I\!\!R} \to \tilde{M})}(q_{1}, q_{2}) =
i \, \frac{g_{\omega_{I\!\!R} \omega_{I\!\!R} \tilde{M}}}{2 M_{0}} \,
\varepsilon_{\mu \nu \rho \sigma}q_{1}^{\rho} q_{2}^{\sigma} \, 
F_{\omega_{I\!\!R} \omega_{I\!\!R} \tilde{M}}(q_{1}^{2}, q_{2}^{2}) \,,
\label{vertex_omeRomeRPS}
\end{eqnarray}
where $g_{\omega_{I\!\!R} \omega_{I\!\!R} \tilde{M}}$ is a dimensionless coupling constant.

Using (\ref{prop_omega}) to (\ref{vertex_omeRomeRPS})
the Born amplitude for the $\omega_{I\!\!R} \omega_{I\!\!R}$-fusion 
giving a pseudoscalar meson $\tilde{M}$ can be parametrized as
\begin{eqnarray}
{\cal M}^{2 \to 3}_{\lambda_{a}\lambda_{b} \to \lambda_{1}\lambda_{2} \tilde{M}} 
\mid_{\omega_{I\!\!R} \omega_{I\!\!R}} =\; &&
 (g_{\omega_{I\!\!R} pp})^{2}\,F_{1}(t_{1})\,F_{1}(t_{2})\, 
 F_{\omega_{I\!\!R} \omega_{I\!\!R} \tilde{M}}(t_{1},t_{2}) \,
  \frac{g_{\omega_{I\!\!R} \omega_{I\!\!R} \tilde{M}}}{2 M_{0}}
\nonumber \\ && \times 
 \bar{u}(p_{1},\lambda_{1}) \gamma^{\mu_{1}} u(p_{a}, \lambda_{a})
\nonumber \\ && \times 
 g_{\mu_{1} \nu_{1}}\, (M_{-})^{-2}\, \left(-i s_{13} \alpha'_{I\!\!R}\right)^{\alpha_{I\!\!R}(t_{1})-1} 
\nonumber \\ && \times 
 \epsilon^{\nu_{1} \nu_{2} \rho \sigma} \, q_{1 \rho} q_{2 \sigma}
\nonumber \\ && \times 
 g_{\nu_{2} \mu_{2}}\, (M_{-})^{-2}\, \left(-i s_{23} \alpha'_{I\!\!R}\right)^{\alpha_{I\!\!R}(t_{2})-1}
\nonumber \\ && \times 
 \bar{u}(p_{2},\lambda_{2}) \gamma^{\mu_{2}} u(p_{b}, \lambda_{b}) \,.
\label{omegaR_omegaR_fusion}
\end{eqnarray}

At even lower energies, for $s_{13}$ and $s_{23}$ near the threshold value
$s_{thr} = (m_{p} + m_{\tilde{M}})^{2}$, respectively $s_{thr} = (m_{p} + m_{M})^{2}$
for a $0^{++}$ meson $M$,
the exchange of reggeons in Fig.~\ref{fig:pp_pompom_ppmes}
should be replaced by particle exchanges.
As an example we give the amplitudes for $\eta$ and $\eta'$ production
at low energies $\sqrt{s_{13}}$ and $\sqrt{s_{23}}$.
It is known from the low energy phenomenology that both $\rho \rho$
and $\omega \omega$ mesons couple to $\eta$ and $\eta'$ mesons.
The $\omega \omega \tilde{M}$ vertex required for constructing 
the meson-exchange current is derived from the Lagrangian densities
\footnote{The Lagrangian (\ref{Lagrangian_pseudoscalar_omega})
is as given in (2.11) of \cite{KK08} and (A.11b) of \cite{NYH11}
taking into account that we use the opposite sign convention for 
$\varepsilon_{\mu \nu \rho \sigma}$; see after (\ref{2to3_kinematics}).
}
\begin{eqnarray}
{\cal L}_{\omega \omega \tilde{M}}(x) = 
\frac{g_{\omega \omega \tilde{M}}}{2 m_{\omega}} \, 
\left[ \partial_{\mu} \omega_{\nu}(x) \right] \left[ \partial_{\rho} \omega_{\sigma}(x) \right] \, 
\varepsilon^{\mu \nu \rho \sigma} \,
\tilde\chi(x) \,
\label{Lagrangian_pseudoscalar_omega}
\end{eqnarray}
and reads
\begin{eqnarray}
i\Gamma_{\mu \nu}^{(\omega \omega \to \tilde{M})}(q_{1}, q_{2}) =
i \, \frac{g_{\omega \omega \tilde{M}}}{m_{\omega}} \, 
\varepsilon_{\mu \nu \rho \sigma} q_{1}^{\rho} q_{2}^{\sigma}\,
F_{\omega \omega \tilde{M}}(q_{1}^{2}, q_{2}^{2}) \,.
\label{vertex_omeomePS}
\end{eqnarray}
The Born amplitude for the $\omega \omega$-fusion 
giving $\tilde{M} = \eta'$~or~$\eta$ can be written as
\begin{eqnarray}
{\cal M}^{2 \to 3}_{\lambda_{a}\lambda_{b} \to \lambda_{1}\lambda_{2} \tilde{M}}
\mid_{\omega \omega} =\; &&
 (g_{\omega pp})^{2}\, F_{\omega}(t_{1})\, F_{\omega}(t_{2})\,
 F_{\omega \omega \tilde{M}}(t_{1},t_{2})\,
  \frac{g_{\omega \omega \tilde{M}}}{m_{\omega}}  
\nonumber \\ && \times 
 \bar{u}(p_{1},\lambda_{1}) \gamma^{\mu_{1}} u(p_{a}, \lambda_{a})
\nonumber \\ && \times 
 \frac{-g_{\mu_{1} \nu_{1}} + q_{1 \mu_{1}} q_{1 \nu_{1}} / m_{\omega}^2}{t_{1} - m_{\omega}^2} \,
 \epsilon^{\nu_{1} \nu_{2} \rho \sigma} \, q_{1 \rho} q_{2 \sigma} \,
 \frac{-g_{\nu_{2} \mu_{2}} + q_{2 \nu_{2}} q_{2 \mu_{2}} / m_{\omega}^2}{t_{2} - m_{\omega}^2}
\nonumber \\ && \times 
 \bar{u}(p_{2},\lambda_{2}) \gamma^{\mu_{2}} u(p_{b}, \lambda_{b}) \,.
\label{omega_omega_fusion}
\end{eqnarray}
%
The coupling constants 
$g_{\omega \omega \eta'} = 4.9$ \cite{NH04,KK08},
$g_{\omega \omega \eta} = 4.84$ \cite{NSL02,NYH11} are known from low energy phenomenology.
In the present calculations we take the $\omega pp$ coupling constant $g_{\omega pp} = 10$.
Here we use form factors 
$F_{\omega \omega \tilde{M}}(t_{1},t_{2}) = F_{\omega}(t_{1})\, F_{\omega}(t_{2})$
for both exponential (\ref{F_exp_formfactor}) 
or monopole (\ref{F_monopol_formfactor}) approaches.
At larger subsystem energies squared, $s_{13}, s_{23} \gg s_{thr}$, one should
use reggeons rather than mesons.
The ``reggeization'' of the amplitude 
given in Eq.~(\ref{omega_omega_fusion}) is included here only approximately 
by a factor assuring asymptotically correct high energy dependence
\begin{eqnarray}
{\cal F} = \left( \frac{s_{13}}{s_{thr}} \right)^{\frac{2}{\pi} 
\arctan[(s_{13}-s_{thr})/\Lambda_{thr}^{2}]
(\alpha_{I\!\!R}(t_{1})-1)}
\left( \frac{s_{23}}{s_{thr}} \right)^{\frac{2}{\pi} 
\arctan[(s_{23}-s_{thr})/\Lambda_{thr}^{2}]
(\alpha_{I\!\!R}(t_{2})-1)} \,,
\label{aux_omeome}
\end{eqnarray}
where $\Lambda_{thr} = 1$~GeV
and $\alpha_{I\!\!R}(0) = 0.5$ and $\alpha'_{I\!\!R} = 0.9$~GeV$^{-2}$.

\subsection{Existing experimental data}
\label{subsection:Existing_experimental_data}

A big step in the investigation of central meson production process (\ref{2to3})
has been taken by the WA91 and WA102 Collaborations,
which have reported remarkable kinematical dependences and different effects; 
see Ref.~\cite{WA91_PLB388, WA102_PLB397, WA102_PLB427, WA102_PLB462, WA102_PLB467, WA102_PLB474, kirk00}.
The WA102 experiment at CERN was the first to discover
a strong dependence of the cross section
on the azimuthal angle between the momenta transferred to the two protons,
a feature that was not expected from standard pomeron phenomenology.
This result inspired some phenomenological works \cite{Close}
pointing to a possible analogy between the pomeron and vector particles
as had been suggested in \cite{DL} (see also chapter 3.7 of \cite{DDLN}).

Close and his collaborators have even proposed to use
transverse momenta correlations of outgoing protons
as tool to discriminate different intrinsic structures 
of the centrally produced object (``glueball filter''); see \cite{Close,CK97}.
In particular, the production of scalar mesons such as
$f_{0}(980)$, $f_{0}(1500)$, $f_{0}(1710)$ 
was found to be considerably enhanced at small $dP_{\perp}$,
while the production of pseudoscalars such as $\eta$, $\eta'$ at large $dP_{\perp}$;
see Fig.3 of \cite{kirk00}.
Here $dP_{\perp} = |d\vec{P}_{\perp}|$ with $d\vec{P}_{\perp}$
the difference of the transverse momenta of the two outgoing protons in (\ref{2to3}); see (\ref{D_4a}).
In Ref.~\cite{WA102_PLB462, kirk00} a study was performed
of resonance production rates as a function of $dP_{\perp}$.
It was observed that all the undisputed $q\bar{q}$ states
(i.e. $\eta$, $\eta'$, $f_{1}(1285)$ etc.) are suppressed as $dP_{\perp} \to 0$,
whereas the glueball candidates, e.g. $f_{0}(1500)$, $f_{2}(1950)$ are prominent.
It is also interesting that the $f_{1}(1420)$ state
disappears at small $dP_{\perp}$ relative to large $dP_{\perp}$.
As can be seen from \cite{kirk00} the mesons $\rho^{0}(770)$, $f_{2}(1270)$, 
and $f_{2}'(1525)$ are produced preferentially at large $dP_{\perp}$
and their cross sections peak at $\phi_{pp} = \pi$,
i.e. the outgoing protons are on opposite sides of the beam.
\footnote{Here $\phi_{pp}$ is the azimuthal angle between 
the momentum vectors of the outgoing protons; see (\ref{D_4}).}
In contrast, for the 'enigmatic' $f_{0}(980)$, $f_{0}(1500)$ and $f_{0}(1710)$ states
the cross sections peak at $\phi_{pp} = 0$.
So far, no dynamical explanation of this empirical observation
has been suggested, so the challenge for theory
is to understand the dynamics behind this ``glueball filter''.

In Ref.~\cite{WA76_ZPC51} the study of the $|t| = |t_{1} + t_{2}|$ dependence
of the resonances observed in the $\pi^{+} \pi^{-}$ and $K^{+}K^{-}$ mass spectra
at $\sqrt{s} = 23.8$~GeV was considered.
It has been observed that 
$\rho(770)$, $\phi(1020)$, $f_{2}(1270)$ and $f'_{2}(1525)$ resonances
are produced more at the high-$|t|$ region ($|t| > 0.3$~GeV$^{2}$)
and at low $|t|$ their signals are suppressed.
The suppression of the $\rho$ and $f_{2}(1270)$ signals
in the low-$|t|$ region is also present at $\sqrt{s} = 12.7$~GeV
for the $\pi^{+} p \to \pi^{+} (\pi^{+} \pi^{-}) p$ reaction; see \cite{WA76_ZPC51}.
In addition, the $dP_{\perp}$, $\phi_{pp}$ and $|t|$ distributions observed in 
the analysis of the $\pi^{+} \pi^{-}$ final state
for the $f_{0}(1370)$ and $f_{0}(1500)$ mesons are similar 
to what was found in the $\pi^{+} \pi^{-} \pi^{+} \pi^{-}$ channel \cite{WA102_PLB474}.

\begin{table}
\caption{Experimental results for the ratios of the cross sections for
the different mesons at $\sqrt{s} = 29.1 $~GeV and 12.7~GeV.}
\label{tab:ratio}
\begin{tabular}{|c|c|c|c|c|c|}
\hline
& $\eta'$ 
& $\rho(770)$ 
& $f_{0}(980)$ 
& $f_{0}(1500)$ 
& $f_{2}(1270)$ \\
\hline
$\frac{\sigma(\sqrt{s} = 29.1 \, GeV)}{\sigma(\sqrt{s} = 12.7 \, GeV)}$
& 0.72 $\pm$ 0.16 \cite{WA102_PLB467}
& 0.36 $\pm$ 0.05 \cite{WA102_PLB462}
& 1.28 $\pm$ 0.21 \cite{WA102_PLB462}
& 1.07 $\pm$ 0.14 \cite{WA102_PLB462}
& 0.98 $\pm$ 0.13 \cite{WA102_PLB462}\\
\hline
\end{tabular}
\end{table}
The ratios of experimental the cross sections for the different mesons 
at $\sqrt{s} = 29.1$~GeV and 12.7~GeV
has also been determined, see Table \ref{tab:ratio}.
Moreover, the WA76 Collaboration reported that the ratio of the $\rho^{0}(770)$
cross section at 23.8~GeV and 12.7~GeV is $0.44 \pm 0.07$; cf.~\cite{kirk00}.
Since the $I = 1$ states cannot be produced by pomeron-pomeron fusion,
the $\rho$ meson signal decreases at high energy.
However, large enhancement of the $\rho$ signal at $\sqrt{s} = 29.1$~GeV
and strong correlation between the directions of the outgoing protons
have been observed \cite{WA91_PLB388, WA102_PLB397}.
Similarly, in the case of the $\omega$ meson production,
where some 'non-central' mechanisms are possible \cite{CLSS},
the cross section is more than twice larger than for the $f_{0}(1500)$ meson,
the lightest scalar glueball candidate \cite{kirk00, SL}.

We turn now to our present calculations of cross sections
and distributions for the central production reaction (\ref{2to3})
with scalar and pseudoscalar mesons.


\section{Results}
\label{section:Results}
Now we wish to compare results of our calculations with existing experimental data.
Theoretical predictions for production of various $J^{PC}$ mesonic states 
for RHIC, Tevatron and LHC, with parameters fixed from the fit to the WA102 experimental data,
can then be easily done.

\subsection{Scalar meson production}
\label{subsection:Scalar_mesons}
We start with discussing the WA102 data at $\sqrt{s} = 29.1$~GeV
where total cross sections are given in Table~1 of Ref.~\cite{kirk00}.
We show these cross sections for the mesons of interest to us in Table~\ref{tab:mesons}.
\begin{table}
\caption{Experimental results for total cross sections
of various mesons in $pp$ collisions at $\sqrt{s} = 29.1 $~GeV; 
from Table~1 of Ref.~\cite{kirk00}.}
\label{tab:mesons}
\begin{tabular}{|c|c|c|c|c|c|c|c|c|}
\hline
& $\eta$ 
& $\eta'$ 
& $f_{0}(980)$ 
& $f_{0}(1370)$ 
& $f_{0}(1500)$ 
& $f_{0}(1710)$
& $f_{0}(2000)$ \\
\hline
$\sigma$($\mu$b)
& 3.86 $\pm$ 0.37 
& 1.72 $\pm$ 0.18 
& 5.71 $\pm$ 0.45 
& 1.75 $\pm$ 0.58
& 2.91 $\pm$ 0.30
& 0.25 $\pm$ 0.07
& 3.14 $\pm$ 0.48 \\
\hline
\end{tabular}
\end{table}
We assume that here the energy is high enough such that we have to consider only 
pomeron-pomeron-meson ($I\!\!P I\!\!P M$) fusion.
We have then determined the corresponding $I\!\!P I\!\!P M$ coupling constants
by approximately fitting the results of our calculations to the
total cross sections given in Table~\ref{tab:mesons}
and the shapes of experimental differential distributions 
(specific details will be given when discussing differential distributions below).
The results depend also on
the pomeron-pomeron-meson form factors (\ref{Fpompommeson_pion}),
as discussed in Appendix~\ref{section:Tensorial_Pomeron},
which are not well known, in particular for larger values of $t$.
In Table~\ref{tab:couplings_S} we show our results for these $I\!\!P I\!\!P M$ coupling constants
for the tensorial and vectorial pomeron ans\"atze.
The figures in bold face represent our ``best'' fit.
We show the resulting total cross sections,
from the coupling $g_{I\!\!P I\!\!P M}'$ alone, 
from $g_{I\!\!P I\!\!P M}''$ alone,
and from the total which includes, of course, the interference term between the two couplings.
The column ``no cuts, total'' has to be compared
to the experimental results shown in Table~\ref{tab:mesons}.
For the cross section with the cuts in $t_{1}t_{2}$
only normalised differential distributions are available; see below.
Thus, our results for the corresponding cross sections there are predictions
to be checked in future experiments.
\begin{table}
\caption{The values of the pomeron-pomeron-meson coupling constants
of the two models of the pomeron exchanges
which are approximately fitted to reproduce the 
experimental total cross sections from Table~\ref{tab:mesons}
and shapes of differential distributions of the WA102 data as discussed below.
The resulting cross sections (in $\mu$b) for scalar meson
central production at $\sqrt{s} = 29.1$~GeV 
without cuts and with cuts in $t_{1}t_{2}$ are also shown.
The figures in bold face represent our ``best fit'' values for
the $I\!\!P I\!\!P M$ coupling constants; 
see the discussion of figures \ref{fig:dsig_dphi_1370}, \ref{fig:dsig_dphi_980}, 
and \ref{fig:dsig_dphi_1500} in the text.}
\label{tab:couplings_S}
\begin{tabular}{|c|c|c|l|l|l|l|l|l|l|l|l|}
\hline
 &  &  &
\multicolumn{9}{c|}{$\sigma$ ($\mu$b) at $\sqrt{s} = 29.1$~GeV} \\
\cline{4-12}
Vertex & $g_{I\!\!P I\!\!P M}'$ & $g_{I\!\!P I\!\!P M}''$ &
\multicolumn{3}{c|}{no cuts}      &  
\multicolumn{3}{c|}{$|t_{1}t_{2}| \leqslant 0.01$~GeV$^{4}$} &  
\multicolumn{3}{c|}{$|t_{1}t_{2}| \geqslant 0.08$~GeV$^{4}$} \\
\cline{4-12}
& $(0,0)$ term &   $(2,2)$ term
& $(0,0)$ & $(2,2)$ & total
& $(0,0)$ & $(2,2)$ & total
& $(0,0)$ & $(2,2)$ & total\\
\hline
$I\!\!P_{T} I\!\!P_{T} f_{0}(980)$ & \textbf{0.788} & \textbf{4} &
5.73 & 1.16 & 5.71 & 
3.56 & 0.12 & 3.51 & 
0.21 & 0.41 & 0.3
\\
      & {\color{blue}\textbf{0.75}} & {\color{blue}\textbf{5.5}} &
5.19 & 2.19 & 5.83 & 
3.22 & 0.23 & 3.23 &
0.19 & 0.77 & 0.55 
\\
$I\!\!P_{V} I\!\!P_{V} f_{0}(980)$ & \textbf{0.27} & \textbf{0.8} &
5.37 & 0.48 & 5.72 & 
2.85 & 0.04 & 2.87 & 
0.34 & 0.2  & 0.49
\\
 & 0.26 & 1.1 &
4.98 & 0.9  & 5.71 & 
2.64 & 0.07 & 2.69 & 
0.31 & 0.38 & 0.63
\\
 & {\color{blue}\textbf{0.24}} & {\color{blue}\textbf{1.5}} &
4.24 & 1.67 & 5.7  & 
2.25 & 0.12 & 2.36 & 
0.27 & 0.71 & 0.9
\\
 & 0.2 & 2 &
2.94 & 2.97 & 5.69 & 
1.56 & 0.22 & 1.76 & 
0.19 & 1.27 & 1.36
\\
\hline
$I\!\!P_{T} I\!\!P_{T} f_{0}(1500)$ & \textbf{1.22} & \textbf{6} &
2.69 & 0.53 & 2.9 & 
1.55 & 0.05 & 1.56 & 
0.12 & 0.19 & 0.21
\\
& {\color{blue}\textbf{1}} & {\color{blue}\textbf{10}} &
1.81 & 1.47 & 2.83 & 
1.04 & 0.14 & 1.13 & 
0.08 & 0.53 & 0.47
\\
$I\!\!P_{V} I\!\!P_{V} f_{0}(1500)$ & \textbf{0.208} & \textbf{0.725} &
2.64 & 0.32 & 2.9 & 
1.37 & 0.02 & 1.39 & 
0.17 & 0.13 & 0.28
\\
  & 0.185 & 1.22 &
2.08 & 0.89 & 2.91 & 
1.09 & 0.06 & 1.14 & 
0.13 & 0.38 & 0.48
\\  
  & {\color{blue}\textbf{0.164}} & {\color{blue}\textbf{1.5}} &
1.64 & 1.35 & 2.91 & 
0.85 & 0.1  & 0.94 & 
0.1  & 0.57 & 0.64
\\
\hline
$I\!\!P_{T} I\!\!P_{T} f_{0}(1370)$ & \textbf{0.81} & -- &
1.75  & -- & -- & 
1.02  & -- & --  & 
0.07  & -- & --
\\
$I\!\!P_{V} I\!\!P_{V} f_{0}(1370)$ & \textbf{0.165} & -- &
1.75  & -- & -- & 
0.91  & -- & --  & 
0.11  & -- & --
\\
\hline
\end{tabular}
\end{table}

\begin{figure}[!ht]
(a)\includegraphics[width = 0.45\textwidth]{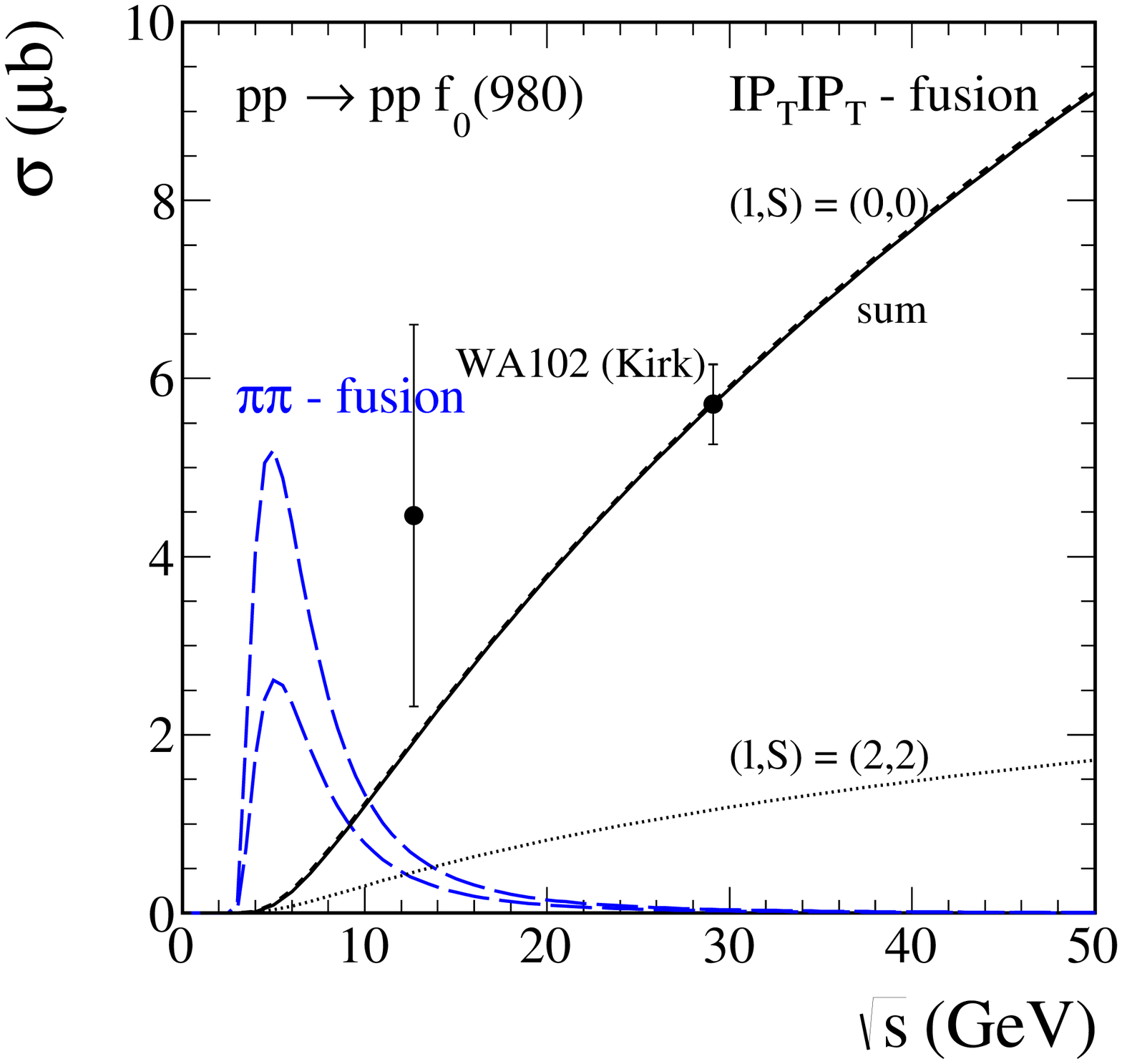}
(b)\includegraphics[width = 0.45\textwidth]{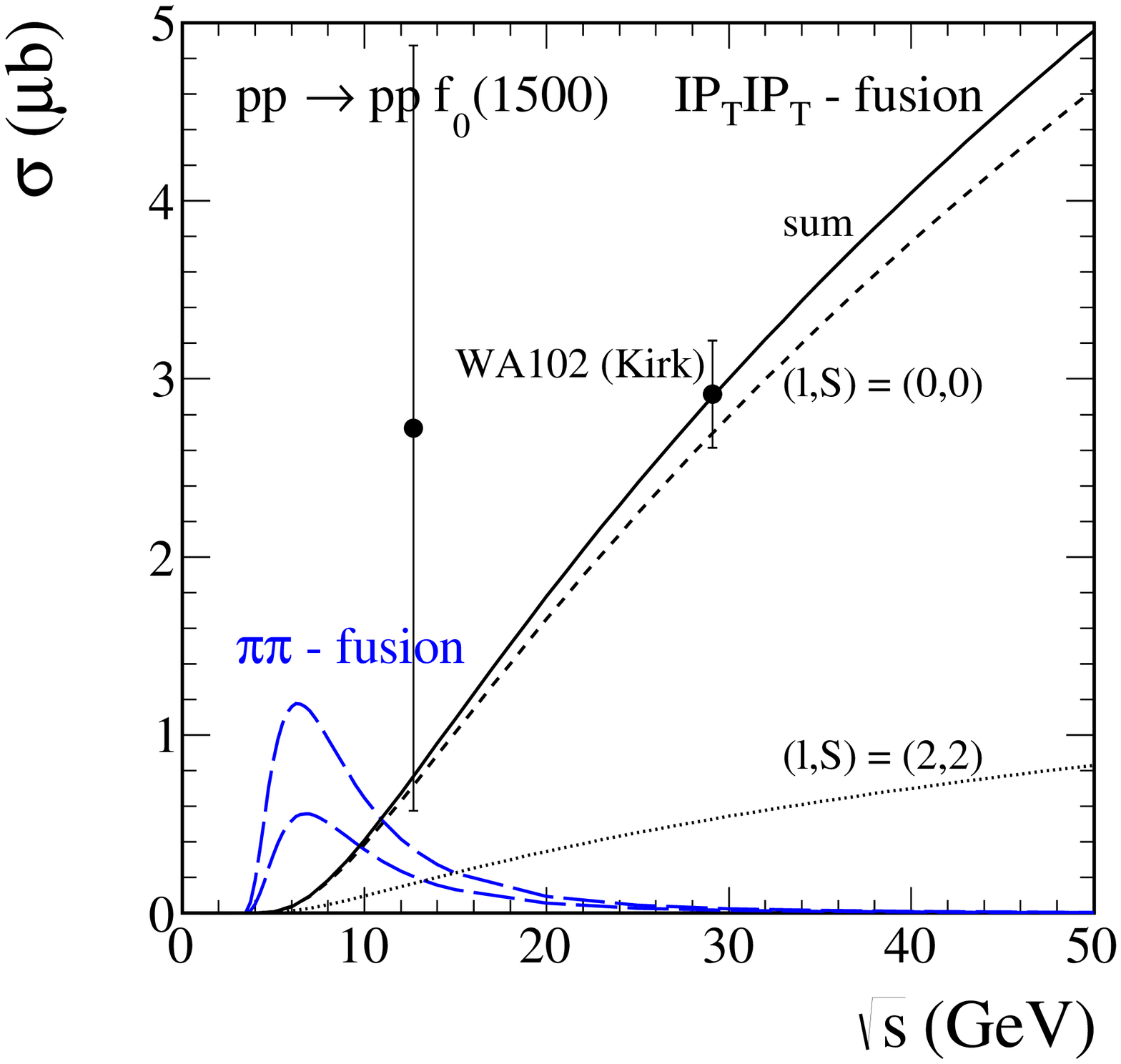}\\
(c)\includegraphics[width = 0.45\textwidth]{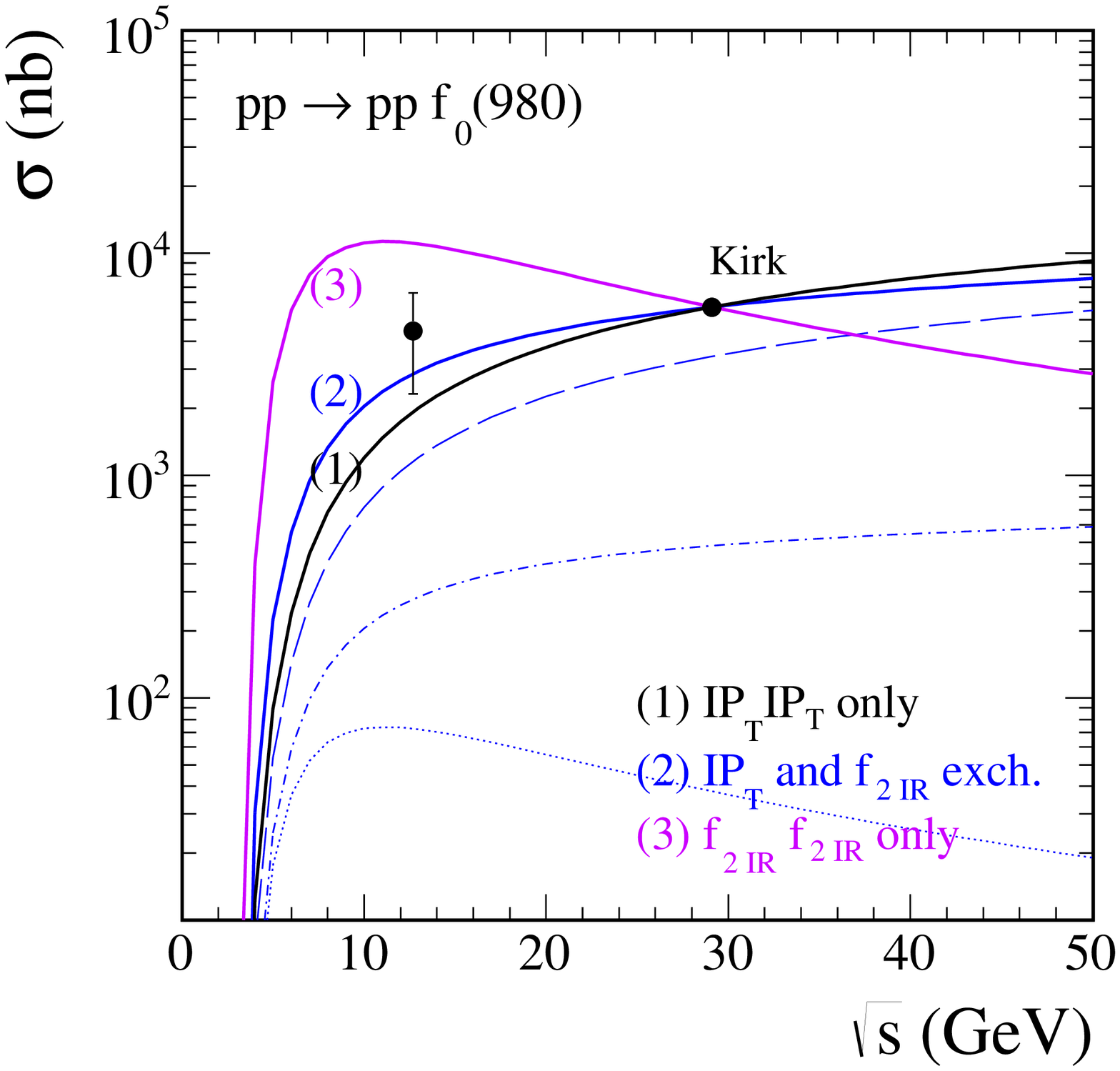}
  \caption{\label{fig:sig_scalars}
  \small
The integrated cross section as a function of the proton-proton center-of-mass energy
for the $pp \to pp f_{0}(980)$ (panel (a))
and $pp \to pp f_{0}(1500)$ (panel (b)) reactions.
We show data points obtained by the WA102 experiment~\cite{kirk00, WA102_PLB462}.
The two long-dashed lines peaked at $\sqrt{s} \approx$~5-7~GeV 
correspond to pion-pion fusion contribution.
The pomeron-pomeron fusion dominates at higher energies.
In panels (a) and (b) we show the individual contributions to the cross section
with $(l,S) = (0,0)$ (short-dashed line) and $(l,S) = (2,2)$ (dotted line).
In panels (a) and (b) we show results when only the $I\!\!P_{T} I\!\!P_{T}$-fusion is included.
In panel (c) the black solid line (1) presents the $I\!\!P_{T} I\!\!P_{T}$-fusion,
the blue solid line (2) correspond to the results with tensor pomeron and $f_{2 I\!\!R}$ exchanges
(the long-dashed, dash-dotted and dotted lines present 
the $I\!\!P_{T} I\!\!P_{T}$, $I\!\!P_{T} f_{2 I\!\!R}$ and $f_{2 I\!\!R} f_{2 I\!\!R}$
contributions, respectively).
The violet solid line (3) presents the $f_{2 I\!\!R} f_{2 I\!\!R}$-fusion alone
normalized to the total cross section from \cite{kirk00} as given in our Table~\ref{tab:mesons}.
}
\end{figure}

In Fig.~\ref{fig:sig_scalars} we present our result for the integrated cross sections
of the exclusive $f_{0}(980)$ (left panel)
and $f_{0}(1500)$ (right panel) scalar meson production
as a function of centre-of-mass energy $\sqrt{s}$.
For this calculation we have taken into account pion-pion fusion and pomeron-pomeron fusion;
see Fig.~\ref{fig:pp_pompom_ppmes}.
We see that at low energy the pion-pion fusion dominates.
The pion-pion contribution grows quickly from the threshold, 
has a maximum at $\sqrt{s} \approx$~5-7~GeV 
and then slowly drops with increasing energy.
This contribution was calculated with monopole vertex form factor (\ref{F_monopol_formfactor}) 
with parameters $\Lambda_{M} = 0.8$~GeV (lower line) and $\Lambda_{M} = 1.2$~GeV (upper line).
See \cite{SL} for more details of the $\pi\pi$-fusion mechanism.
The difference between the lower and upper curves represents the uncertainties
on the pion-pion component.
At intermediate energies other exchange processes such as the pomeron-$f_{2 I\!\!R}$,
$f_{2 I\!\!R}$-pomeron and $f_{2 I\!\!R}$-$f_{2 I\!\!R}$ exchanges are possible.
For the $f_{2 I\!\!R} pp$ vertex and the $f_{2 I\!\!R}$ exchange effective propagator
we shall make an ansatz in complete analogy to (\ref{vertex_pomNN}) and (\ref{prop_pom})
for the tensorial pomeron, respectively, 
with the coupling constant $g_{f_{2 I\!\!R} pp}=11.04$
and the trajectory as (\ref{reggeon_trajectory}); see \cite{EMN13}.
The $f_{2 I\!\!R} f_{2 I\!\!R} f_{0}(980)$
and $f_{2 I\!\!R} I\!\!P_{T} f_{0}(980)$ vertices should have the general structure
of the $I\!\!P_{T} I\!\!P_{T} f_{0}(980)$ vertex (\ref{vertex_pompomS}), but, of course,
with different and independent coupling constants.
In panel (c) we show results with $I\!\!P_{T} I\!\!P_{T}$ (black solid line (1))
and $f_{2 I\!\!R} f_{2 I\!\!R}$ (violet solid line (3)) exchanges,
obtained for the coupling constants $(g_{I\!\!P I\!\!P M}', g_{I\!\!P I\!\!P M}'')=(0.788, 4)$ 
and $(g_{f_{2 I\!\!R} f_{2 I\!\!R} M}', g_{f_{2 I\!\!R} f_{2 I\!\!R} M}'')=(9.5, 80)$, respectively.
We see that fixing the $I\!\!P_{T}$ or $f_{2 I\!\!R}$ contributions
to the point at $\sqrt{s} = 29.1$~GeV
the $I\!\!P_{T}$ curve is below, the $f_{2 I\!\!R}$ curve above 
the experimental point at $\sqrt{s} = 12.7$~GeV.
Clearly, we have to include all $I\!\!P_{T}$ and $f_{2 I\!\!R}$ exchanges.
The corresponding curve (2) reproduces the experiment.
The individual contributions are also shown in Fig.~\ref{fig:sig_scalars}(c),
corresponding to
$(g_{I\!\!P I\!\!P M}', g_{I\!\!P I\!\!P M}'')=(0.47, 2.4)$,
$(g_{I\!\!P f_{2 I\!\!R} M}', g_{I\!\!P f_{2 I\!\!R} M}'')=
 (g_{f_{2 I\!\!R} I\!\!P M}', g_{f_{2 I\!\!R} I\!\!P M}'')= (0.63, 3.2)$,
$(g_{f_{2 I\!\!R} f_{2 I\!\!R} M}', g_{f_{2 I\!\!R} f_{2 I\!\!R} M}'')=(0.79, 3.9)$.

\begin{figure}[!ht]
\includegraphics[width = 0.45\textwidth]{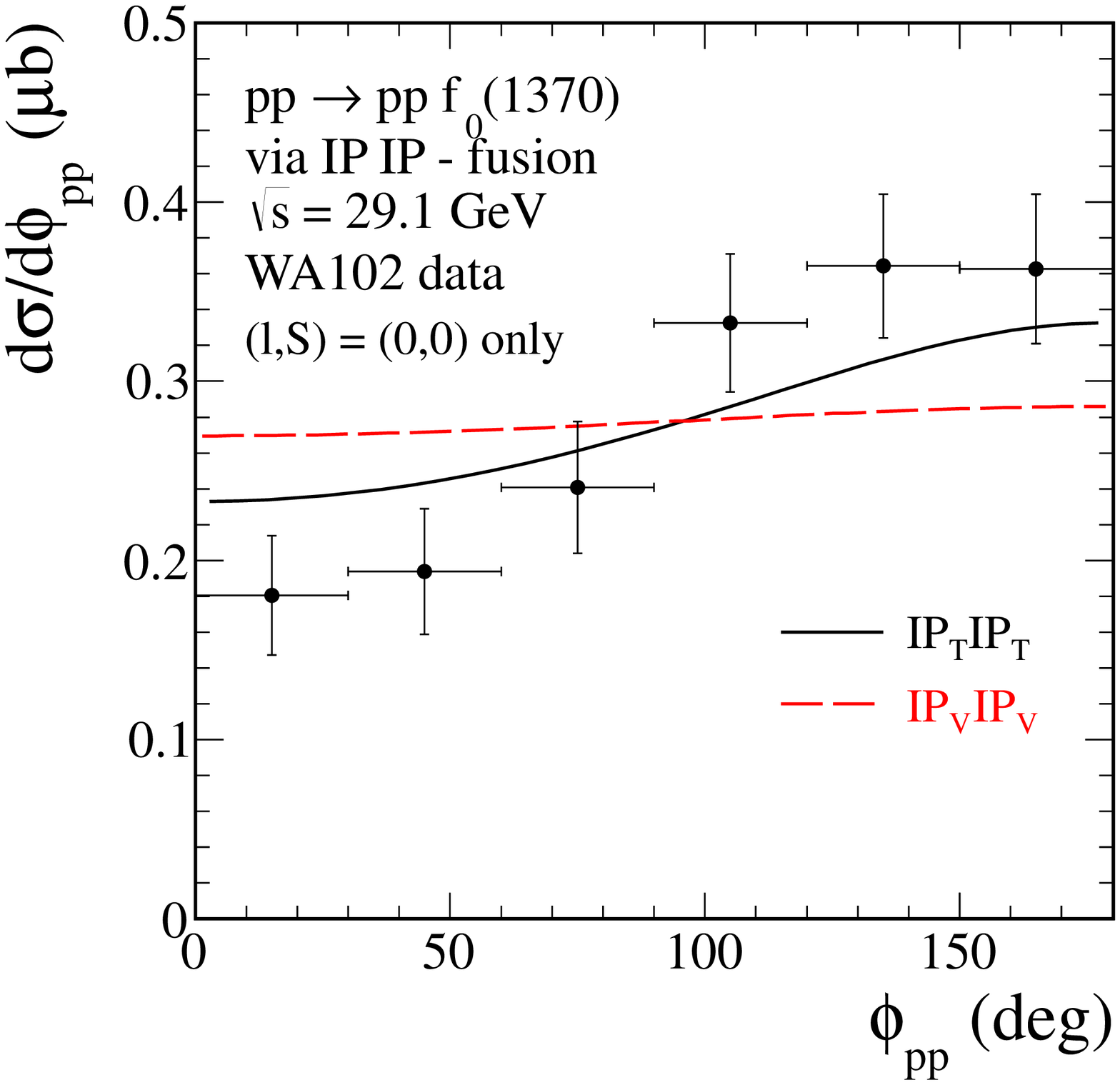}
  \caption{\label{fig:dsig_dphi_1370}
  \small
The distribution in azimuthal angle between the outgoing protons
for central exclusive $f_{0}(1370)$ meson production
by a fusion of two tensor (solid line) and vector (long-dashed line) pomerons at $\sqrt{s} = 29.1$~GeV.
The WA102 experimental data points from \cite{WA102_PLB462}
have been normalized to the total cross section from \cite{kirk00} as given in our Table~\ref{tab:mesons}.
The corresponding $I\!\!P I\!\!P M$ coupling constants are as given in Table~\ref{tab:couplings_S}.
}
\end{figure}

In Fig.~\ref{fig:dsig_dphi_1370} we show the distribution in azimuthal angle $\phi_{pp}$
between outgoing protons for central exclusive $f_{0}(1370)$ meson production
by the fusion of two tensor (solid line) and vector (long-dashed line) pomerons at $\sqrt{s} = 29.1$~GeV.
The results of the two models of pomeron exchanges are compared with the WA102 data.
The tensorial pomeron with the $(l,S) = (0,0)$
coupling alone already describes the azimuthal angular correlation
for $f_{0}(1370)$ meson reasonable well. 
The vectorial pomeron with the $(l,S) = (0,0)$ term alone is disfavoured here.
The preference of the $f_{0}(1370)$ for the $\phi_{pp} \approx \pi$ domain in contrast to
the enigmatic $f_{0}(980)$ and $f_{0}(1500)$ scalars has been observed by 
the WA102 Collaboration \cite{WA102_PLB462}.

\begin{figure}[!ht]
(a)\includegraphics[width = 0.29\textwidth]{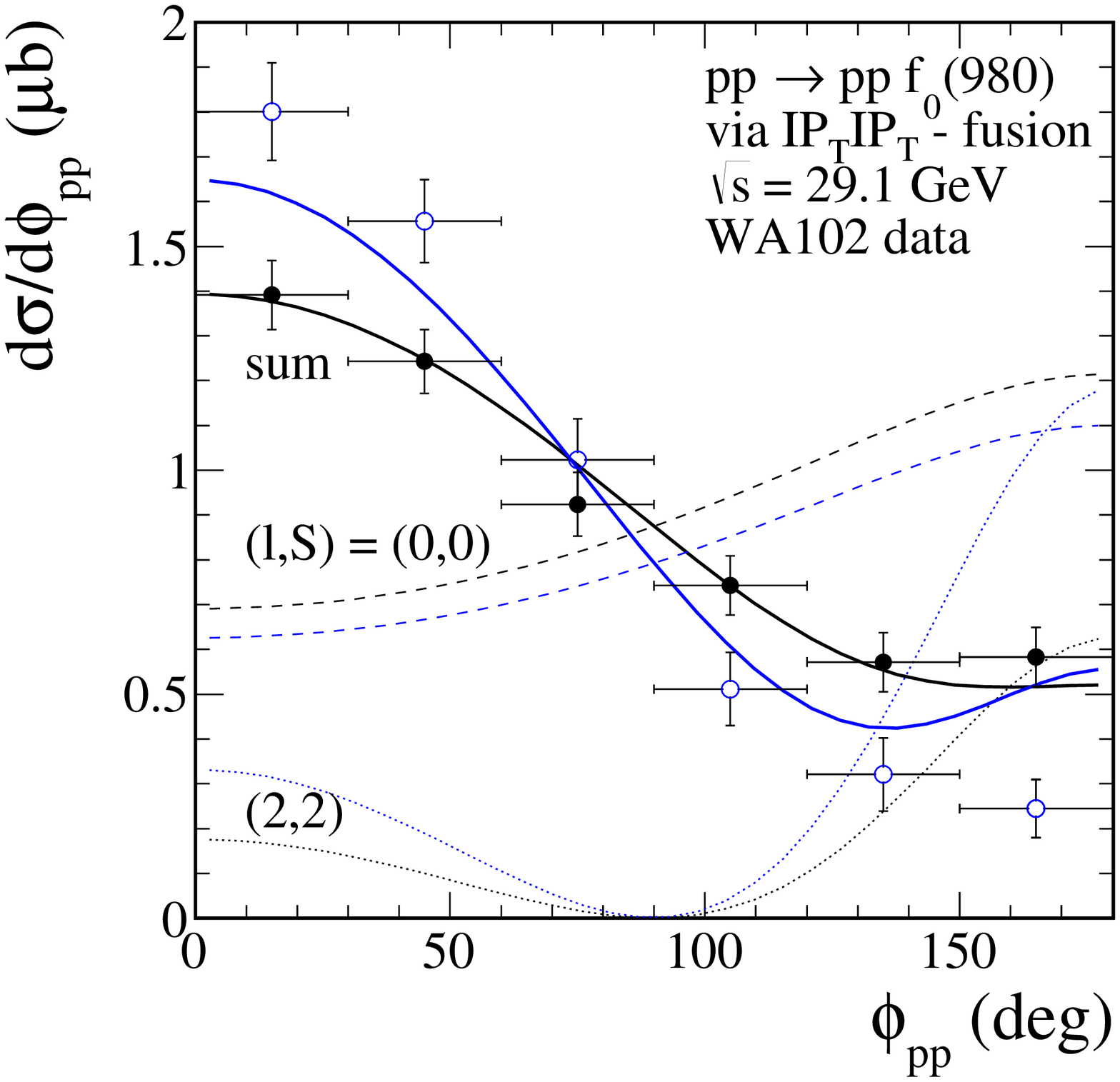}
(b)\includegraphics[width = 0.29\textwidth]{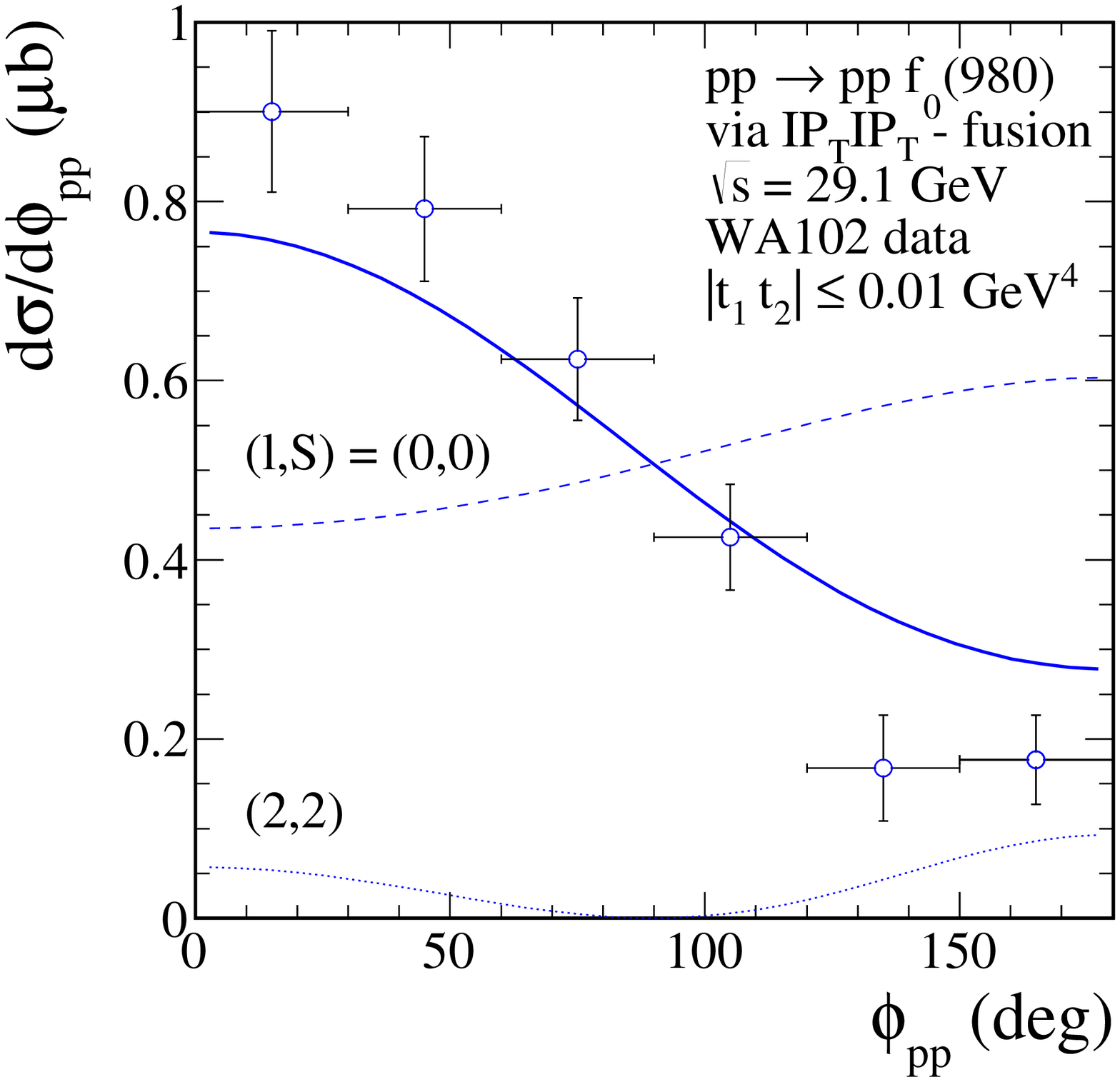}
(c)\includegraphics[width = 0.29\textwidth]{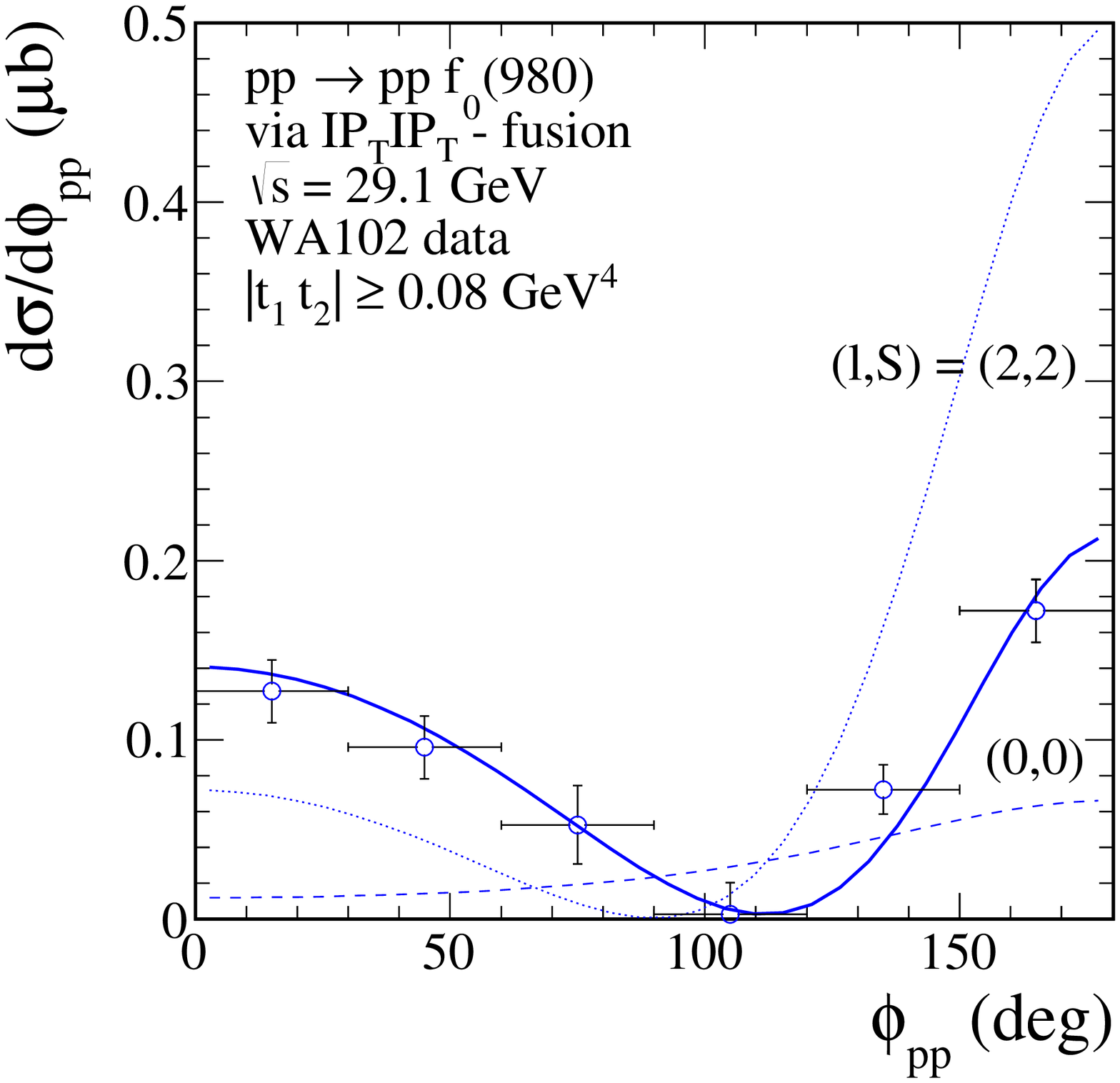}\\
(d)\includegraphics[width = 0.29\textwidth]{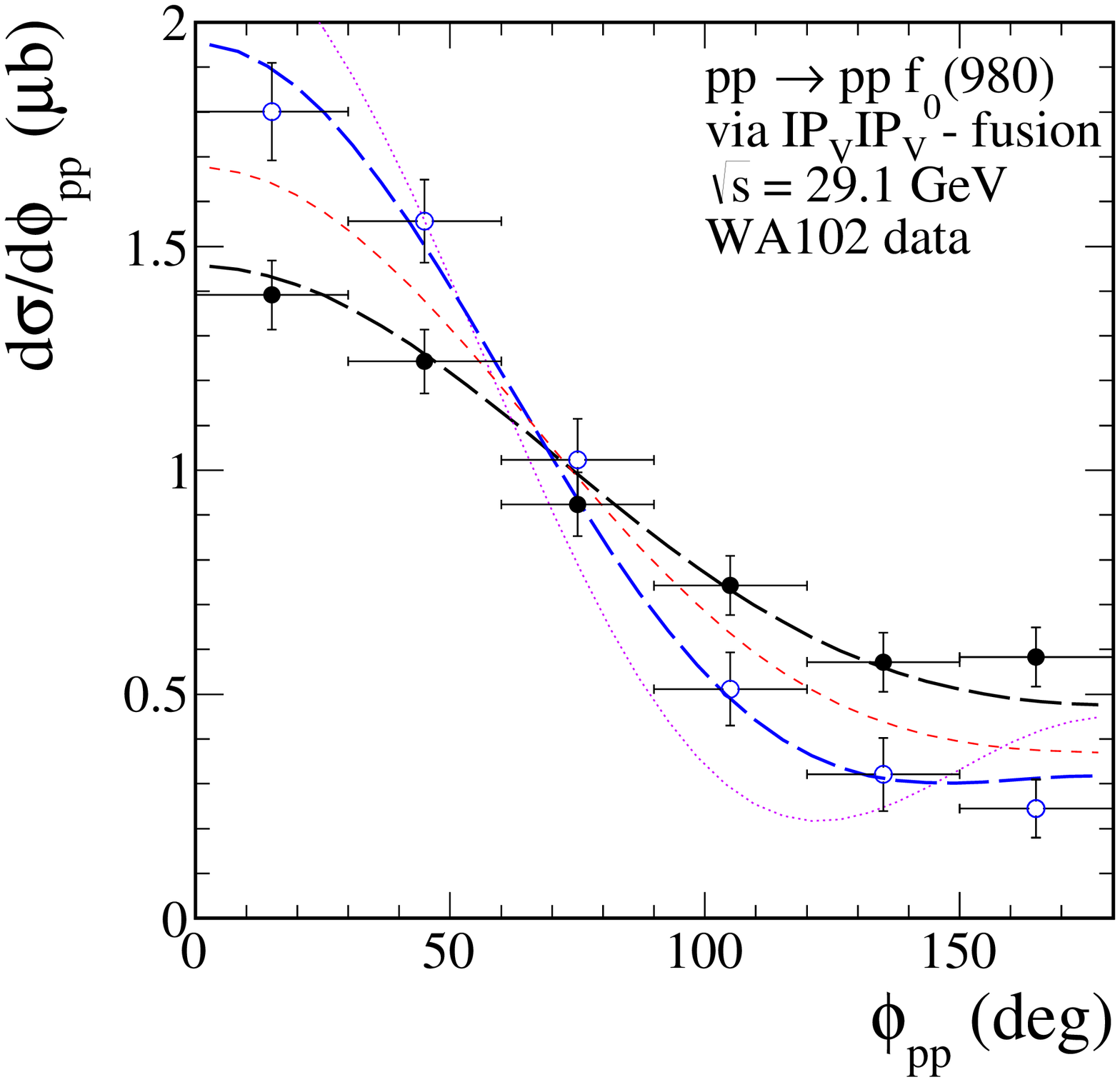}
(e)\includegraphics[width = 0.29\textwidth]{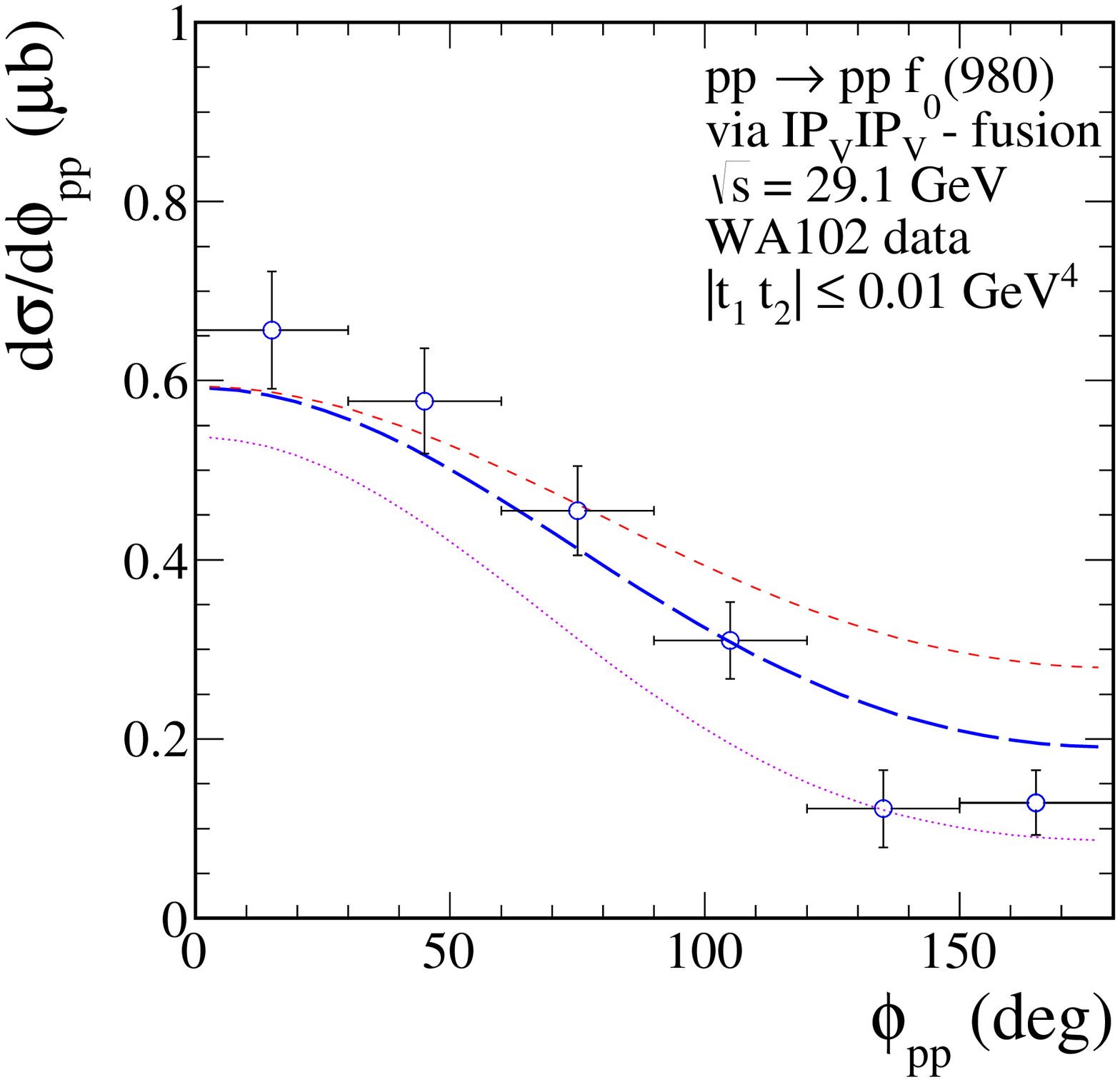}
(f)\includegraphics[width = 0.29\textwidth]{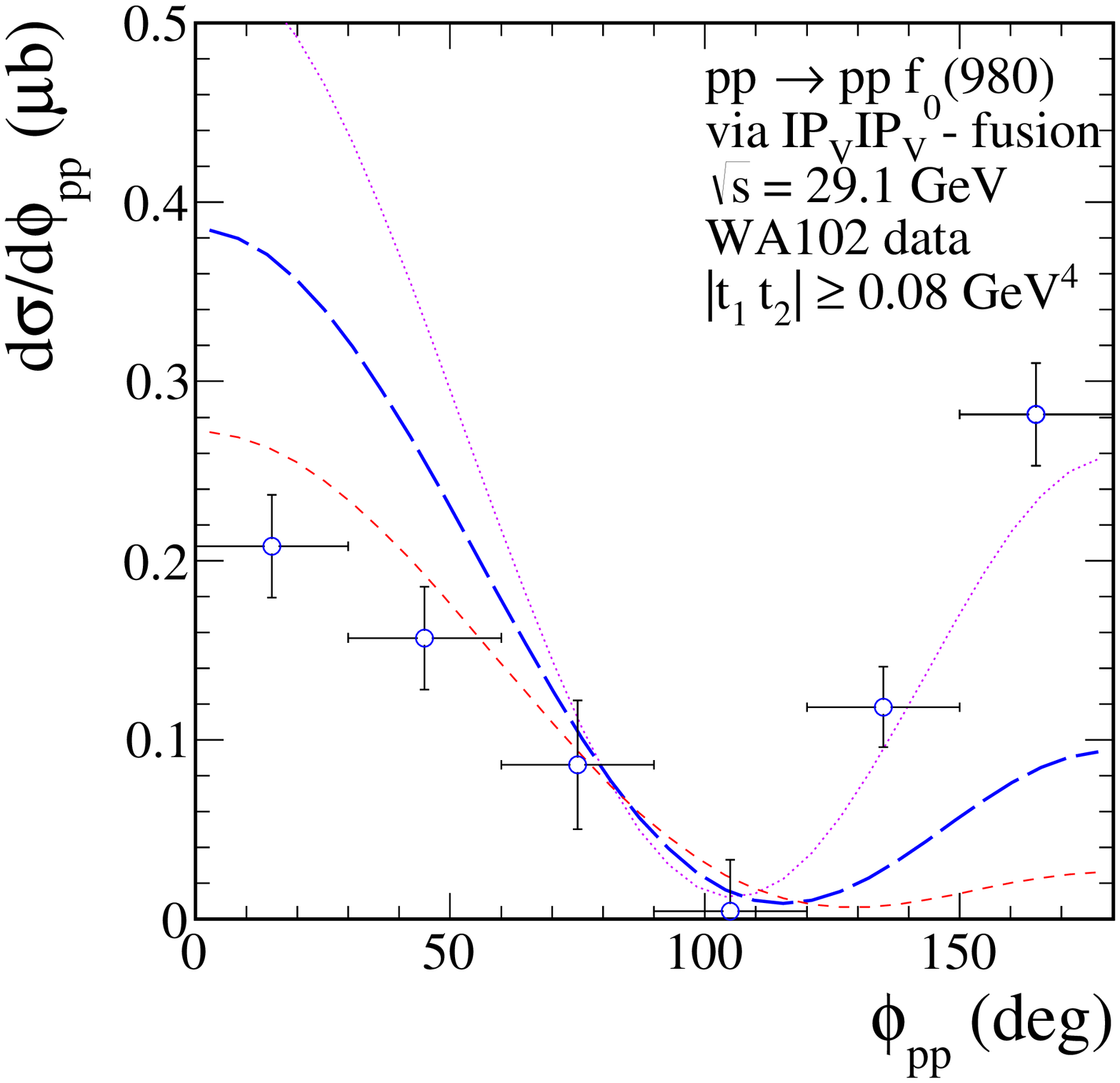}\\
(g)\includegraphics[width = 0.29\textwidth]{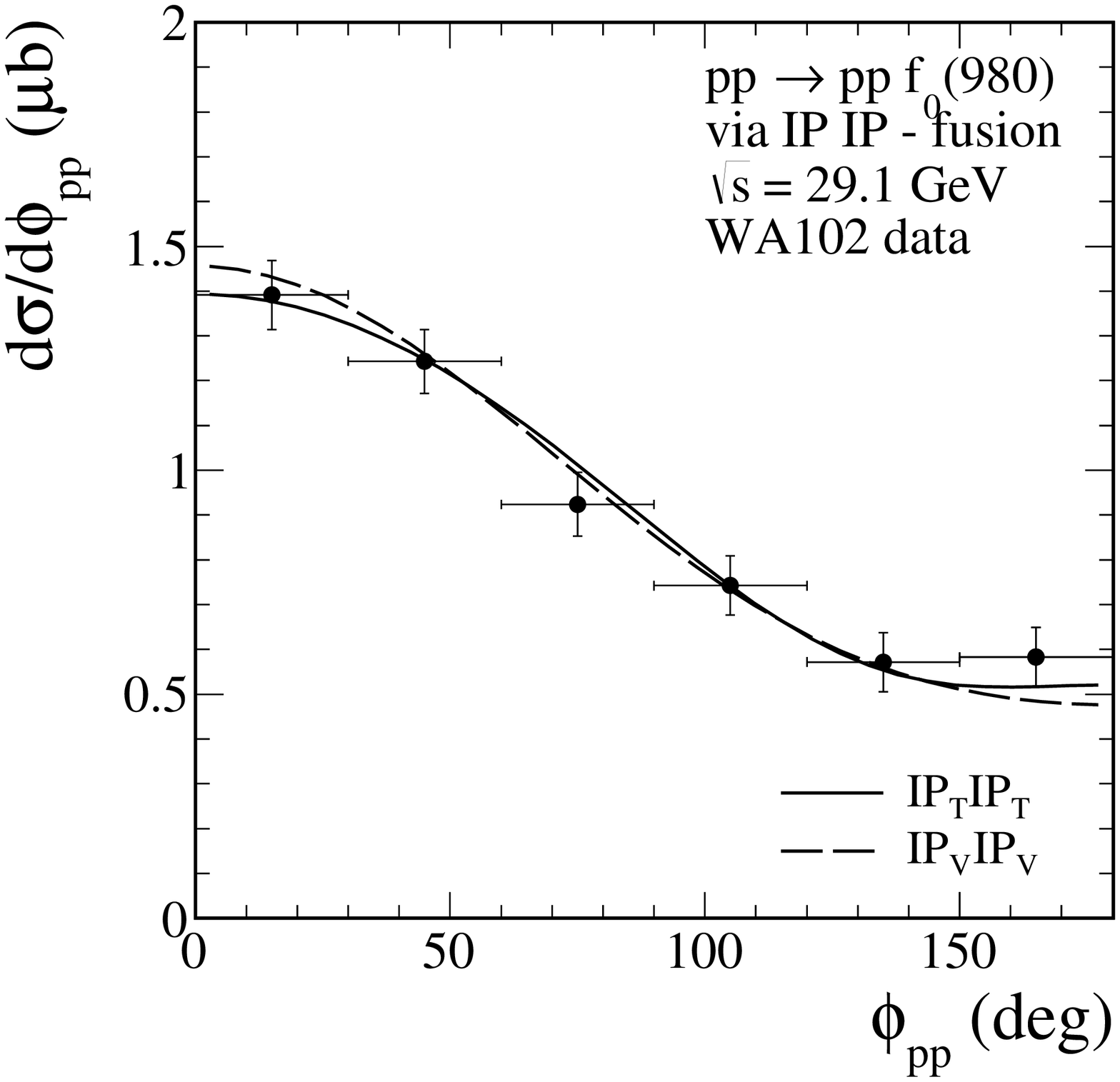}
(h)\includegraphics[width = 0.29\textwidth]{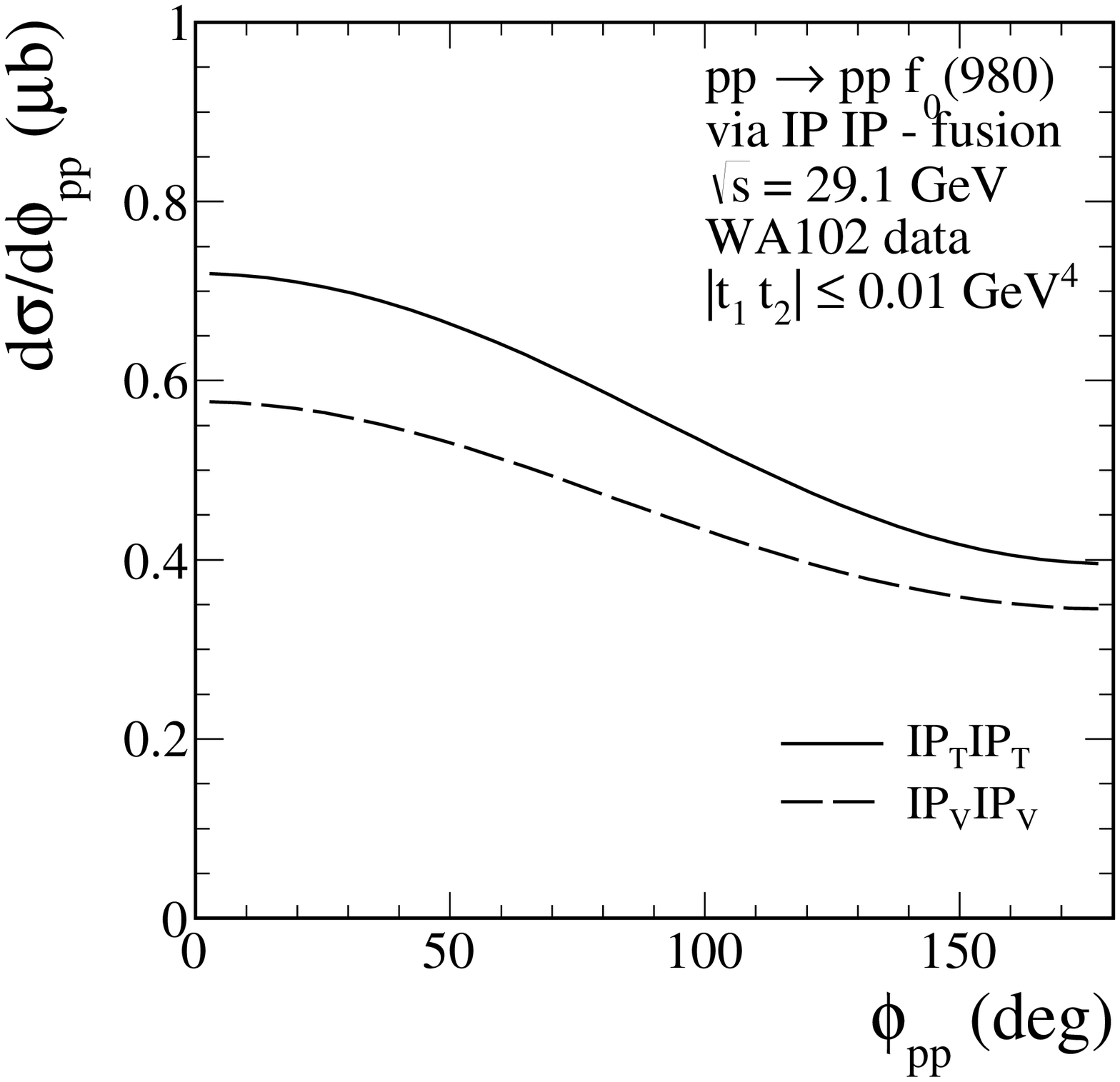}
(i)\includegraphics[width = 0.29\textwidth]{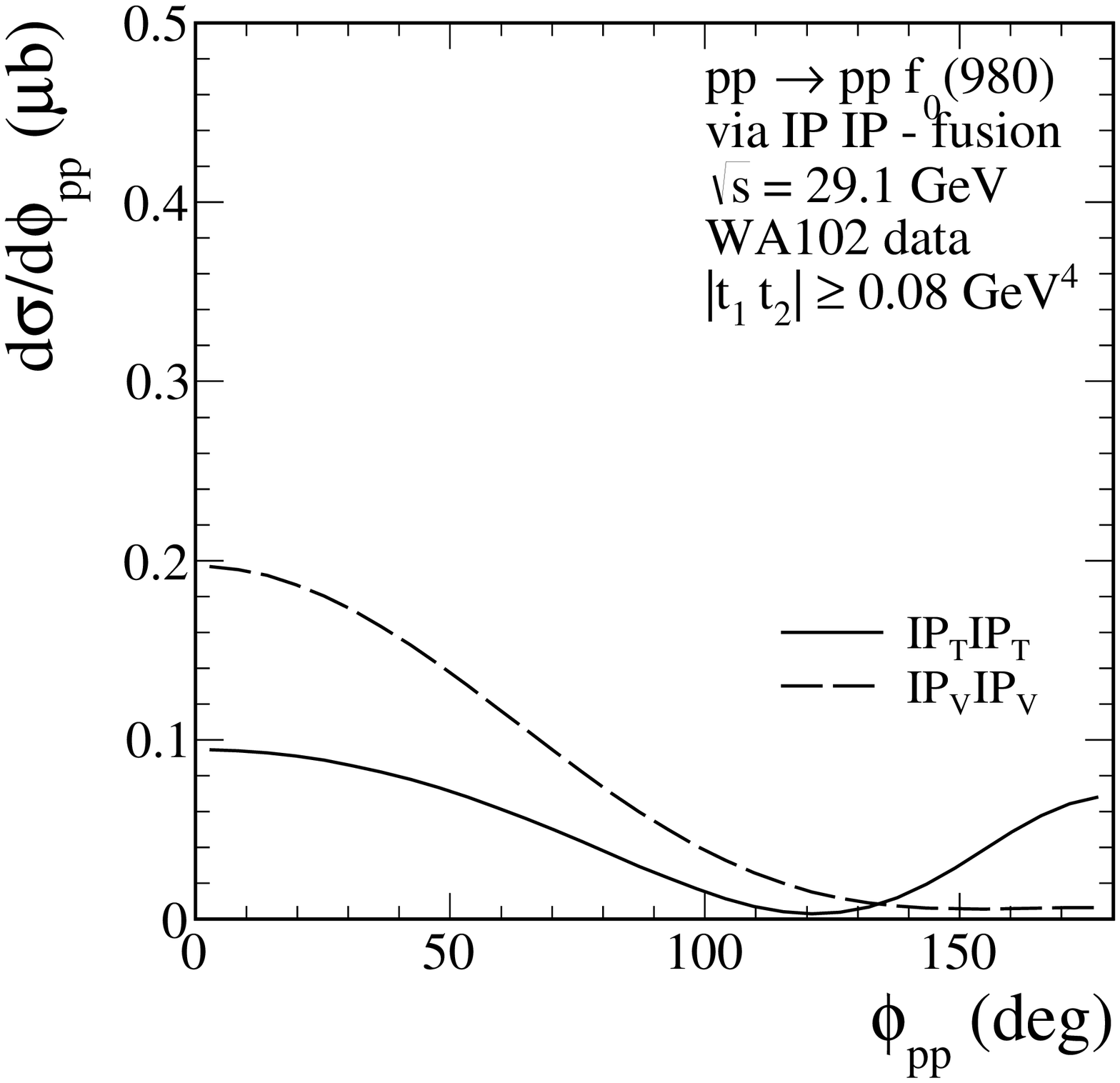}
  \caption{\label{fig:dsig_dphi_980}
    \small
The distribution in azimuthal angle between outgoing protons
for the central exclusive $f_{0}(980)$ meson production 
by the fusion of both tensor (panels (a) - (c)) 
and vector (panels (d) - (f)) pomerons at $\sqrt{s} = 29.1$~GeV.
Results in the left panels and the WA102 data points
from \cite{WA102_PLB462} (black points) and from \cite{WA102_PLB467} (blue points)
have been normalized to the mean value of the total cross section given in 
Table~\ref{tab:mesons}, obtained from Ref.~\cite{kirk00}.
The $\phi_{pp}$ distributions have also been analysed in two intervals of $|t_{1}t_{2}|$
and compared with experimental data .
These data are obtained from \cite{WA102_PLB467} with the normalisation
calculated in the tensorial and vectorial pomeron models themselves.
We show in panels (a) - (c) the results in the tensorial pomeron model.
For tensorial pomeron the individual contributions to the cross sections 
with $(l,S) = (0,0)$ (short-dashed line) and $(l,S) = (2,2)$ (dotted line) are also shown.
Panels (d) - (f) show the results obtained for the vectorial pomeron model.
In panels (g) - (i) the results obtained in the two models of pomeron are compared.
}
\end{figure}

\begin{figure}[!ht]
(a)\includegraphics[width = 0.29\textwidth]{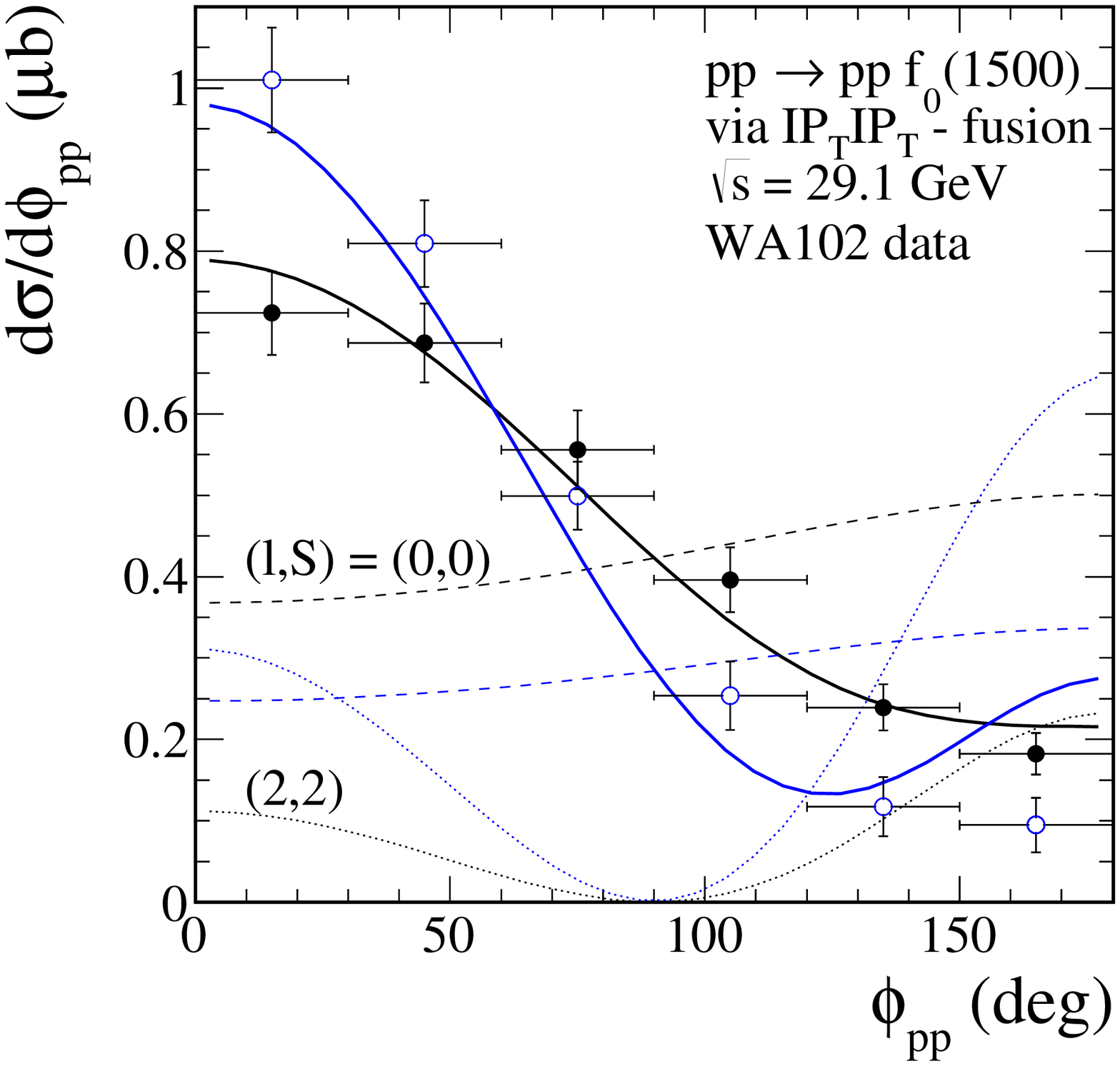}
(b)\includegraphics[width = 0.29\textwidth]{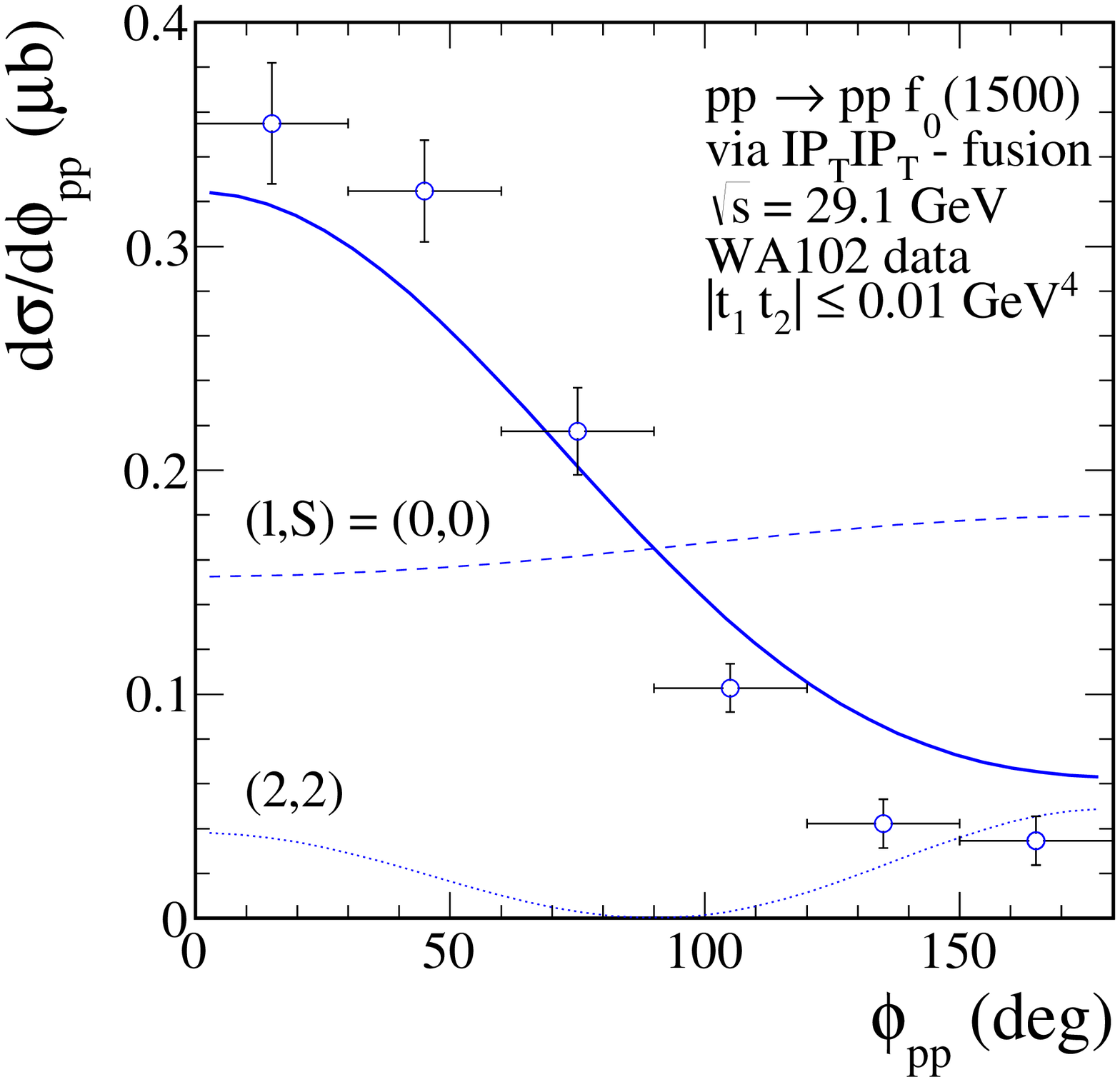}
(c)\includegraphics[width = 0.29\textwidth]{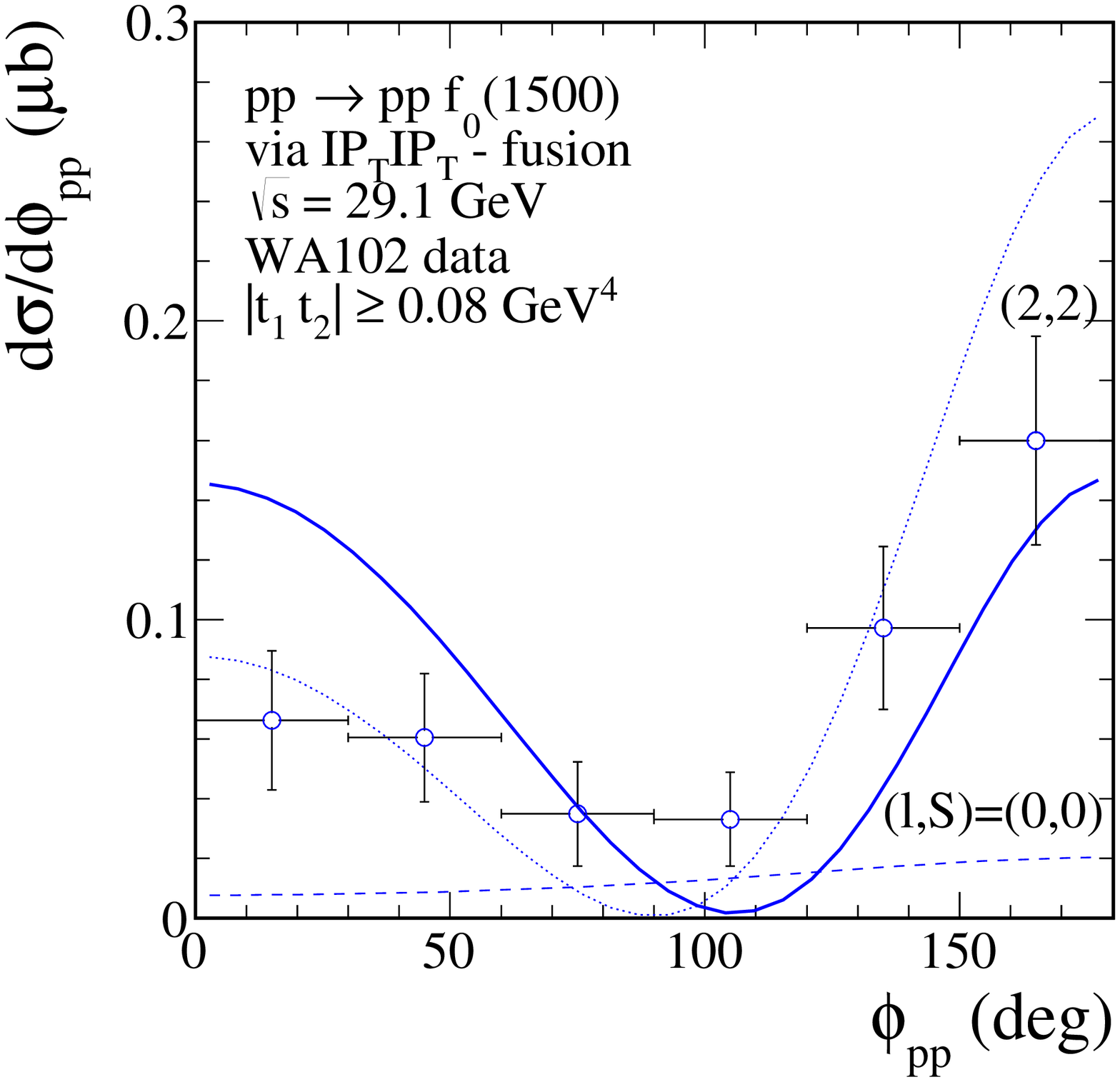}\\
(d)\includegraphics[width = 0.29\textwidth]{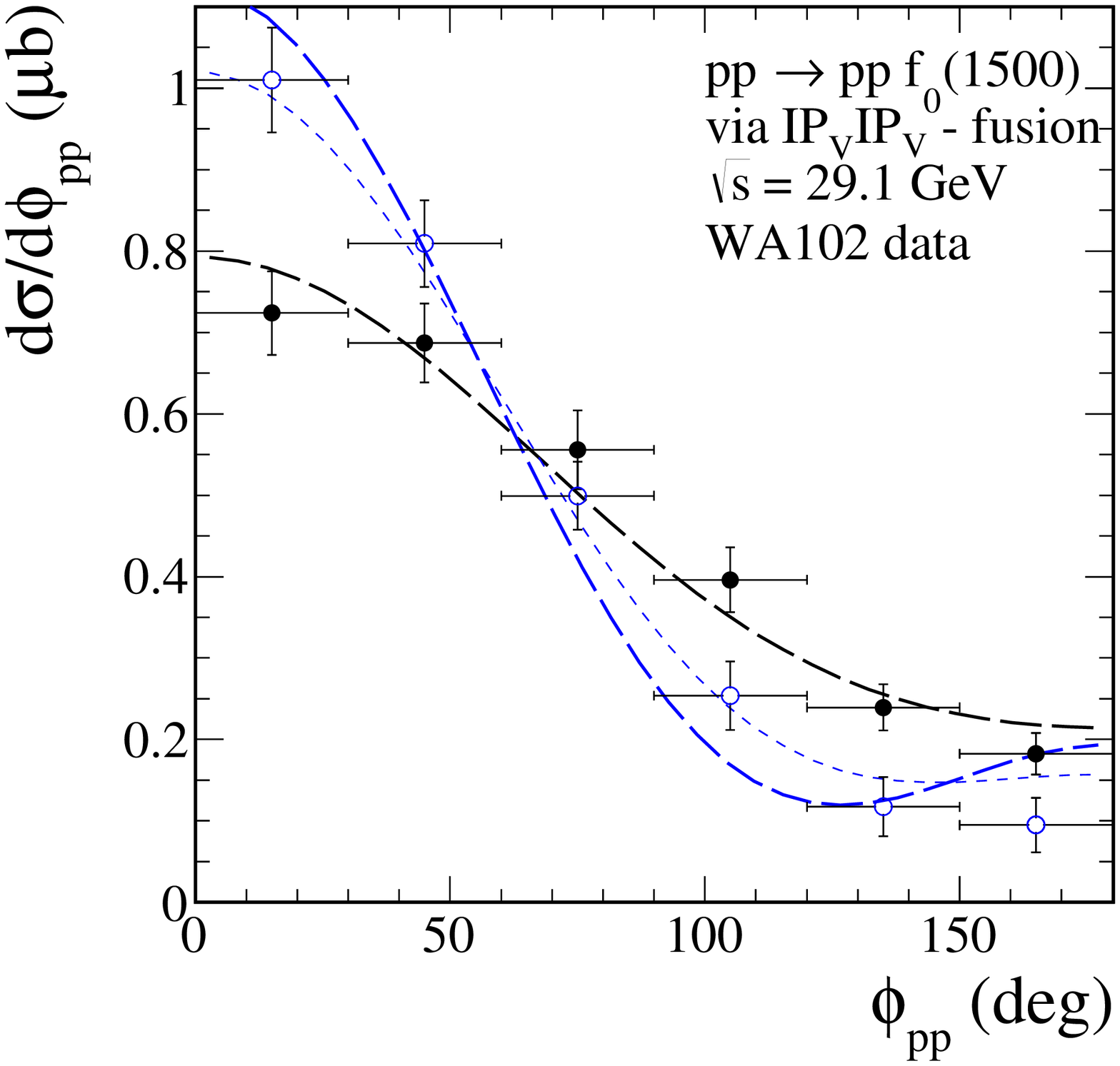}
(e)\includegraphics[width = 0.29\textwidth]{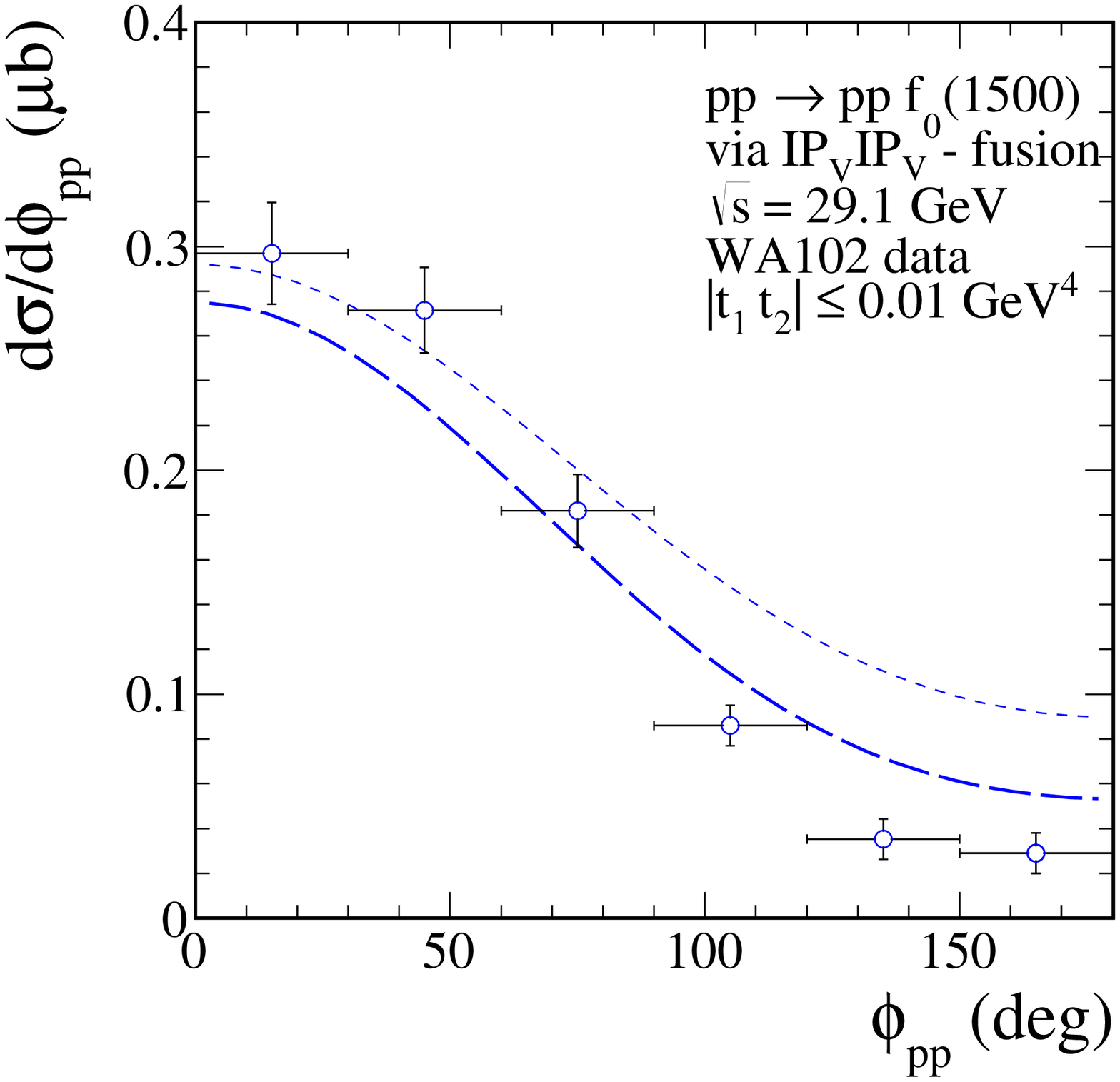}
(f)\includegraphics[width = 0.29\textwidth]{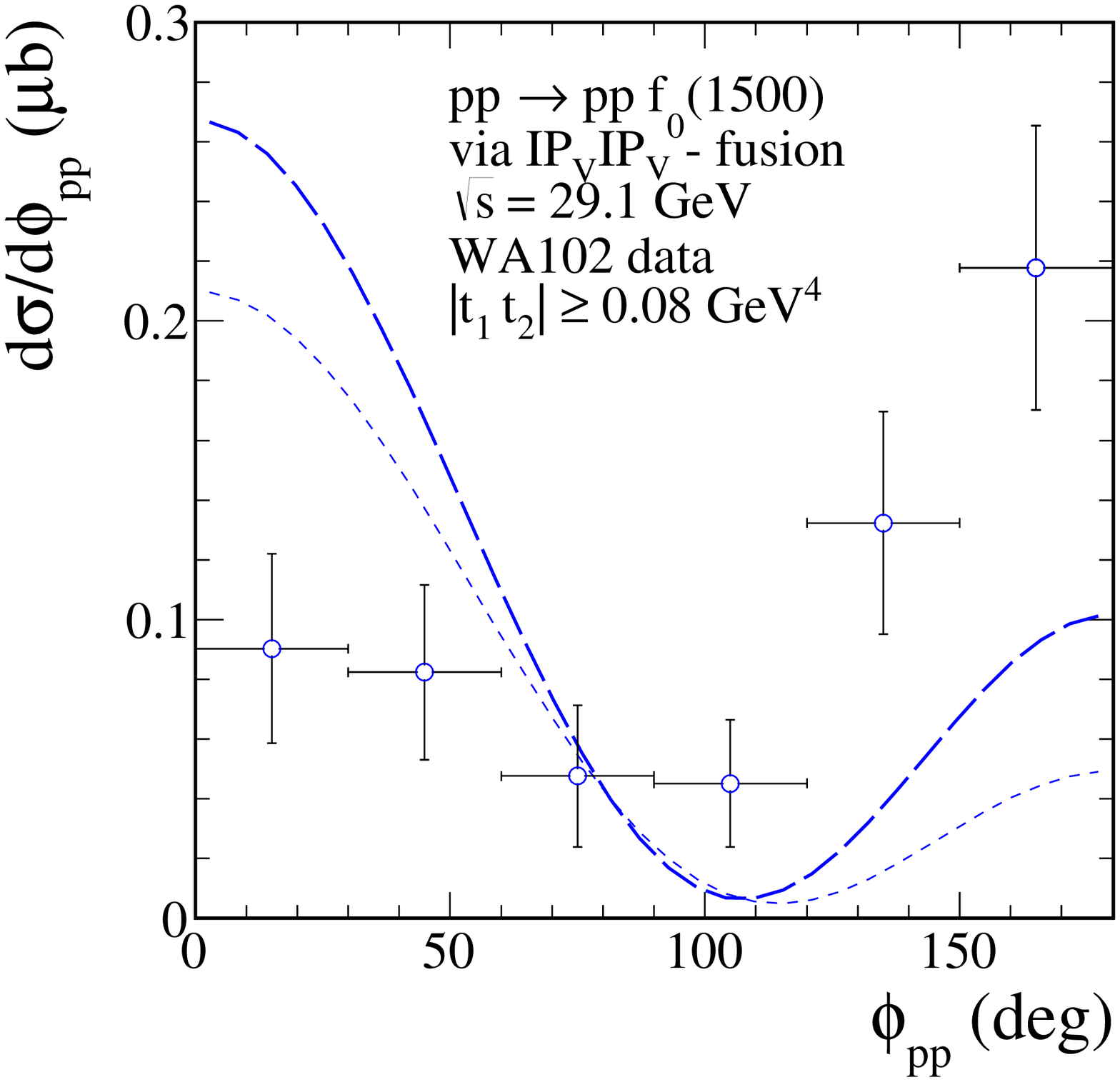}\\
(g)\includegraphics[width = 0.29\textwidth]{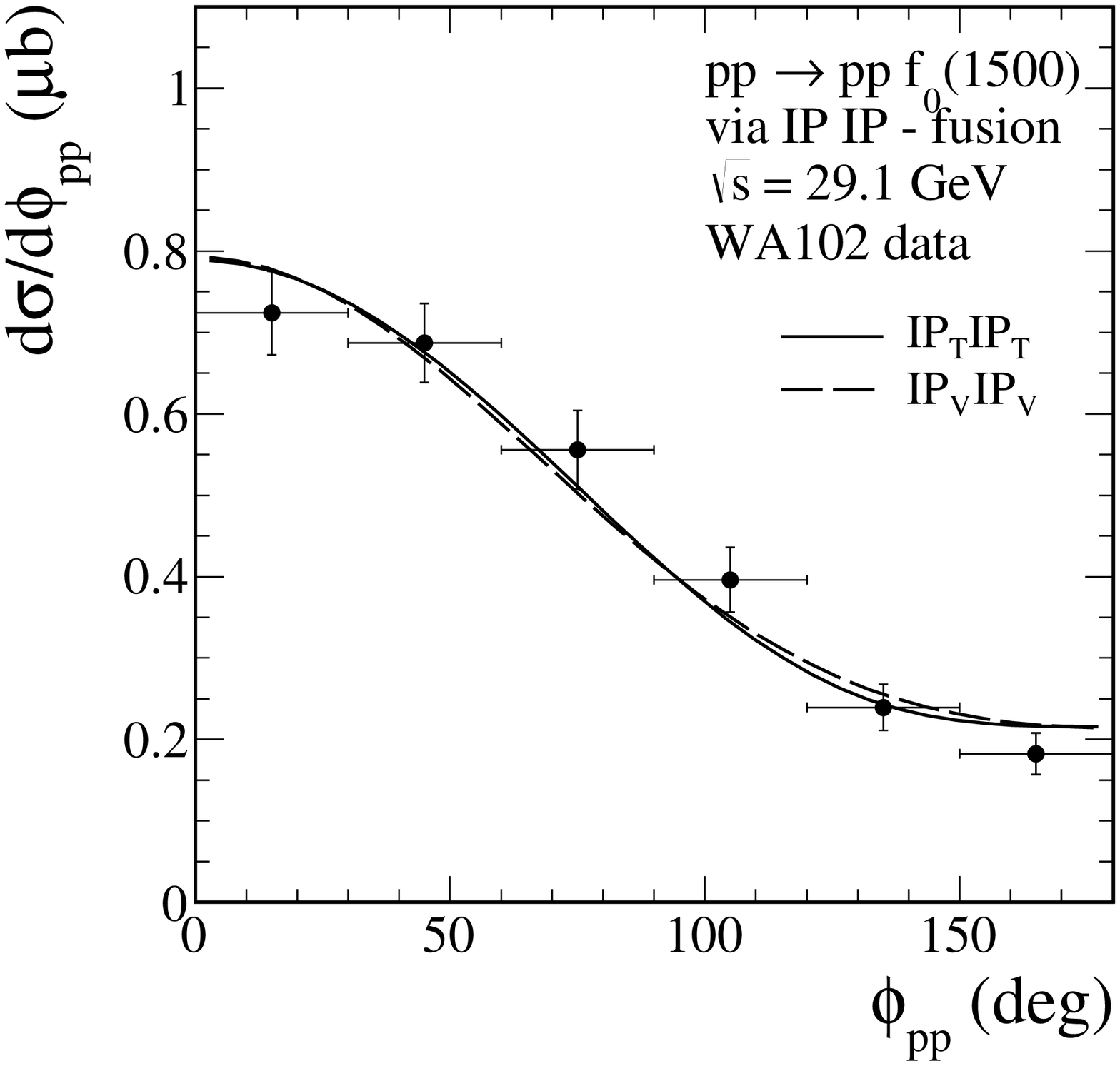}
(h)\includegraphics[width = 0.29\textwidth]{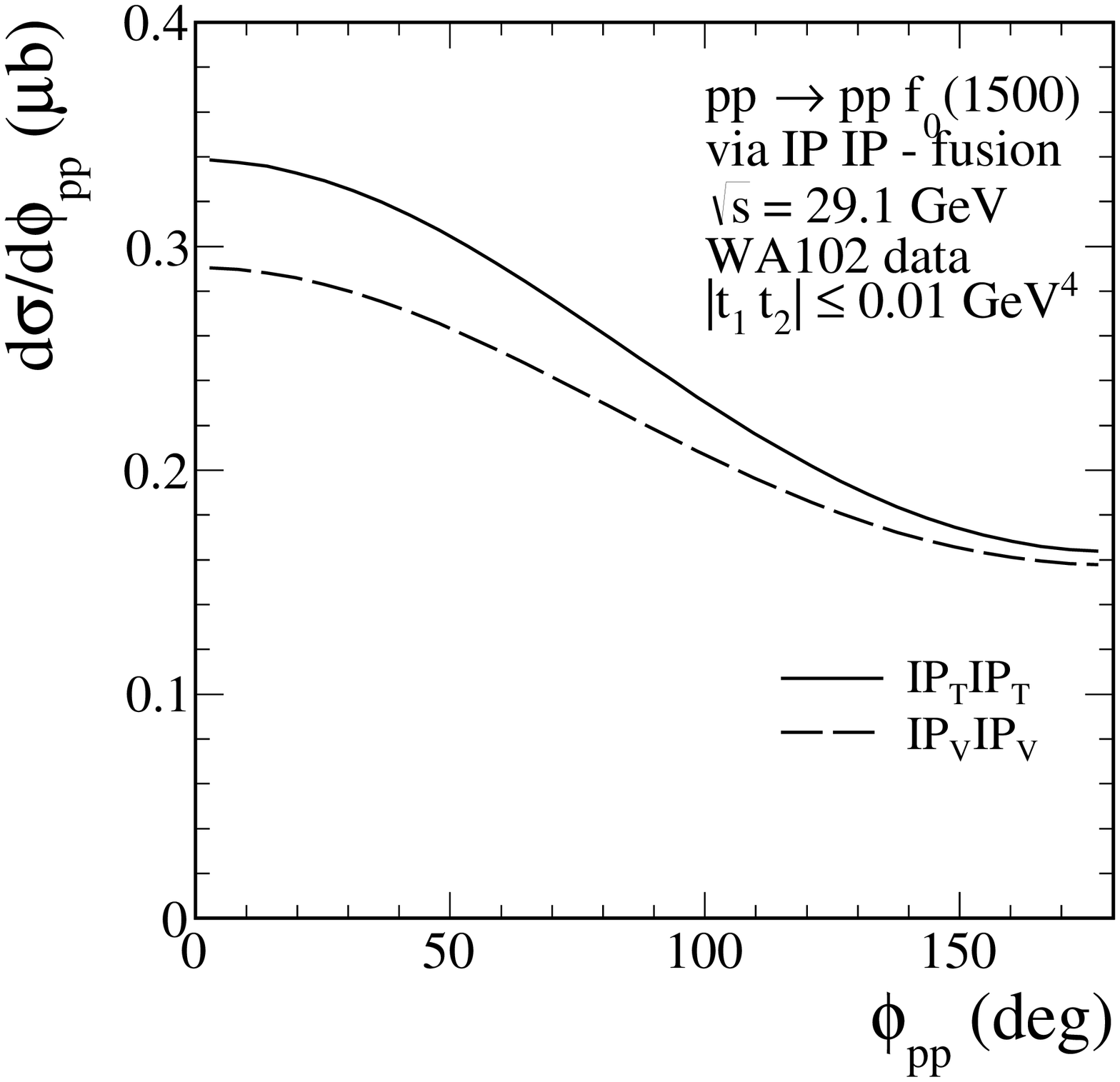}
(i)\includegraphics[width = 0.29\textwidth]{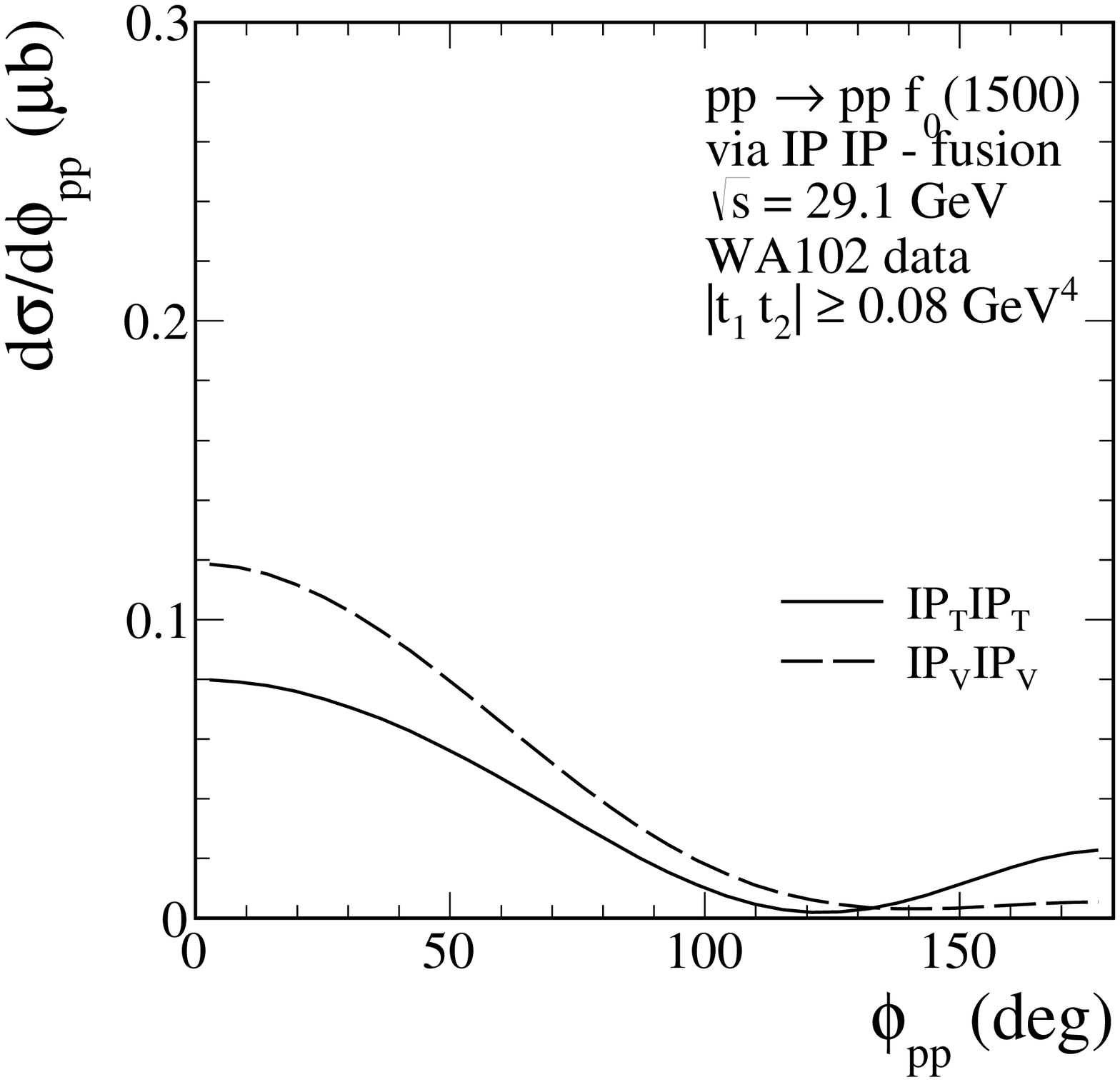}
  \caption{\label{fig:dsig_dphi_1500}
  \small
Same as Fig.~\ref{fig:dsig_dphi_980}, but for the central exclusive $f_{0}(1500)$ meson production.
}
\end{figure}

The distributions in azimuthal angle $\phi_{pp}$ between the outgoing protons 
for the central exclusive production of the mesons $f_{0}(980)$ and $f_{0}(1500)$ 
at $\sqrt{s} = 29.1$~GeV
are shown in Figs.~\ref{fig:dsig_dphi_980} and \ref{fig:dsig_dphi_1500}, respectively.
We compare results obtained by the fusion of two pomerons
(the tensor pomeron exchanges are shown in panels (a) - (c) 
and the vector pomeron exchanges are shown in panels (d) - (f))
with the data measured by the WA102 Collaboration in
\cite{WA102_PLB462} (the black filled points) and \cite{WA102_PLB467} (the blue circle points).
In the left panels we show the $\phi_{pp}$ distribution without experimental cuts,
the middle panels show the $\phi_{pp}$ distribution for $|t_{1}t_{2}| \leqslant 0.01$~GeV$^{4}$
and the right panels show the corresponding distribution for $|t_{1}t_{2}| \geqslant 0.08$~GeV$^{4}$.
Note that in \cite{WA102_PLB462} and \cite{WA102_PLB467} only normalised distributions are given. 
We have multiplied these distributions with the mean value of the total cross sections 
from Table~\ref{tab:mesons} for panels (a) and (d). 
For panels (b), (c), (e), and (f) we have multiplied
the normalised data distributions given in \cite{WA102_PLB467}
with the cross sections obtained from our calculations in the tensorial
and vectorial pomeron models, respectively; see Table~\ref{tab:couplings_S}.
These normalisation factors are different for the $I\!\!P_{T}$ and $I\!\!P_{V}$ cases.
Therefore, also the ``data'' shown in panels (b) and (e), as well as in (c) and (f), are different. 
Also note that the difference in the data from \cite{WA102_PLB462} and \cite{WA102_PLB467}
shown in panels (a) and (d) has an experimental origin, as far as the authors can tell.
Correspondingly, in the panels (a) the black filled and the blue circle experimental points
are described by the tensorial pomeron exchanges for different values of the two $(l,S)$ contributions.
For the $f_{0}(980)$ (Fig.~\ref{fig:dsig_dphi_980}(a)) we obtain these coupling constants as
$(g_{I\!\!P I\!\!P M}', g_{I\!\!P I\!\!P M}'') = (0.788, 4)$ (the black solid line) and
$(g_{I\!\!P I\!\!P M}', g_{I\!\!P I\!\!P M}'') = (0.75, 5.5)$ (the blue solid line), respectively.
The values of the couplings for $f_{0}(1500)$ production shown in Fig.~\ref{fig:dsig_dphi_1500}(a) are
$(g_{I\!\!P I\!\!P M}', g_{I\!\!P I\!\!P M}'') = (1.22, 6)$ (the black solid line) and
$(g_{I\!\!P I\!\!P M}', g_{I\!\!P I\!\!P M}'') = (1, 10)$ (the blue solid line), respectively.
From our results we conclude that both $(l,S)$ contributions 
are necessary if the distributions in azimuthal angle are to be described accurately.
The $(l,S) = (2,2)$ contribution increases the cross section at large $\phi_{pp}$
while decreasing it for small $\phi_{pp}$.
The panels (d) - (f) show the results obtained for two vector pomerons coupling to the mesons.
The curves present contributions from different $(l,S)$ couplings collected in Table~\ref{tab:couplings_S}.
In the panel (d) of Fig.~\ref{fig:dsig_dphi_980} ($f_{0}(980)$ production)
the black long-dashed line corresponds to
$(g_{I\!\!P I\!\!P M}', g_{I\!\!P I\!\!P M}'') = (0.27, 0.8)$ 
and the blue long-dashed line to
$(g_{I\!\!P I\!\!P M}', g_{I\!\!P I\!\!P M}'') = (0.24, 1.5)$.
For $f_{0}(1500)$ production shown in panel (d) of Fig.~\ref{fig:dsig_dphi_1500}
the black long-dashed line corresponds to
$(g_{I\!\!P I\!\!P M}', g_{I\!\!P I\!\!P M}'') = (0.208, 0.725)$,
the blue long-dashed line to
$(g_{I\!\!P I\!\!P M}', g_{I\!\!P I\!\!P M}'') = (0.164, 1.5)$.
With these values we are able to describe well 
the black filled and blue circle experimental points, respectively.
For panels (e) and (f) we have multiplied the normalised data from \cite{WA102_PLB467}
with the cross sections obtained from our calculations.
In panels (g) - (i) the results obtained with the two models of pomeron are compared.
From Figs.~\ref{fig:dsig_dphi_980} and \ref{fig:dsig_dphi_1500}
we conclude that, especially for $|t_{1}t_{2}| \geqslant 0.08$~GeV$^{4}$,
the tensorial pomeron ansatz is in better (qualitative)
agreement with the data than the vectorial ansatz.
But let us recall that for panels (b), (c), (e), and (f) the normalisation is
taken from the models themselves for lack of experimental information.

At present we have calculated only so-called bare amplitudes
which are subjected to absorption corrections.
The absorption effects lead usually to a weak energy dependent damping of the cross sections. 
At the energy of the WA102 experiment ($\sqrt{s} = 29.1$~GeV)
the damping factor is expected to be at most of the order of 2
and should increase with rising collision energy.
The absorption effects both in initial and final states have been considered in Ref.~\cite{PRSG05}.
It was stressed in Ref.~\cite{PRSG05} that at the WA102 energies
absorptive effects are not so significant and the azimuthal angle dependence 
looks like the ``bare'' one.

\begin{figure}[!ht]
(a)\includegraphics[width = 0.29\textwidth]{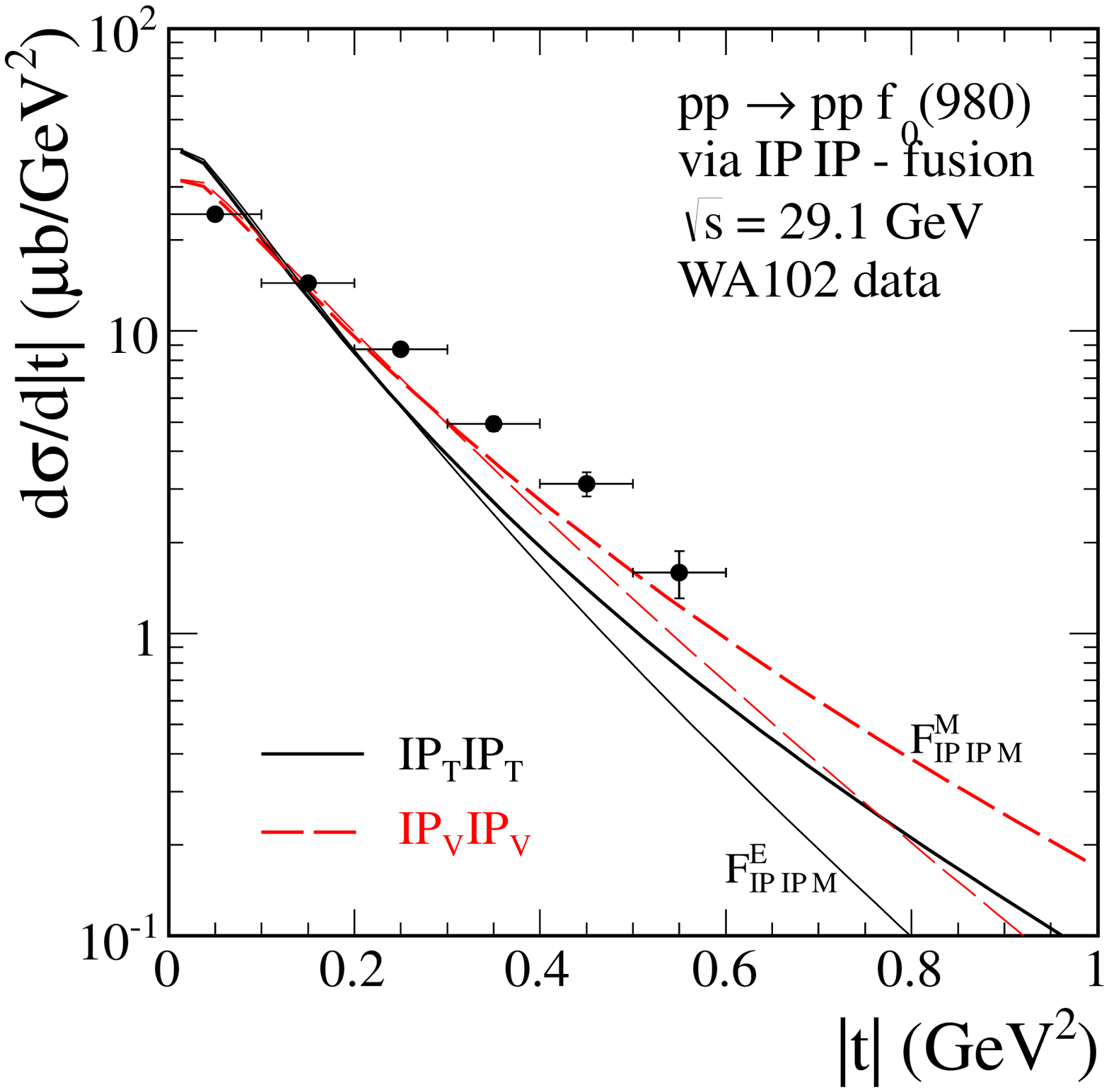}
(b)\includegraphics[width = 0.29\textwidth]{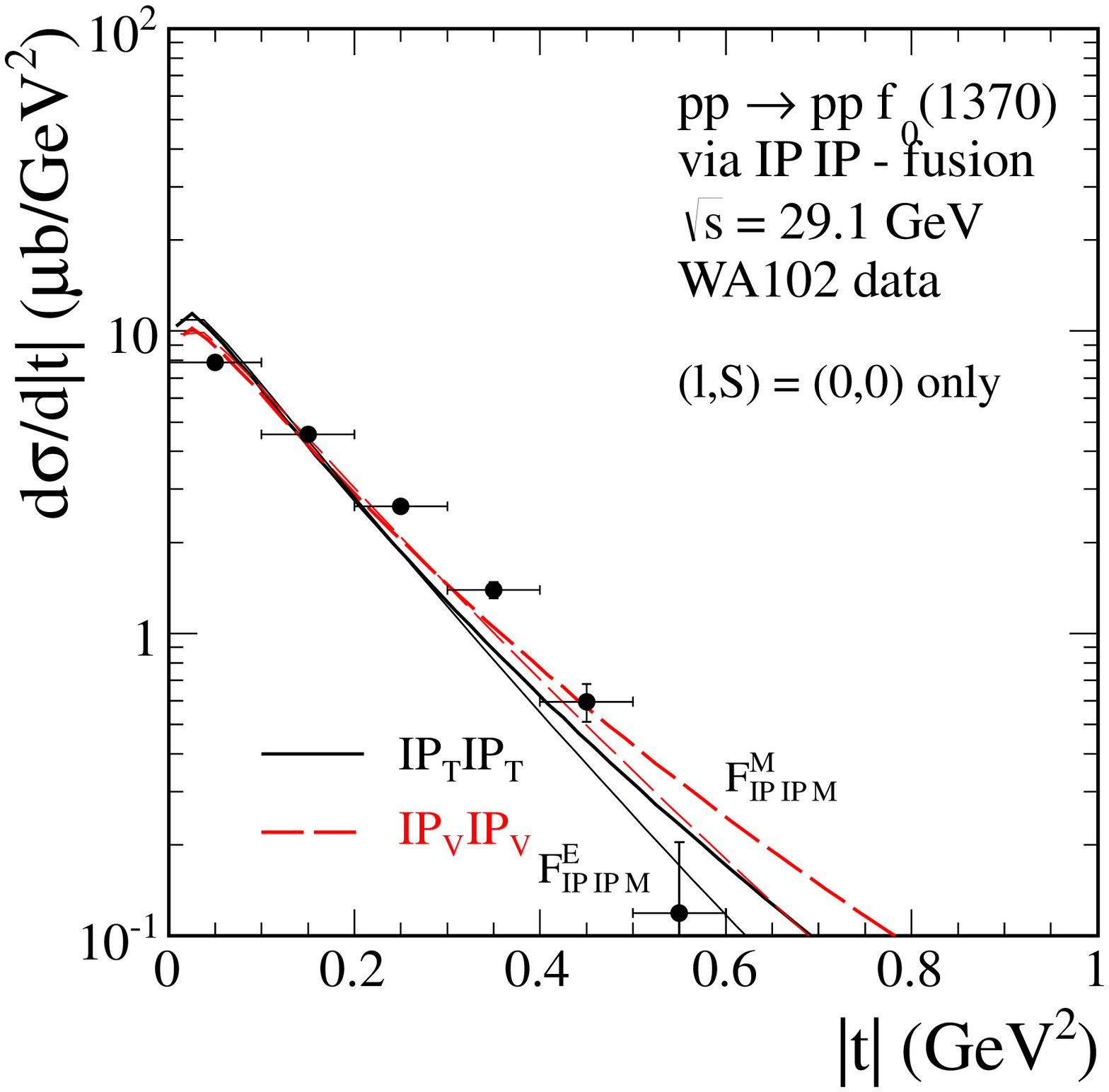}
(c)\includegraphics[width = 0.29\textwidth]{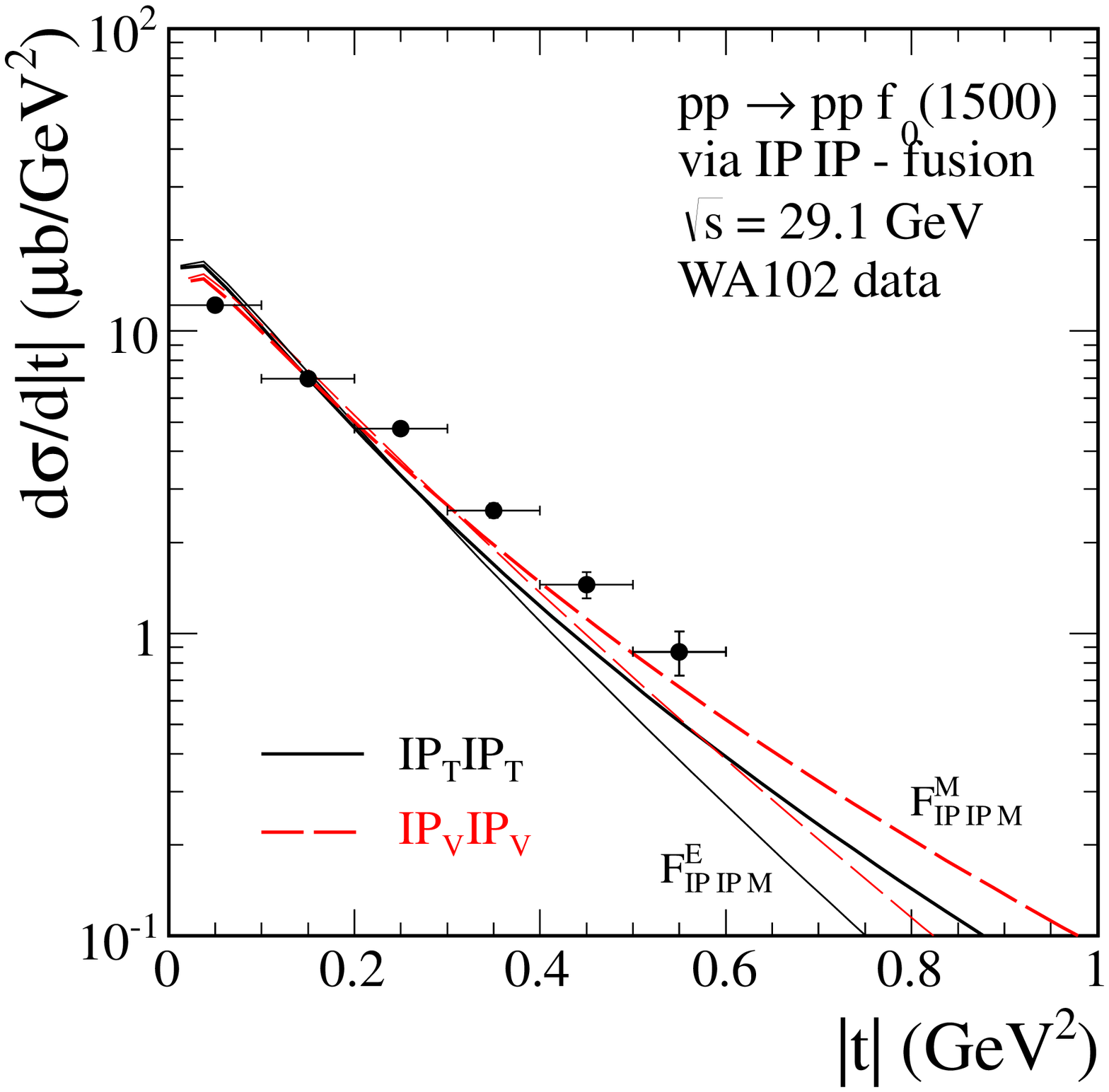}\\
(d)\includegraphics[width = 0.29\textwidth]{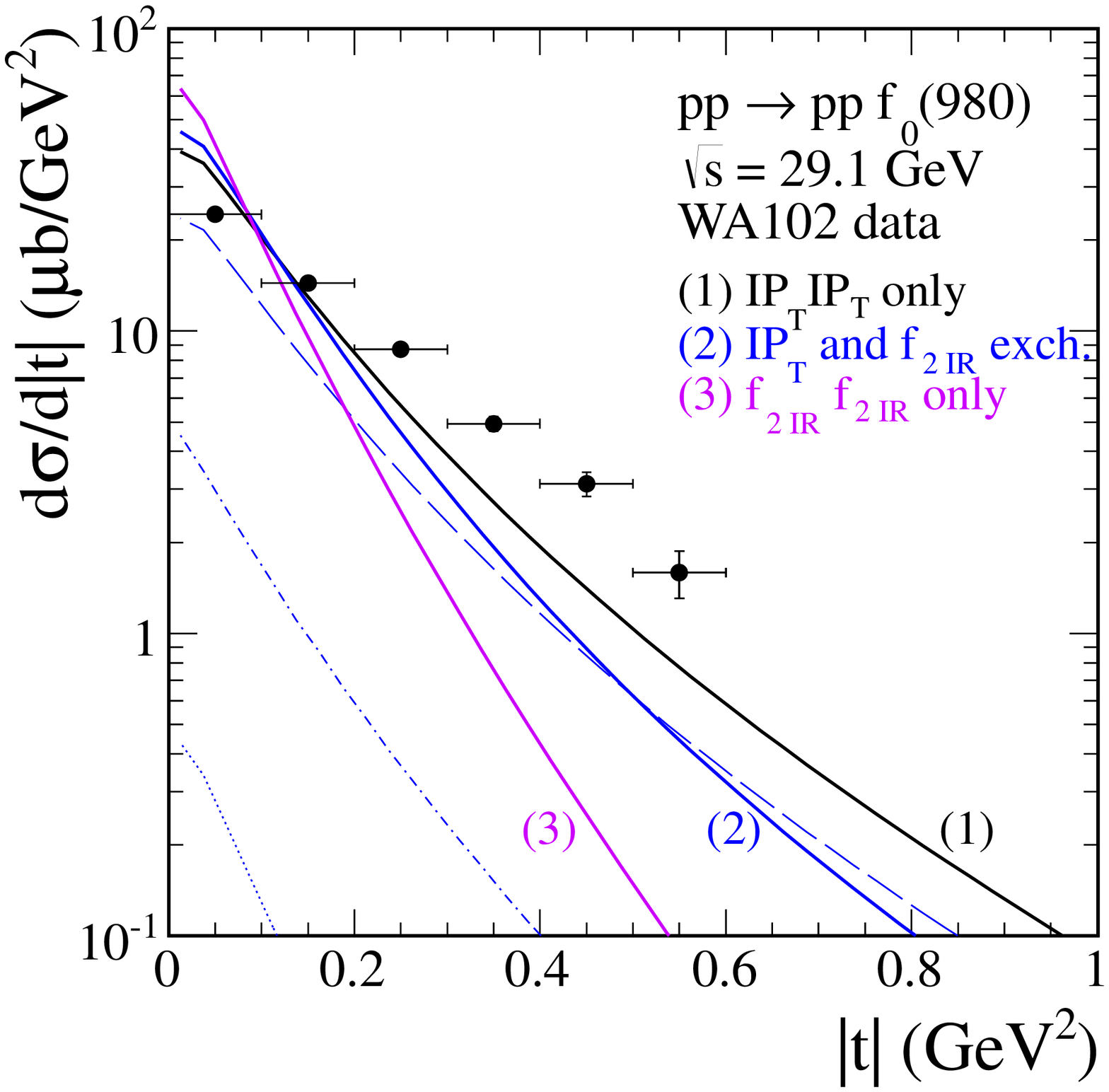}
(e)\includegraphics[width = 0.29\textwidth]{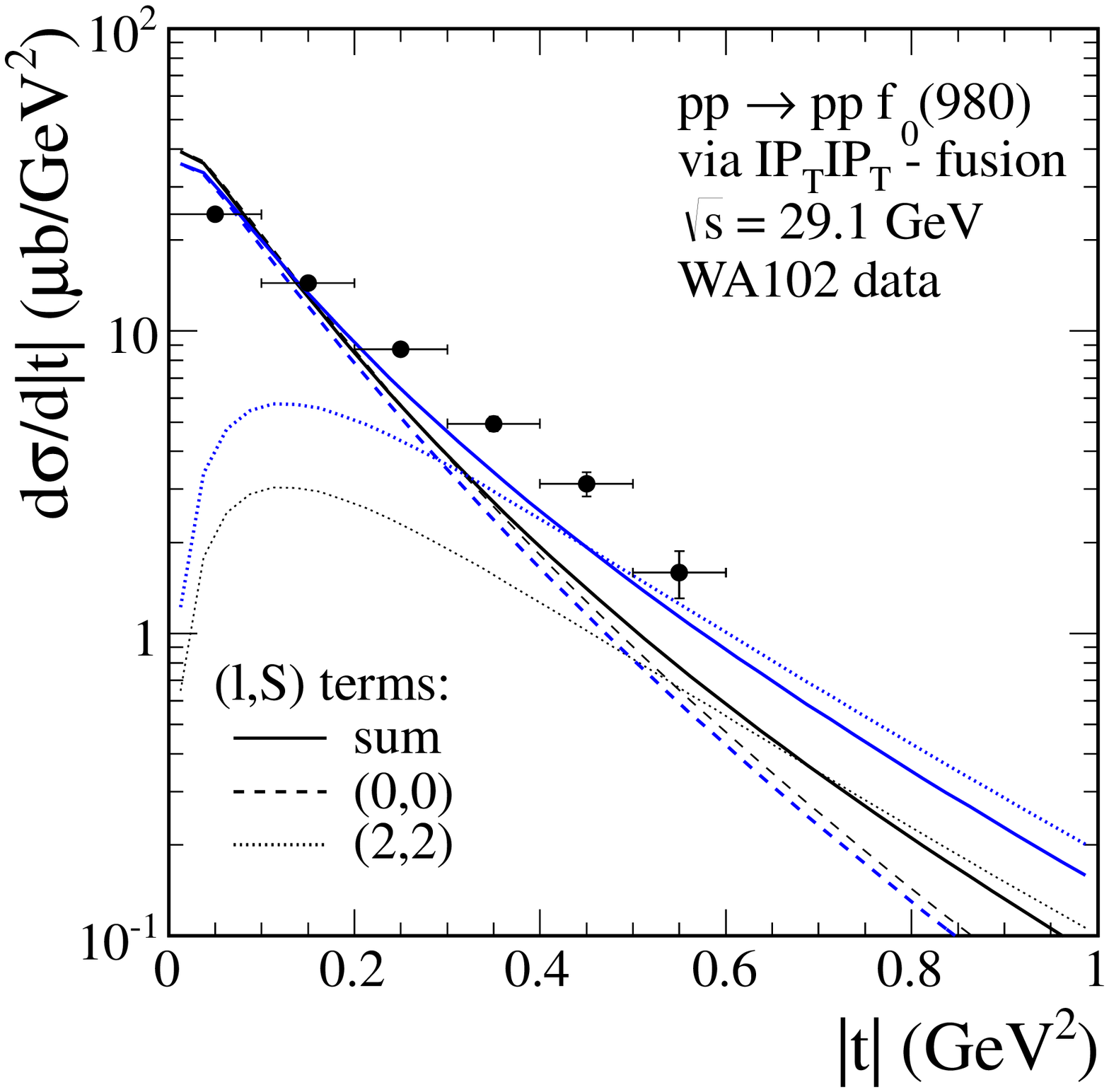}
(f)\includegraphics[width = 0.29\textwidth]{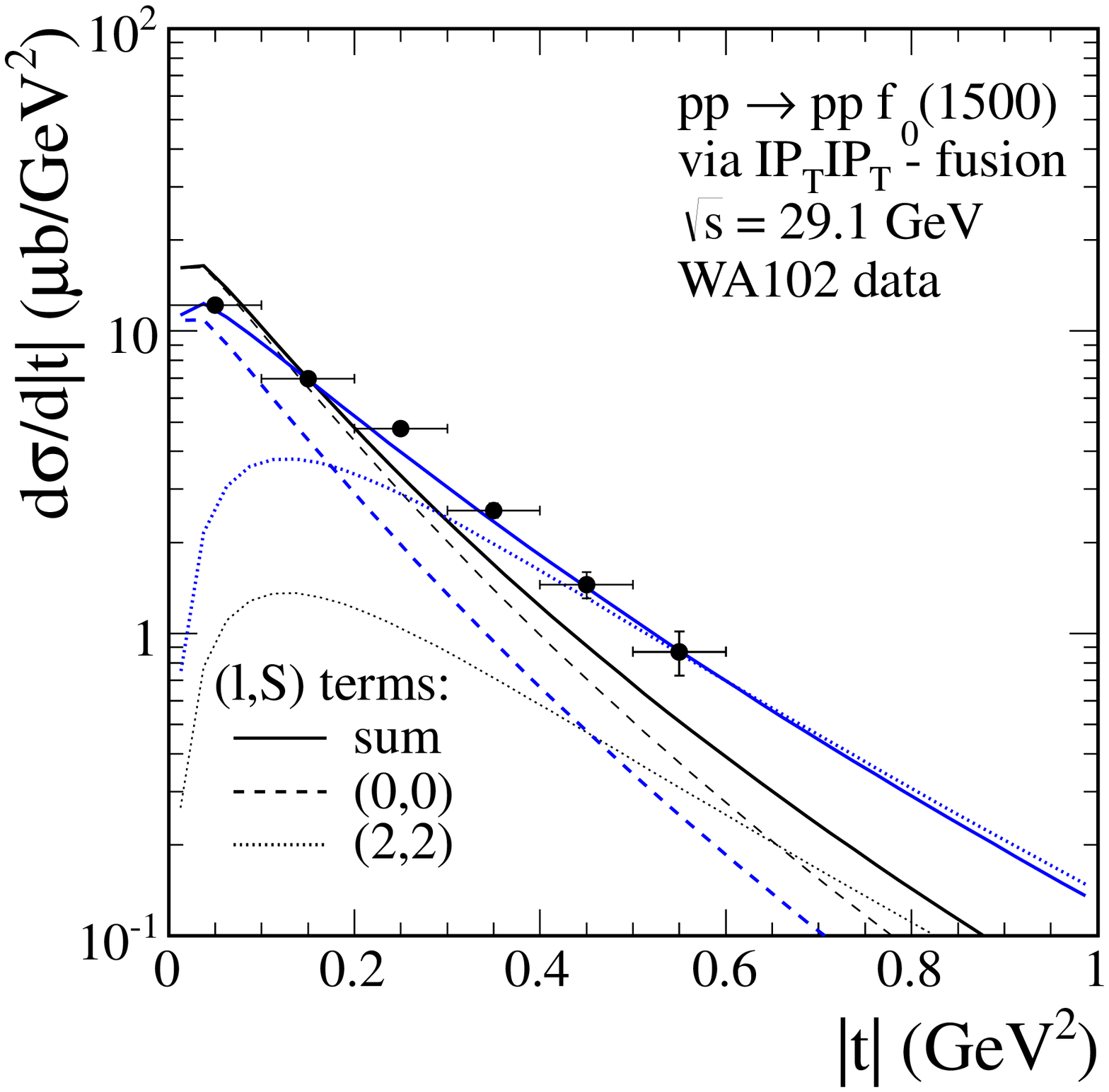}
  \caption{\label{fig:dsig_dt}
  \small
The $t$ distributions for $f_{0}(980)$ (panels (a), (d) and (e)),
$f_{0}(1370)$ (panel (b)), 
and $f_{0}(1500)$ (panels (c) and (f)) meson production at $\sqrt{s} = 29.1$~GeV.
The WA102 experimental data points from \cite{WA102_PLB462}
have been normalized to the mean value of the total cross sections 
given in Table~\ref{tab:mesons} as obtained from \cite{kirk00}.
In panels (a) - (c) the results for the fusion of two tensor (solid line) 
and vector (long-dashed line) pomerons are shown.
The lower lines correspond to calculations with the exponential form factor (\ref{Fpompommeson_exp}) 
for the cut-off parameter $\Lambda_{E}^{2} = 0.6$~GeV$^{2}$, 
the upper lines to calculations with the monopole form factor (\ref{Fpompommeson_pion})
for $\Lambda_{0}^{2} = 0.5$~GeV$^{2}$.
In panel (d) the black solid line (1) corresponds to the $I\!\!P_{T} I\!\!P_{T}$-fusion only,
the blue solid line (2) corresponds to the tensor pomeron and $f_{2 I\!\!R}$ exchanges
(the long-dashed, dash-dotted and dotted lines present 
the $I\!\!P_{T} I\!\!P_{T}$, $I\!\!P_{T} f_{2 I\!\!R}$ ($f_{2 I\!\!R} I\!\!P_{T}$) 
and $f_{2 I\!\!R} f_{2 I\!\!R}$ contributions, respectively),
and the violet solid line (3) presents the $f_{2 I\!\!R} f_{2 I\!\!R}$-fusion alone
normalized to the integrated cross section from \cite{kirk00}; see Table~\ref{tab:mesons}.
In panels (e) and (f) we show the individual spin contributions to the cross sections
with $(l,S) = (0,0)$ (short-dashed line) and $(l,S) = (2,2)$ (dotted line)
as well as lines for the two sets of couplings
fixed previously by comparison with the experimental azimuthal angular correlations
(see panels (a) in Figs.~\ref{fig:dsig_dphi_980} and \ref{fig:dsig_dphi_1500}).
}
\end{figure}

In Fig.~\ref{fig:dsig_dt} we show the distributions
in transferred four-momentum squared $t$ between the initial and final protons
at $\sqrt{s} = 29.1$~GeV for $f_0(980)$, $f_0(1370)$, and $f_0(1500)$ mesons.
While for $f_0(1370)$ the $(l,S) = (0,0)$ coupling is sufficient 
(see discussion of azimuthal correlations in Fig.~\ref{fig:dsig_dphi_1370}) 
for $f_0(980)$ and $f_0(1500)$ both the $(0,0)$ and $(2,2)$ couplings are included.
A different structure of the central vertex for vector and tensor
leads to a difference in $t$ distribution; see panels (a) - (c).
The difference seems, however, too small to be verified experimentally.
In addition, in panels (a) - (c) we compare distributions 
obtained for two types of pomeron-pomeron-meson form factors
of the exponential form (\ref{Fpompommeson_exp}) 
and the monopole form (\ref{Fpompommeson_pion}).
The calculations with the exponential form factor (\ref{Fpompommeson_exp}) 
and for the cut-off parameter $\Lambda_{E}^{2} = 0.6$~GeV$^{2}$ give a sizeable decrease
of the cross sections at large $|t|$.
In panel (d) we show contributions for two tensor pomerons (the line (1))
and $f_{2 I\!\!R}$ reggeons (the line (3)) exchanges alone, 
since the contribution with tensorial pomeron and $f_{2 I\!\!R}$ reggeon 
is included as well (the line (2)).
We conclude that the $f_{2 I\!\!R} f_{2 I\!\!R}$ component alone
does not describe the WA102 data.
In panels (e) and (f) we show a decomposition of the $t$-distribution into $(0,0)$ and $(2,2)$
components for the tensor pomeron exchanges.
At $t = 0$ the $(2,2)$ component vanishes, in contrast 
to the $(0,0)$ component. Therefore, the latter dominates at small $|t|$. 
As previously, we show lines for the two parameter sets obtained from the fits 
to the two different experimental azimuthal angular correlations
(see panels (a) in Figs.~\ref{fig:dsig_dphi_980} and \ref{fig:dsig_dphi_1500}).

\begin{figure}[!ht]
\includegraphics[width = 0.32\textwidth]{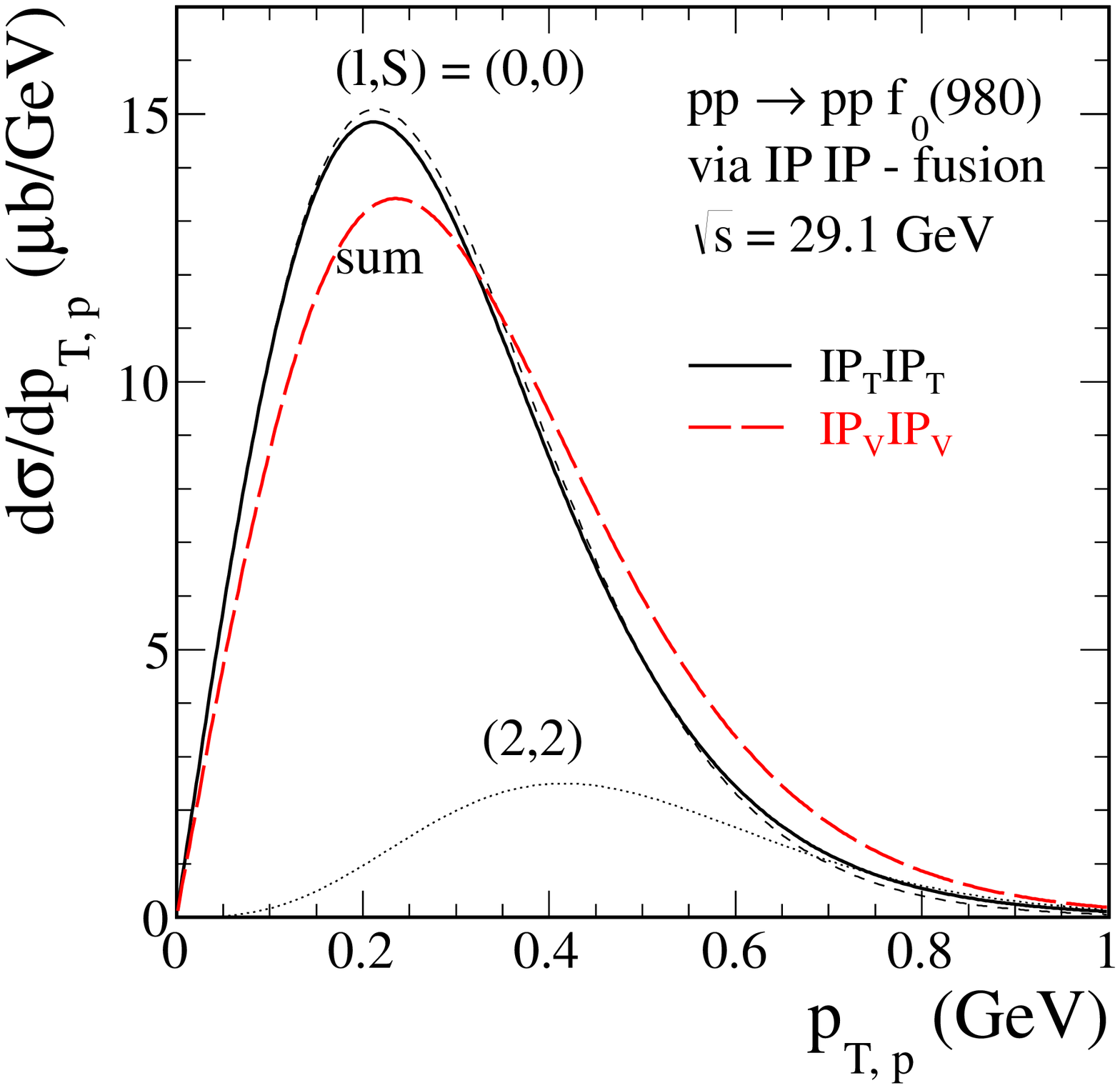}
\includegraphics[width = 0.32\textwidth]{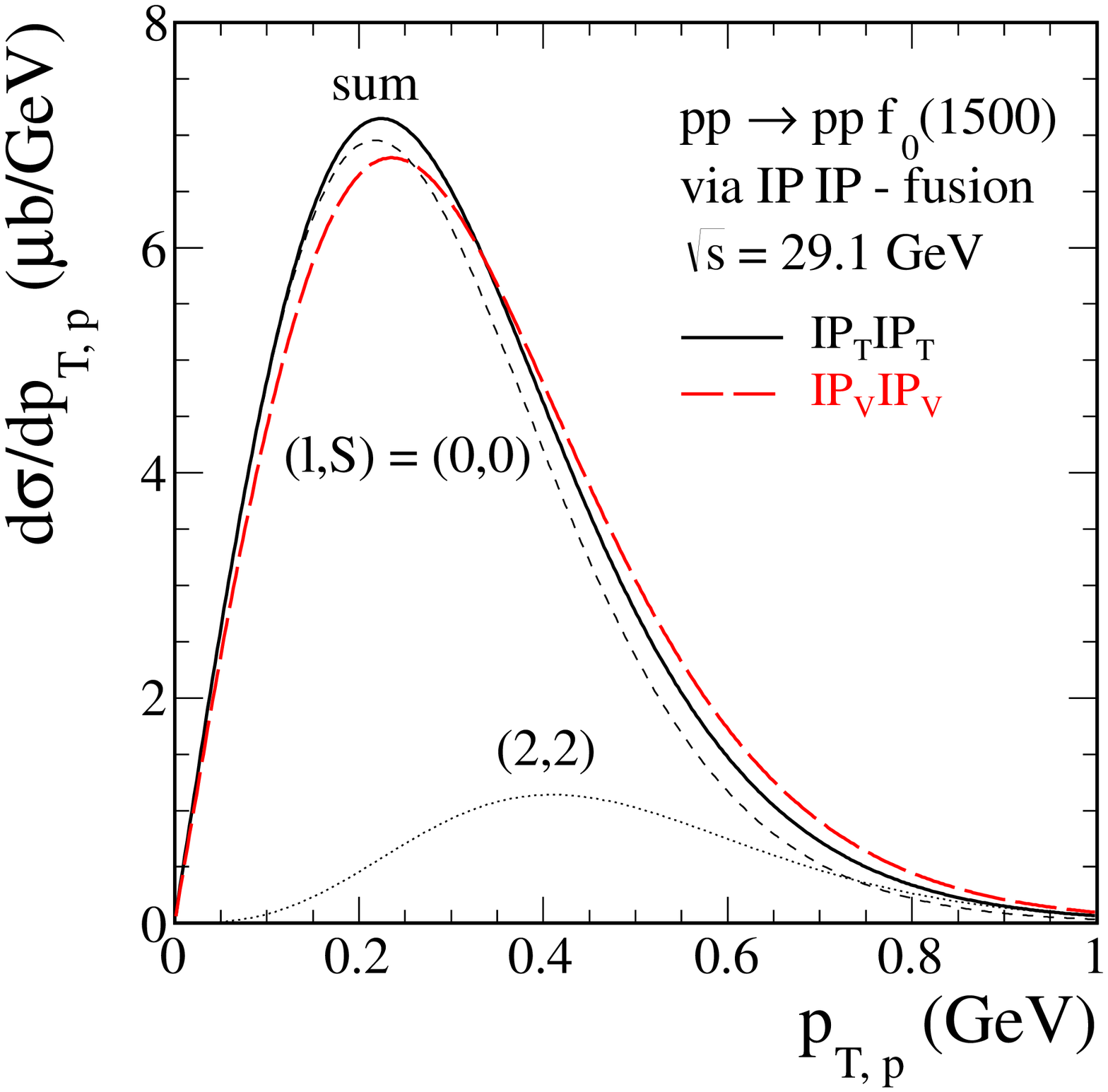}
\includegraphics[width = 0.32\textwidth]{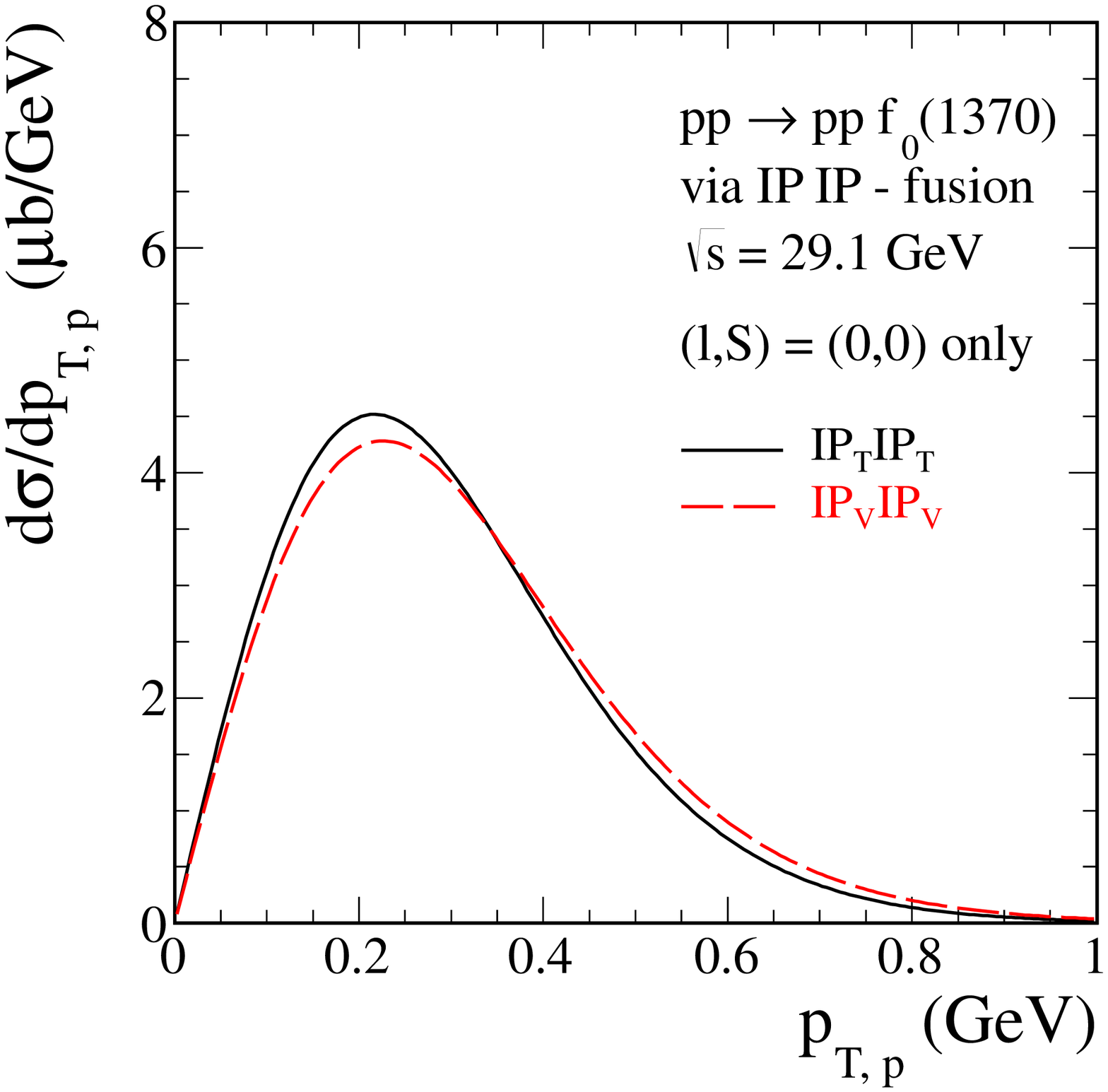}\\
\includegraphics[width = 0.32\textwidth]{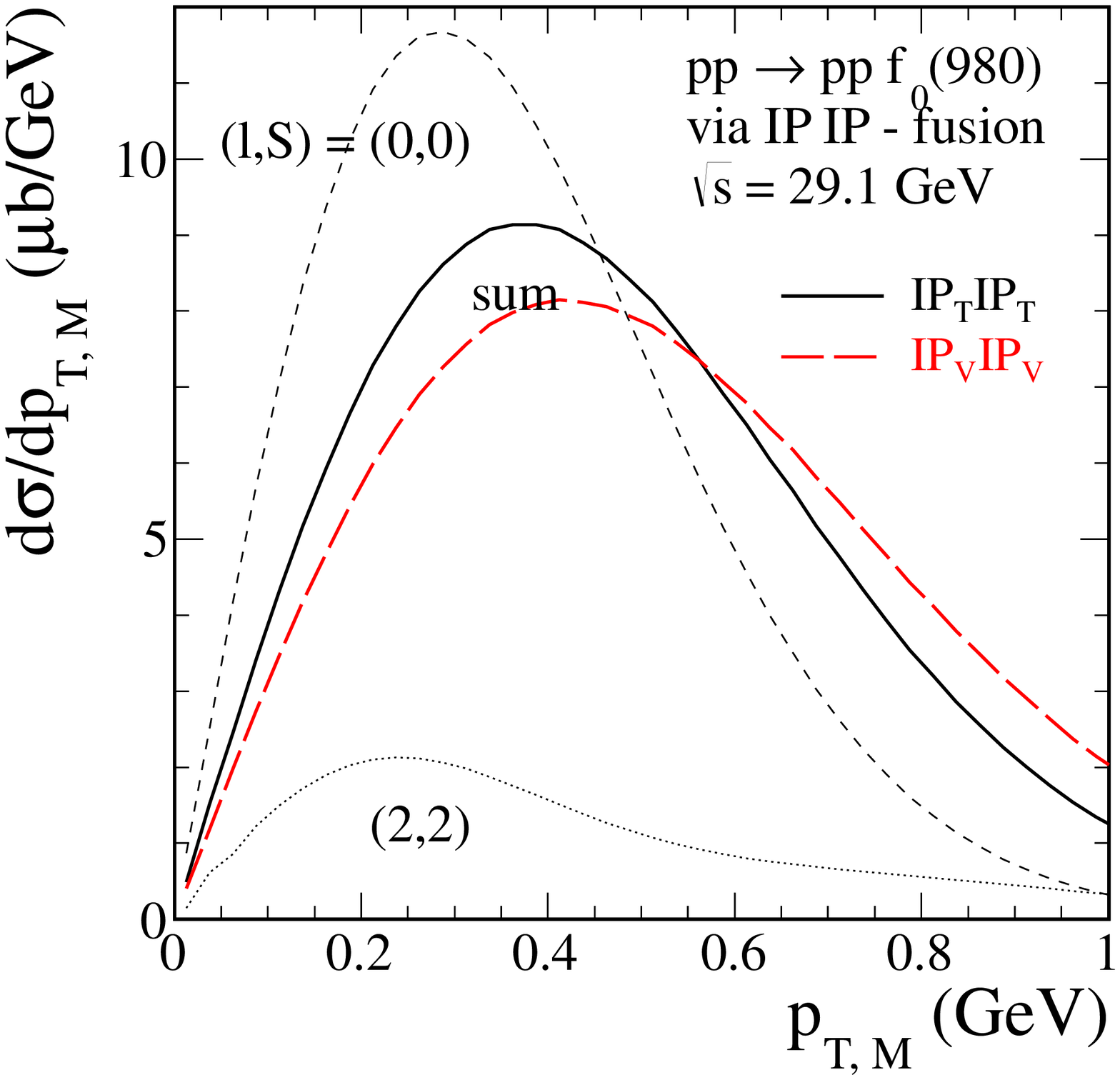}
\includegraphics[width = 0.32\textwidth]{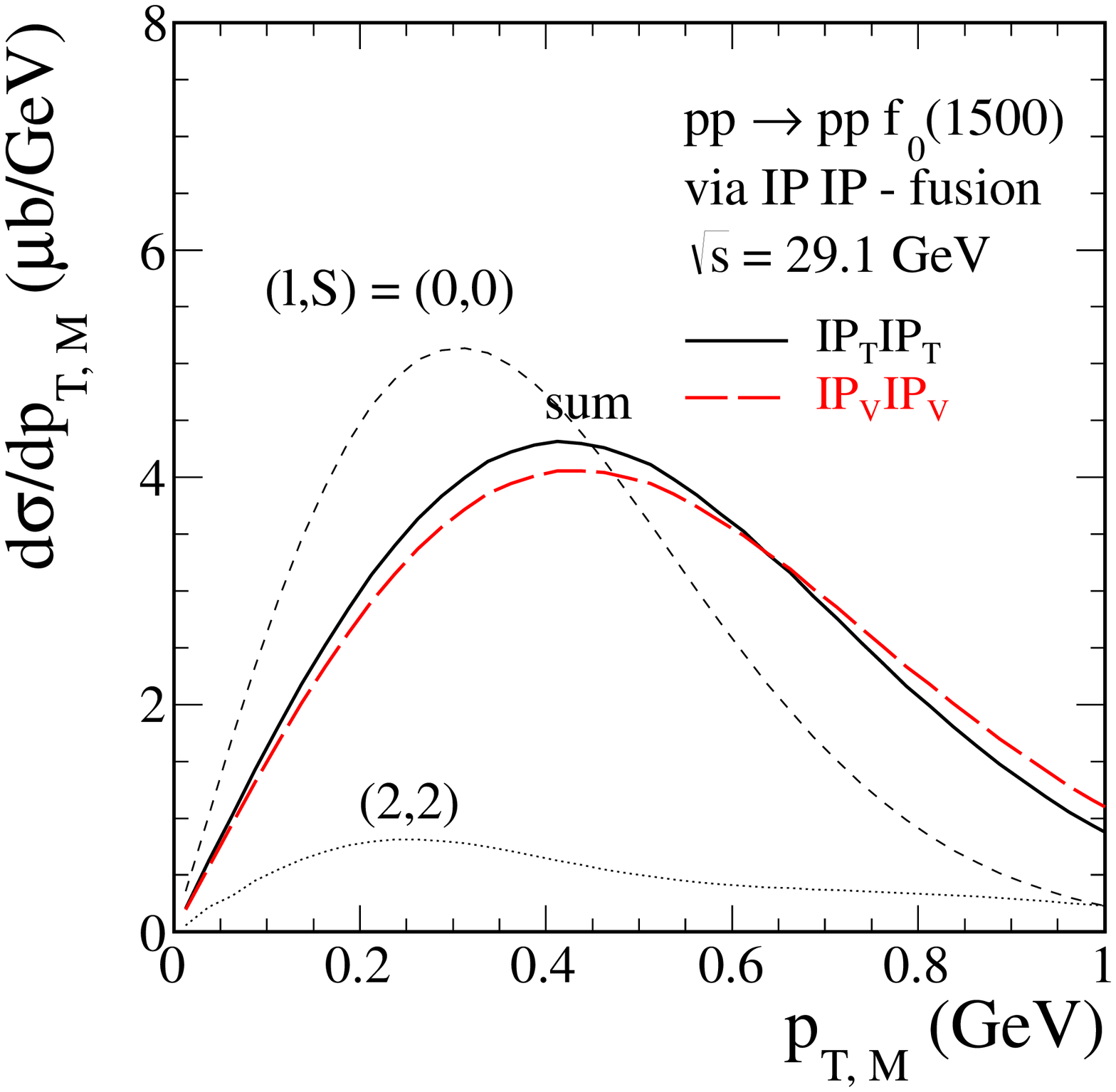}
\includegraphics[width = 0.32\textwidth]{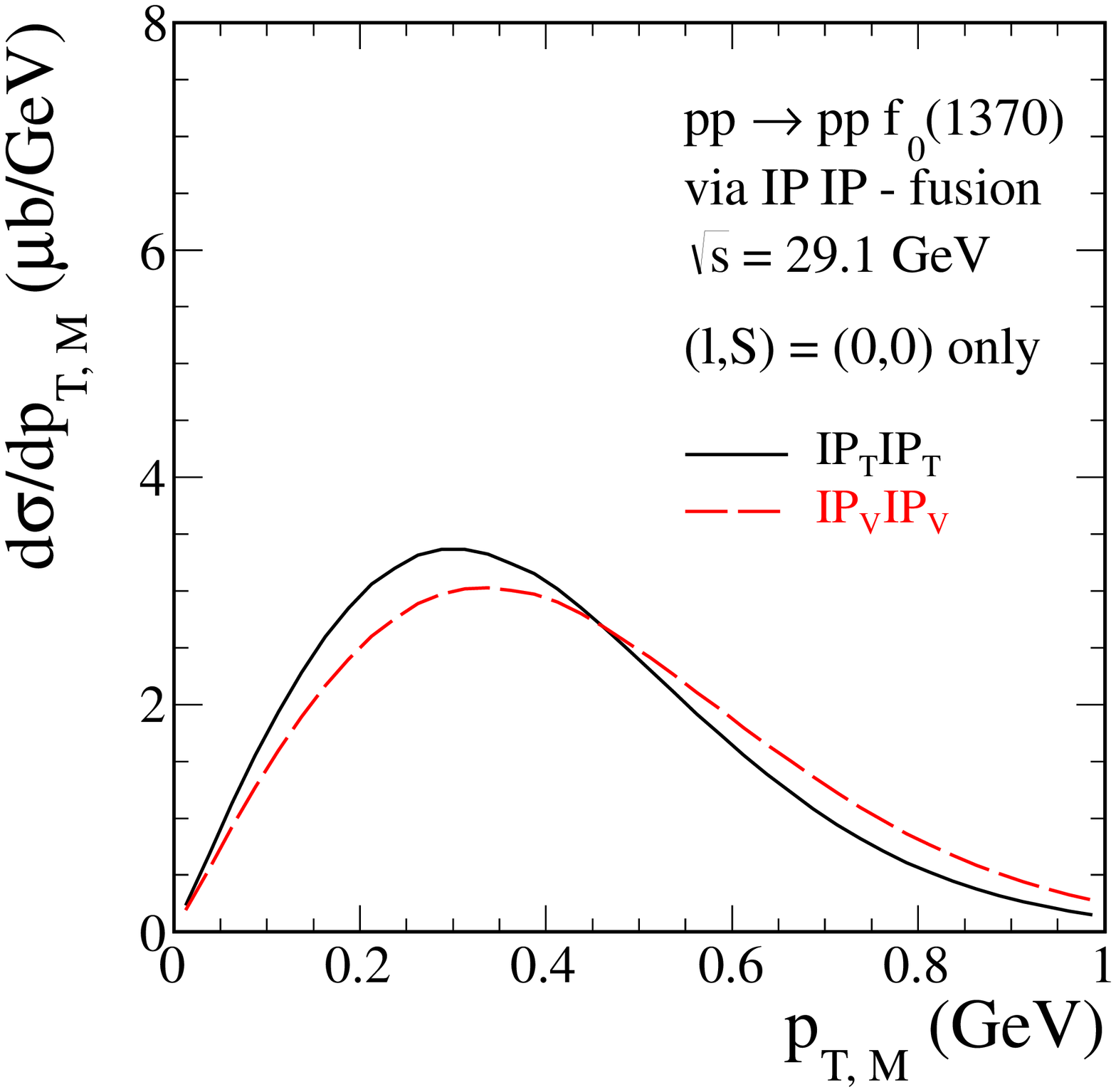}\\
\includegraphics[width = 0.32\textwidth]{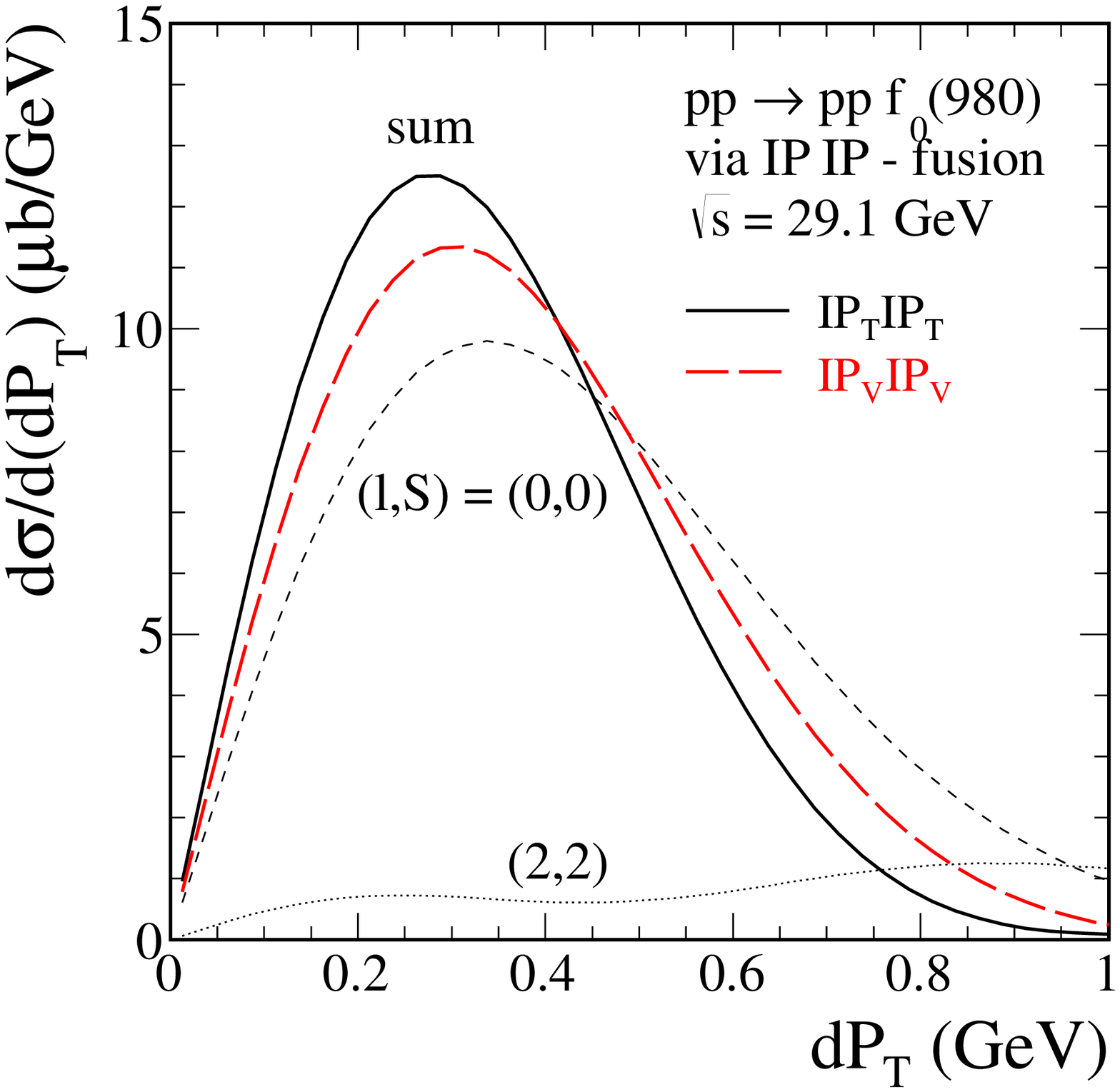}
\includegraphics[width = 0.32\textwidth]{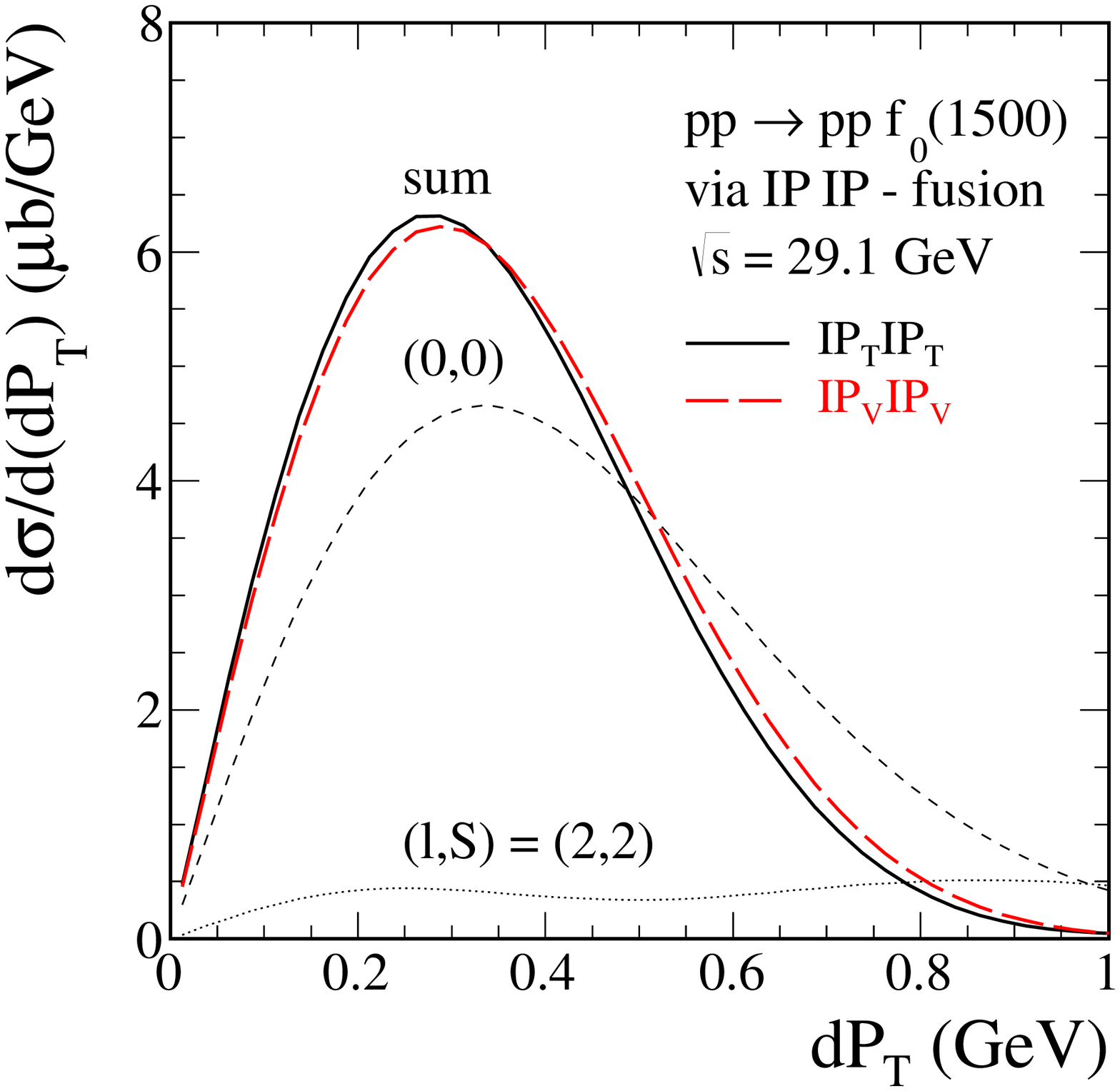}
\includegraphics[width = 0.32\textwidth]{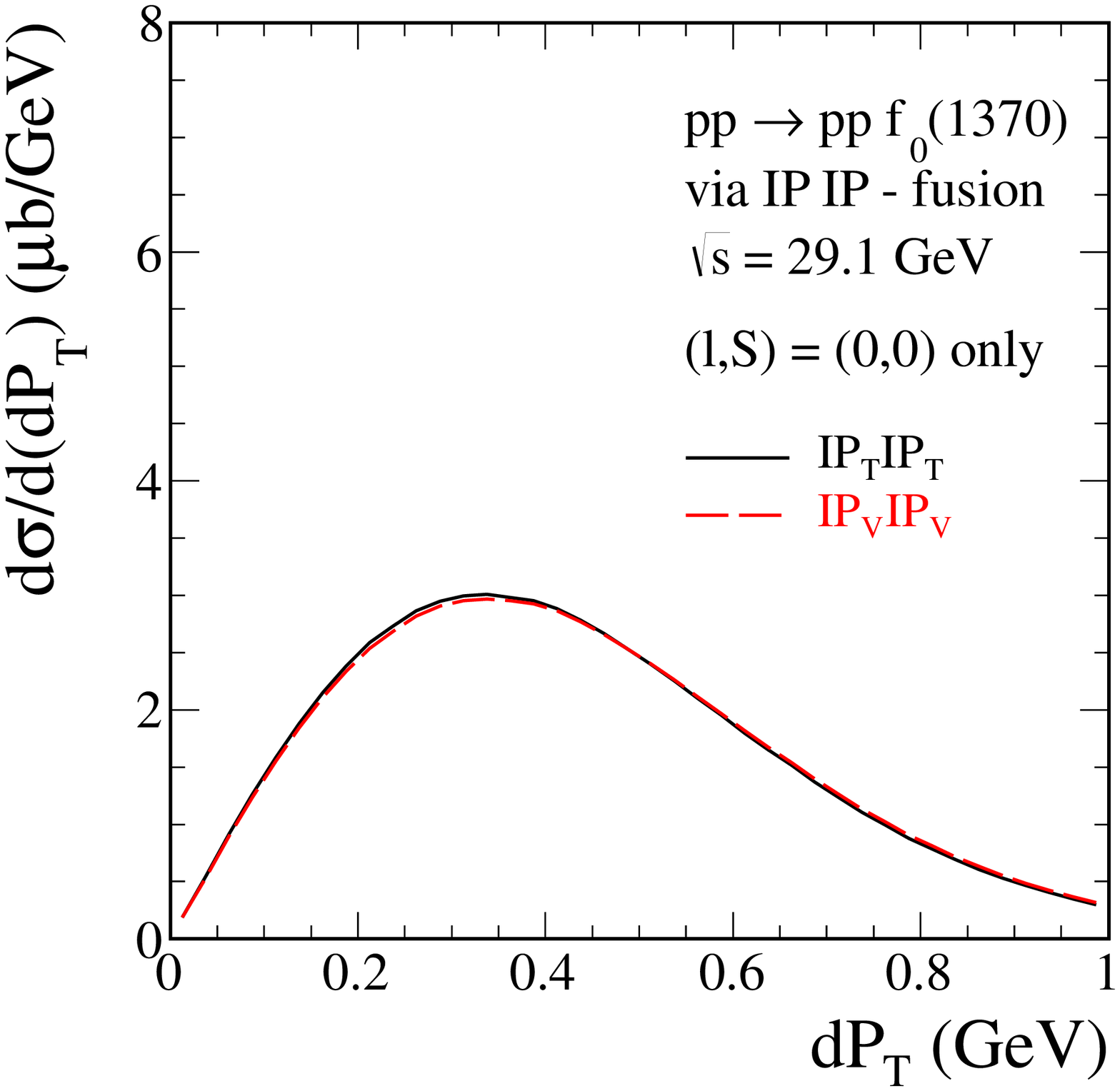}
  \caption{\label{fig:dsig_ds2}
  \small
The different differential observables for the central exclusive production of $f_{0}(980)$ (left panel),
$f_{0}(1500)$ (central panel) and $f_{0}(1370)$ (right panel) mesons
by the fusion of two tensor (solid line) and vector (long-dashed line) pomerons at $\sqrt{s} = 29.1$~GeV.
The results have been normalized to the mean value 
of the total cross sections given in Table~\ref{tab:mesons}.
For the tensorial pomeron case we show the individual spin contributions to the cross sections with 
$(l,S) = (0,0)$ (short-dashed line) and $(l,S) = (2,2)$ (dotted line).
}
\end{figure}

In Fig.~\ref{fig:dsig_ds2} we present different differential observables
(in proton and meson transverse momenta 
as well as in the so-called ``glueball filter variable'' $dP_{\perp}$) 
at $\sqrt{s} = 29.1$~GeV
for the central exclusive production of three different scalar mesons,
$f_{0}(980)$ (left panel), $f_{0}(1500)$ (middle panel) and $f_0(1370)$ (right panel).
As explained in the figure caption we show results for both tensor (solid line) and 
vector (long-dashed line) pomerons 
as well as the individual spin $(l,S)$ contributions for tensor pomeron only.
The coherent sum of the $(0,0)$ and $(2,2)$ components is shifted to smaller $dP_{\perp}$
with respect to the $(0,0)$ component alone.
This seems to be qualitatively consistent with the WA102 Collaboration result
presented in Table~2 of Ref.~\cite{kirk00}.
Further studies how different scalar mesons
are produced as a function of $dP_{\perp}$
will be presented in the next section; see discussion of Fig.~\ref{fig:4a}.
For meson transverse momentum one can see a shift in the opposite direction.

\begin{figure}[!ht]
\includegraphics[width = 0.32\textwidth]{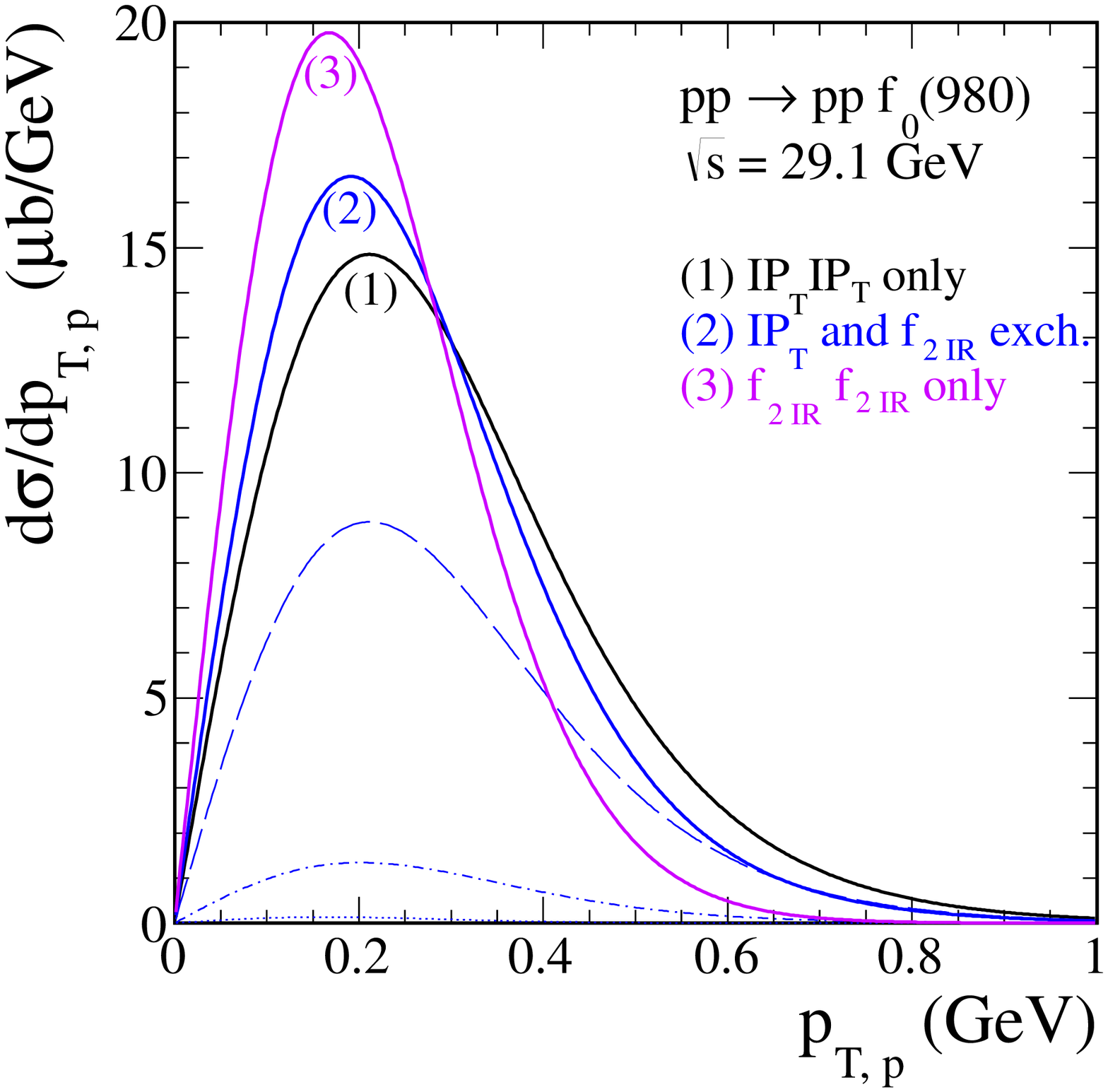}
\includegraphics[width = 0.32\textwidth]{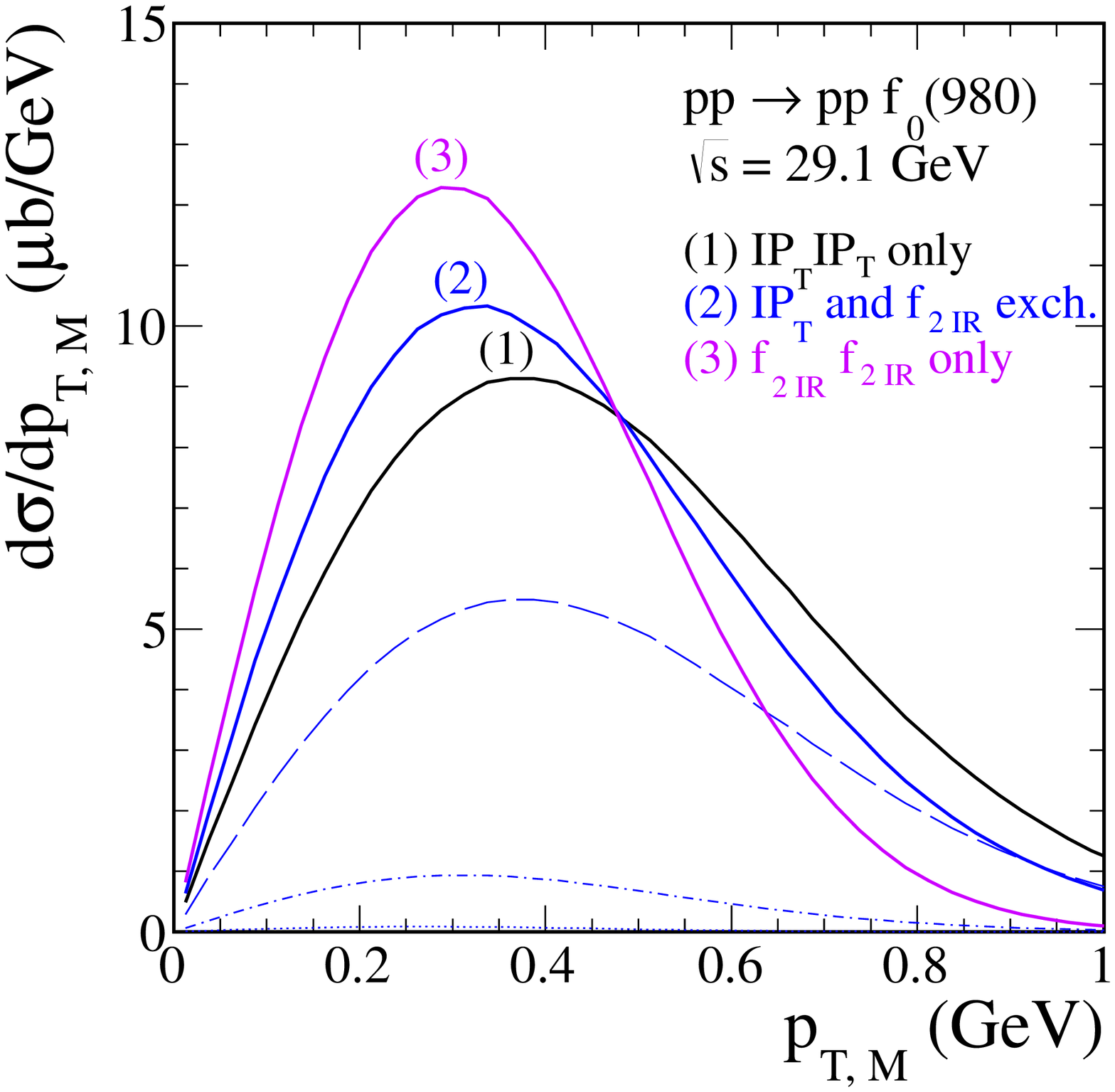}
\includegraphics[width = 0.32\textwidth]{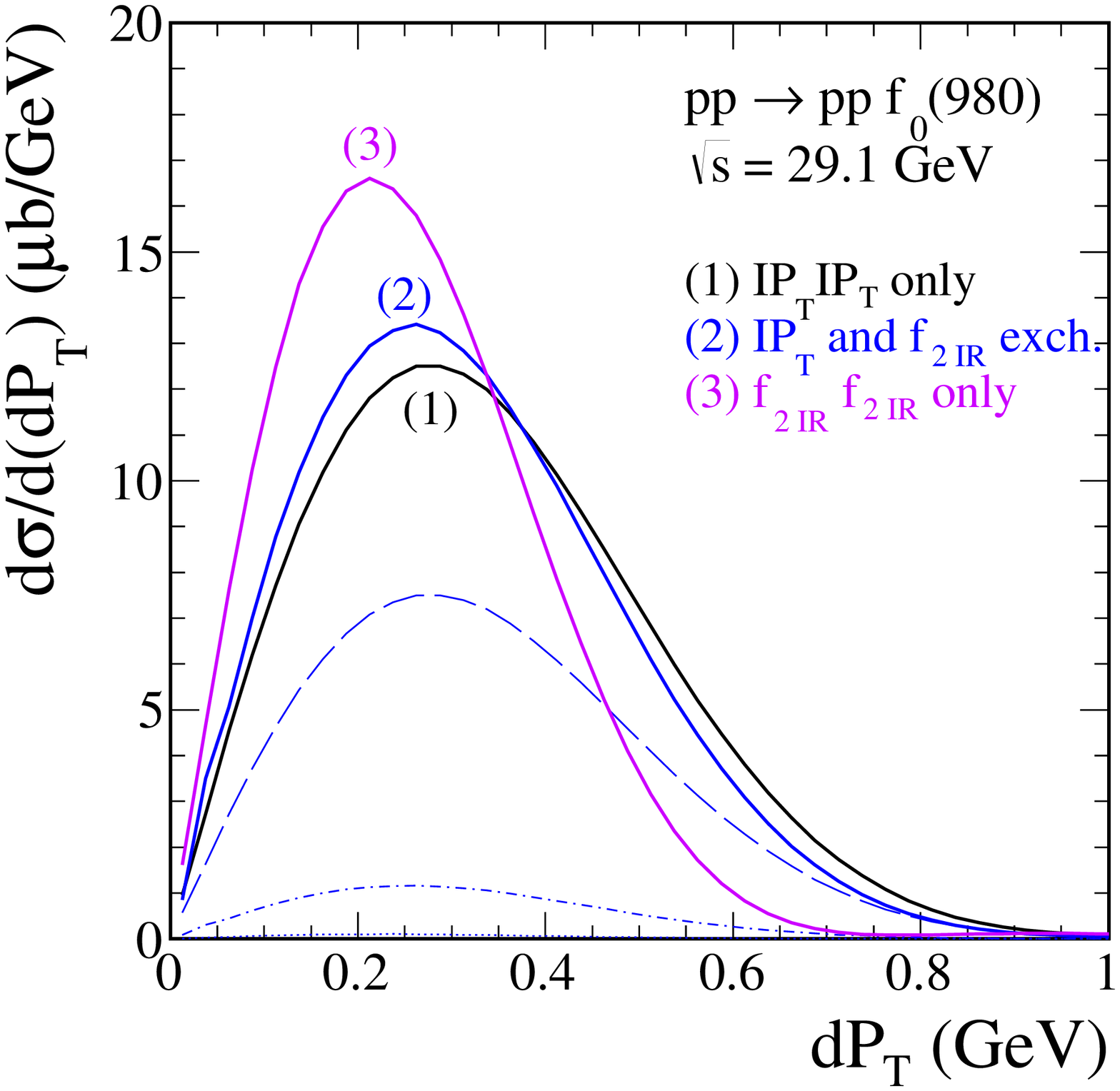}
  \caption{\label{fig:dsig_ds3}
  \small
Different differential observables 
for the central exclusive production of $f_{0}(980)$ meson at $\sqrt{s} = 29.1$~GeV.
The results have been normalized to the mean value 
of the total cross section given in Table~\ref{tab:mesons}.
The black solid line (1) corresponds to the $I\!\!P_{T} I\!\!P_{T}$-fusion,
the blue solid line (2) to the results with tensor pomeron and $f_{2 I\!\!R}$ exchanges
(the long-dashed, dash-dotted and dotted lines present 
the $I\!\!P_{T} I\!\!P_{T}$, $I\!\!P_{T} f_{2 I\!\!R}$, 
and $f_{2 I\!\!R} f_{2 I\!\!R}$ contributions, respectively),
and the violet solid line (3) presents the $f_{2 I\!\!R} f_{2 I\!\!R}$-fusion alone
normalized to the integrated cross section from \cite{kirk00}; see Table~\ref{tab:mesons}.
}
\end{figure}

In Fig.~\ref{fig:dsig_ds3} we show distributions in transverse momenta of protons, mesons
and in the $dP_{\perp}$ for the $f_{0}(980)$ meson production.
The three tensorial scenarios of meson production, as in Fig.~\ref{fig:dsig_dt}~(b), are presented.
One conclusion is that the $f_{2 I\!\!R} f_{2 I\!\!R}$ contribution,
indicated in the figure as curve (3),
does not give the expected $dP_{\perp}$ distribution as in Table~2 of Ref.~\cite{kirk00}.

\begin{figure}[!ht]
\includegraphics[width = 0.32\textwidth]{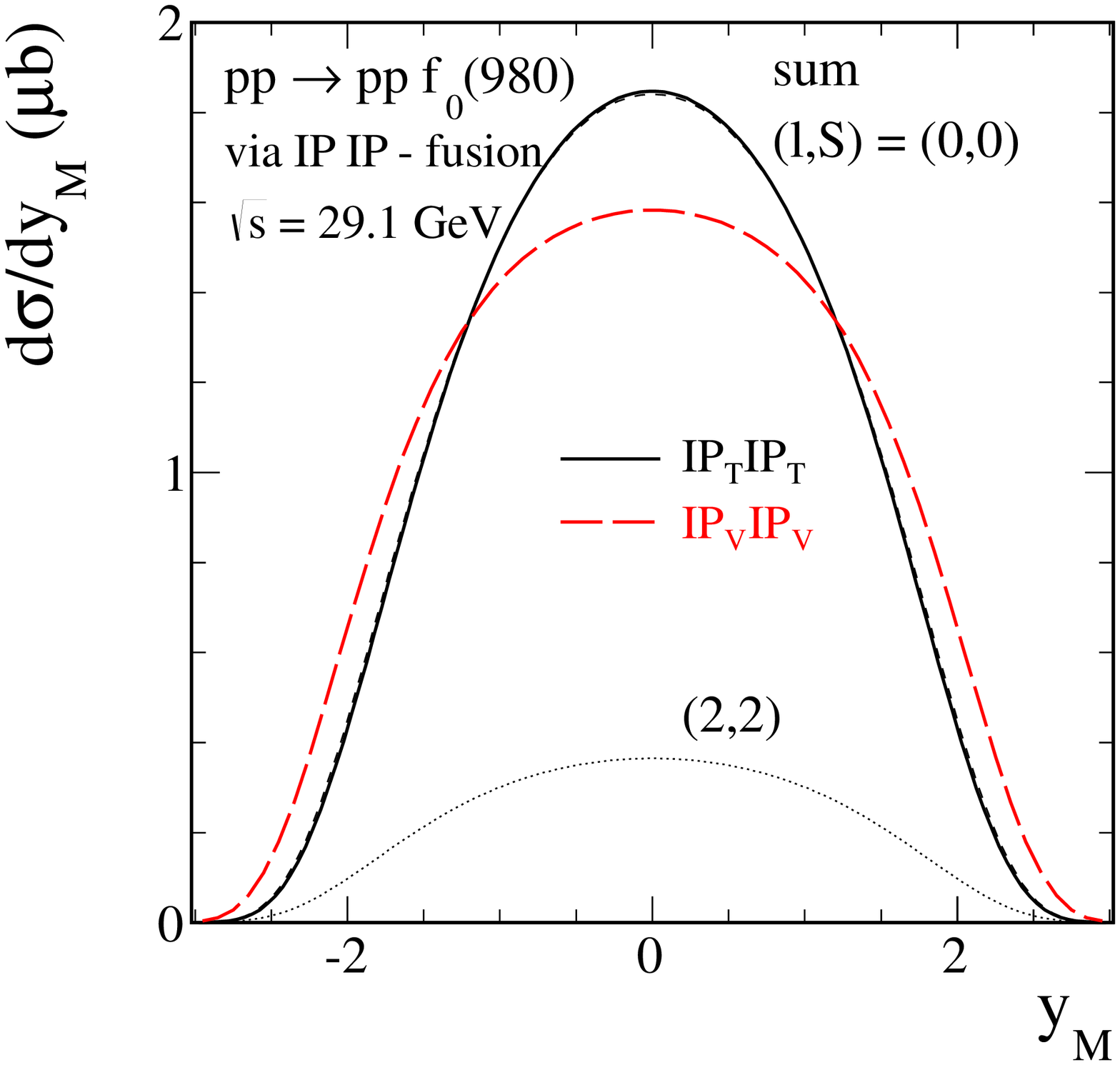}
\includegraphics[width = 0.32\textwidth]{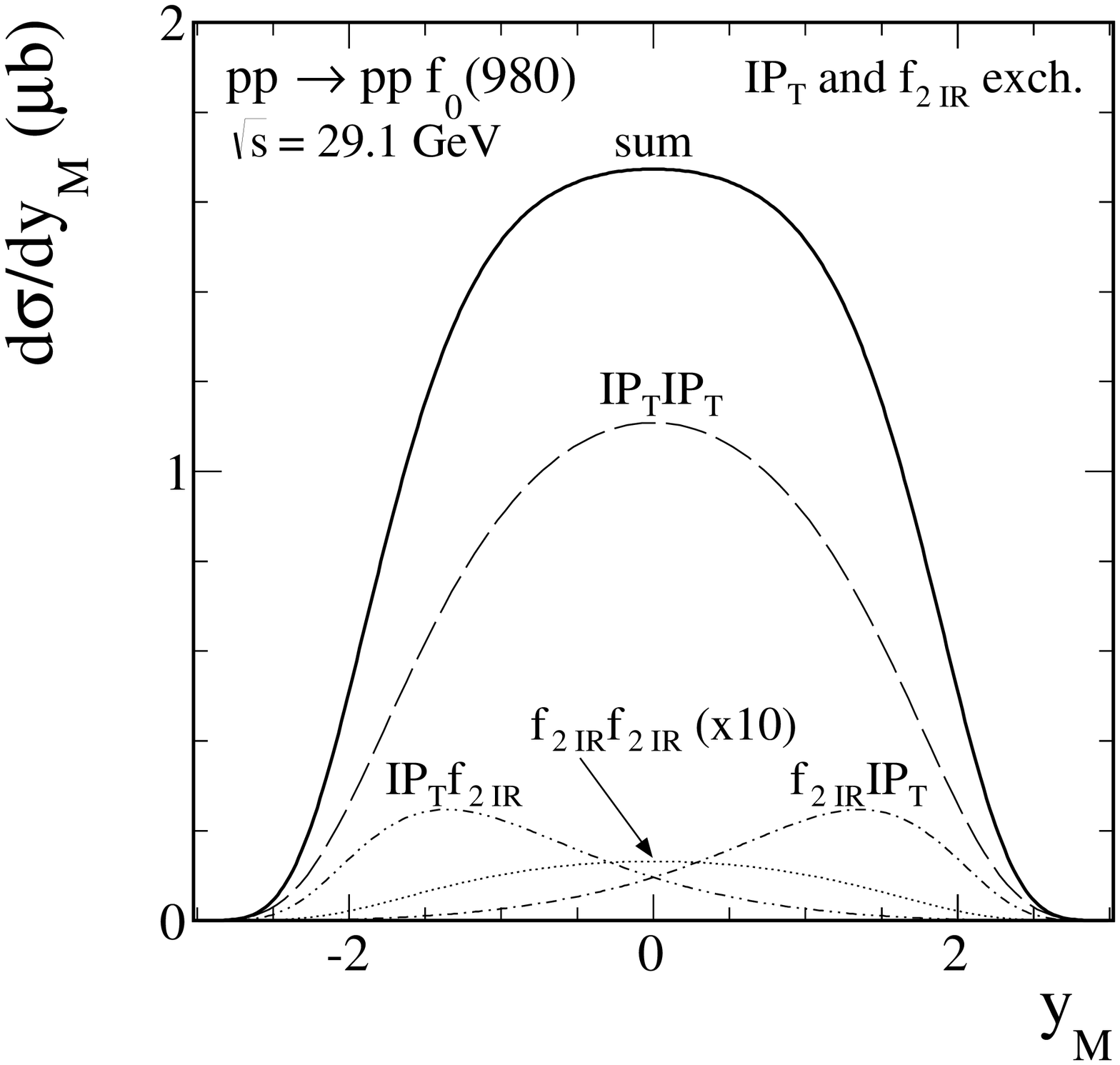}
\includegraphics[width = 0.32\textwidth]{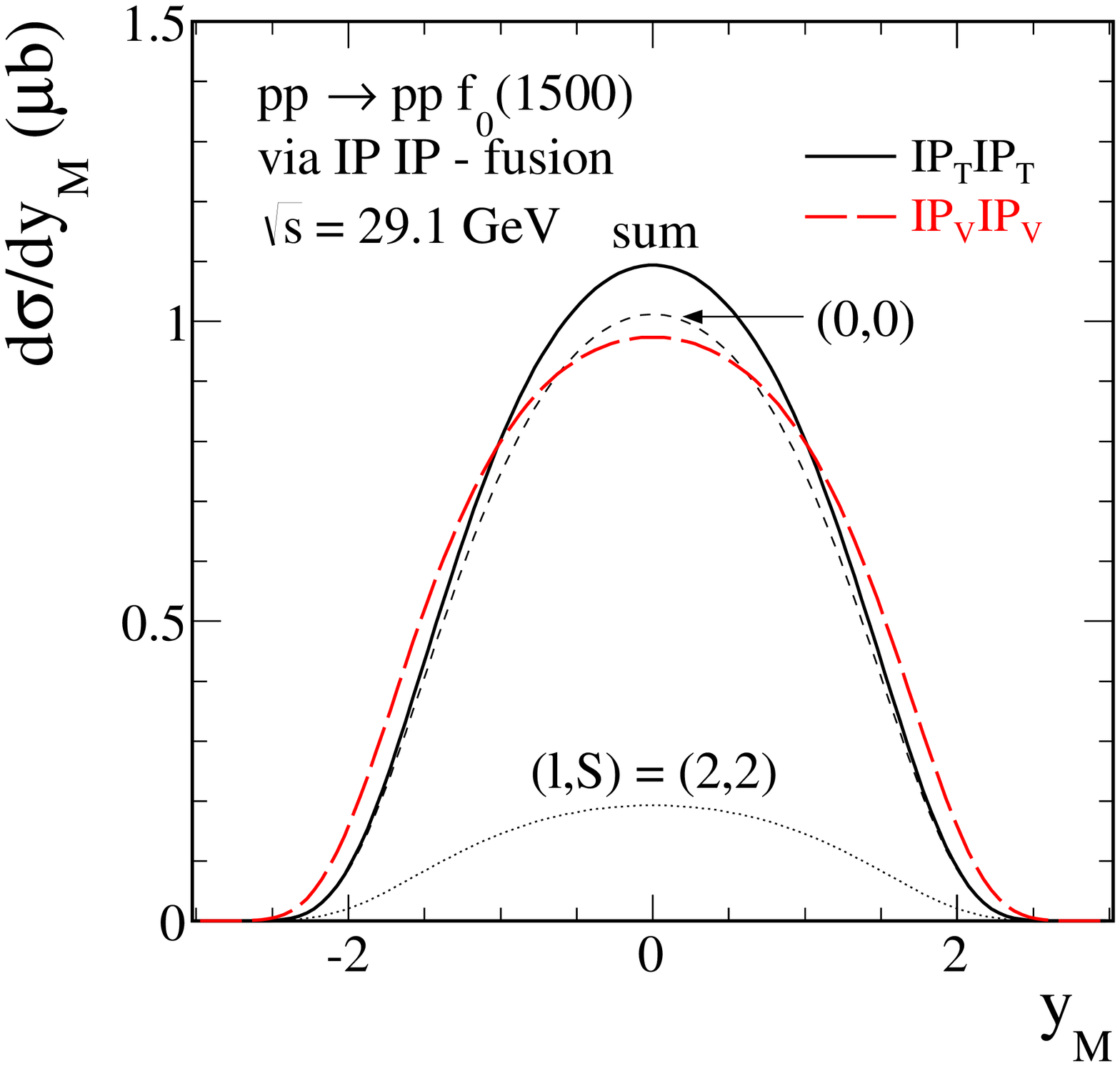}\\
\includegraphics[width = 0.32\textwidth]{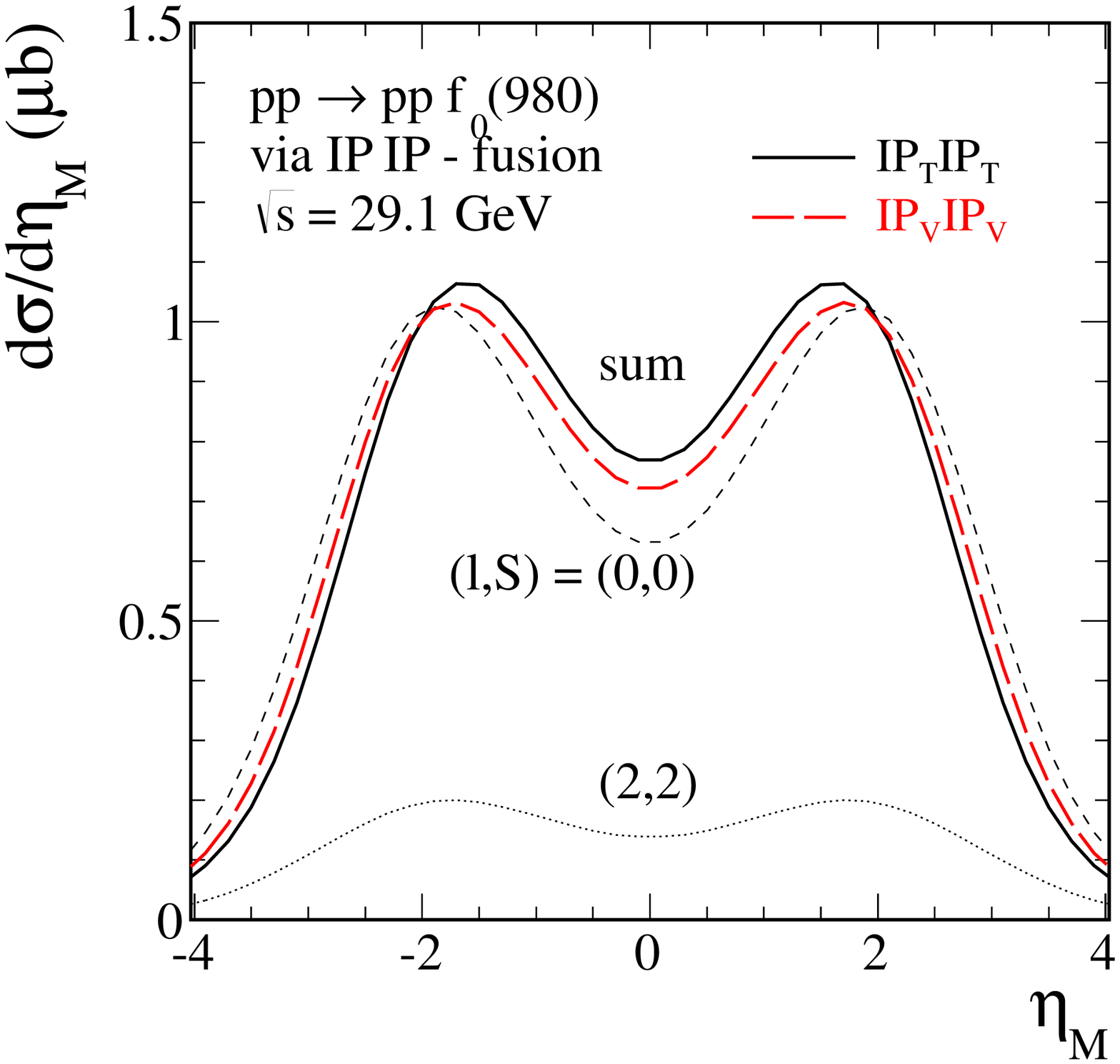}
\includegraphics[width = 0.32\textwidth]{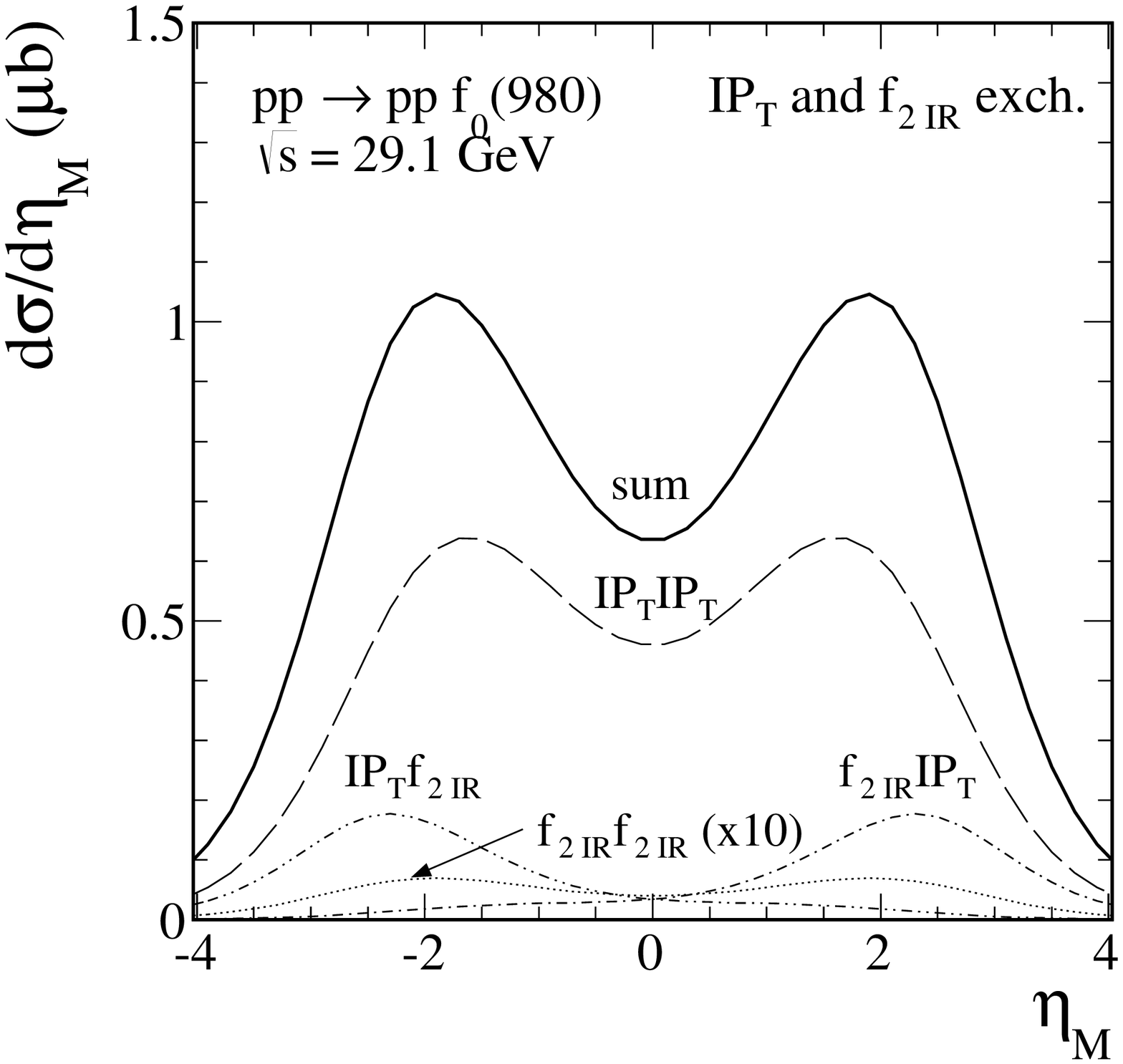}
\includegraphics[width = 0.32\textwidth]{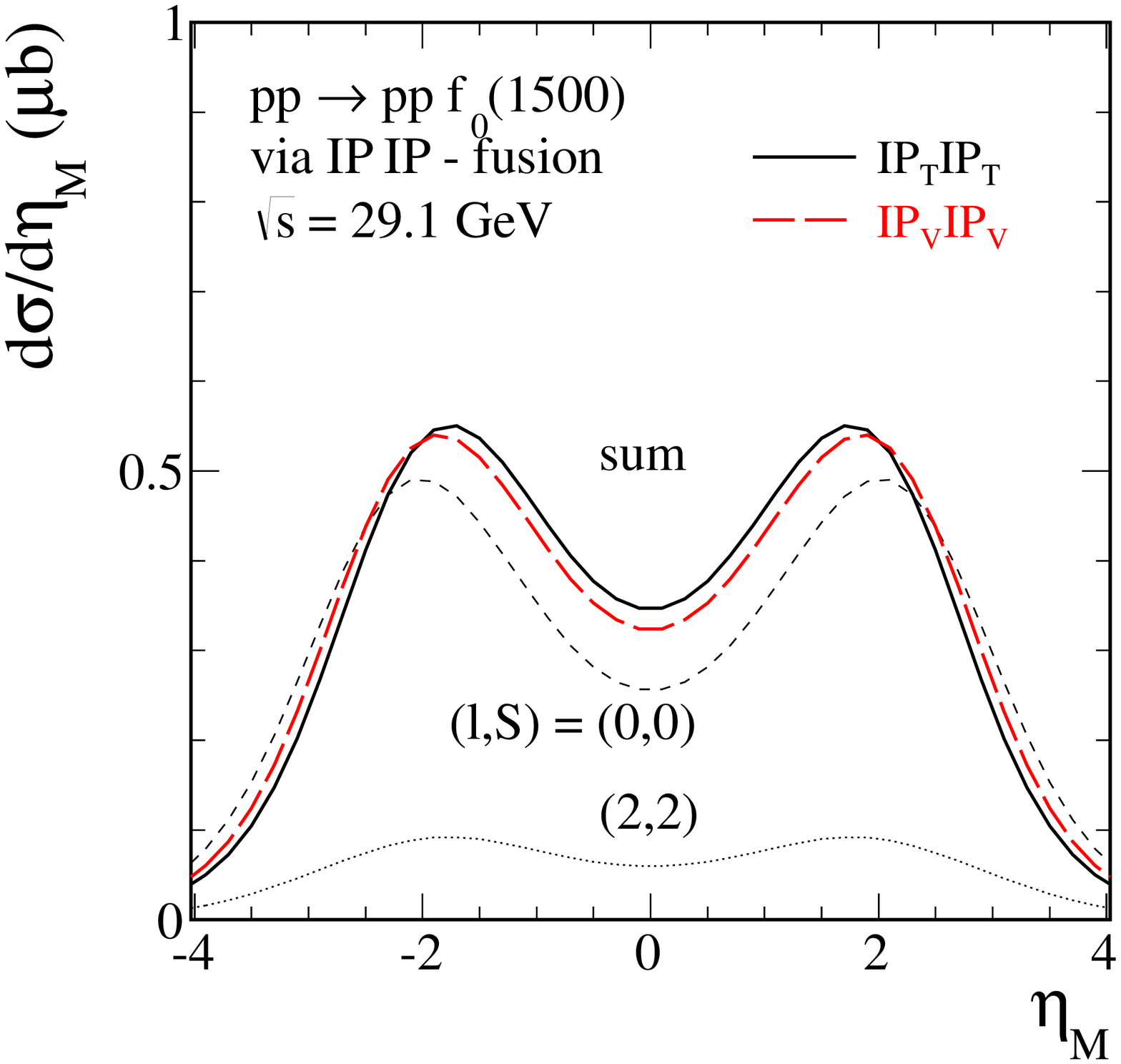}
  \caption{\label{fig:dsig_dy}
  \small
Rapidity and pseudorapidity distributions of $f_{0}(980)$
and $f_{0}(1500)$ produced by the fusion of two tensor (solid line)
and vector (long-dashed line) pomerons at $\sqrt{s} = 29.1$~GeV.
The results have been normalized to the mean value 
of the total cross sections given in Table~\ref{tab:mesons}.
For tensorial pomeron the individual contributions of $(l,S) = (0,0)$ (short-dashed line),
$(l,S) = (2,2)$ (dotted line), and their coherent sum (solid line) are shown.
In the center panels we show the results for the $f_{0}(980)$ meson production 
with the tensorial $f_{2 I\!\!R}$ contributions included.
}
\end{figure}

In Fig.~\ref{fig:dsig_dy} we show distributions 
in rapidity $\mathrm{y}_{M}$ of $f_{0}(980)$ and $f_{0}(1500)$ mesons
and the corresponding distributions in pseudorapidity $\eta_{M}$ at $\sqrt{s} = 29.1$~GeV. 
In these observables both $(l,S)$ components and their coherent sum have similar shape. 
The minimum in the pseudorapidity distributions
can be understood as a kinematic effect; 
see Appendix~\ref{section:Kinematic_Relations}.
In addition, for the $f_{0}(980)$ meson production 
we have included the tensorial $f_{2 I\!\!R}$ contributions; see the central panels.
The $I\!\!P_{T} I\!\!P_{T}$ and the $f_{2 I\!\!R} f_{2 I\!\!R}$ exchanges 
contribute at midrapidity of the meson, while
the $I\!\!P_{T} f_{2 I\!\!R}$ and $f_{2 I\!\!R} I\!\!P_{T}$ exchanges
at backward and forward meson rapidity, respectively.
The interference of these components in the amplitude produces
enhancements of the cross section at large $|\mathrm{y}_{M}|$ and $|\eta_{M}|$.

\begin{figure}[!ht]
\includegraphics[width = 0.45\textwidth]{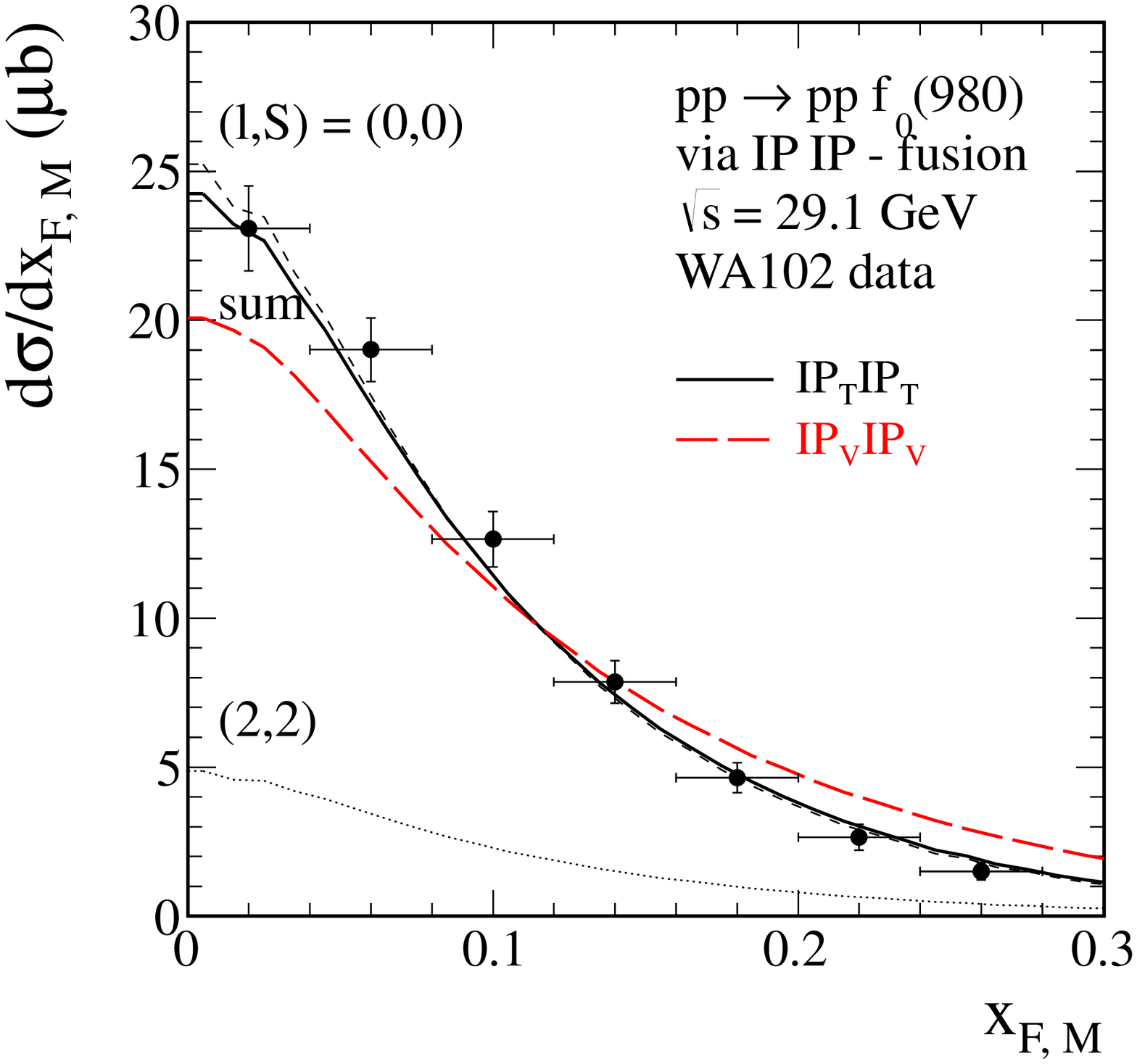}
\includegraphics[width = 0.45\textwidth]{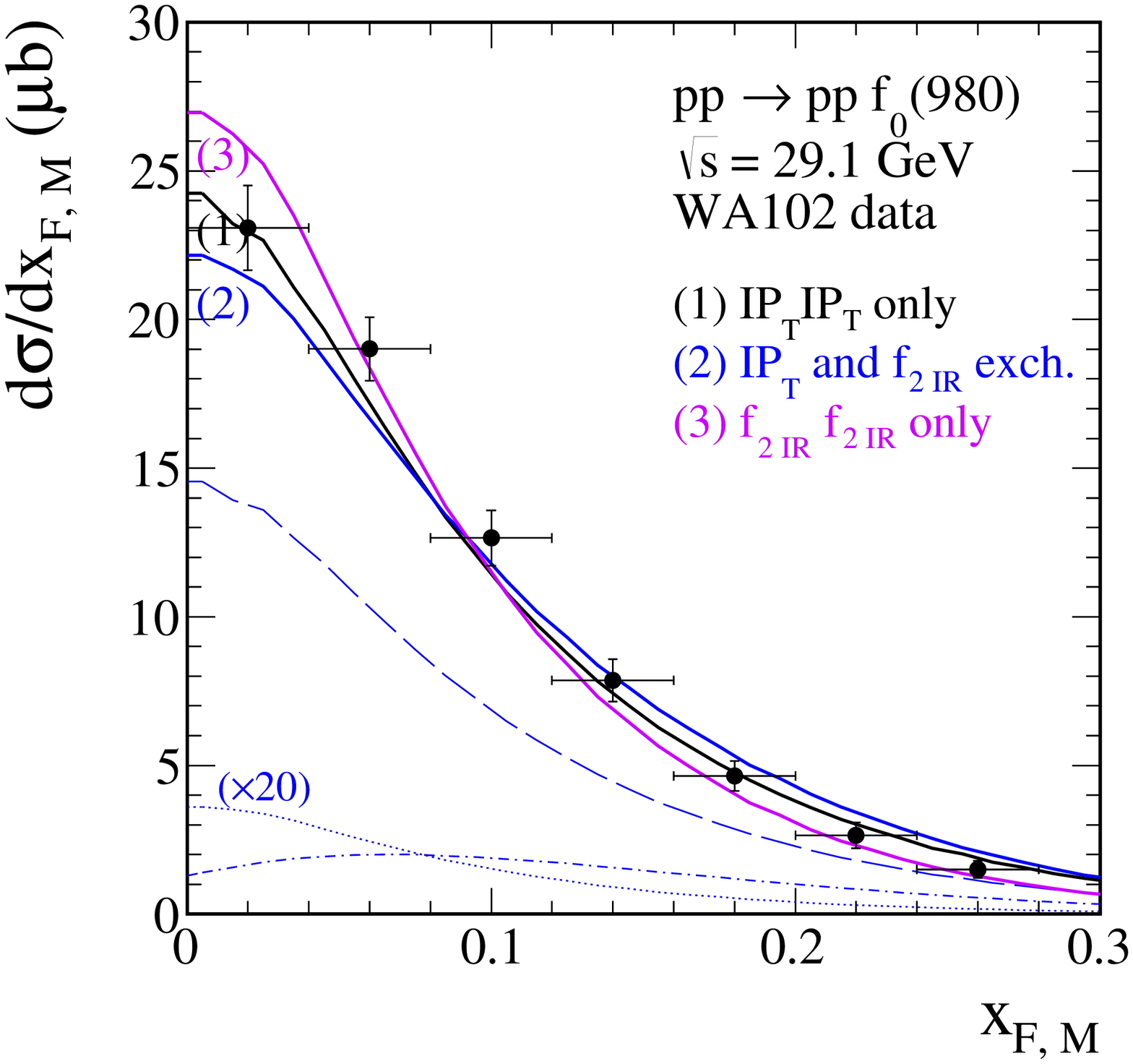}
  \caption{\label{fig:dsig_dxf}
  \small
The $x_{F,M}$ distribution for the central exclusive $f_{0}(980)$
meson production at $\sqrt{s} = 29.1$~GeV.
The WA102 experimental data points from \cite{CloseKirk2000}
have been normalized 
to the mean values of the total cross section given in Table~\ref{tab:mesons}.
In the left panel we show the results obtained 
by the fusion of two tensor pomerons. 
In addition, the individual $(l,S) = (0,0)$ and $(2,2)$ contributions denoted
by the short-dashed and dotted lines, respectively, are presented.
In the right panel the black solid line (1) corresponds to the $I\!\!P_{T} I\!\!P_{T}$-fusion,
the blue solid line (2) to the results with tensor pomeron and $f_{2 I\!\!R}$ exchanges
(the long-dashed, dash-dotted and dotted lines present 
the $I\!\!P_{T} I\!\!P_{T}$, $I\!\!P_{T} f_{2 I\!\!R}$ 
and $f_{2 I\!\!R} f_{2 I\!\!R}$ (enlarged by a factor $20$)
contributions, respectively),
and the violet solid line (3) presents the $f_{2 I\!\!R} f_{2 I\!\!R}$-fusion contribution
alone normalized to the mean value of the total cross section given in Table~\ref{tab:mesons}.
}
\end{figure}

In Fig.~\ref{fig:dsig_dxf} we show the distribution in Feynman-$x_{F}$
for the central exclusive $f_{0}(980)$ meson (the only available experimentally) 
production at $\sqrt{s} = 29.1$~GeV. 
The good agreement of the $I\!\!P_{T} I\!\!P_{T}$-fusion result 
(see the solid line in the left panel) with the WA102 data suggests that
for the tensor pomeron model the pomeron-reggeon and reggeon-reggeon contributions are small.

\begin{figure}[!ht]
(a)\includegraphics[width = 0.43\textwidth]{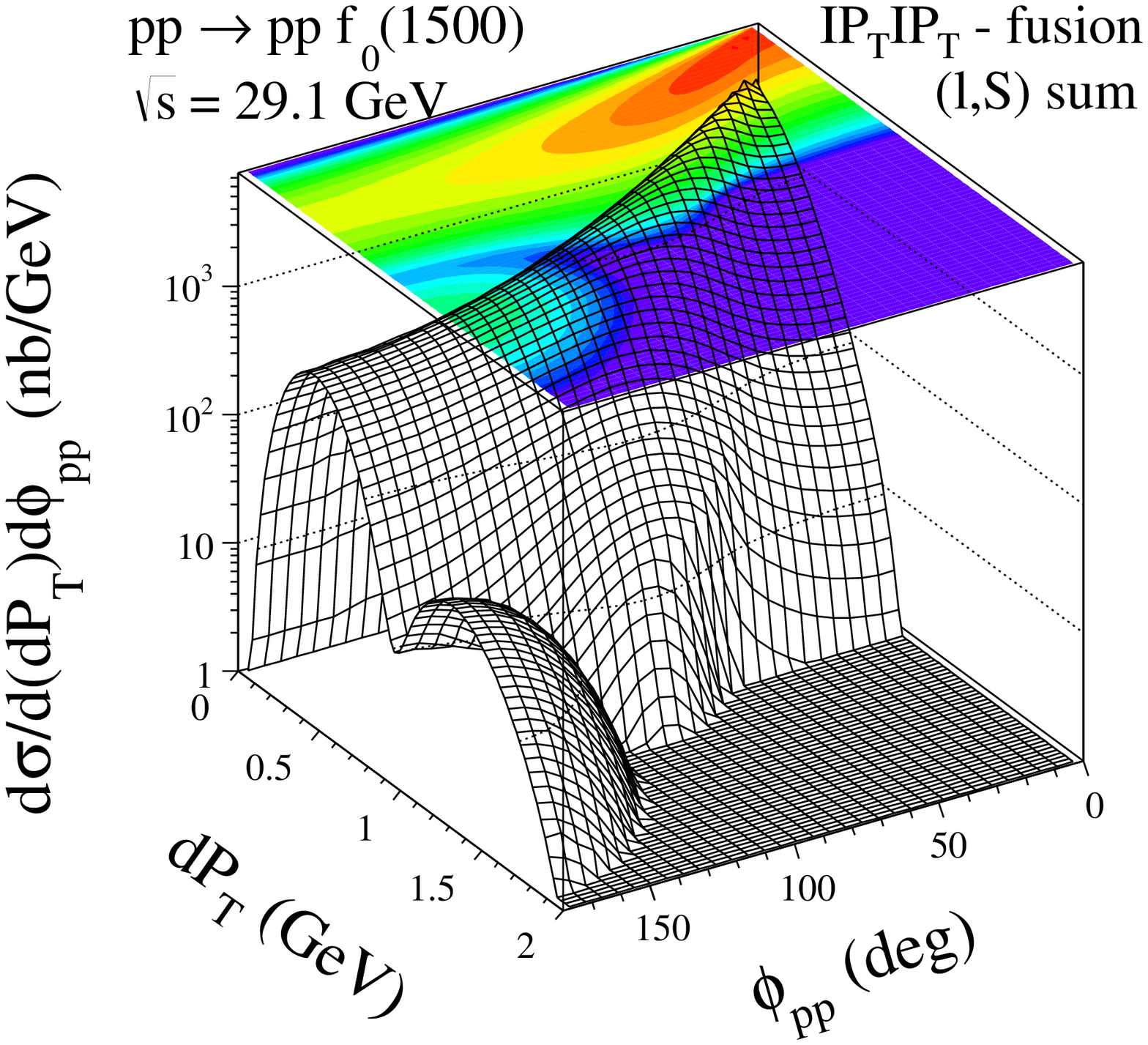}
(b)\includegraphics[width = 0.43\textwidth]{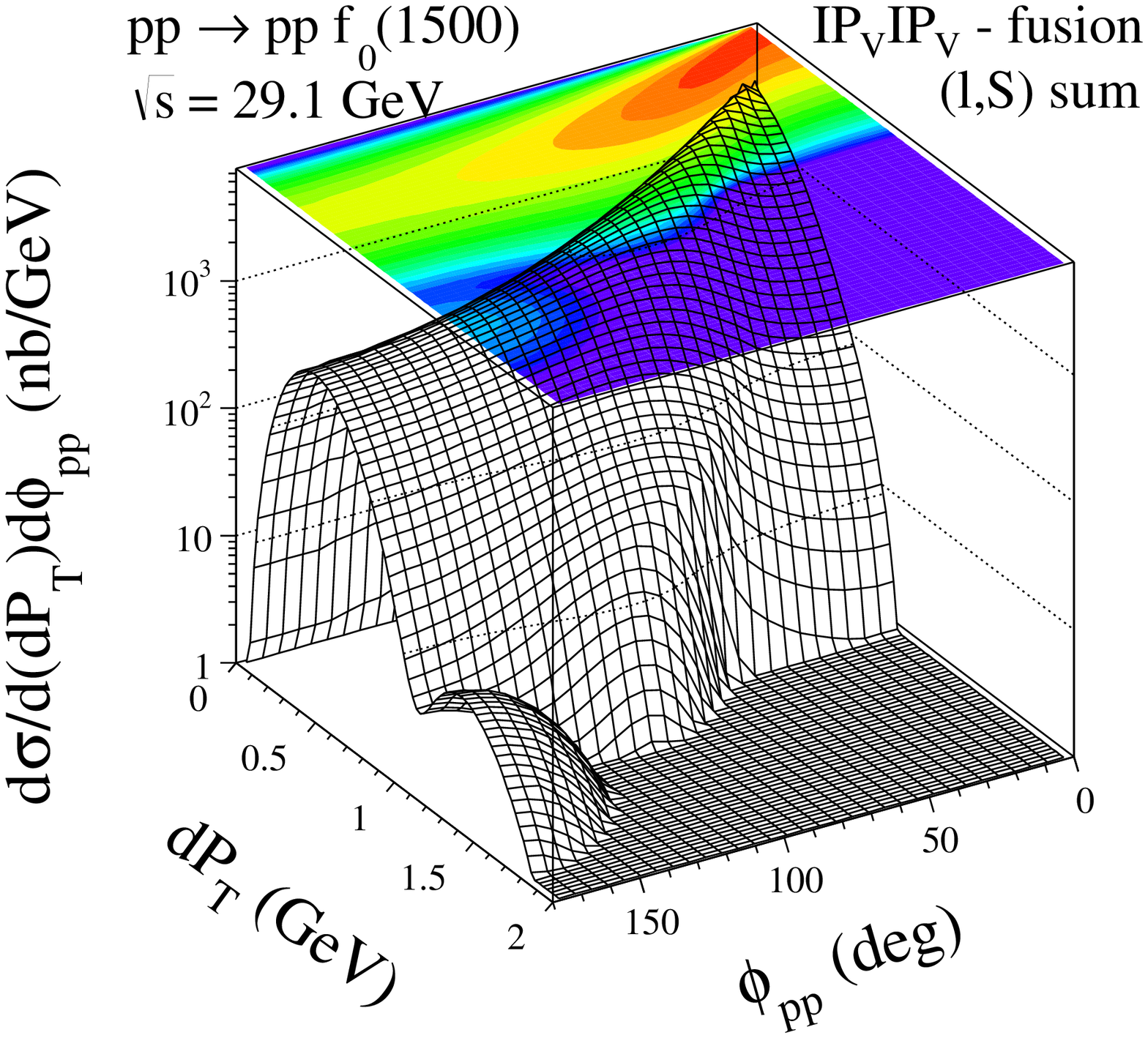}
(c)\includegraphics[width = 0.43\textwidth]{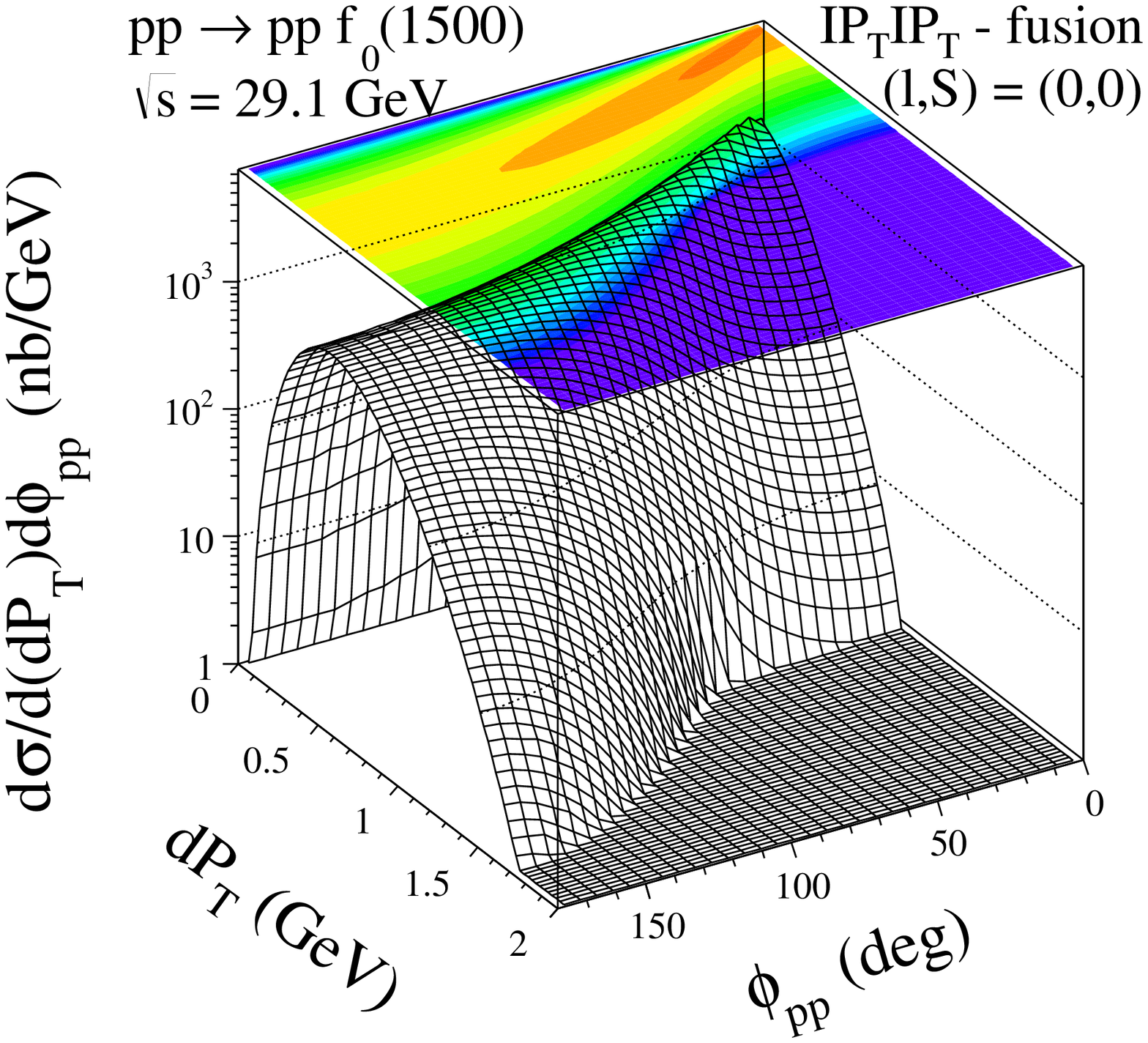}
(d)\includegraphics[width = 0.43\textwidth]{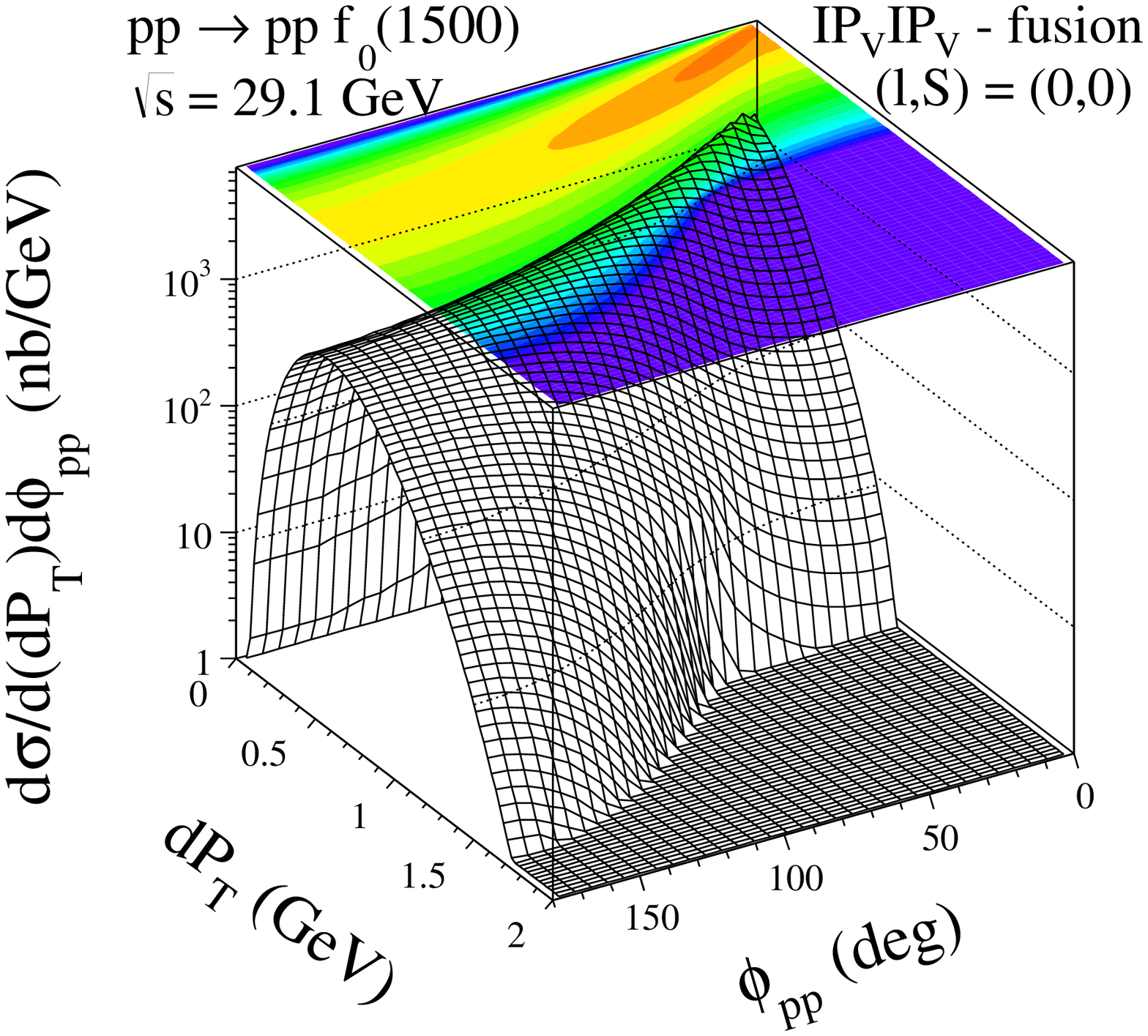}
(e)\includegraphics[width = 0.43\textwidth]{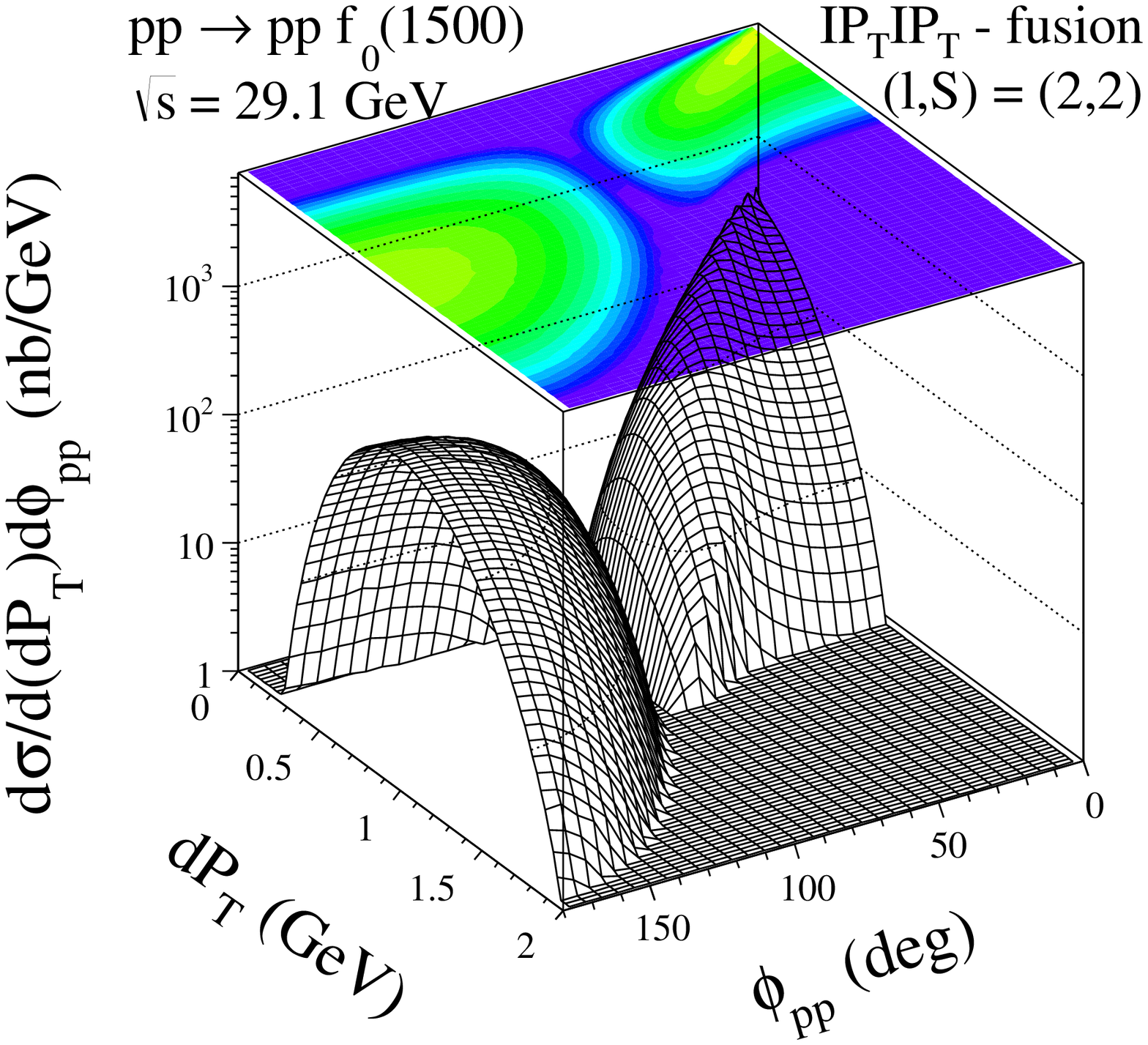}
(f)\includegraphics[width = 0.43\textwidth]{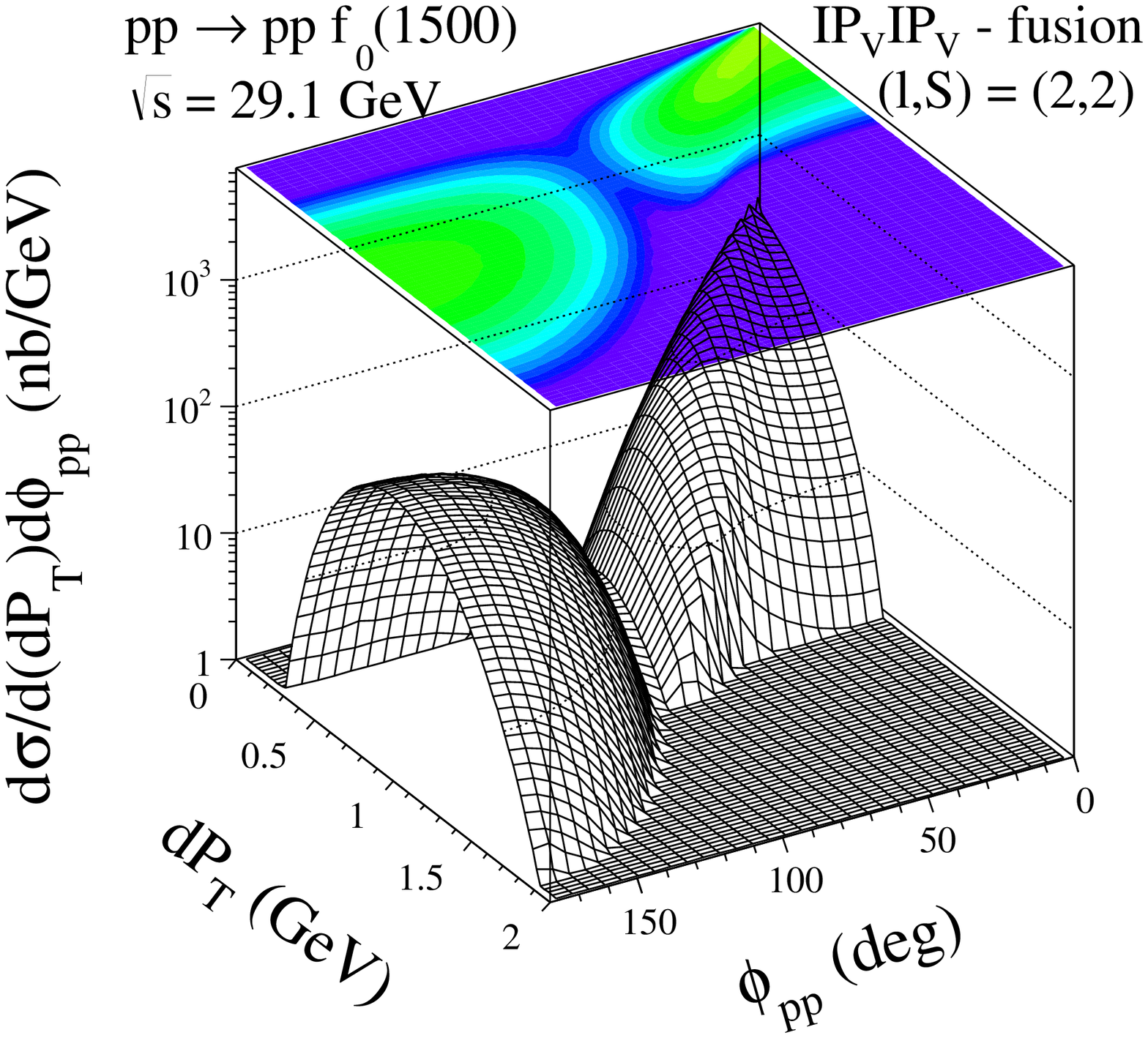}
  \caption{\label{fig:map_ptperpphi}
  \small
Distributions in ($dP_{\perp}, \phi_{pp}$)
for the central exclusive $f_{0}(1500)$ meson production
via the tensorial (left panels) and vectorial (right panels) pomeron exchanges at $\sqrt{s} = 29.1$~GeV.
The individual contributions of $(l,S) = (0,0)$ (panels (c) and (d))
and $(l,S) = (2,2)$ (panels (e) and (f)) are shown separately.
}
\end{figure}

Up to now we have observed some differences of the results 
for $(l,S) = (0,0)$ and $(2,2)$ couplings. 
The differences can be made better visible in two-dimensional distributions.
In Fig.~\ref{fig:map_ptperpphi} we show, as an example, two-dimensional
distributions in ($dP_{\perp}, \phi_{pp}$).
We show results for the fusion of two tensor (left panels) and two vector (right panels) pomerons.
In panels (a) and (b) we show the results for both $(l,S)$ components added coherently.
In panels (c, d) and (e, f) we show the individual components for $(l,S) = (0,0)$ and $(2,2)$, respectively. 
The distributions for both cases are very different. 
By comparing panels (a) and (b) to panels (c, e) and (d, f), respectively,
we see that the interference effects are rather large.

\subsection{Pseudoscalar meson production}
\label{subsection:Pseudocalar_mesons}
We turn now to the presentation of our results for pseudoscalar mesons. 
It is known that the $\eta$ and $\eta'$ mesons, 
the isoscalar members of the nonet of the lightest pseudoscalar mesons,
play an important role in the understanding of various aspects of nonperturbative effects of QCD;
see for instance \cite{DRB12}.
The $\eta'$-meson being dominantly a ($\alpha \ket{s \bar{s}} + \beta \ket{gg}$) state,
with presence of a sizeable gluonic component \cite{Kroll},
is particularly interesting for our study as here the pomeron-pomeron fusion 
should be the dominant mechanism in central production.
For central production of the $\eta$ meson the situation may be more complicated 
and requires consideration of additional $f_{2 I\!\!R}$ reggeon exchanges \cite{KMV99, KMV00}.
In contrast to $\eta'$ production, no good fit with 
(tensorial or vectorial) pomeron-pomeron
component only is possible for the $\eta$ meson production.
Therefore we have decided to include in addition $f_{2 I\!\!R} I\!\!P$, 
$I\!\!P f_{2 I\!\!R}$ and $f_{2 I\!\!R} f_{2 I\!\!R}$ contributions
into our analysis.
\footnote{In addition some other 'non-central' mechanisms are possible \cite{CLSS, LS13}.
One of them is diffractive excitation of $N(1535)$~$J^{P} = 1/2^{-}$
which decays into the $p + \eta$ channel with branching fraction of about 50~$\%$ \cite{PDG}.
The issue of diffractive excitation of nucleon resonances is so far not well understood
and goes beyond the scope of present paper.}
The corresponding coupling constants were roughly fitted to existing
experimental differential distributions (some specific details
will be given when discussing differential distributions); see Table \ref{tab:couplings_PS}.
We recall from the discussion in Section~\ref{subsection:Scalar_and_pseudoscalar_meson_production}
that for the tensorial pomeron two $I\!\!P I\!\!P \tilde{M}$ couplings,
$(l,S) = (1,1)$ and $(3,3)$, are possible.
For the vectorial pomeron we have only $(l,S) = (1,1)$.
As will be discussed below in addition to pomeron-pomeron fusion
the inclusion of secondary reggeons
is required for a simultaneous description of $d\sigma/d\phi_{pp}$,
$d\sigma/dt$ and $d\sigma/dx_{F}$ experimental data for the $\eta$ production.
\begin{table}
\caption{The values of the pomeron-pomeron-meson $\tilde{M}$ coupling constants
of the two models of the pomeron exchanges
which are approximately fitted to reproduce the correct normalization, see Table~\ref{tab:mesons}, and 
shapes of differential distributions of the WA102 experiment.
In addition, the cross sections (in $\mu$b) for the individual $(l,S)$ contributions 
at $\sqrt{s} = 29.1$~GeV are shown.}
\label{tab:couplings_PS}
\begin{tabular}{|c|c|c|c|c|c|c|}
\hline
Meson & Exchanges & $g_{I\!\!P I\!\!P \tilde{M}}'$ & $g_{I\!\!P I\!\!P \tilde{M}}''$ &
\multicolumn{3}{c|}{$\sigma$ ($\mu$b) at $\sqrt{s} = 29.1$~GeV}      \\
\cline{5-7}
$\tilde{M}$& & $(1,1)$ term &   $(3,3)$ term
& $(1,1)$ & $(3,3)$ & total
\\
\hline
$\eta$ & 
$I\!\!P_{T} I\!\!P_{T}$, 
$I\!\!P_{T} f_{2 I\!\!R}$, 
$f_{2 I\!\!R} I\!\!P_{T}$, 
$f_{2 I\!\!R} f_{2 I\!\!R}$ 
& 0.8, 2.45, 2.45, 2 & 1.4, 4.29, 4.29, 3.5 &
5.05 & 0.85 & 3.85
\\
 & $I\!\!P_{T} I\!\!P_{T}$ & 2 & 2.25 &
4.83 & 0.55 & 3.85
\\
 & $I\!\!P_{V} I\!\!P_{V}$ & 8.47 & - &
3.86 & -- & --
\\
\hline
$\eta'$ & $I\!\!P_{T} I\!\!P_{T}$ & 2.61 & 1.5 &
1.86 & 0.05 & 1.71 
\\
 & $I\!\!P_{V} I\!\!P_{V}$ & 6.08 & - &
1.72 & -- & --
\\
\hline
\end{tabular}
\end{table}

\begin{figure}[!ht]
(a)\includegraphics[width = 0.45\textwidth]{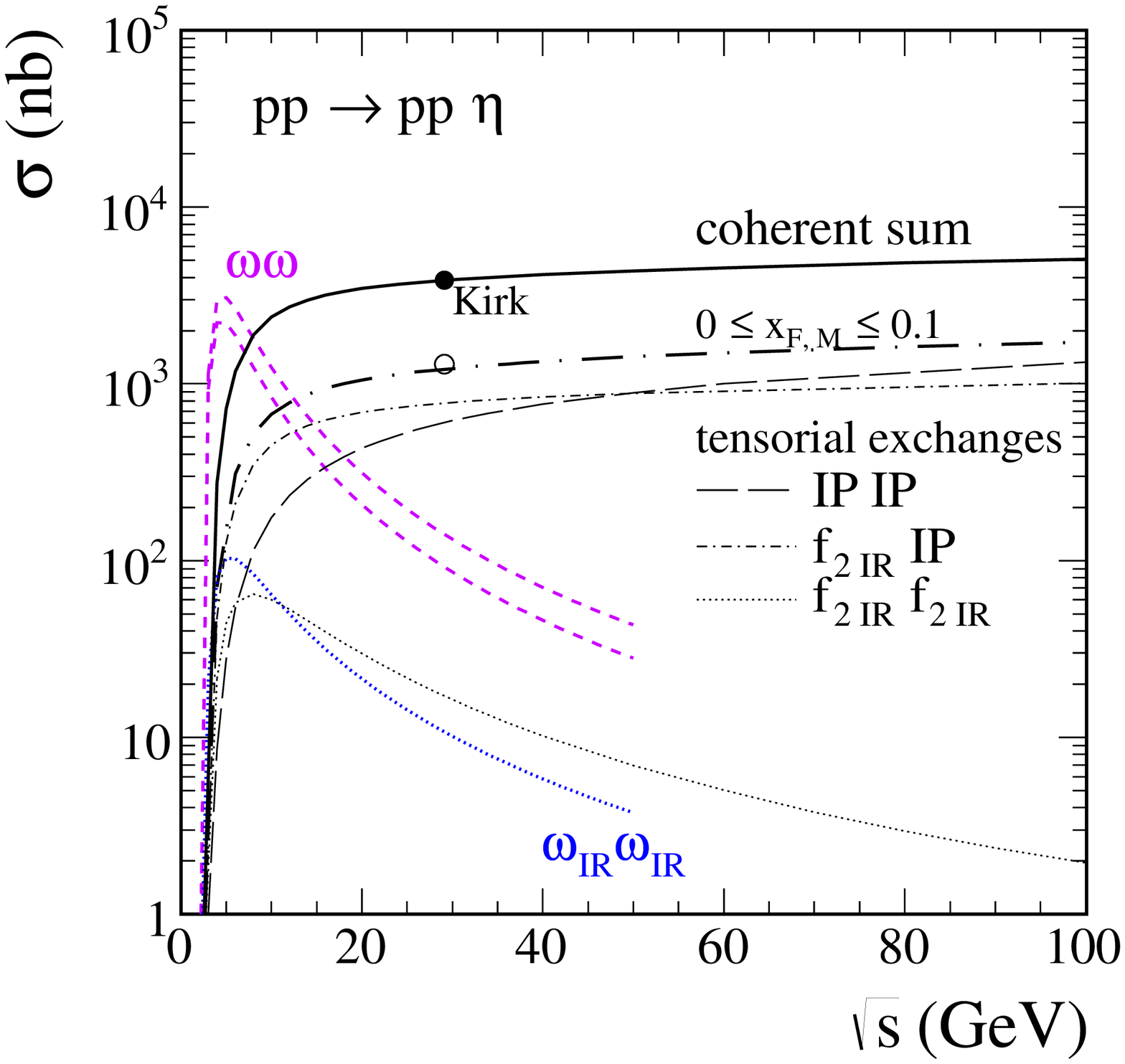}
(b)\includegraphics[width = 0.45\textwidth]{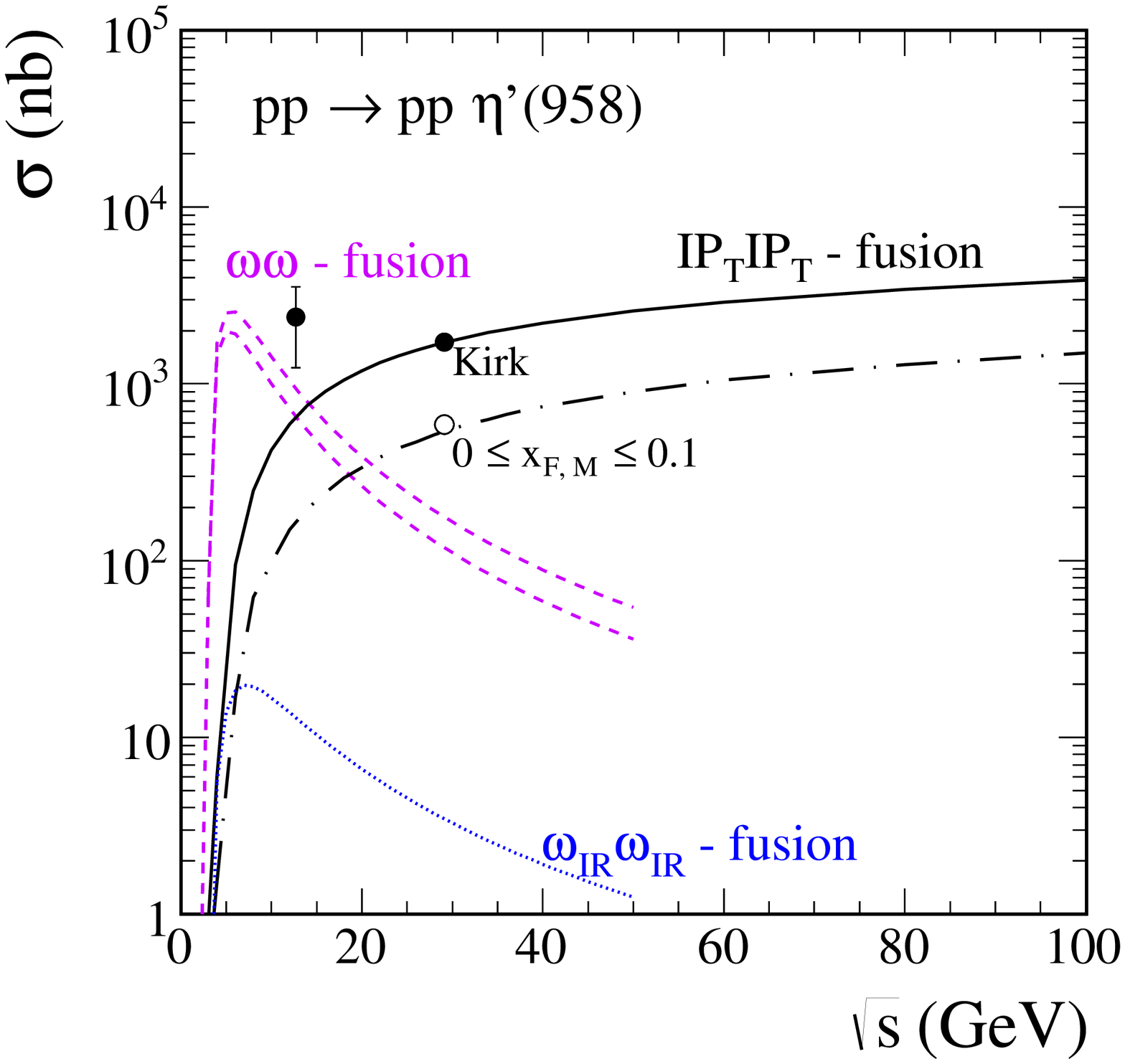}\\
(c)\includegraphics[width = 0.45\textwidth]{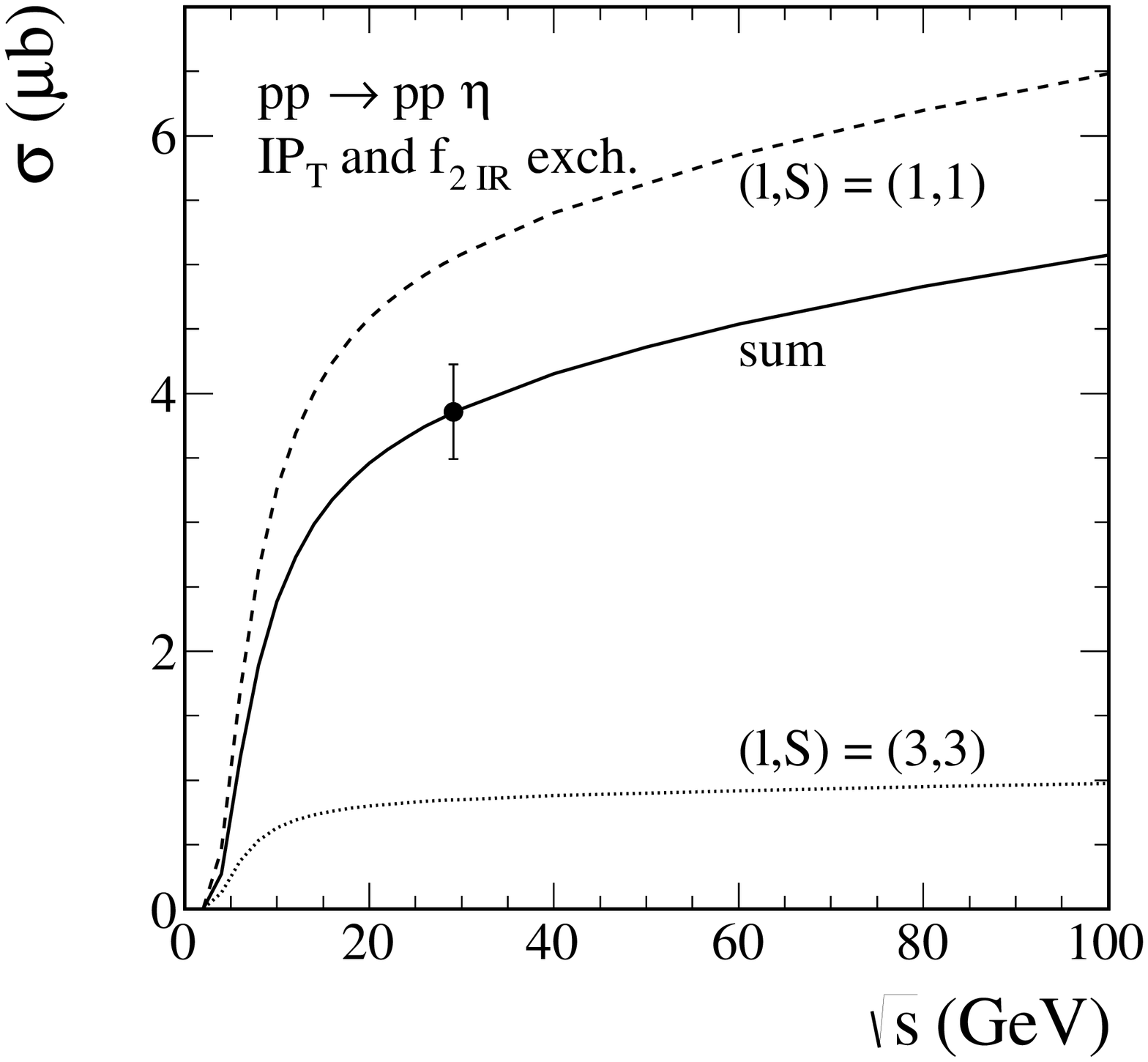}
(d)\includegraphics[width = 0.45\textwidth]{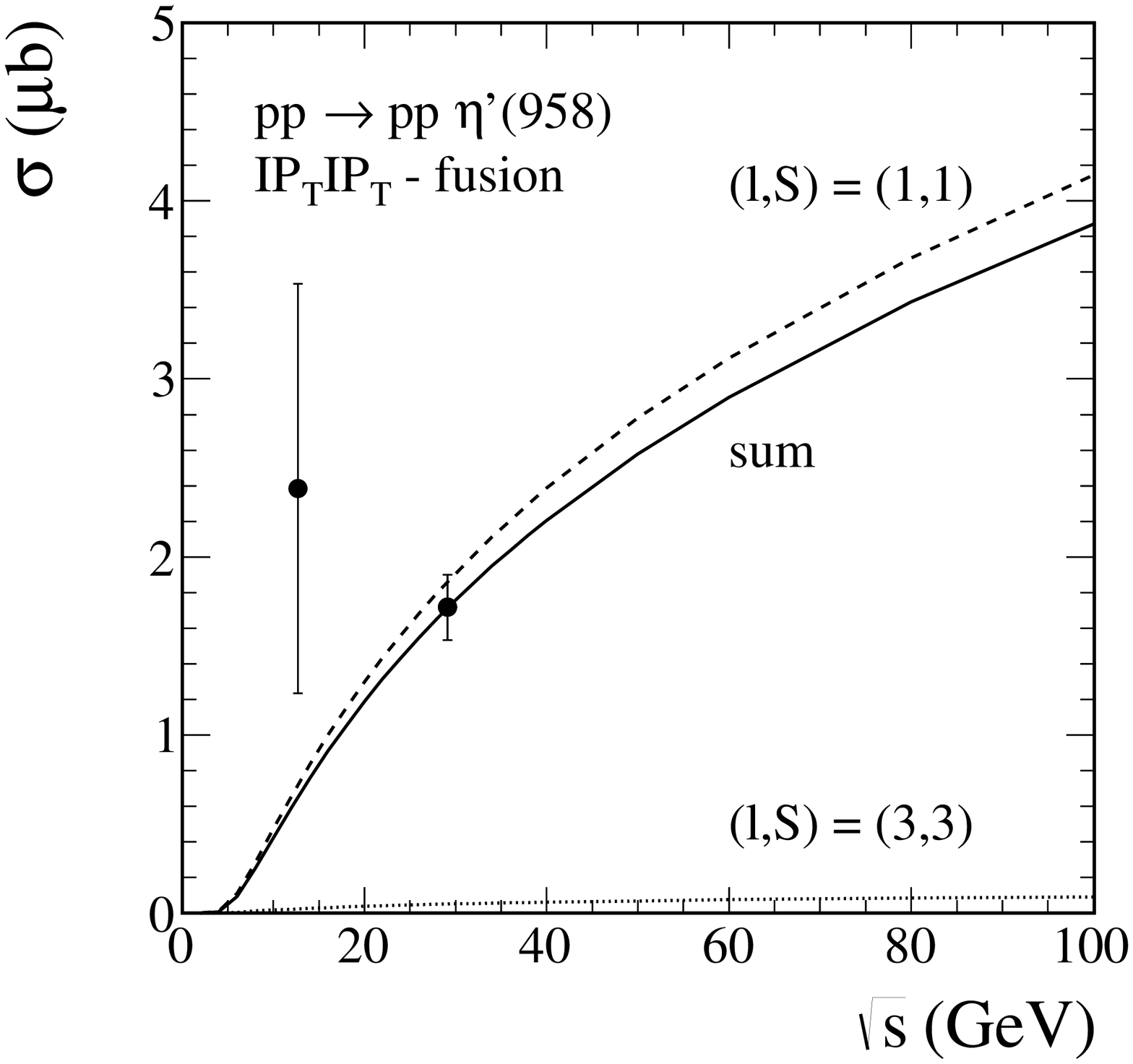}
  \caption{\label{fig:sig_pseudoscalars}
  \small
Cross section for the $pp \to pp \eta$ (panel a) 
and $pp \to pp \eta'(958)$ (panel b) reaction
as a function of proton-proton center-of-mass energy $\sqrt{s}$.
The experimental data are from the WA102 experiment at $\sqrt{s} = 29.1 $~GeV; 
see Table~\ref{tab:mesons} obtained from \cite{kirk00},
and for the Feynman-$x_{F}$ interval $0 \leqslant x_{F,M} \leqslant 0.1$ \cite{WA102_PLB427}.
There is also a data point at $\sqrt{s} = 12.7$~GeV obtained from Table~\ref{tab:ratio}.
The $\omega\omega$-fusion contribution is important only at lower energies
while tensorial pomeron fusion contribution dominates at higher energies.
In the diffractive mechanism we use vertex form factor 
(\ref{Fpompommeson_pion}) and the value of coupling constants collected in Table~\ref{tab:couplings_PS}.
For the $\eta$ meson production the tensorial contributions of
$I\!\!P I\!\!P$, $f_{2 I\!\!R} I\!\!P$ ($I\!\!P f_{2 I\!\!R}$) 
and $f_{2 I\!\!R} f_{2 I\!\!R}$ exchanges were included. 
Their coherent sum is shown by the solid line. 
For the $\eta'$ meson production the solid line 
represents the cross section obtained via tensor pomeron exchanges only.
For comparison, in the panels (c) and (d), 
we show the individual contributions to the cross section
with $(l,S) = (1,1)$ (short-dashed line) and $(l,S) = (3,3)$ (dotted line).
}
\end{figure}

In Fig.~\ref{fig:sig_pseudoscalars} we present energy dependences of the cross sections for
$\eta$ (panels (a) and (c)) and $\eta'$ (panels (b) and (d)) meson production.
It was argued in Ref.~\cite{KMV00} that $f_{2 I\!\!R}$-pomeron and pomeron-$f_{2 I\!\!R}$ exchanges
could be important for both $\eta$ and $\eta'$ central production.
For comparison, we show the results where $f_{2 I\!\!R}$ exchanges are included for $\eta$ production.
We observe a large interference of different components in the amplitude
(the long-dashed line denotes the pomeron-pomeron component,
the dash-dotted line -- $f_{2 I\!\!R}$-pomeron (or pomeron-$f_{2 I\!\!R}$) component,
and the dotted line -- $f_{2 I\!\!R} f_{2 I\!\!R}$ component).
In the diffractive mechanism we use vertex form factor given by 
Eqs.~(\ref{Fpompommeson_pion}) and (\ref{Fpion}).
Our results have been normalized to the experimental total cross sections 
given in Table~\ref{tab:mesons}
and take into account (see the dash-dotted line in panels (a) and (b))
the limited Feynman-$x_{F}$ domain $0 \leqslant x_{F,M} \leqslant 0.1$ 
for the corresponding data points; see \cite{WA102_PLB427}.
Moreover, at lower energies we can expect large contributions from
$\omega$-$\omega$ exchanges due to the large coupling of the $\omega$ meson to the nucleon.
The dashed bottom and upper lines at low energies represent the $\omega \omega$-contribution
calculated with the monopole (\ref{F_monopol_formfactor}) 
and exponential (\ref{F_exp_formfactor}) form factors, respectively.
In the case of meson exchanges we use 
values of the cut-off parameters $\Lambda_{E} = \Lambda_{M} = 1.4$~GeV.
We have taken rather maximal $\Lambda_{E}$ and $\Lambda_{M}$ in order
to obtain an upper limit for this contribution.
As explained in Section \ref{subsection:Scalar_and_pseudoscalar_meson_production}
at higher subsystem squared energies $s_{13}$ and $s_{23}$
the meson exchanges are corrected to obtain the high energy behaviour
appropriate for reggeon exchange, cf. Eq.~(\ref{aux_omeome}).

In both panels (a) and (b) the dotted line represents
the $\omega_{I\!\!R} \omega_{I\!\!R}$-contribution
calculated with coupling constant $g_{\omega_{I\!\!R} \omega_{I\!\!R} \tilde{M}} = 60$.
Due to charge-conjugation invariance the $\eta$ and $\eta'$ cannot be produced
by $\omega$-pomeron exchange and isospin conservation forbids $\rho$-pomeron exchange.
In the region of small momentum transfer squared the contribution from other processes
such as photon-(vector meson) and photon-photon fusion is possible \cite{CEF98},
but the cross section is expected to be several orders of magnitude smaller \cite{SPT07,KN98}
than for the double pomeron processes.
\footnote{
In Ref.~\cite{SZ03} the authors considered glueballs and $\eta'$ production
in semiclassical theory based on interrupted tunneling (instantons) or QCD
sphaleron production and predicted cross section
(with the cut $0 \leqslant x_{F, M} \leqslant 0.1$)
$\sigma(\eta') \approx$~255~nb in comparison to the $588 \pm 63$~nb
observed empirically \cite{WA102_PLB427}.}

\begin{figure}[!ht]
(a)\includegraphics[width = 0.45\textwidth]{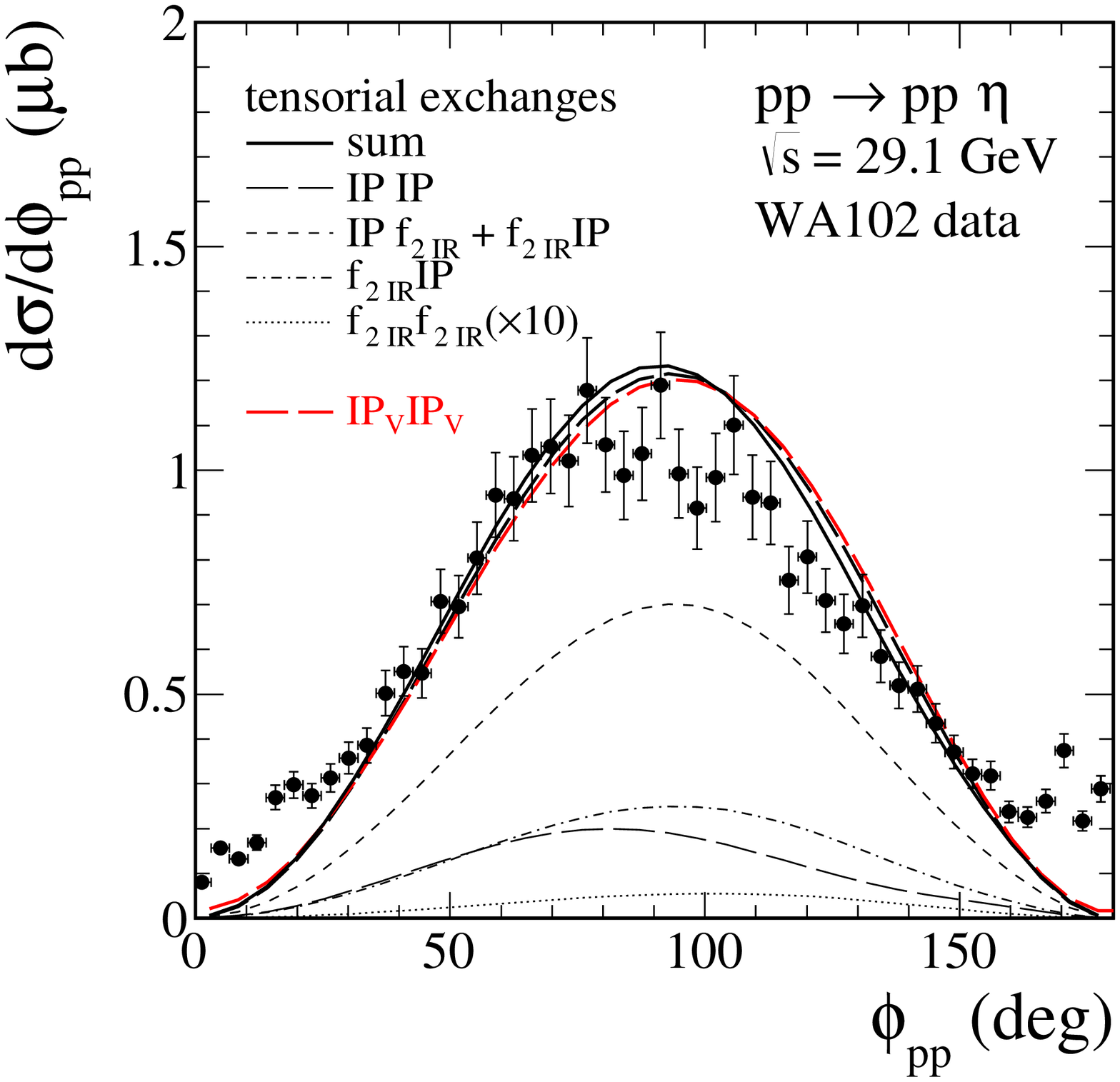}
(b)\includegraphics[width = 0.45\textwidth]{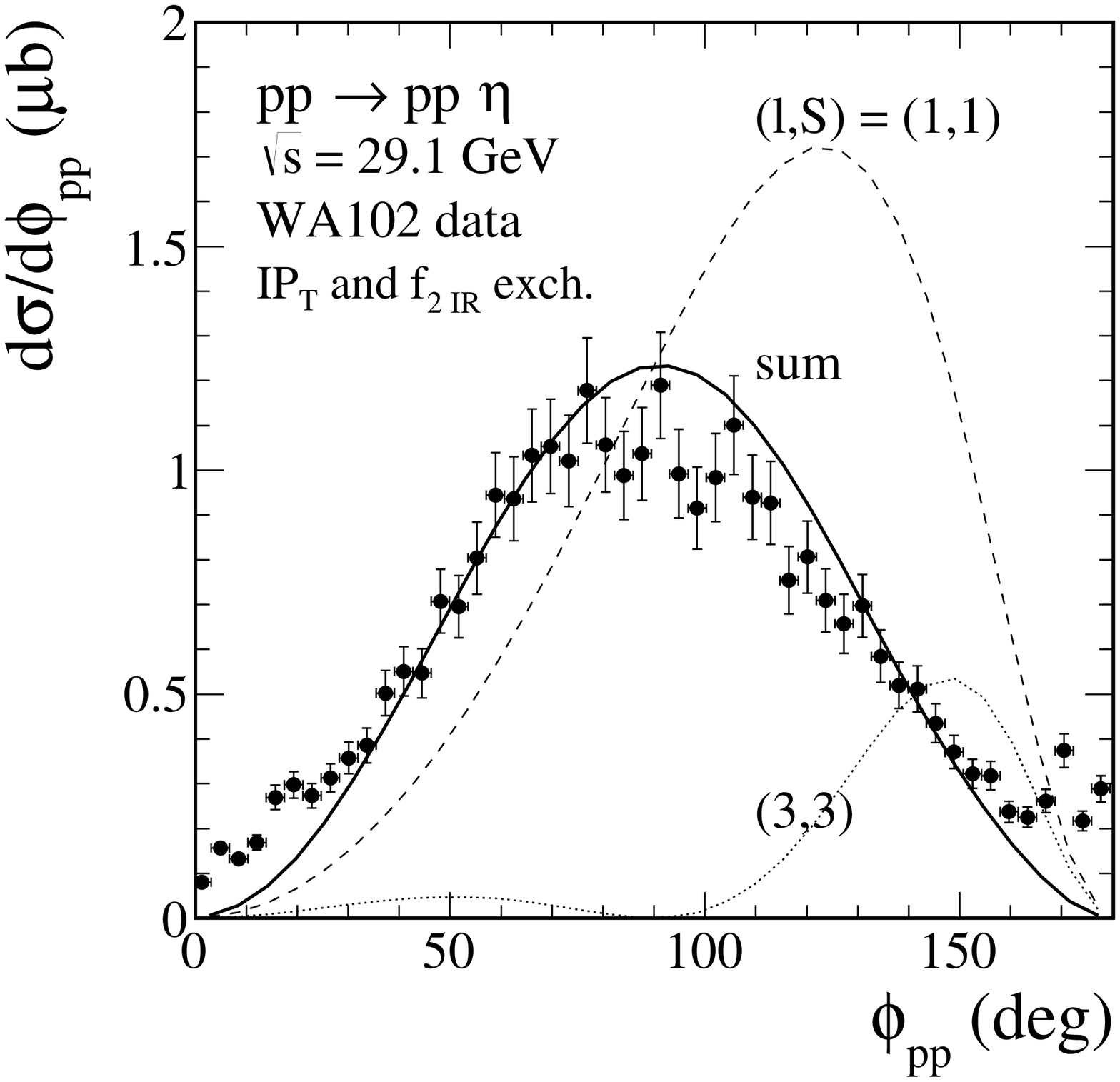}\\
(c)\includegraphics[width = 0.45\textwidth]{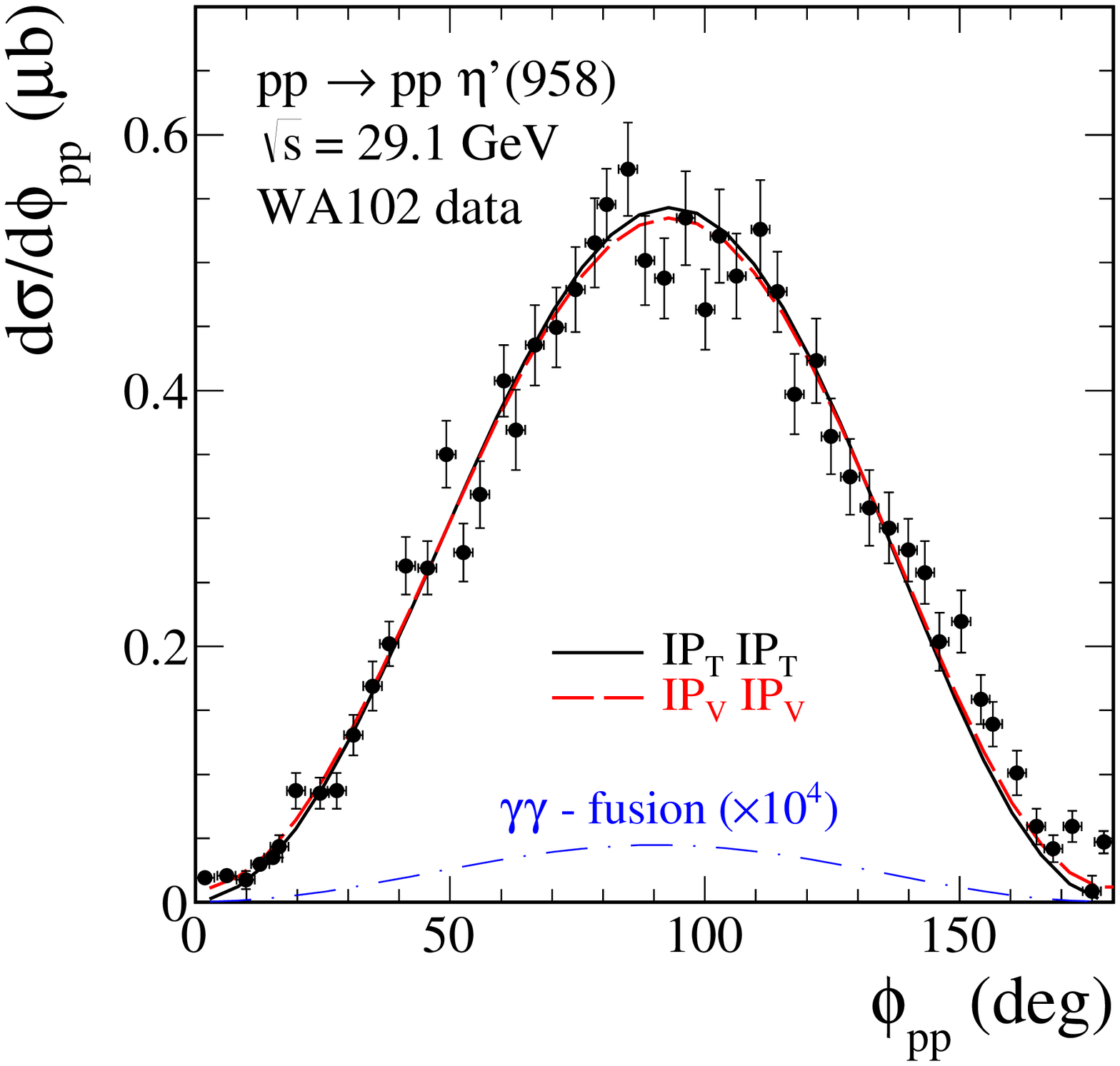}
(d)\includegraphics[width = 0.45\textwidth]{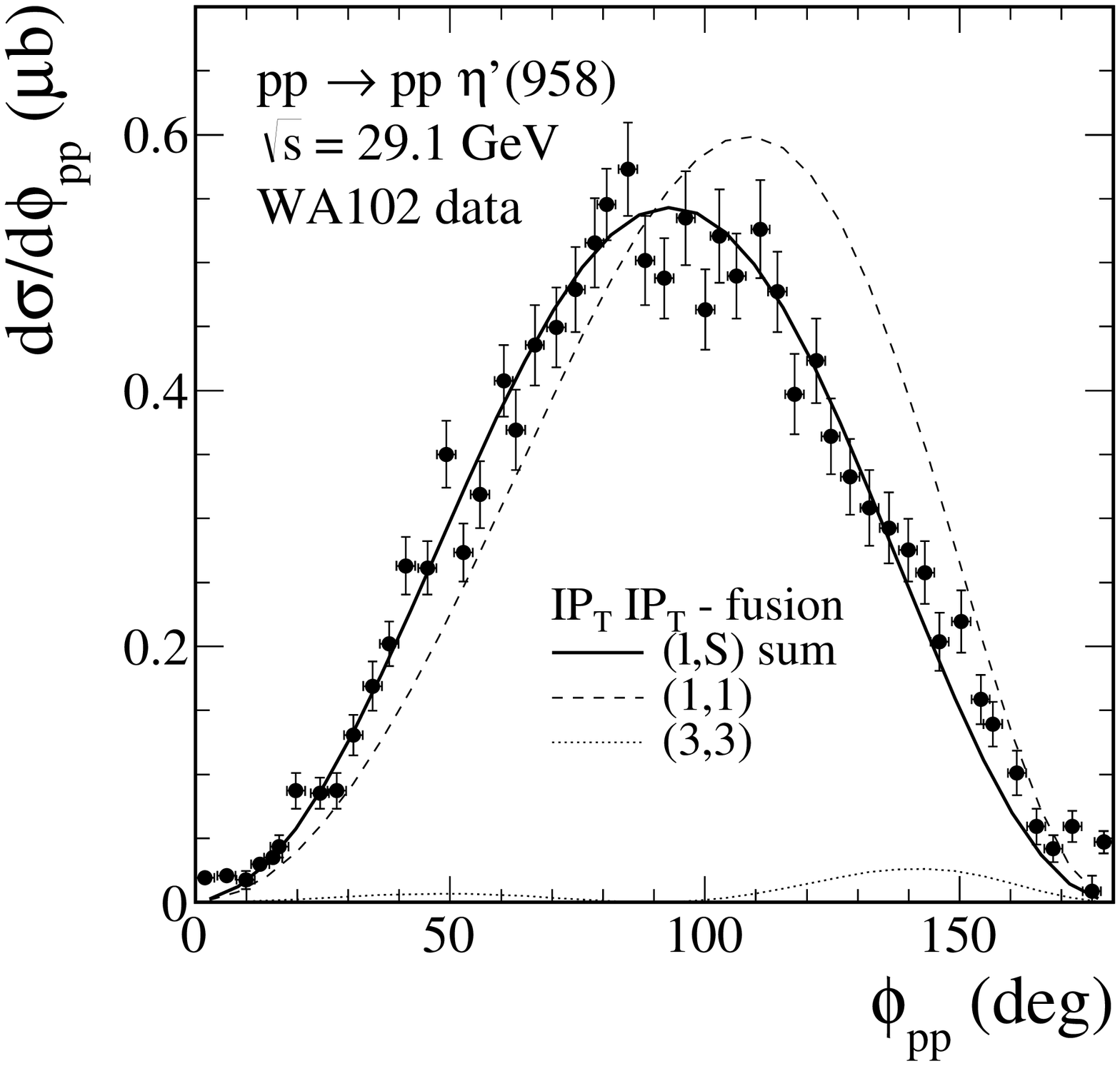}
  \caption{\label{fig:1}
  \small
Differential cross section $d\sigma/d\phi_{pp}$
for the $pp \to pp \eta$ and $pp \to pp \eta'(958)$ reactions
at $\sqrt{s} = 29.1$~GeV.
The WA102 experimental data from \cite{WA102_PLB427}
have been normalized to the mean values of the total cross sections 
given in Table~\ref{tab:mesons}.
Panel (a) shows the results for $\eta$ production.
The solid line is the result for the tensorial pomeron including two $(l,S)$ couplings
as well as $f_{2 I\!\!R} I\!\!P$, $I\!\!P f_{2 I\!\!R}$,
and $f_{2 I\!\!R} f_{2 I\!\!R}$ exchanges. 
The red long-dashed line corresponds to vector pomeron exchange only 
and $(l,S) = (1,1)$ coupling.
In panel (b) the two $(l,S)$ contributions from the tensorial pomeron exchanges 
and their total are shown.
In panel (c) we show the results for $\eta'$ production 
for the case of tensor and vector pomeron exchanges
as well as the $\gamma \gamma$-fusion enlarged by a factor $10^{4}$.
Panel (d) shows the results for $I\!\!P_{T} I\!\!P_{T}$-fusion.
}
\end{figure}

In Fig.~\ref{fig:1} we show the cross section 
as a function of the azimuthal angle $\phi_{pp}$ between 
the transverse momentum vectors of the two outgoing protons; see (\ref{D_4}).
The vertex form factor (\ref{Fpompommeson_pion}) was used in calculations.
For tensor pomeron the strengths of the $(l,S) = (1,1)$ and $(3,3)$
were adjusted to roughly reproduce the azimuthal angle distribution.
The contribution of the $(1,1)$ component alone
is not able to describe the azimuthal angular dependence (see panel (b)).
For both models the theoretical distributions are somewhat skewed 
with respect to a simple $\sin^{2}(\phi_{pp})$ dependence
as obtained e.g. from vector-vector-pseudoscalar coupling alone without phase space effects.
The small deviation in this case is due to phase space angular dependence.
The matrix element squared itself is proportional to $\sin^{2}(\phi_{pp})$.
For comparison, the dash-dotted line in the panel (c)
corresponds to $\gamma \gamma$-fusion for the $\eta'$ production calculated as in \cite{SPT07}.

\begin{figure}[!ht]
(a)\includegraphics[width = 0.45\textwidth]{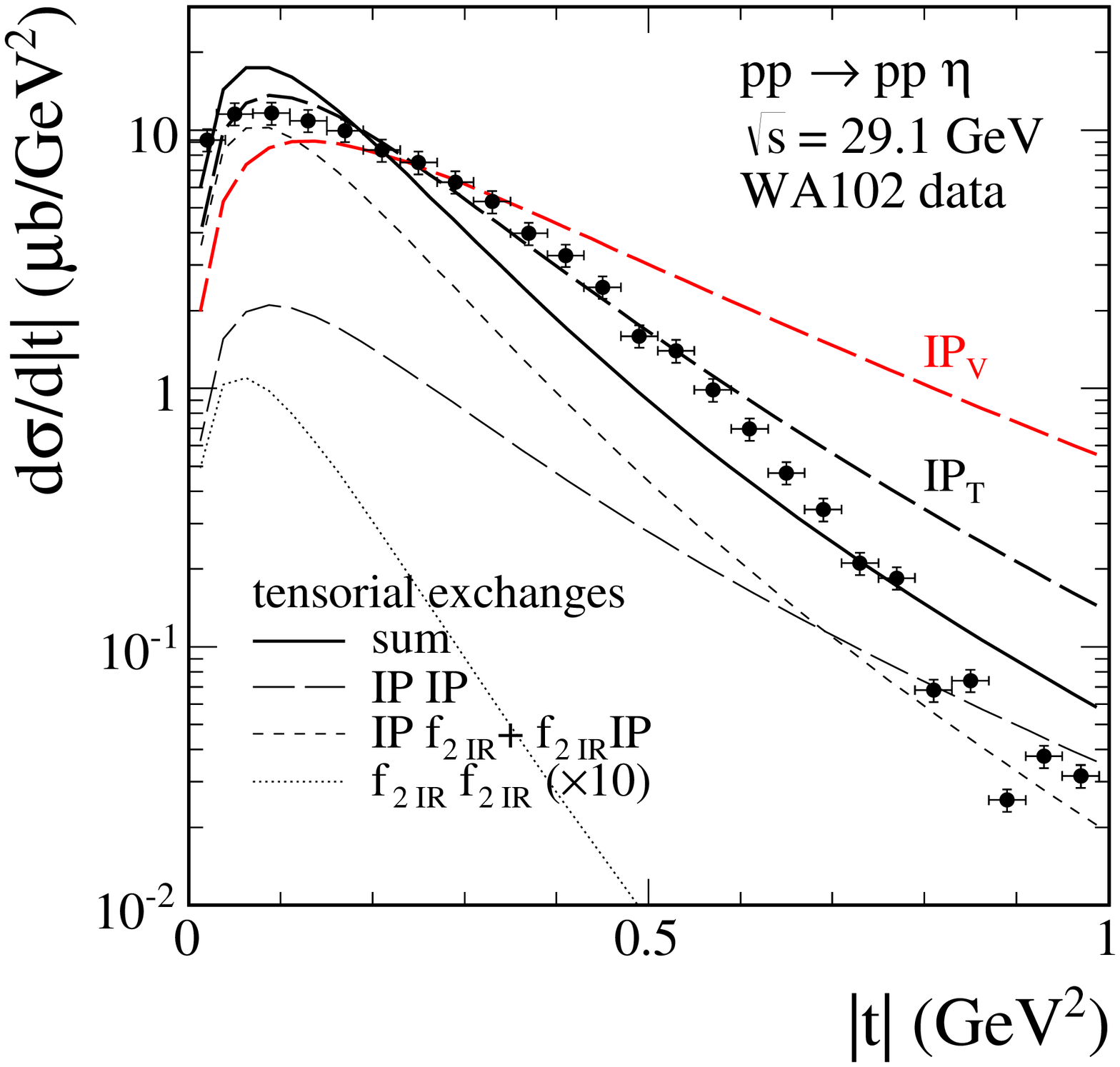}
(b)\includegraphics[width = 0.45\textwidth]{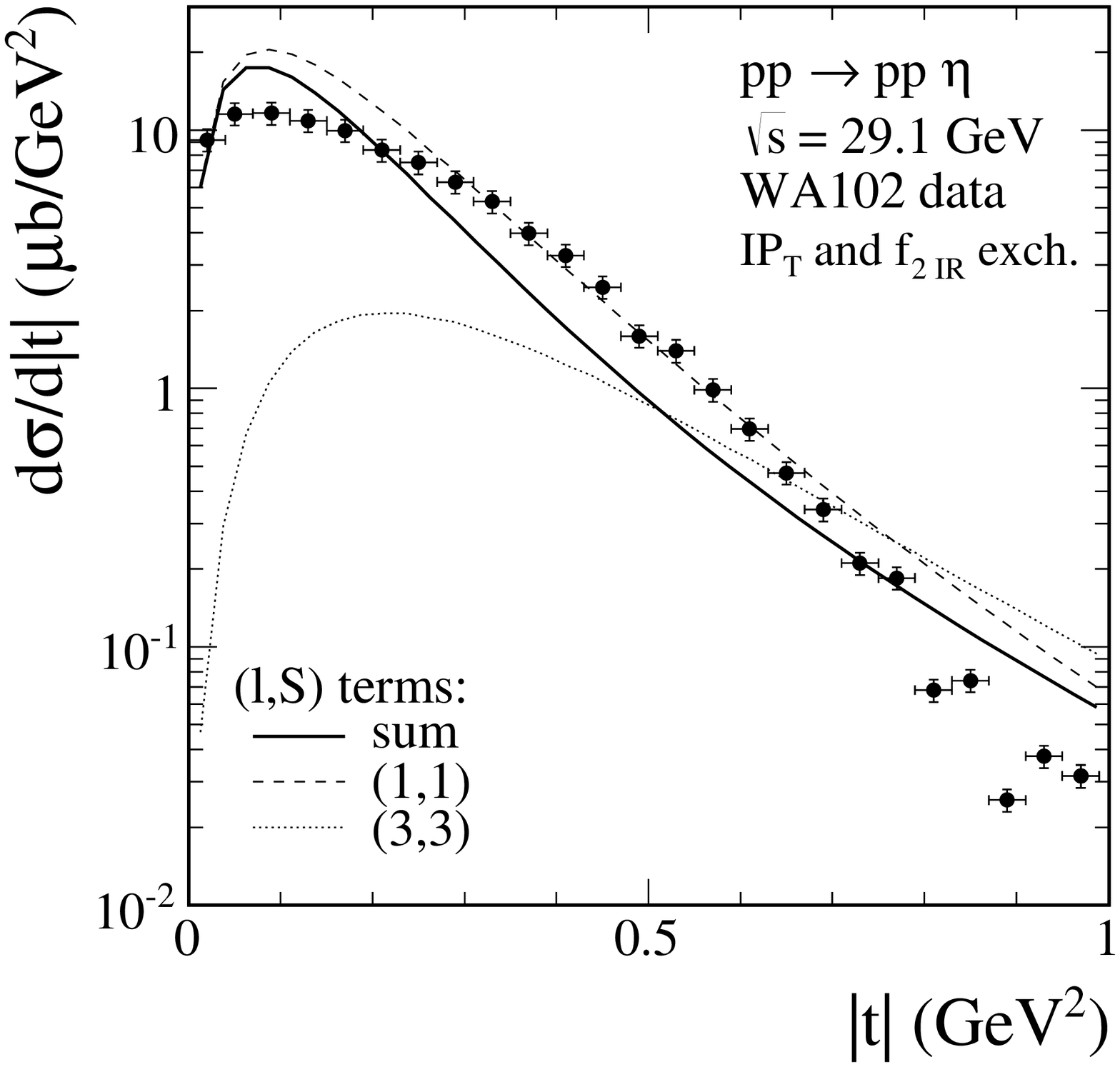}
(c)\includegraphics[width = 0.45\textwidth]{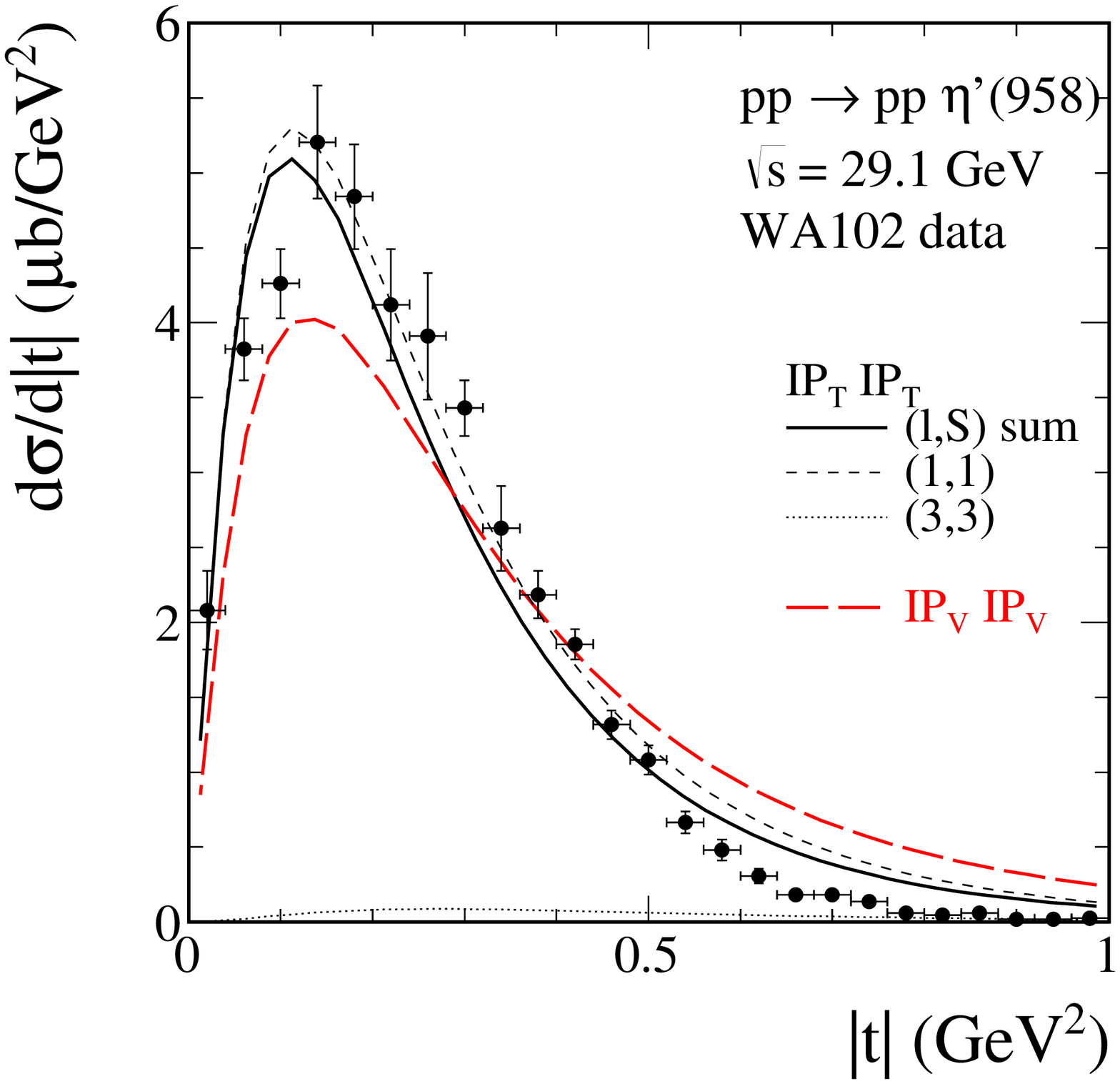}
(d)\includegraphics[width = 0.45\textwidth]{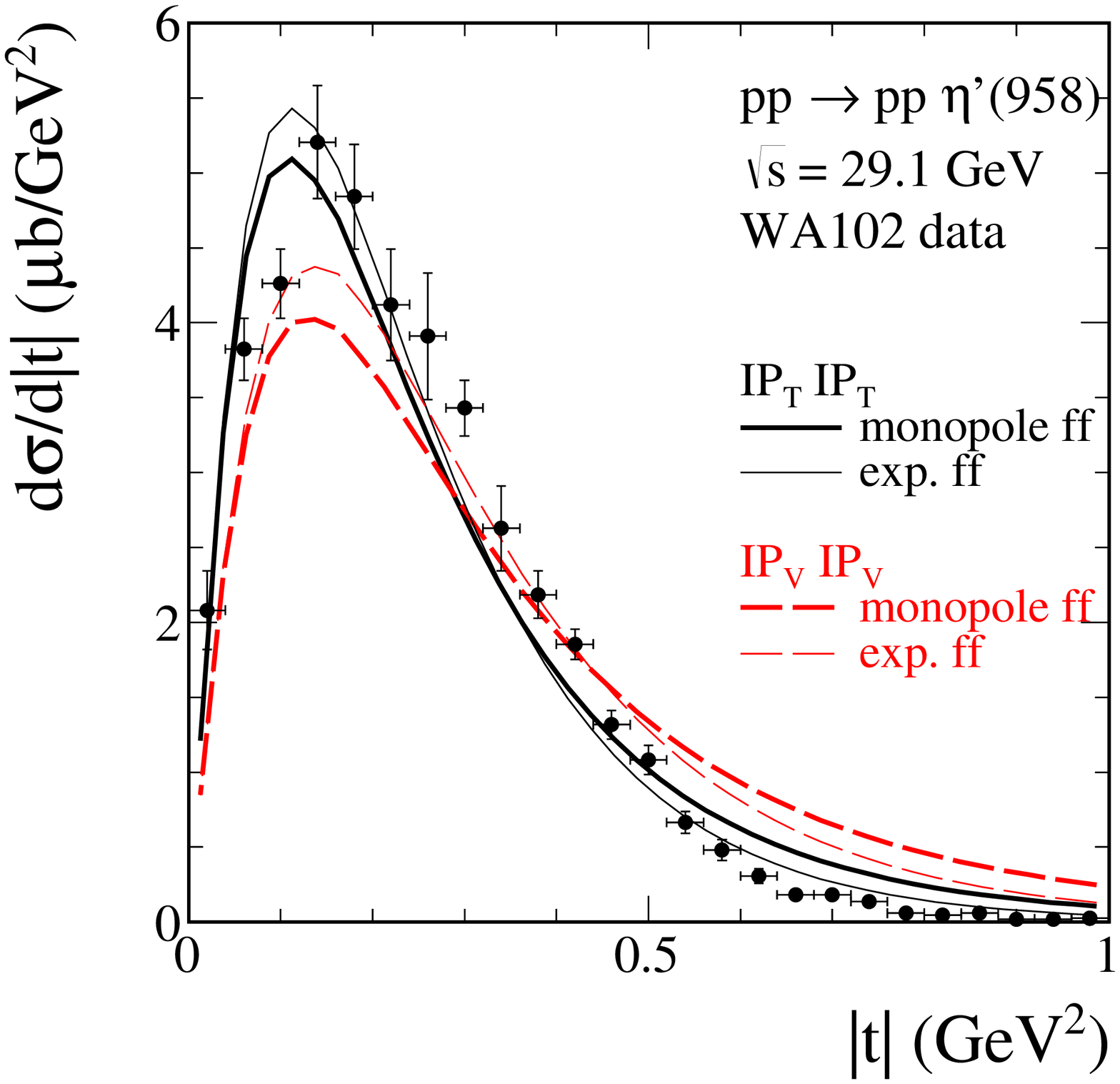}
  \caption{\label{fig:2}
  \small
Differential cross section $d\sigma/d|t|$
for the $pp \to pp \eta$ (panels (a) and (b))
and $pp \to pp \eta'$ (panels (c) and (d)) reactions at $\sqrt{s} = 29.1$~GeV.
The WA102 experimental distributions from \cite{WA102_PLB427}
have been normalized to the mean values of the total cross sections given in Table~\ref{tab:mesons}.
The solid line corresponds to the model with tensorial pomeron
while the dashed line to the model with vectorial pomeron.
For $\eta$ production the $f_{2 I\!\!R}$ exchanges were included in addition.
In the present calculations we use vertex form factor 
given by Eqs.~(\ref{Fpompommeson_pion}) and (\ref{Fpion}).
For comparison, in panel (d), we also show the results for 
exponential form factor (\ref{Fpompommeson_exp}) and for $\Lambda_{E}^{2} = 0.7$~GeV$^{2}$.
}
\end{figure}

In Fig.~\ref{fig:2} we present distribution in $|t_{1}|$ and $|t_{2}|$,
which are, of course identical. Therefore we label them by $|t|$.
As can be seen from panels (a) and (c) 
the results for the tensorial exchanges give a better description of $t$ distribution
than the vector pomeron exchanges.
The $t$-dependence of $\eta$ and $\eta'$ production 
is very sensitive to the form factor $F_{I\!\!P I\!\!P M}(t_{1},t_{2})$, cf.~(\ref{Fpompommeson_pion}),
in the pomeron-pomeron-meson vertex.

\begin{figure}[!ht]
(a)\includegraphics[width = 0.45\textwidth]{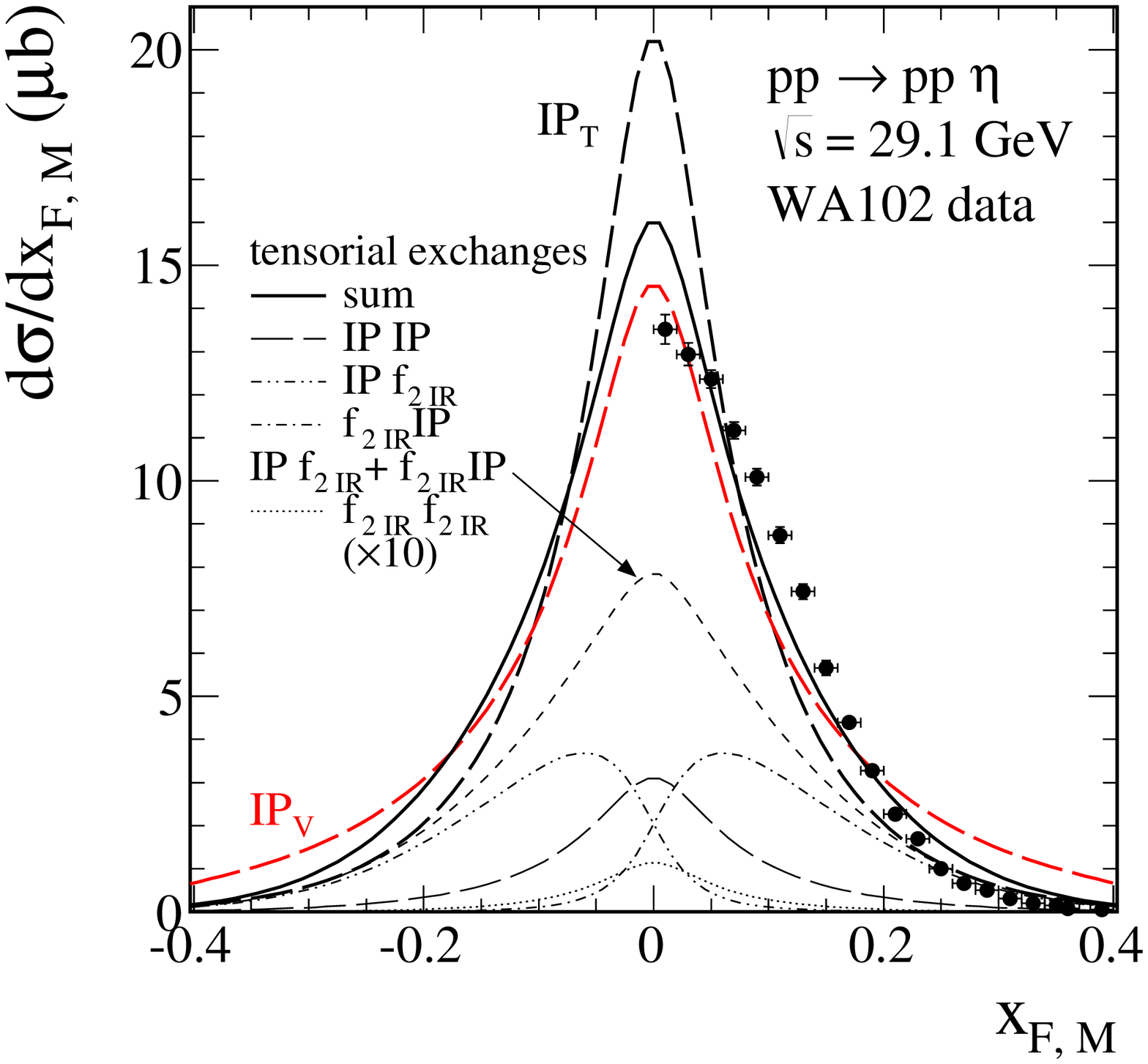}
(b)\includegraphics[width = 0.45\textwidth]{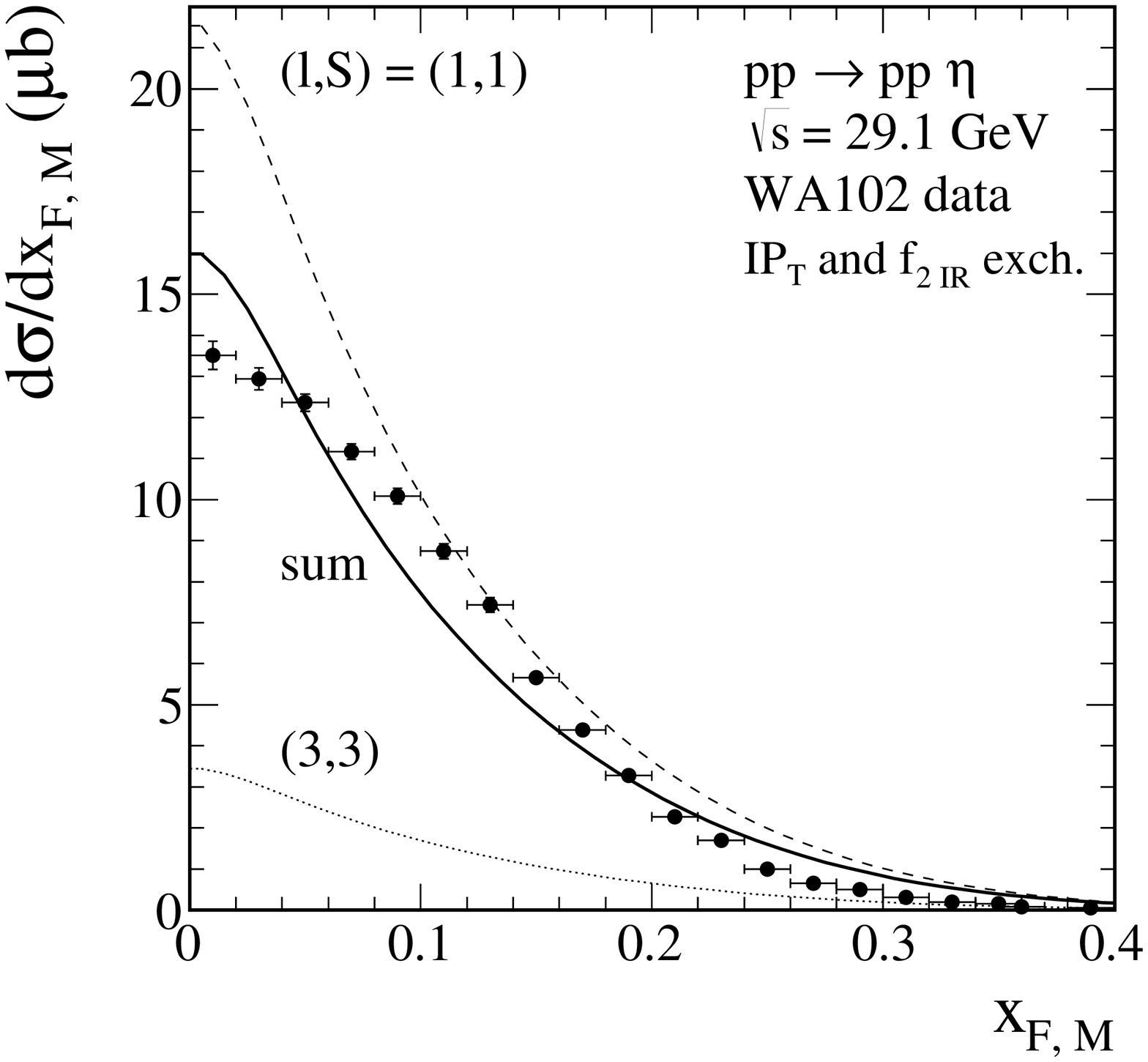}
(c)\includegraphics[width = 0.45\textwidth]{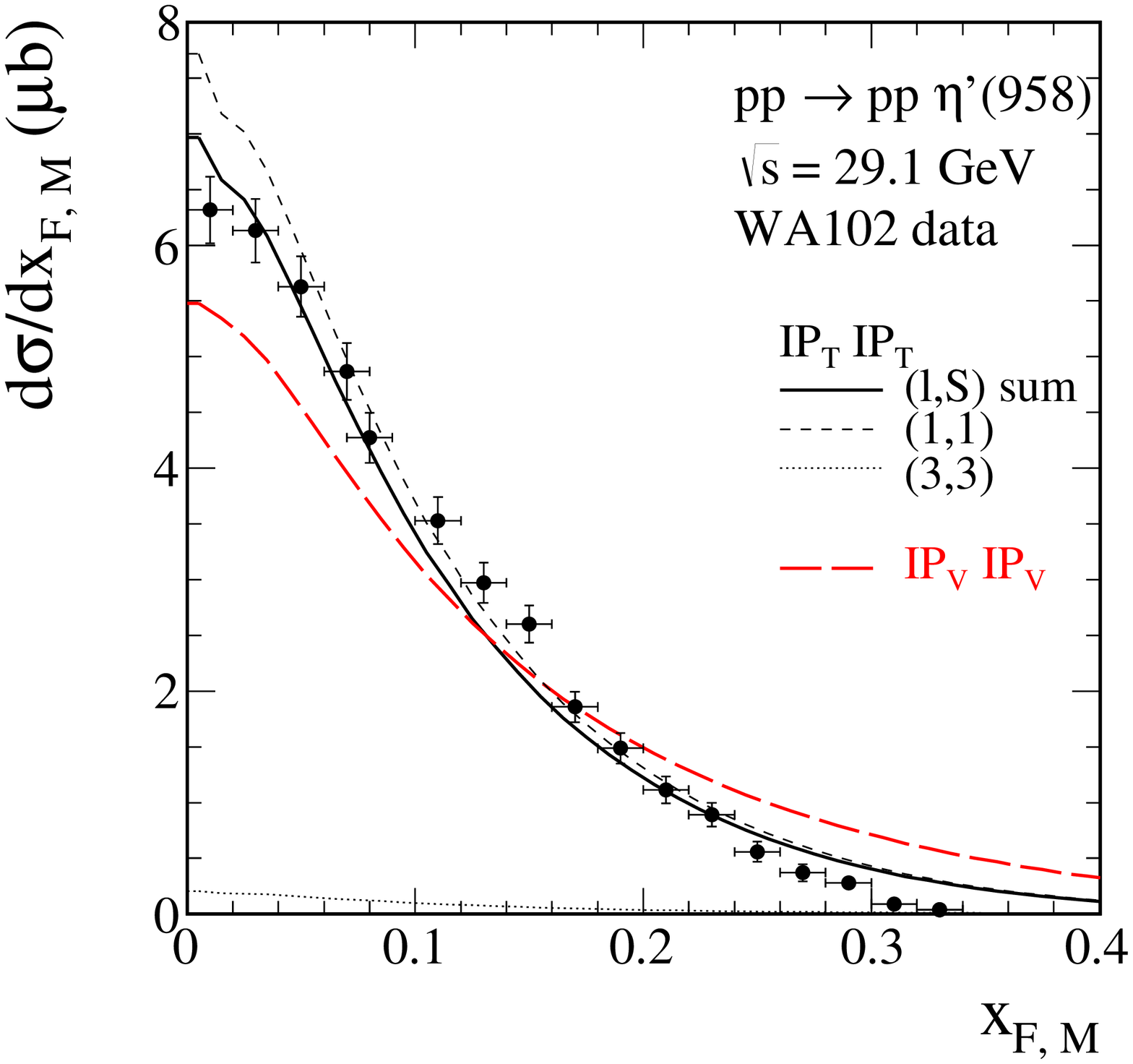}
(d)\includegraphics[width = 0.45\textwidth]{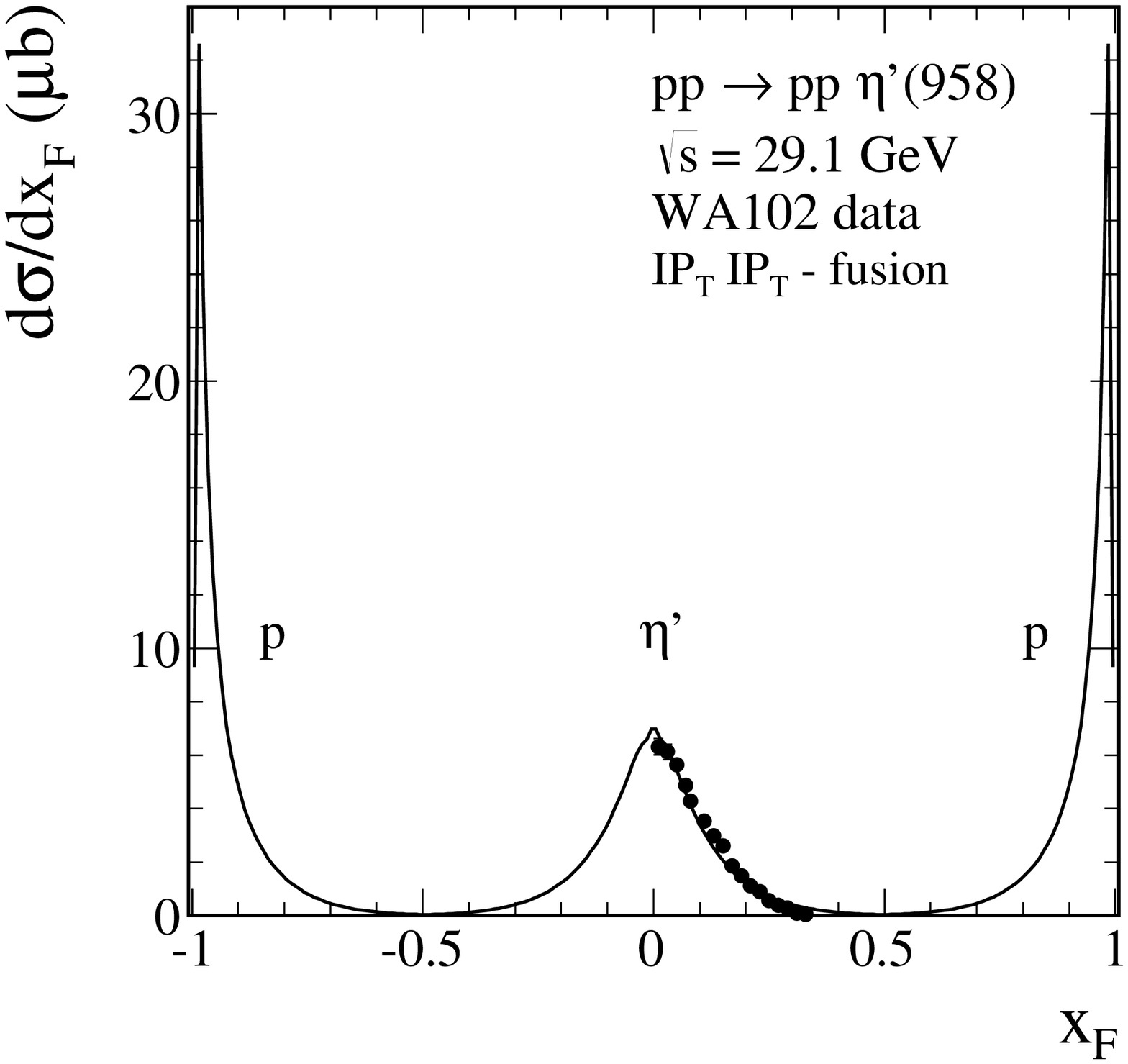}
  \caption{\label{fig:3}
  \small
Differential cross section $d\sigma/dx_{F, M}$
for the $pp \to pp \eta$ (panels (a) and (b))
and $pp \to pp \eta'$ (panels (c) and (d)) reactions at $\sqrt{s} = 29.1$~GeV.
The WA102 experimental data \cite{WA102_PLB427} are shown for comparison
and have been normalized to the mean values of the total cross sections given in Table~\ref{tab:mesons}.
In the present calculations we use vertex form factor (\ref{Fpompommeson_pion})
and two model of pomeron exchanges.
In panel (a) the results for the tensorial pomeron and $f_{2 I\!\!R}$ exchanges are shown;
the pomeron-pomeron component peaks at $x_{F, M} = 0$ (the long-dashed line),
the pomeron-$f_{2 I\!\!R}$ ($f_{2 I\!\!R}$-pomeron) peaks at backward (forward) $x_{F, M}$, respectively,
and the coherent sum of pomeron-$f_{2 I\!\!R}$ and $f_{2 I\!\!R}$-pomeron component
effectively dominates in the central region of $x_{F, M}$ (the short-dashed line).
In panels (b) and (c) we show the individual contributions to the cross section
with $(l,S) = (1,1)$ (the short-dashed line), $(3,3)$ (the dotted line),
and their coherent sum (the solid line).
The long-dashed line in panel (c) corresponds to the model with vectorial pomeron.
In panel (d) the $x_{F}$ distributions for $\eta'$ (at $x_{F} = 0$)
and for the protons (at $x_{F} \to \pm 1$) are shown for $I\!\!P_{T} I\!\!P_{T}$ fusion.
}
\end{figure}

In Fig.~\ref{fig:3} we present the $d\sigma/dx_{F}$ distribution.
We see that $\eta$ (panels~(a) and (b)) and $\eta'$ (panels~(c) and (d)) meson distributions
are peaked at $x_{F, M} \approx 0$,
which is consistent with the dominance of the pomeron-pomeron exchange.
In the calculations we use the pomeron-pomeron-meson couplings
collected in Table \ref{tab:couplings_PS}.
For the description of the $\eta$ production in the case of the tensorial pomeron
the $f_{2 I\!\!R}$ exchanges in the amplitude were included.
In panel (a) the solid line corresponds to the model with tensorial pomeron
plus $f_{2 I\!\!R}$ exchanges and
the long-dashed line to the model with vectorial pomeron.
The enhancement of the $\eta$ distribution at larger values of $x_{F, M}$
can be explained by significant $f_{2 I\!\!R}$-pomeron and pomeron-$f_{2 I\!\!R}$ exchanges.
As can be seen from panel (a) these contributions have maxima at $x_{F, M} \neq 0$.
The corresponding couplings constants were fixed
to differential distributions of the WA102 Collaboration \cite{WA102_PLB427}.
In panel (b) we show for the tensorial pomeron 
the individual contributions to the cross section
with $(l,S) = (1,1)$ (the short-dashed line), $(l,S) = (3,3)$ (the dotted line),
and their coherent sum (the solid line).
In panel (c) we show the Feynman-$x_{F}$ distribution
of the $\eta'$ meson and the theoretical curves for $I\!\!P_{T} I\!\!P_{T}$
and $I\!\!P_{V} I\!\!P_{V}$ fusion, respectively.
The diffractively scattered outgoing protons are placed at
$x_{F} \approx \pm 1$; see panel (d).

\begin{figure}[!ht]
\includegraphics[width = 0.32\textwidth]{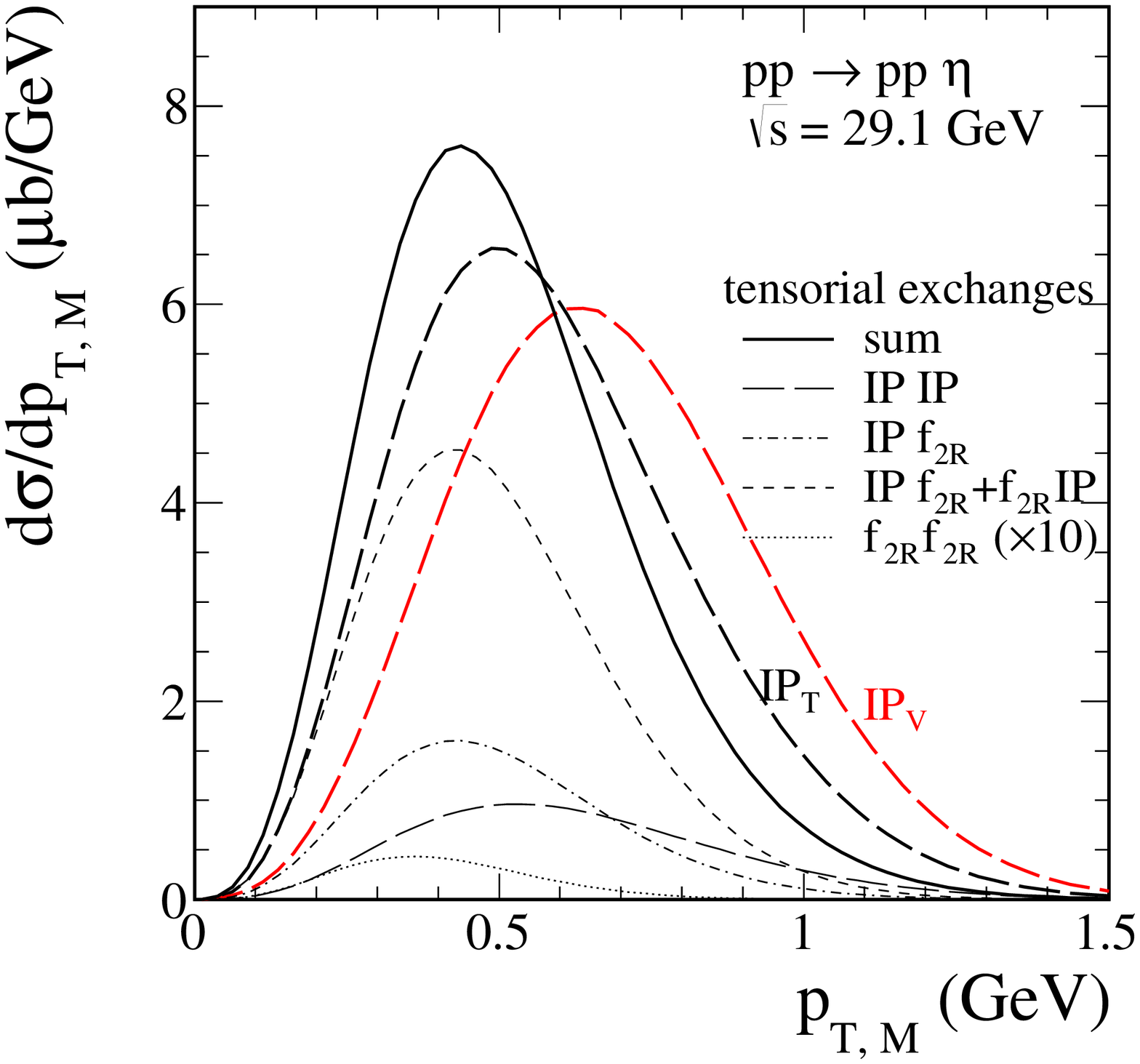}
\includegraphics[width = 0.32\textwidth]{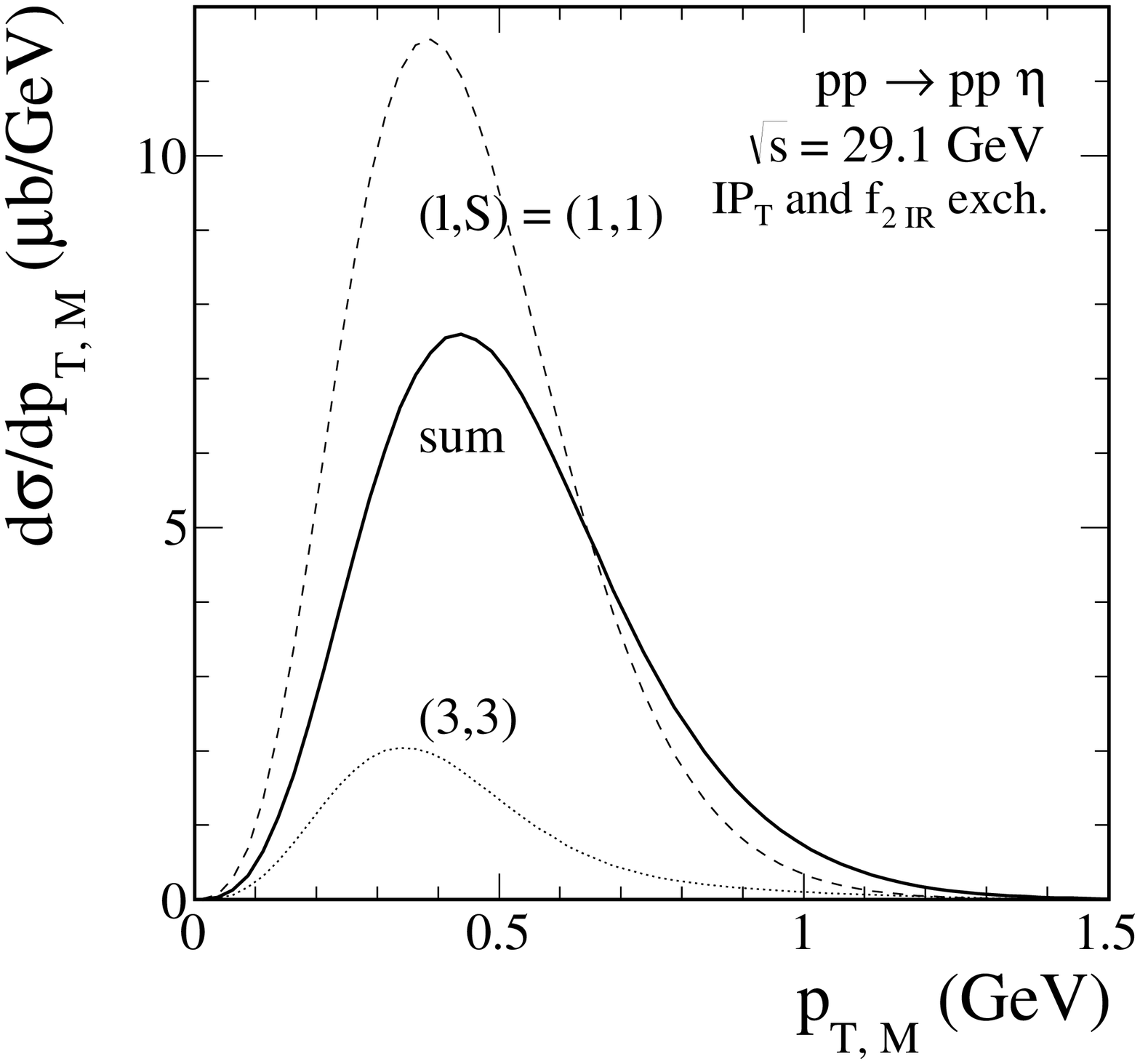}
\includegraphics[width = 0.32\textwidth]{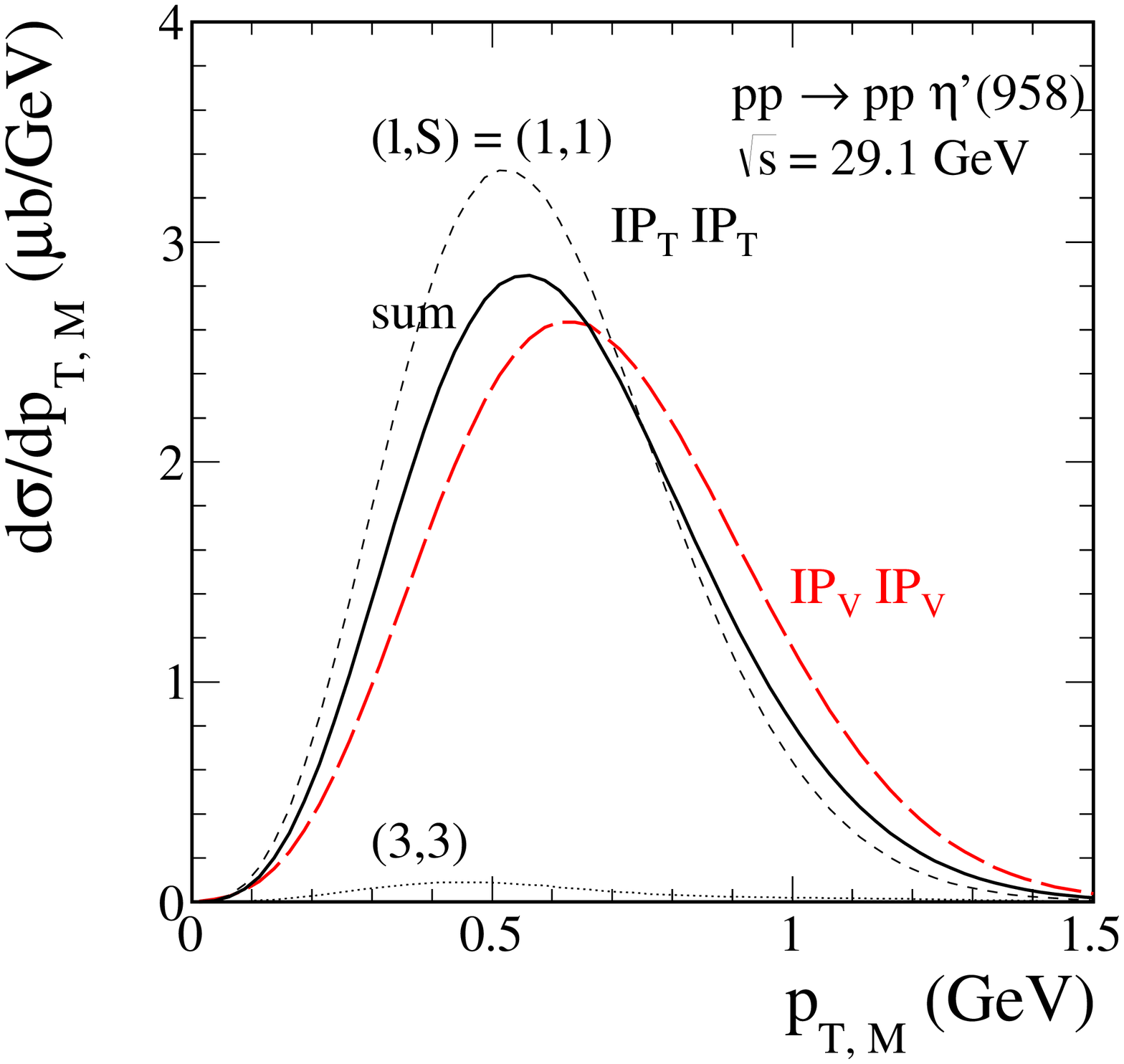}\\
\includegraphics[width = 0.32\textwidth]{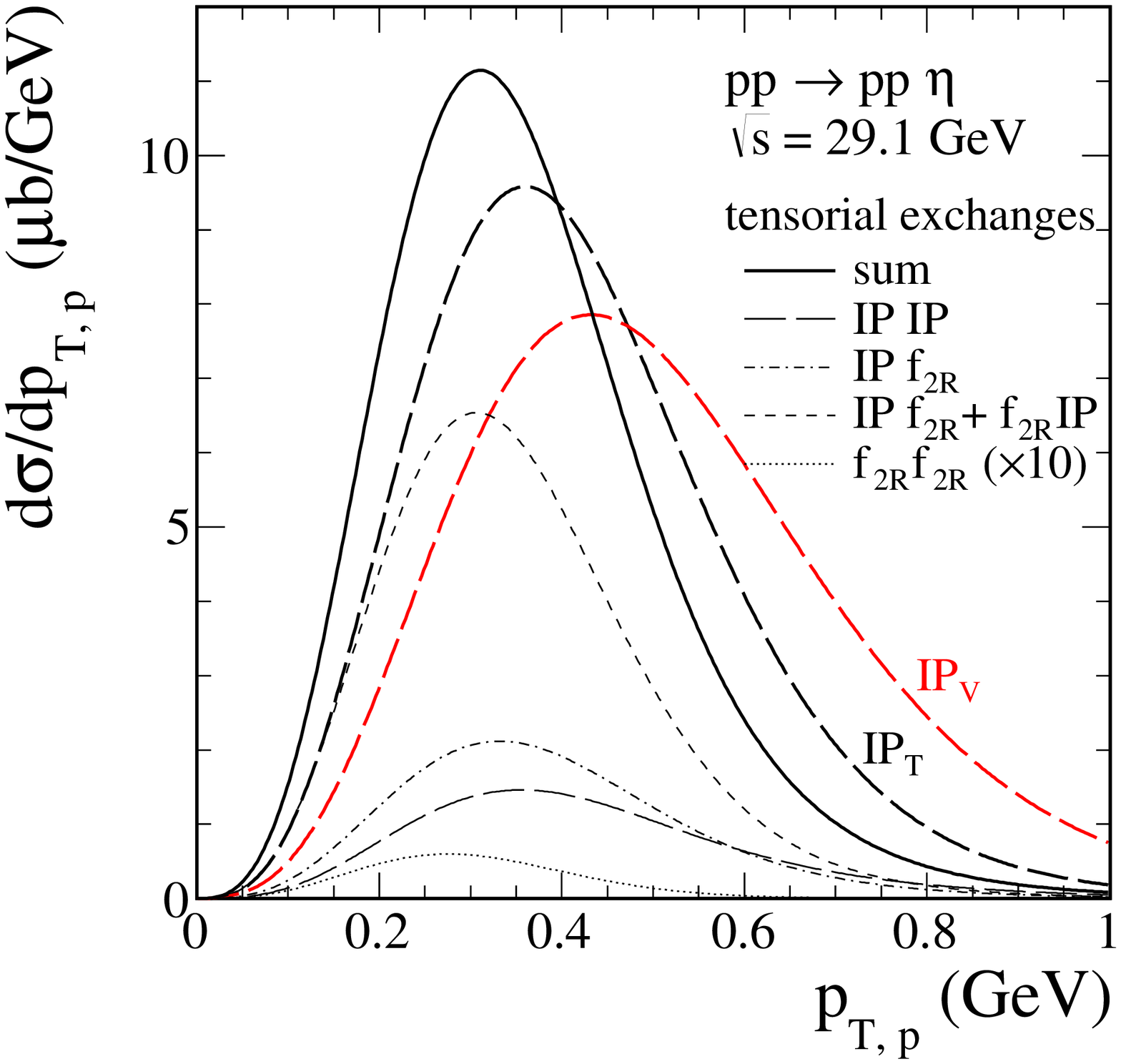}
\includegraphics[width = 0.32\textwidth]{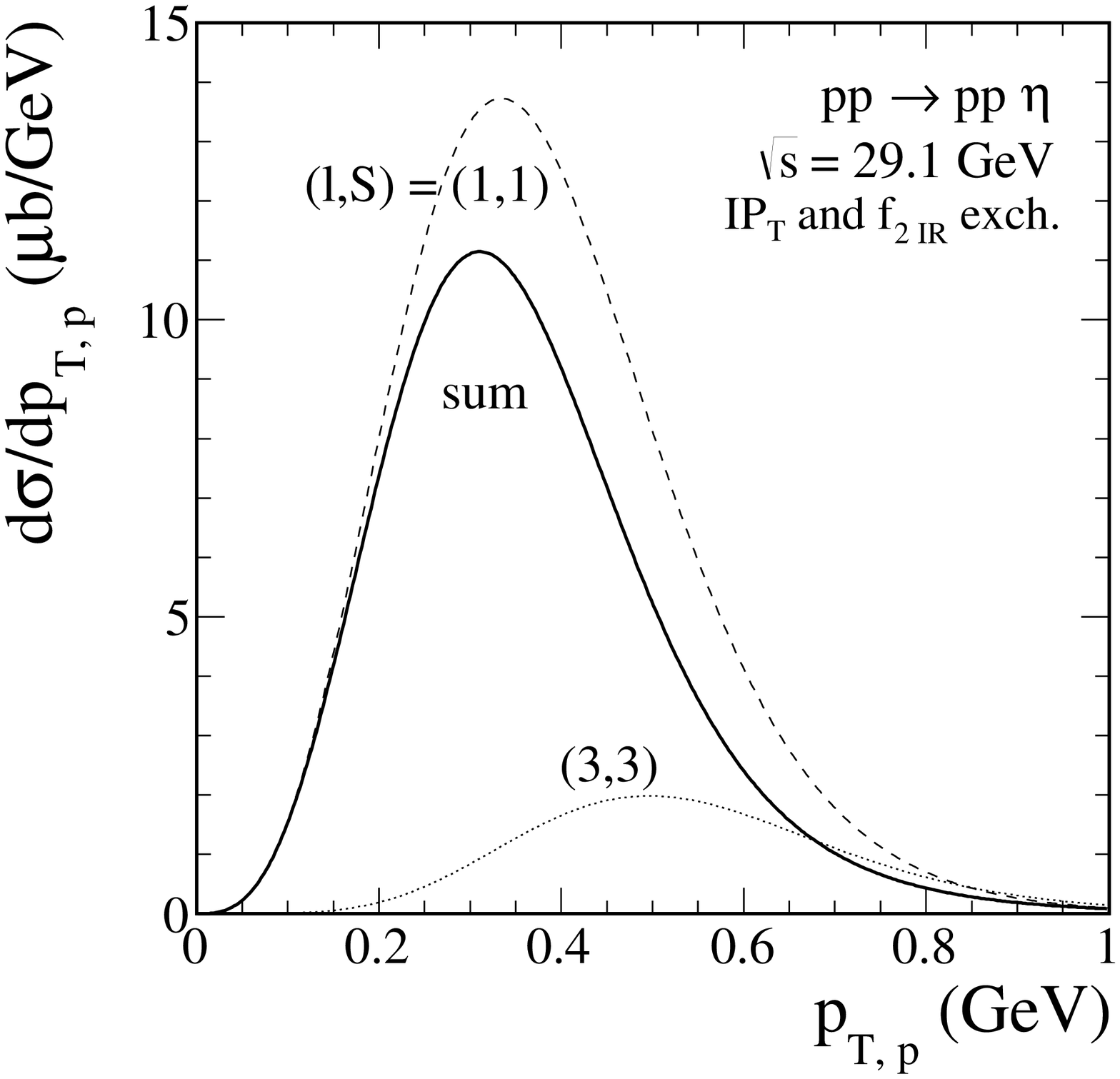}
\includegraphics[width = 0.32\textwidth]{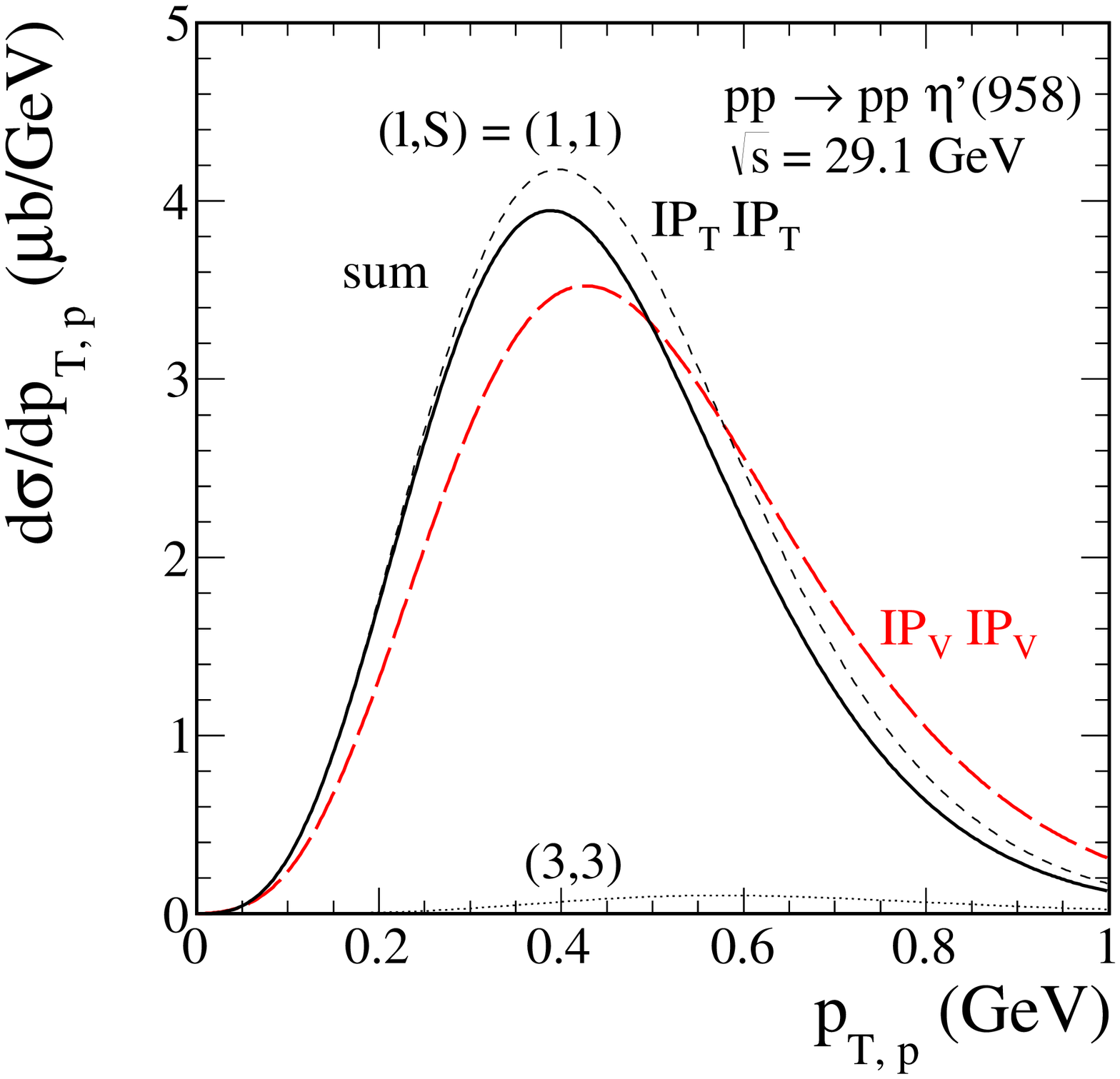}
  \caption{\label{fig:4}
  \small
Differential cross sections $d\sigma/dp_{\perp,M}$ 
and $d\sigma/dp_{\perp,p}$ (the forward proton $p_{1}$)
for the central exclusive $\eta$ and $\eta'$ meson production at $\sqrt{s} = 29.1$~GeV.
The solid line corresponds to the model with tensorial pomeron
while the dashed line to the model with vectorial pomeron.
For $\eta$ production the $f_{2 I\!\!R}$ exchanges in the amplitude were included
in addition as discussed in the text.
We show for the tensorial case also the individual contributions to the cross section
with $(l,S) = (1,1)$ (short-dashed line) and $(l,S) = (3,3)$ (dotted line).
}
\end{figure}

In Fig.~\ref{fig:4} we present distributions
in meson transverse momentum $p_{\perp,M}$ and
proton transverse momentum $p_{\perp,p}$.
As already explained above for $\eta$ meson production we 
include in addition tensorial reggeon exchanges. 
Their individual contributions are shown in the left panels.
In addition, we show the individual spin contributions to the cross section
with $(l,S) = (1,1)$ (short-dashed line) and $(l,S) = (3,3)$ (dotted line).
The coherent sum of $(1,1)$ and $(3,3)$ tensorial components 
is shifted with respect to the $(1,1)$ vectorial component alone.

\begin{figure}[!ht]
(a)\includegraphics[width = 0.45\textwidth]{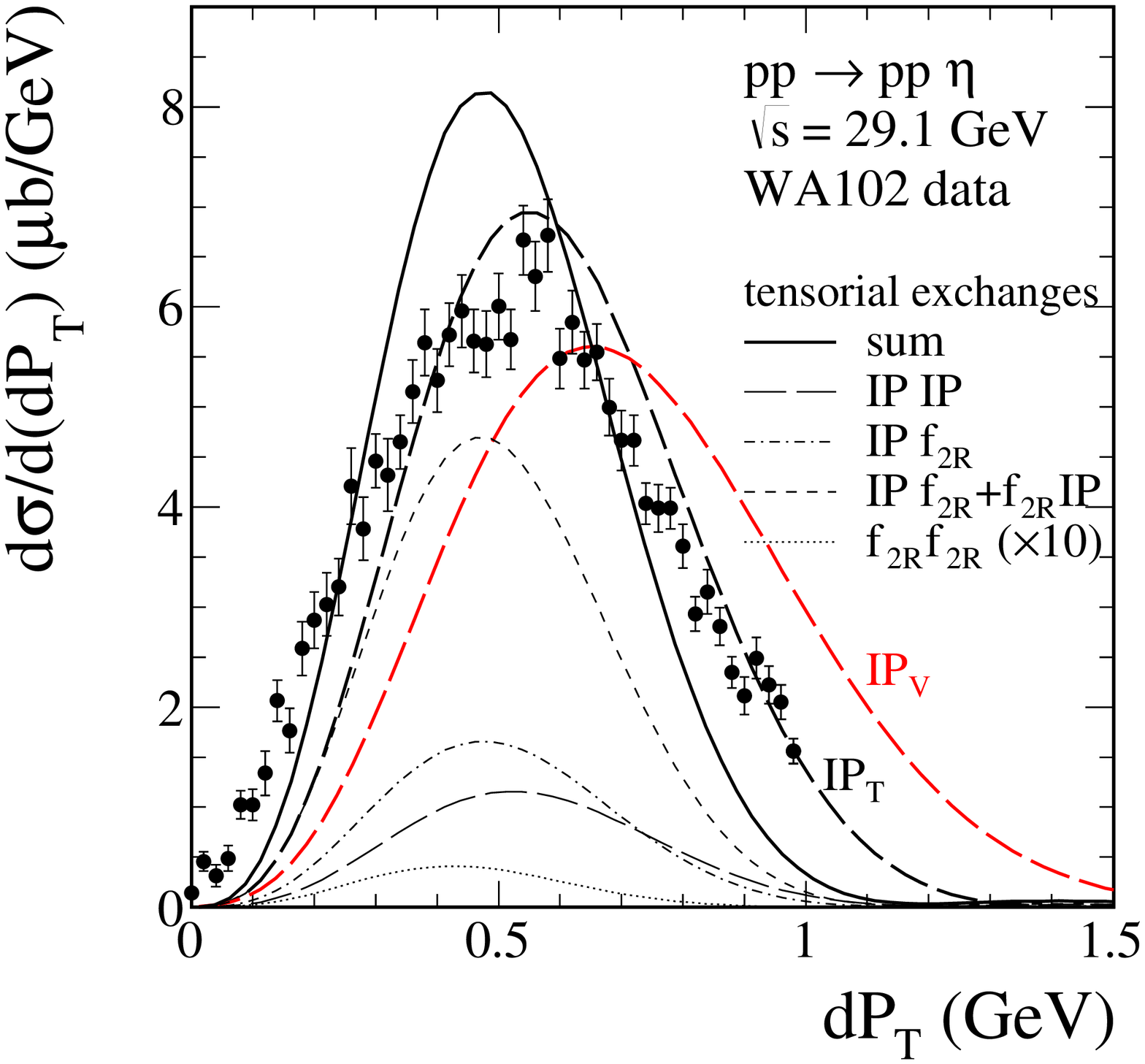}
(b)\includegraphics[width = 0.45\textwidth]{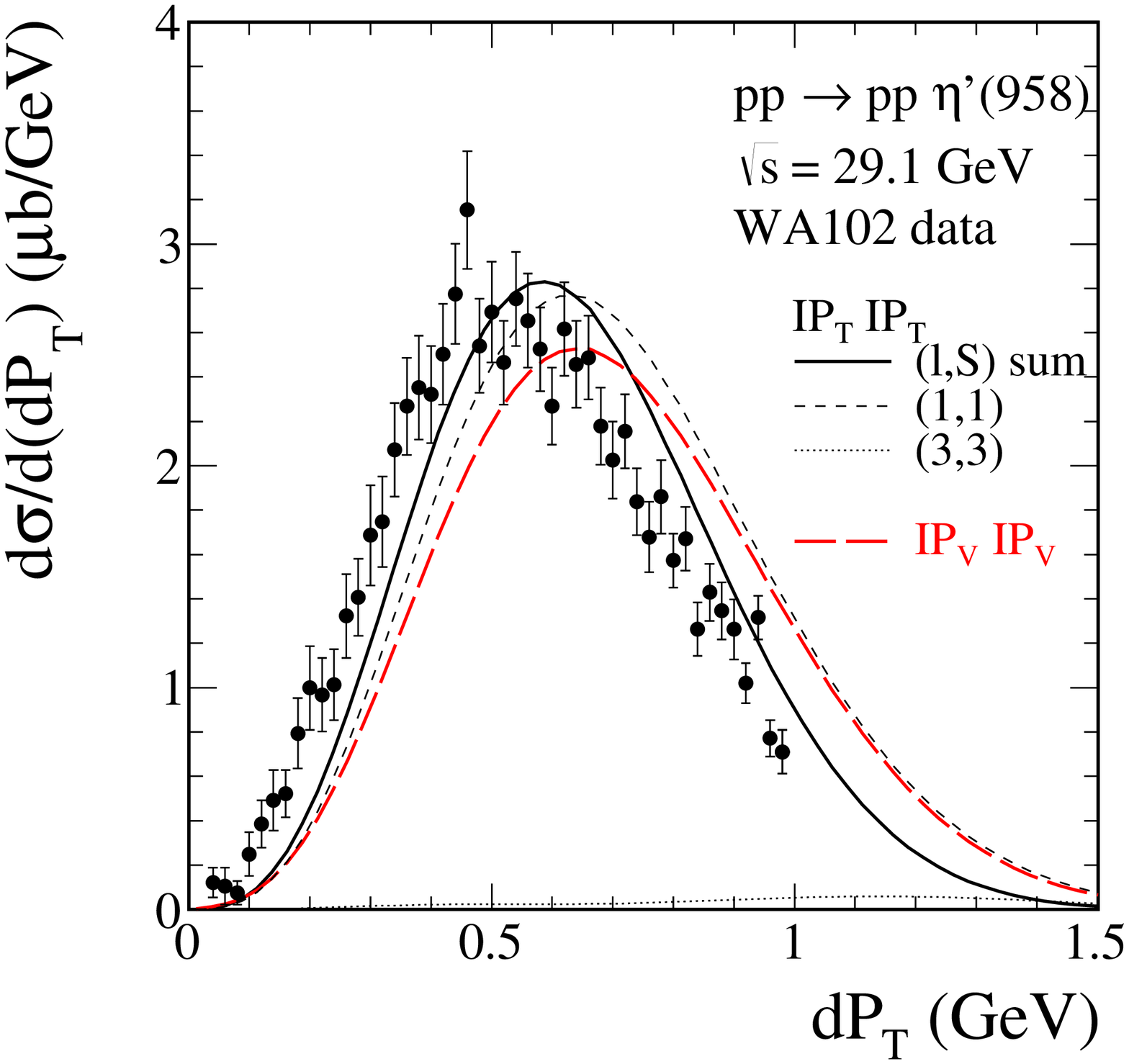}\\
(c)\includegraphics[width = 0.45\textwidth]{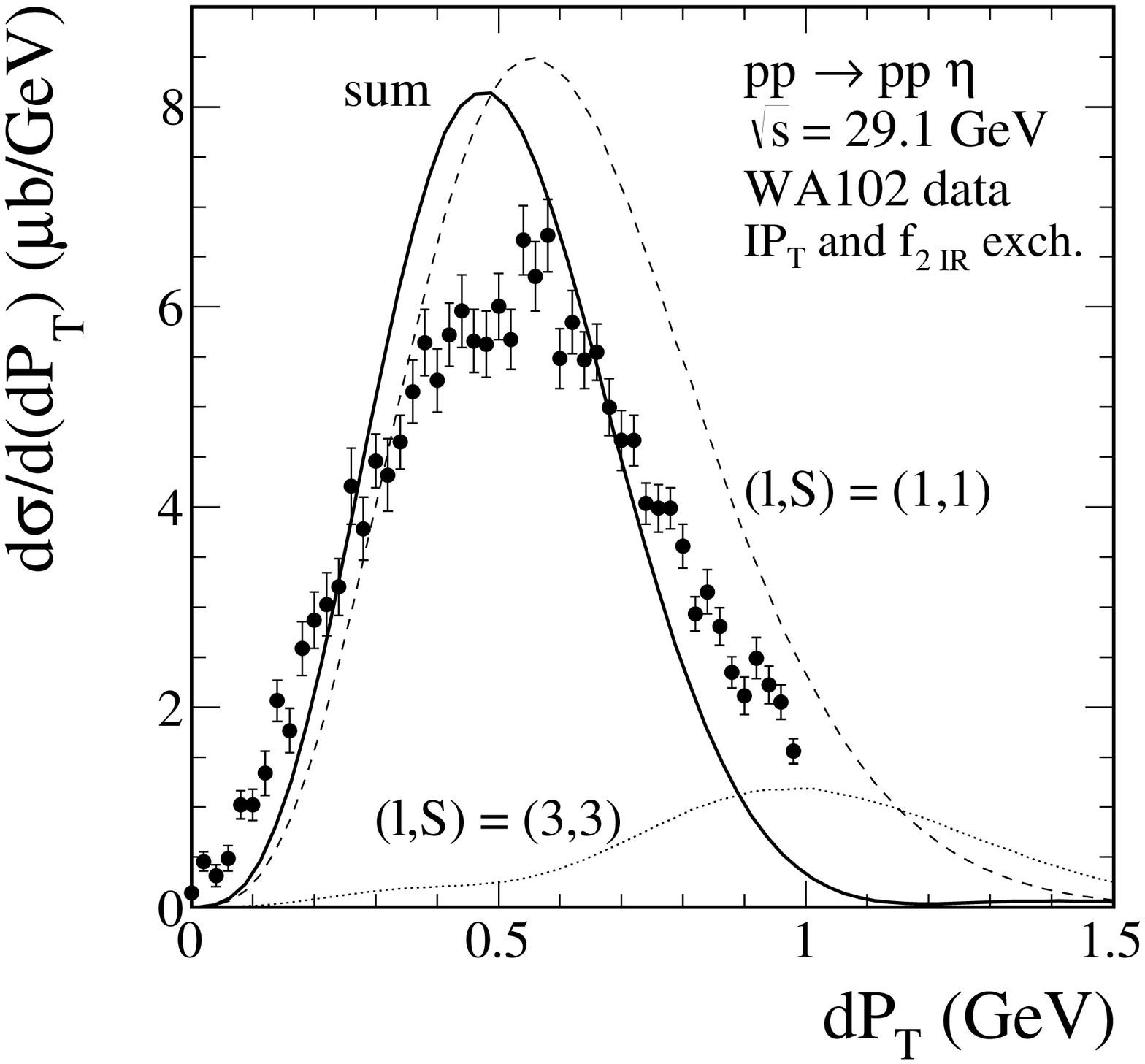}
(d)\includegraphics[width = 0.45\textwidth]{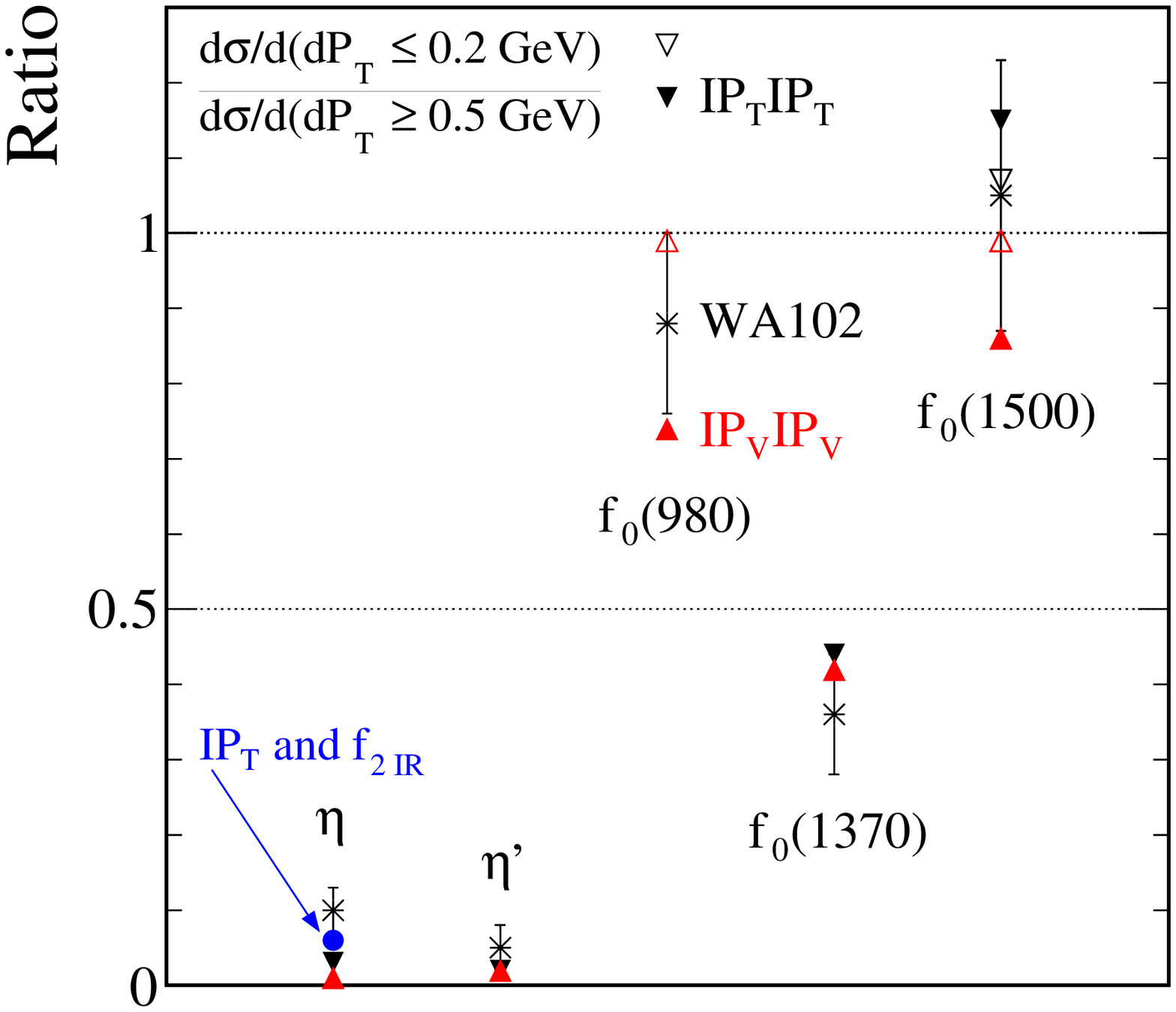}
  \caption{\label{fig:4a}
  \small
Differential cross section $d\sigma/d(dP_{\perp})$
for the central exclusive $\eta$ (panels (a) and (c)) 
and $\eta'$ (panel (b)) mesons production at $\sqrt{s} = 29.1$~GeV.
The WA102 experimental distributions from \cite{WA102_PLB427}
have been normalized to the mean values of the 
total cross sections from Table~\ref{tab:mesons}.
Results for the tensorial and vectorial pomeron models are presented.
For $\eta$ production the $f_{2 I\!\!R}$ exchanges in the amplitude were included in addition.
The $(l,S)$ contributions to the differential cross sections are also shown.
Panel (d) shows the ratio of production
at small $dP_{\perp}$ to large $dP_{\perp}$
for each pseudoscalar and scalar meson discussed in this paper and 
collected in Table~\ref{tab:ratio_dPt}.
Experimental results for the ratio are taken from Table~2 of Ref.~\cite{kirk00}.
For the $f_{0}(980)$ and $f_{0}(1500)$ meson production we show results 
obtained for the two sets of $(l,S)$ contributions
fitted to the experimental azimuthal angular correlations data
shown in Figs.~\ref{fig:dsig_dphi_980} and \ref{fig:dsig_dphi_1500}.
For the $f_{0}(980)$ and $f_{0}(1500)$ the filled points correspond to
$(g_{I\!\!P_{T} I\!\!P_{T} M}', g_{I\!\!P_{T} I\!\!P_{T} M}'') = (0.788, 4)$ and $(1.22, 6)$,
the open points to
$(g_{I\!\!P_{T} I\!\!P_{T} M}', g_{I\!\!P_{T} I\!\!P_{T} M}'') = (0.75, 5.5)$ and $(1, 10)$, respectively,
see Table~\ref{tab:couplings_S}.
}
\end{figure}

In Fig.~\ref{fig:4a} we present the ``glueball variable'' $dP_{\perp}$ distribution.
Theoretical predictions of $dP_{\perp}$ seem
to be qualitatively consistent with the WA102 data
presented in Table~2 of Ref.~\cite{kirk00}.
We show results for the mesons of interest to us in Table~\ref{tab:ratio_dPt}.
In addition, in Fig.~\ref{fig:4a}(d), the ratio of production
at small $dP_{\perp}$ to large $dP_{\perp}$ has been 
compared with the experimental results taken from \cite{kirk00}; see also \cite{WA102_PLB462}.
It can be observed that scalar mesons which could have a large 'gluonic component' 
have a large value for this ratio.
The fact that $f_{0}(1370)$ and $f_{0}(1500)$ have different $\phi_{pp}$
and $dP_{\perp}$ dependences confirms that these are not simply $J$ dependent phenomena. 
This is also true for the $J = 2$ states,
where the $f_{2}(1950)$ has a different $\phi_{pp}$ dependence compared to the $f_{2}(1270)$ 
and $f_{2}'(1525)$ states; see Fig.5 of~\cite{kirk00}.
The $dP_{\perp}$ and $\phi_{pp}$ effects are in our present work understood
as being due to the fact that in general more than one coupling structure,
$I\!\!P I\!\!P M$ respectively $I\!\!P I\!\!P \tilde{M}$, is possible.
It remains a challenge for theory to predict these coupling structures from
calculations in the framework of QCD.
\begin{table}
\caption{Results of meson production as a function of $dP_{\perp}$
expressed as a percentage of its total contribution 
at the WA102 collision energy $\sqrt{s}=29.1$~GeV.
The theoretical numbers quoted for $\eta$, $\eta'$ and $f_{0}(1370)$
correspond to the coupling parameters 
given in Tables~\ref{tab:couplings_PS} and \ref{tab:couplings_S}, respectively.
For the $f_{0}(980)$ and $f_{0}(1500)$ the numbers (those in parentheses)
correspond to the coupling parameters
which fit in Figs.~\ref{fig:dsig_dphi_980}(a) and \ref{fig:dsig_dphi_1500}(a), respectively,
the black filled (blue circle) points.
See the discussion of these figures in the text.
In the last column the ratios of 
$\frac{d\sigma/d(dP_{\perp} \leqslant \,0.2~\mathrm{GeV})}
{d\sigma/d(dP_{\perp}\geqslant \,0.5~\mathrm{GeV})}$ are given.
The experimental numbers are from Table~2 of Ref.~\cite{kirk00}.
}
\label{tab:ratio_dPt}
\begin{tabular}{|c|c|c|c|c|c|}
\hline
Meson & Exchanges & 
$dP_{\perp} \leqslant 0.2$~GeV & 
$0.2 \leqslant dP_{\perp} \leqslant 0.5$~GeV & 
$dP_{\perp}\geqslant 0.5$~GeV & Ratio\\
\hline
$\eta$ & $I\!\!P_{T}$ and $f_{2 I\!\!R}$ &
3.0 & 46.8 & 50.1 & 0.06\\
       & $I\!\!P_{T} I\!\!P_{T}$ &
1.8 & 33.4 & 64.8 & 0.03\\
       & $I\!\!P_{V} I\!\!P_{V}$ &
1.1 & 21.0 & 77.8 & 0.01\\
       & exp. &
$6 \pm 2$ & $34 \pm 2$ & $60 \pm 3$ & $0.10 \pm 0.03$\\
\hline
$\eta'$ & $I\!\!P_{T} I\!\!P_{T}$ &
1.4 & 28.3 & 70.4 & 0.02\\
        & $I\!\!P_{V} I\!\!P_{V}$ &
1.2 & 22.1 & 76.7 & 0.02\\
       & exp. &
$3 \pm 2$ & $32 \pm 2$ & $64 \pm 3$ & $0.05 \pm 0.03$\\
\hline
$f_{0}(980)$ & $I\!\!P_{T}$ and $f_{2 I\!\!R}$ &
25.3 & 59.2 & 15.2 & 1.67\\
       & $I\!\!P_{T} I\!\!P_{T}$ &
22.7 (23.9)& 57.9 (57.0)& 19.3 (19.1)& 1.18 (1.25)\\
       & $I\!\!P_{V} I\!\!P_{V}$ &
19.3 (21.6)& 54.9 (56.4)& 25.9 (21.9)& 0.74 (0.99)\\
       & exp. &
$23 \pm 2$ & $51 \pm 2$ & $26 \pm 3$ & $0.88 \pm 0.12$\\
\hline
$f_{0}(1370)$ & $I\!\!P_{T} I\!\!P_{T}$ &
15.5 & 49.0 & 35.5 & 0.44\\
              & $I\!\!P_{V} I\!\!P_{V}$ &
15.2 & 48.5 & 36.3 & 0.42\\
       & exp. &
$18 \pm 4$ & $32 \pm 2$ & $50 \pm 3$ & $0.36 \pm 0.08$\\
\hline 
$f_{0}(1500)$ & $I\!\!P_{T} I\!\!P_{T}$ &
22.5 (23.7)& 57.8 (54.3)& 19.7 (22.0)& 1.15 (1.07)\\
              & $I\!\!P_{V} I\!\!P_{V}$ &
20.4 (22.4)& 56.0 (54.9)& 23.6 (22.7)& 0.86 (0.99)\\
       & exp. &
$24 \pm 2$ & $54 \pm 3$ & $22 \pm 4$ & $1.05 \pm 0.18$\\
\hline 
\end{tabular}
\end{table}

\begin{figure}[!ht]
(a)\includegraphics[width = 0.43\textwidth]{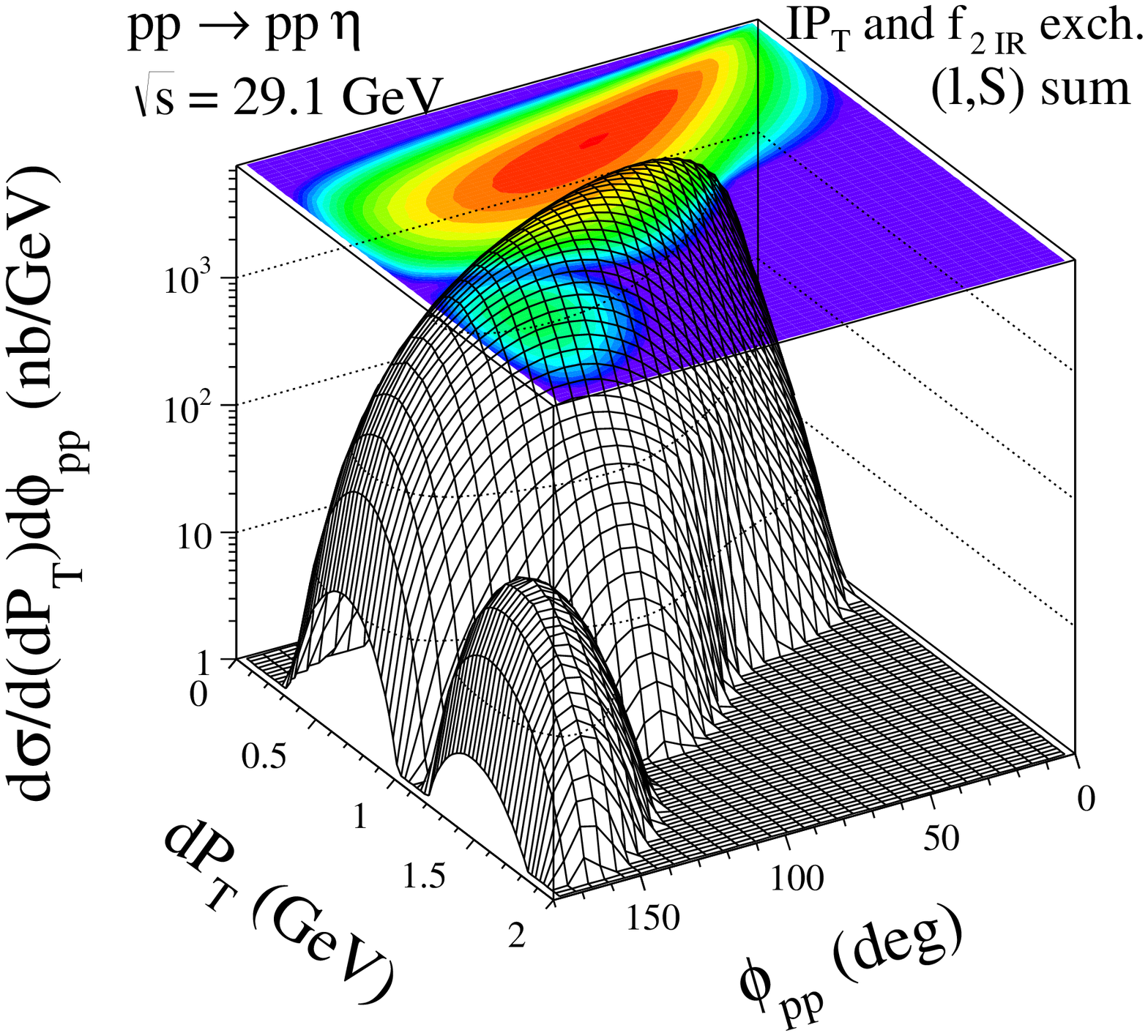}
(b)\includegraphics[width = 0.43\textwidth]{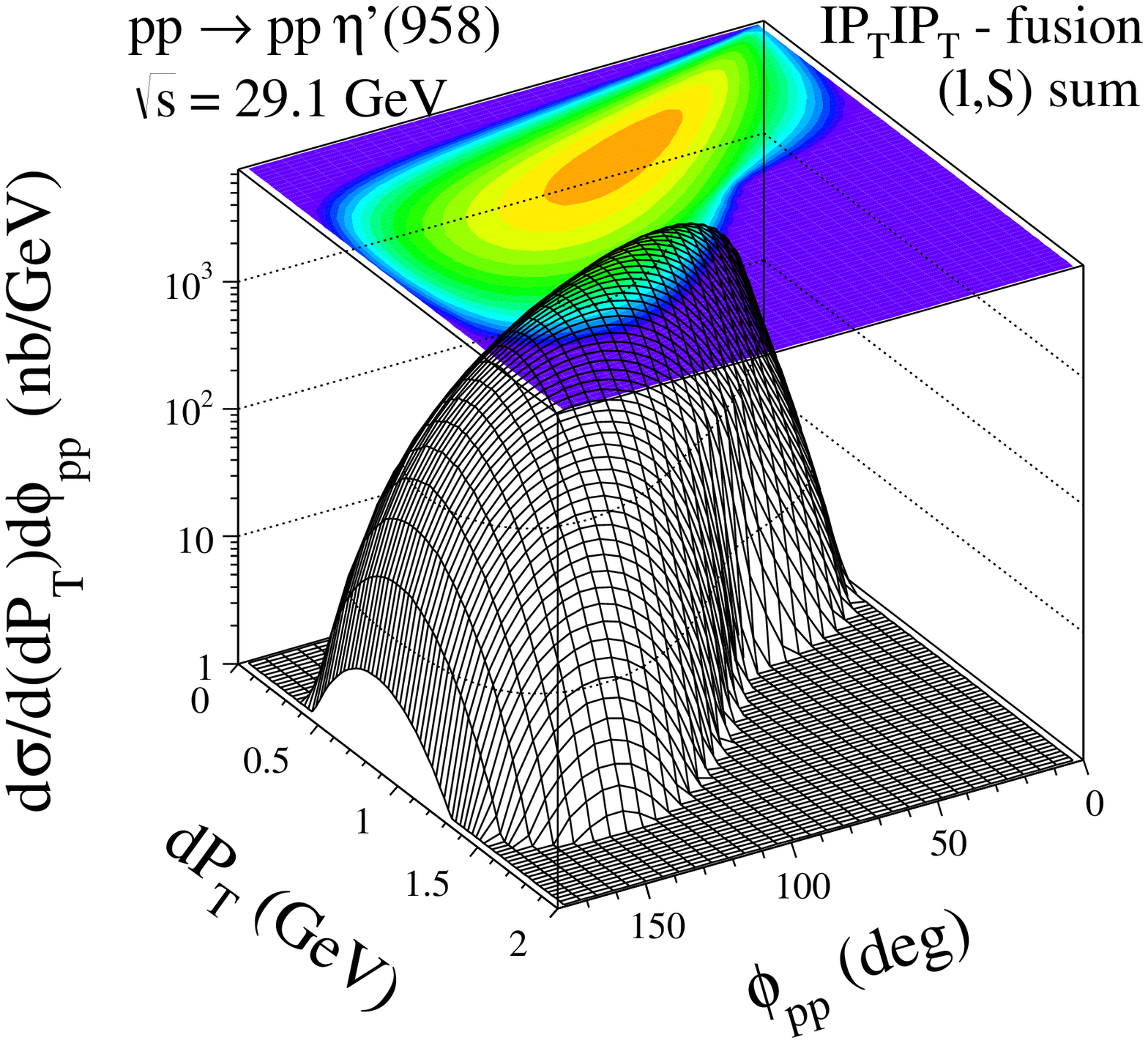}
(c)\includegraphics[width = 0.43\textwidth]{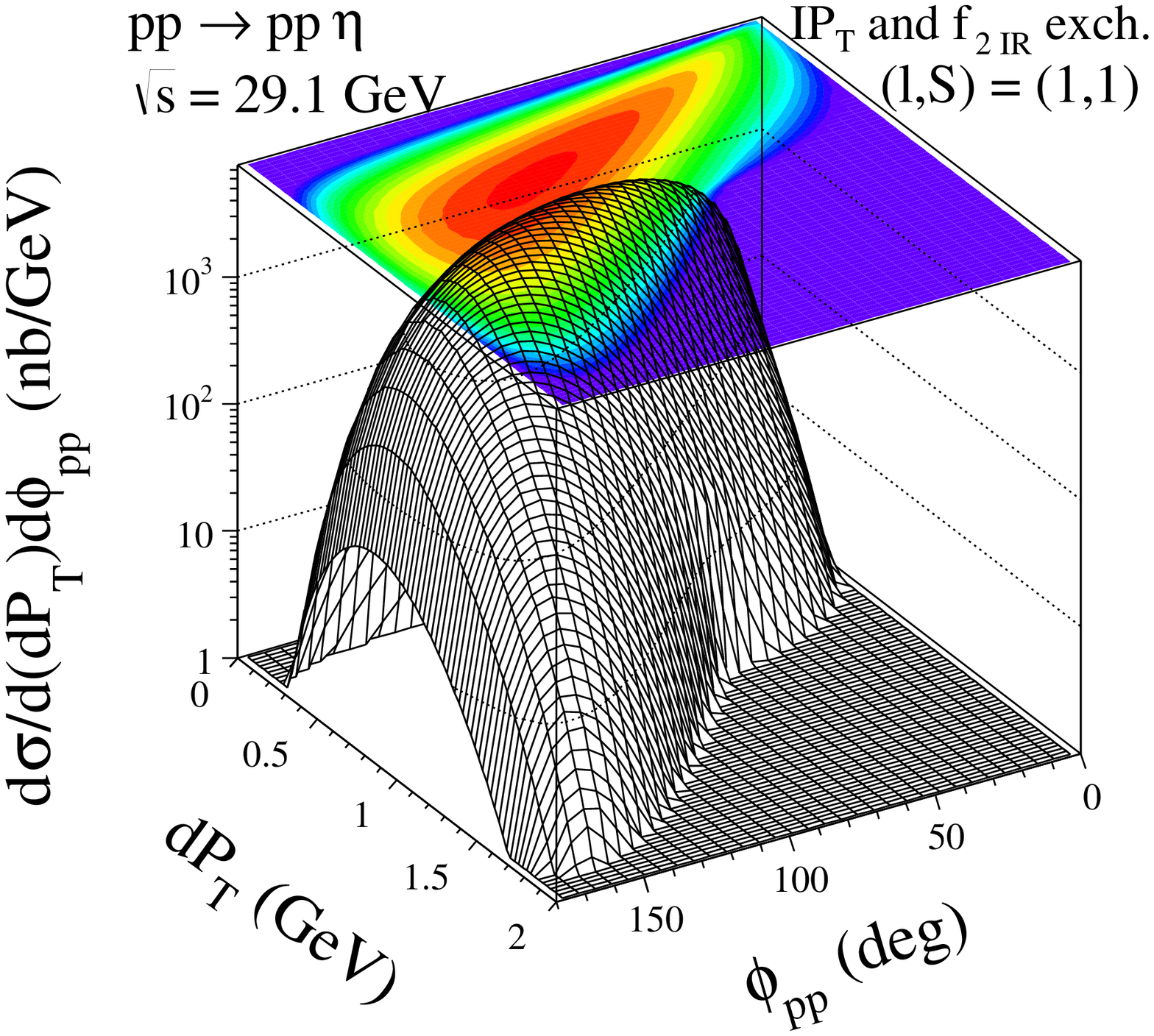}
(d)\includegraphics[width = 0.43\textwidth]{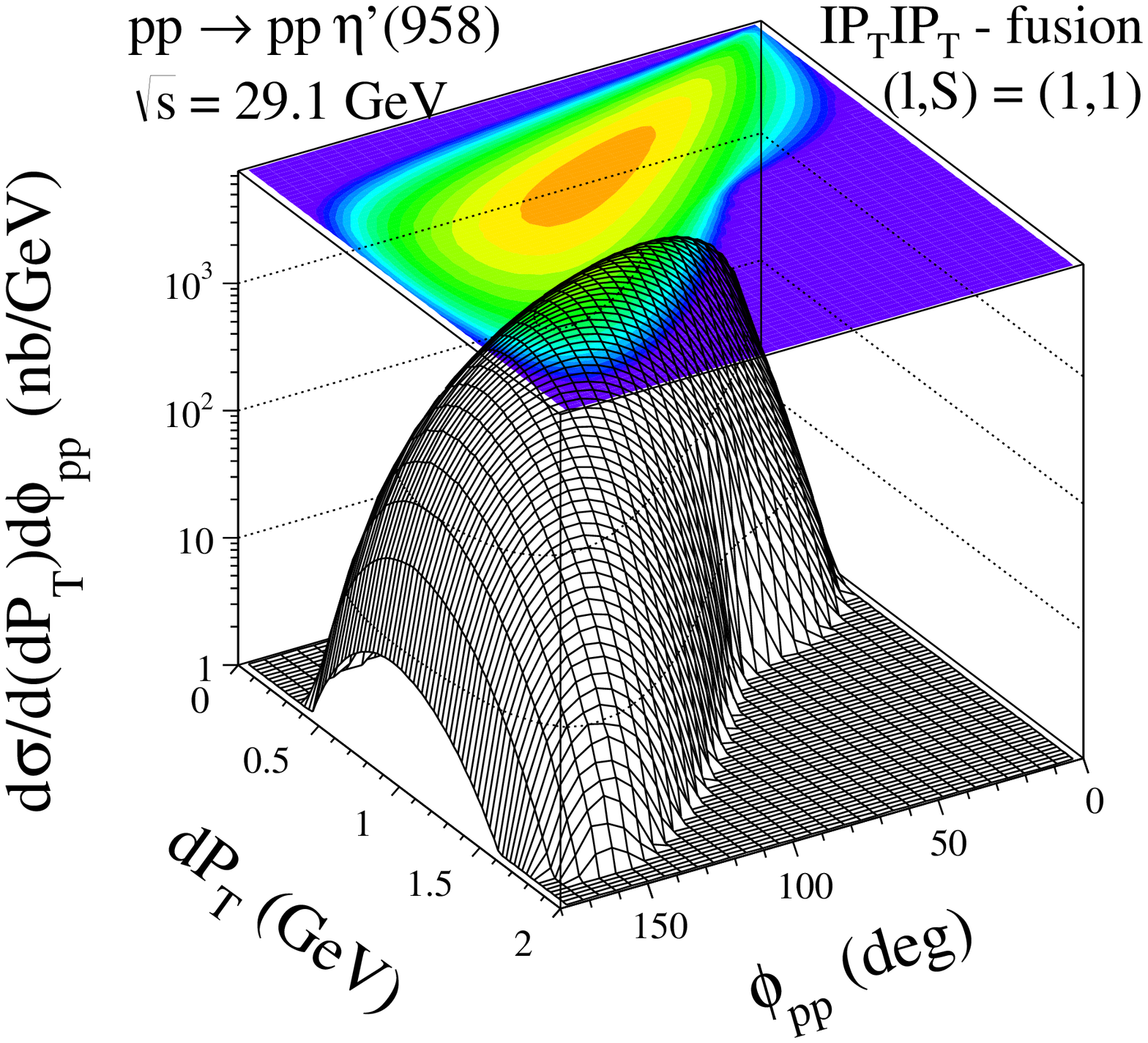}
(e)\includegraphics[width = 0.43\textwidth]{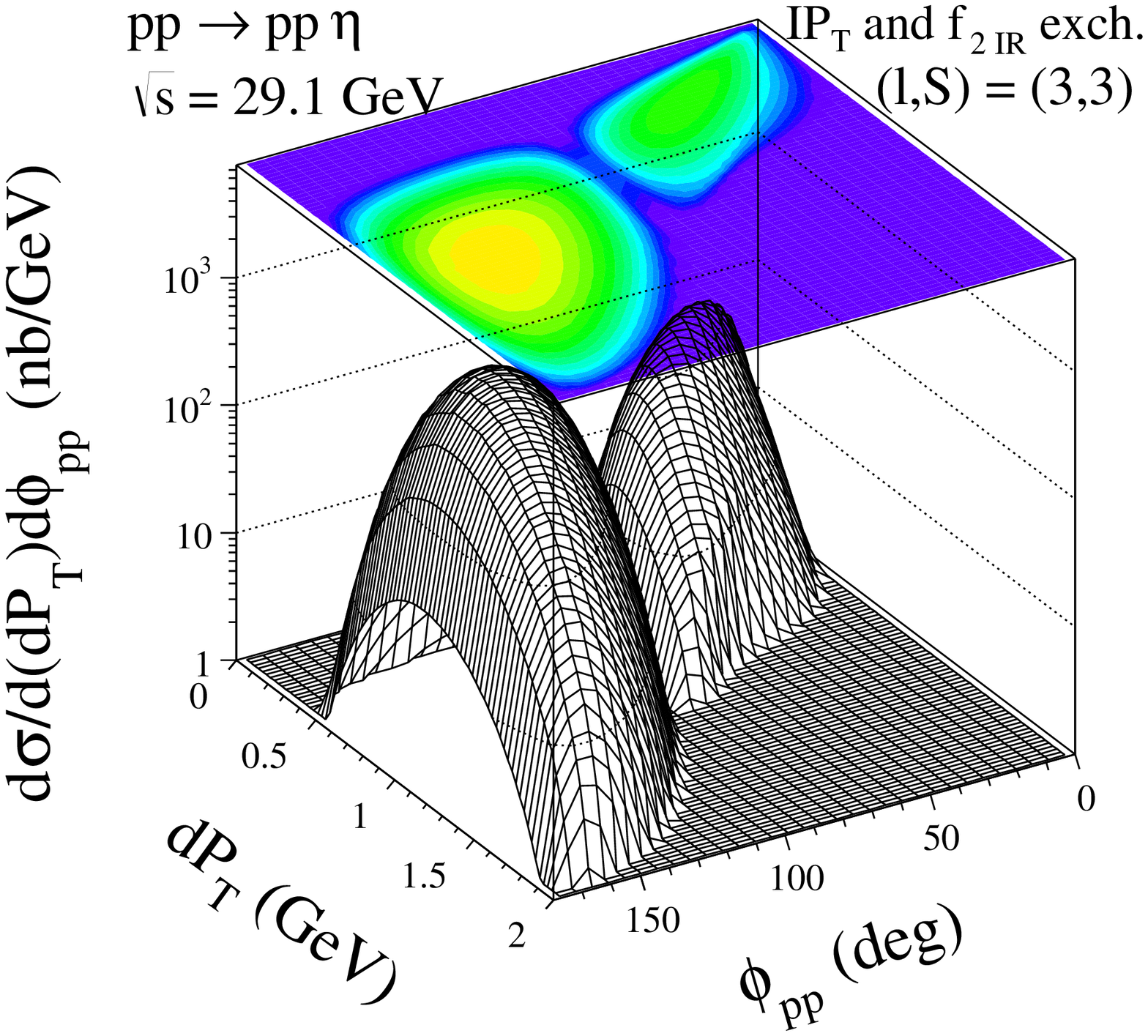}
(f)\includegraphics[width = 0.43\textwidth]{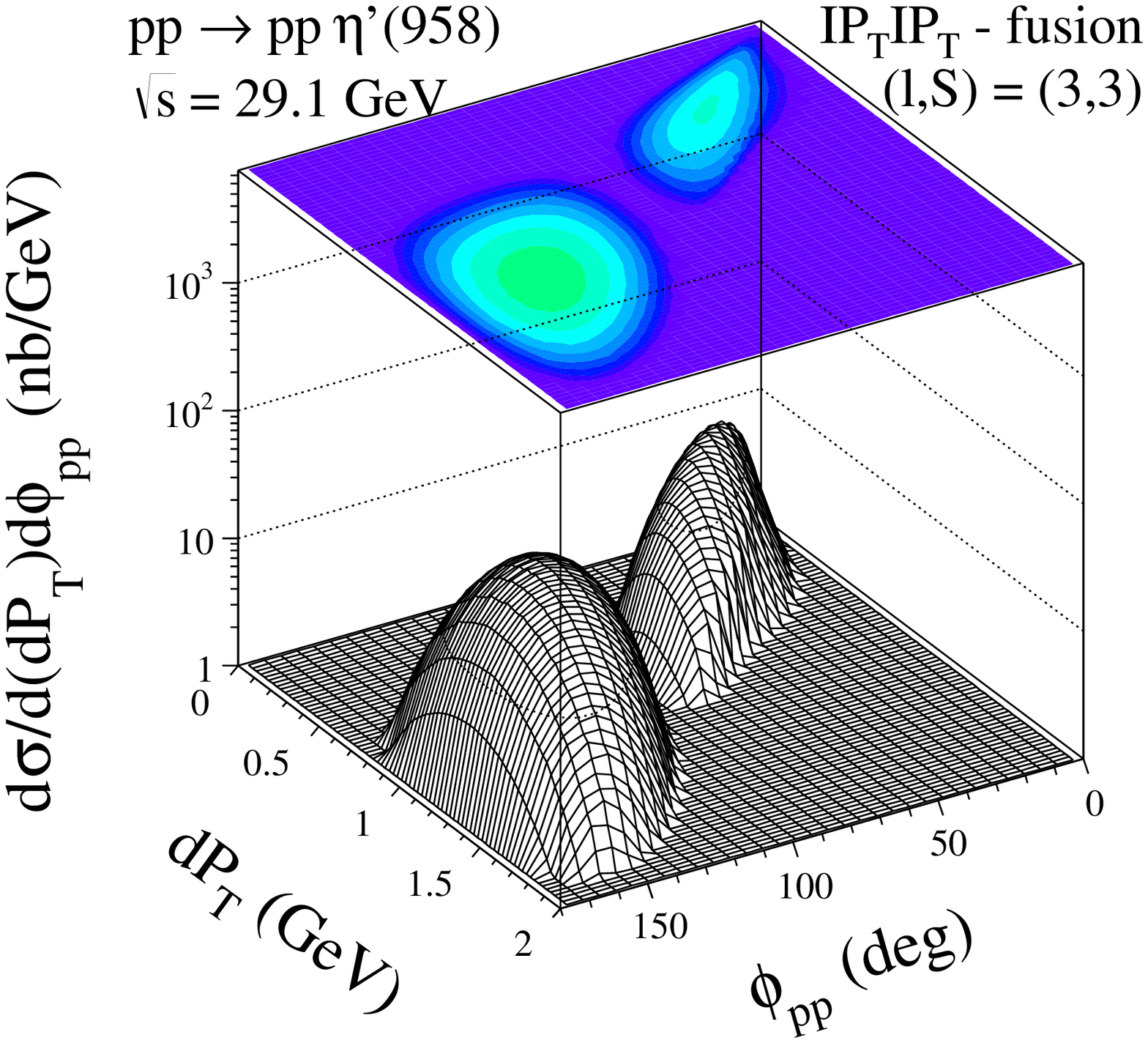}
  \caption{\label{fig:map_ptperpphi_PS}
  \small
Distributions in ($dP_{\perp}, \phi_{pp}$)
for the $\eta$ (left panels) and $\eta'(958)$ (right panels) meson production at $\sqrt{s} = 29.1$~GeV,
Results for $\eta$ meson correspond to the model with the tensor pomeron and $f_{2}$-reggeon exchanges
while $\eta'$ meson production to the model with tensorial pomeron only.
The individual contributions of $(l,S) = (1,1)$ (panels (c) and (d))
and $(l,S) = (3,3)$ (panels (e) and (f)) are shown separately.
}
\end{figure}

In Fig.~\ref{fig:map_ptperpphi_PS} we show two-dimensional
distributions in ($dP_{\perp}, \phi_{pp}$) 
for the $\eta$ (left panels) and $\eta'(958)$ (right panels) meson production
in the fusion of two tensor pomerons.
In panels (a) and (b) we show the result for $(l,S)$ components added coherently.
In panels (c) - (d) and (e) - (f) we show the individual spin components 
for $(l,S) = (1,1)$ and $(3,3)$, respectively. 
By comparing panels (a) - (f) we infer that the interference effects are rather large.

\begin{figure}[!ht]
\includegraphics[width = 0.32\textwidth]{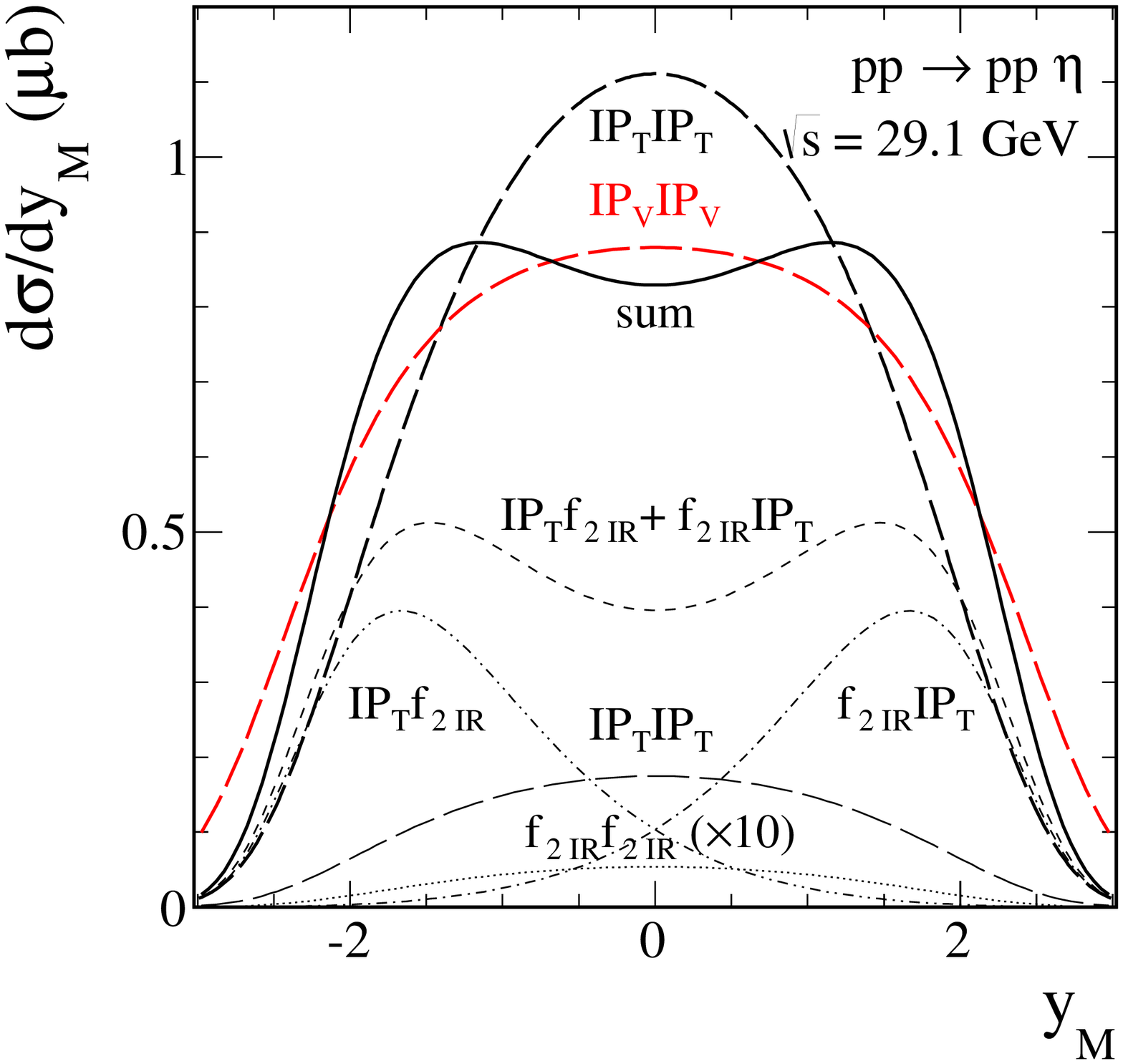}
\includegraphics[width = 0.32\textwidth]{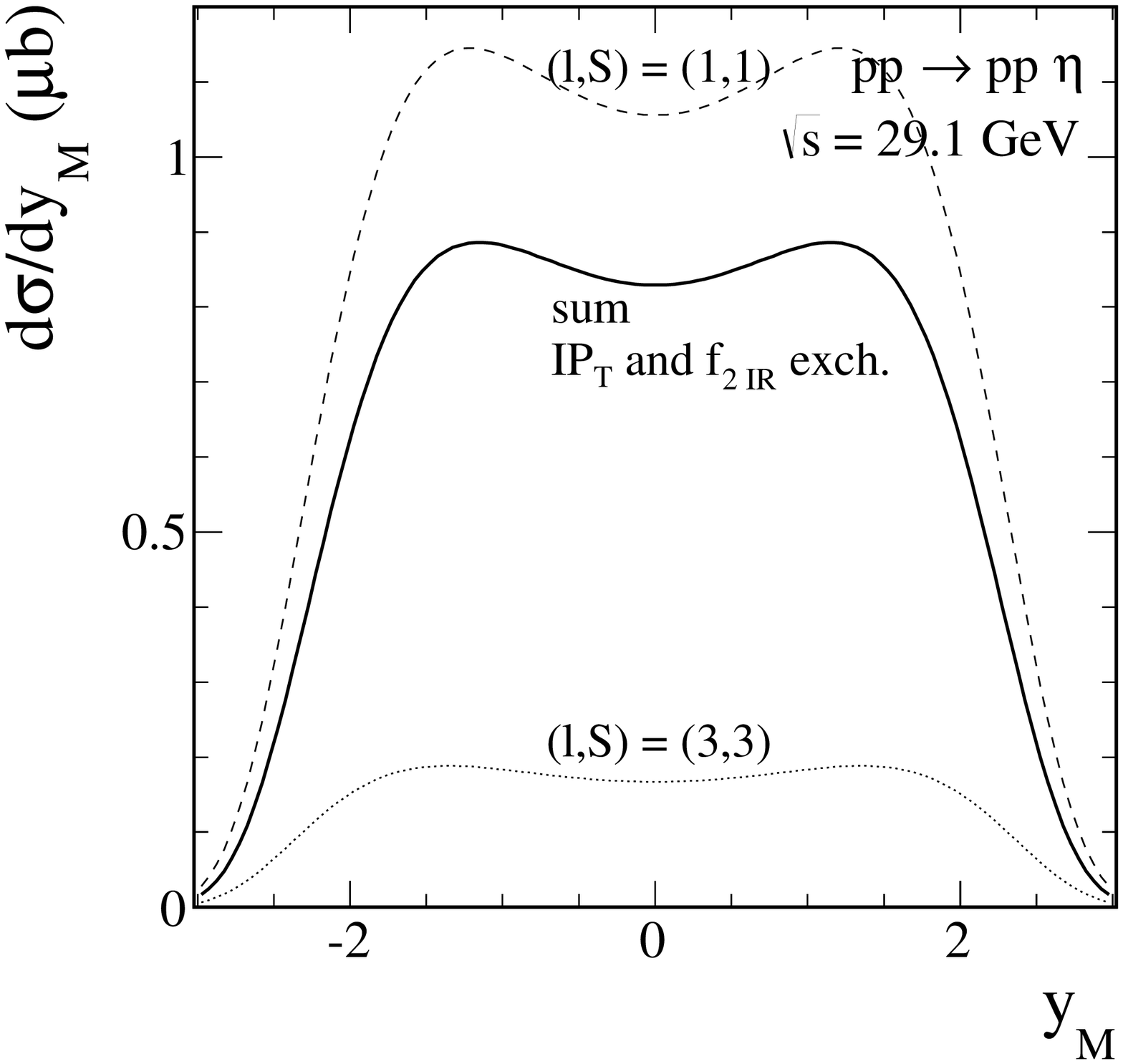}
\includegraphics[width = 0.32\textwidth]{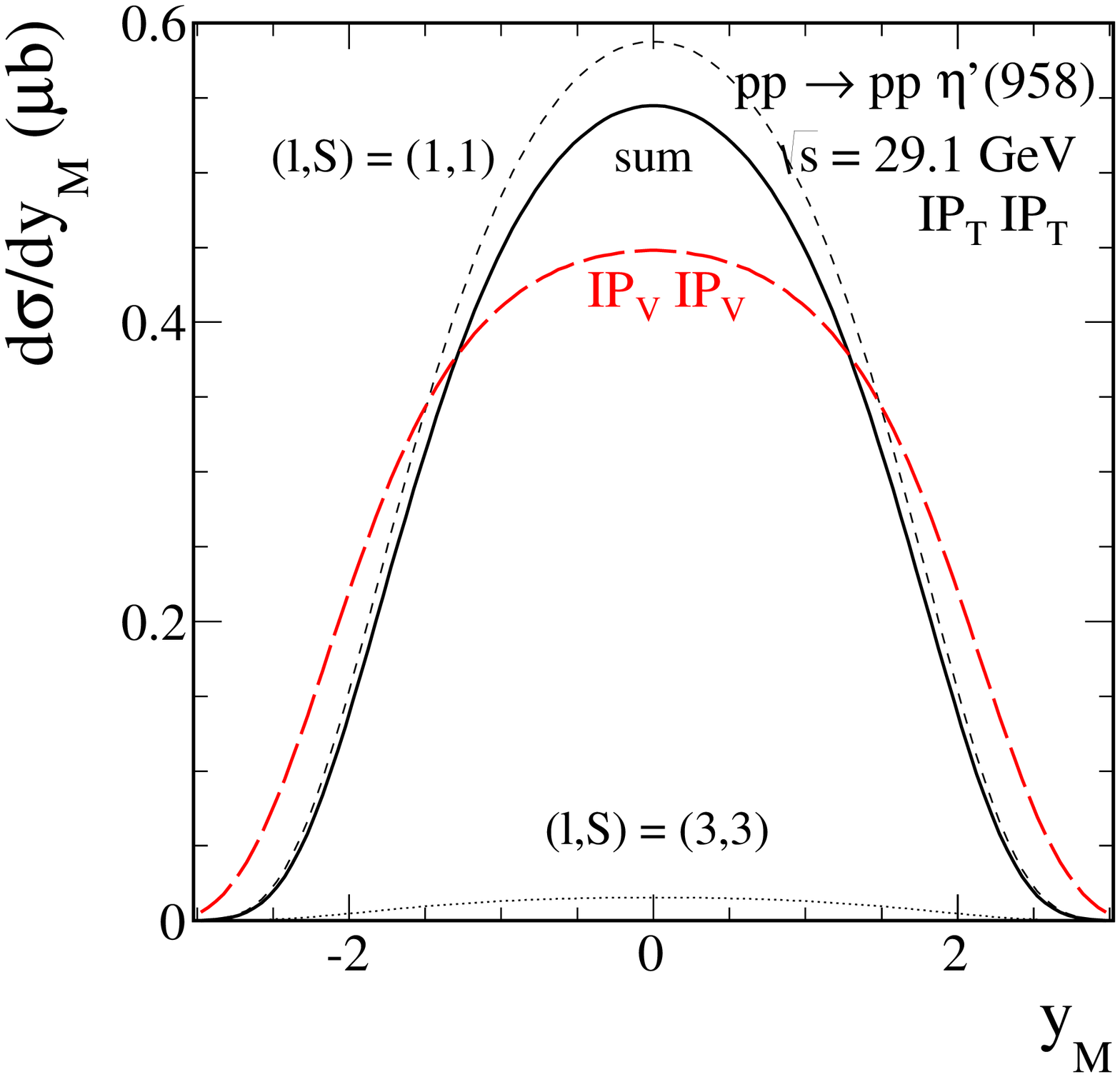}\\
\includegraphics[width = 0.32\textwidth]{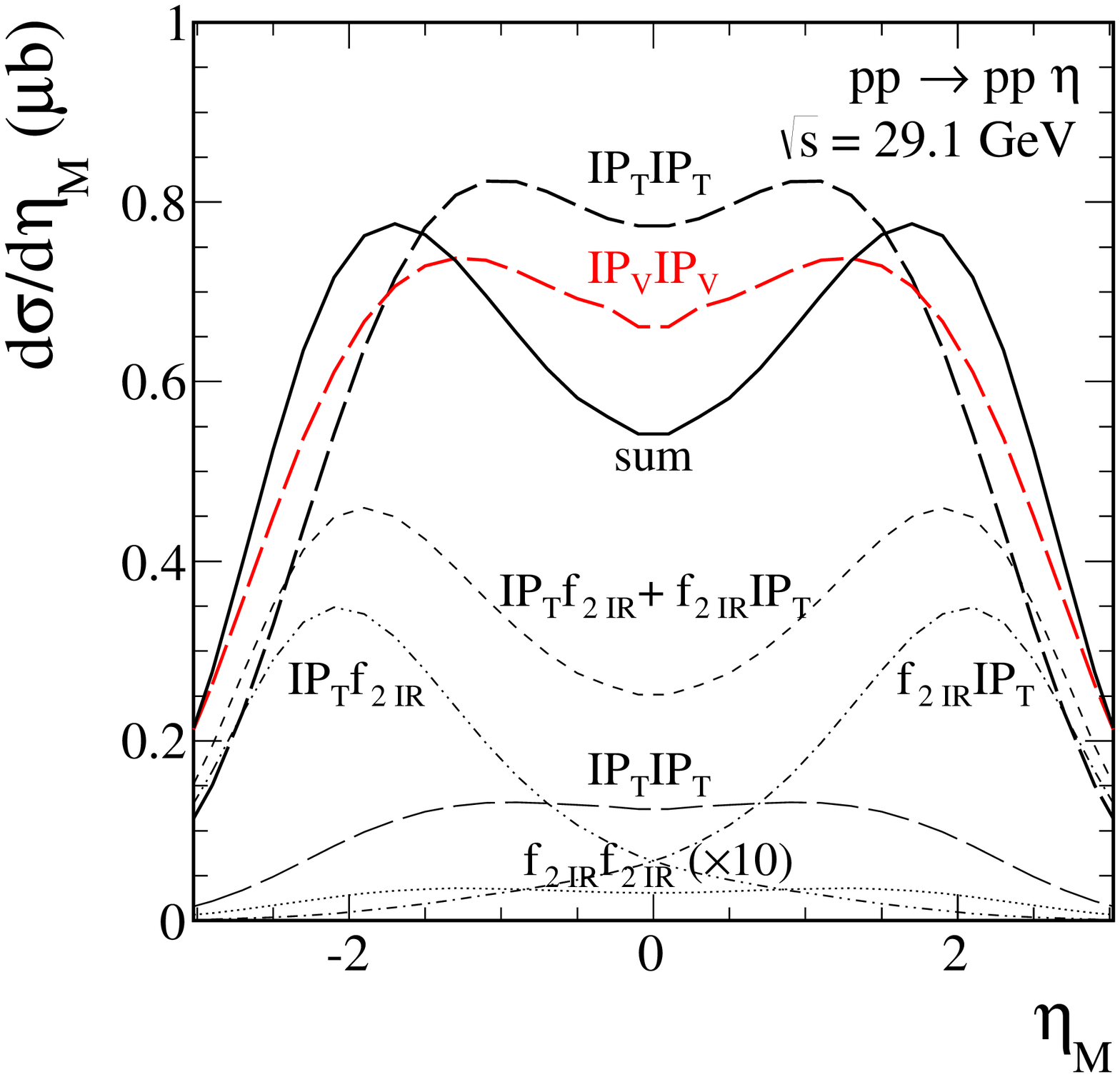}
\includegraphics[width = 0.32\textwidth]{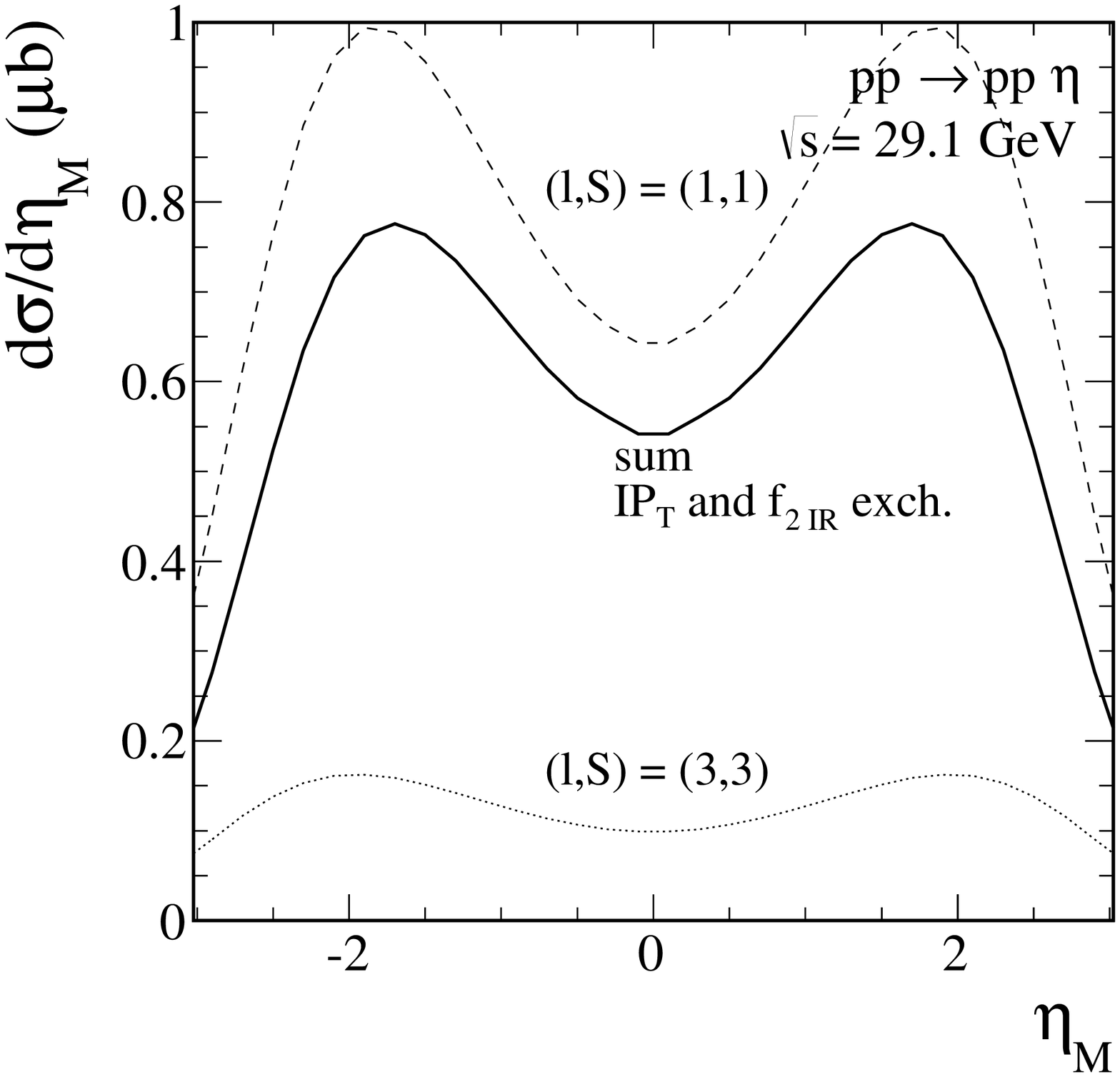}
\includegraphics[width = 0.32\textwidth]{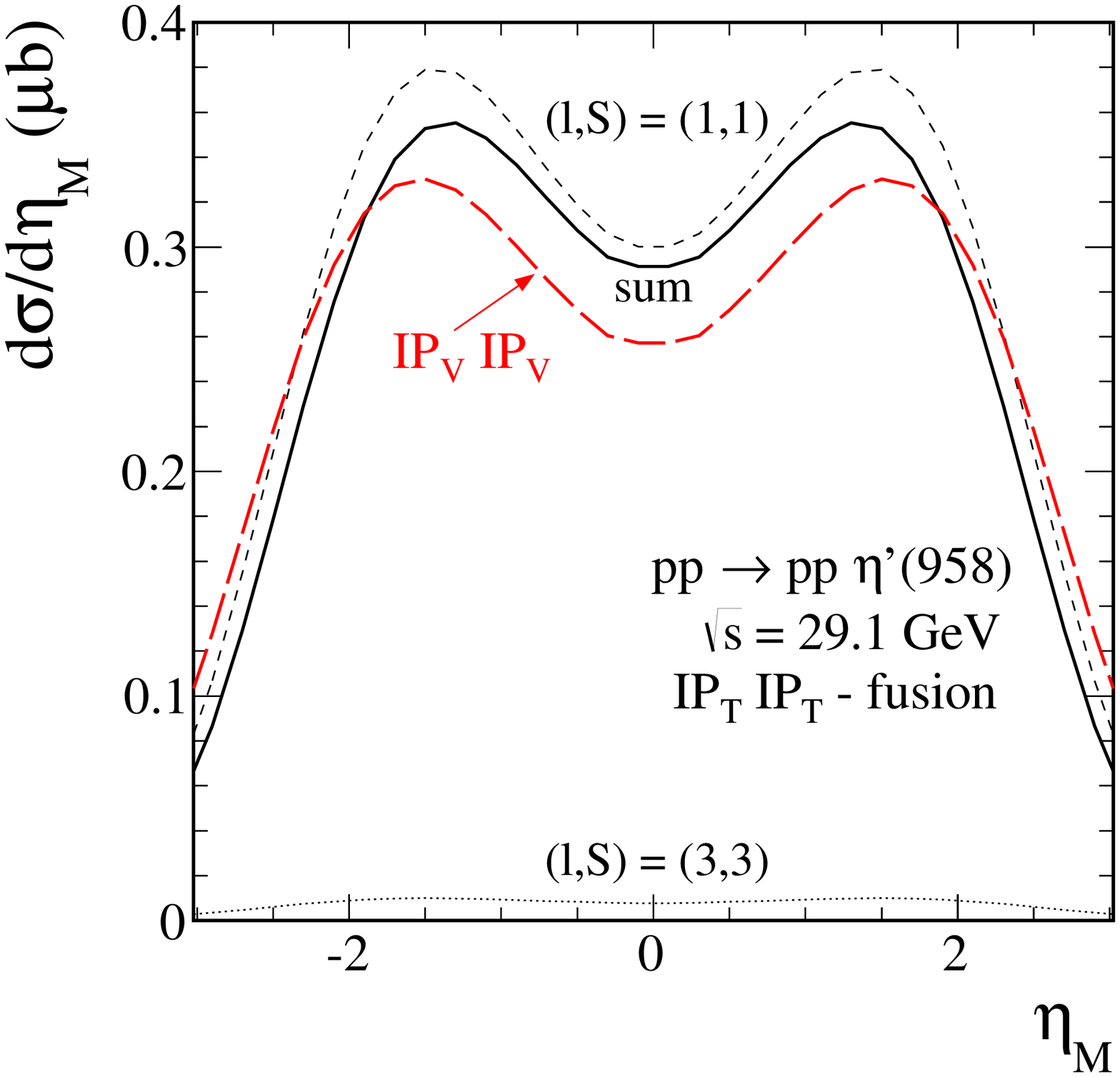}
  \caption{\label{fig:5}
  \small
Differential cross section $d\sigma/d\mathrm{y}_{M}$ (top panels) 
and $d\sigma/d\eta_{M}$ (bottom panels)
for the $\eta$ and $\eta'$ production at $\sqrt{s} = 29.1$~GeV.
The solid line corresponds to the model with tensorial pomeron
while the long-dashed line to the model with vectorial pomeron.
The different lines correspond to the situation when all 
or only some components of the pomeron and $f_{2 I\!\!R}$ exchanges 
in the amplitude are included 
(the pomeron-pomeron component dominates at midrapidities of $\mathrm{y}_{M}$
and the pomeron-reggeon (reggeon-pomeron) 
peaks at backward (forward) rapidities of $\mathrm{y}_{M}$, respectively). 
}
\end{figure}

For completeness, differential distributions in the $\eta$ or $\eta'$ 
rapidity (top panels) and pseudorapidity (bottom panels)
are shown in Fig.~\ref{fig:5} for the two models of the pomeron exchanges.


\section{Conclusions}
\label{section:Conclusions}

We have analyzed proton-proton collisions with the exclusive central production of
scalar and pseudoscalar mesons.
We analyzed the predictions of two different models of the soft pomeron. 
The first one is the commonly used model with vectorial pomeron which is, 
however, difficult to be supported from a theoretical point of view.
The second one is a recently proposed model of tensorial pomeron, which,
in our opinion, has better theoretical foundations.
We have presented formulae for corresponding 
pomeron-pomeron-meson vertices and amplitudes for the $pp \to pMp$ reaction.
In general, different couplings with different orbital angular momentum and spin of 
two ``pomeron particles'' are possible. In most cases one has to
add coherently amplitudes for two couplings. The corresponding
coupling constants are not known and have been fitted to existing experimental data.  

We have performed calculations of several differential distributions.
We wish to emphasize that the tensorial pomeron can - at least - equally well describe
experimental data on the exclusive meson production discussed here as the less 
theoretically justified vectorial pomeron frequently used in the literature. 
This has been illustrated for the production of several scalar and pseudoscalar mesons.
The existing low-energy experimental data do not allow to
clearly distinguish between the two models as the presence of subleading
reggeon exchanges is at low energies very probable for many reactions. 
This seems to be the case for the $\eta$ meson production.
In these cases we have included in our analysis also exchanges of subleading 
trajectories which improve the agreement with experimental data.
Production of $\eta'$ meson seems to be less affected by contributions from subleading exchanges.

Now we list some issues which deserve further studies but are beyond the scope of our present paper.
For the resonances decaying e.g. into the $\pi \pi$ channel an interference
of the resonance signals with the two-pion continuum has to be included in addition. 
This requires a consistent model of the resonances and the non-resonant background.
It would be very interesting to see if the exchange of tensorial pomerons may
modify differential distributions for the $\pi^+ \pi^-$ continuum
compared to the previous calculations \cite{LS10, LPS11}.
Furthermore, absorption effects 
are frequently taken into account by simply multiplying
cross sections with a gap survival factor. 
But absorption effects may also change the shapes
of $t_1/t_2$, $\phi_{pp}$, etc. distributions.
The deviation from ``bare'' distributions probably is more significant at high energies
where the absorptive corrections should be more important.
Consistent inclusion of these effects clearly goes beyond the scope
of the present study where we have limited ourselves to simple Born term calculations
at the WA102 collision energy.
It would clearly be interesting to extend the studies of central meson production
in diffractive processes to higher energies,
where the dominance of the pomeron exchange can be better justified.

To summarise: our study of scalar and pseudoscalar meson production
certainly shows the potential of these reactions for testing the nature of the soft pomeron.
Pseudoscalar meson production could be of particular interest
in this respect since there the distribution in the azimuthal angle $\phi_{pp}$
between the two outgoing protons may contain, for the tensorial pomeron,
a term which is not possible for the vectorial pomeron; see the discussion
after (\ref{vertex_pomVpomVPS}) 
and after (\ref{couplings_pseudoscalar}) 
in Section~\ref{subsection:Scalar_and_pseudoscalar_meson_production}.
Clearly, our study can be extended to the central production of other mesons like
the $f_{2}(1270)$.
We hope to come back to this issue in a future publication.

Our main aim with these studies is to provide detailed models for central
meson production, for both the tensorial and the vectorial pomeron ansatz,
where all measurable distributions of the particles in the final state can be calculated.
The models contain only a few free coupling parameters to be determined by experiment.
The hope is, of course, that future experiments will be able to select
the correct soft pomeron model. In any case, our models should provide good ``targets''
for experimentalists to shoot at.
Supposing that one model survives the experimental tests
we have then the theoretical challenge of deriving the corresponding
$I\!\!P I\!\!P M$ coupling constants from QCD.

Future experimental data on exclusive meson production at high
energies should thus provide good information on the spin structure of 
the pomeron and on its couplings to the nucleon and the mesons.
On the other hand, the low energy data could help in understanding the role of 
subleading trajectories. 
Several experimental groups, 
e.g. COMPASS \cite{COMPASS}, STAR \cite{STAR}, CDF \cite{CDF}, ALICE \cite{ALICE}, ATLAS \cite{SLTCS11} 
have the potential to make very significant contributions to this program aimed at
understanding the coupling and the spin structure of the soft pomeron.

\vspace{0.5cm}
{\bf Acknowledgments}
We are indebted to C. Ewerz, K. Kochelev, and R. Schicker for useful discussions.
Piotr Lebiedowicz is thankful to the Wilhelm and Else Heraeus - Foundation
for warm hospitality during his stay at
WE-Heraeus-Summerschool \textit{Diffractive and electromagnetic processes at high energies}
in Heidelberg when this work was completed.
This work was partially supported by the Polish grants:
DEC-2011/01/B/ST2/04535, DEC-2013/08/T/ST2/00165, and PRO-2011/01/N/ST2/04116.
\appendix

\section{Tensorial pomeron}
\label{section:Tensorial_Pomeron}

For the case of the tensorial pomeron the $I\!\!P p p$ vertex
and the $I\!\!P$ propagator read as follows, see \cite{talkN} and \cite{EMN13},
\begin{eqnarray}
i\Gamma_{\mu \nu}^{(I\!\!P pp)}(p',p)=
-i 3 \beta_{I\!\!P NN} F_{1}\bigl((p'-p)^2\bigr)
\left\lbrace 
\frac{1}{2} 
\left[ \gamma_{\mu}(p'+p)_{\nu} 
     + \gamma_{\nu}(p'+p)_{\mu} \right]
- \frac{1}{4} g_{\mu \nu} ( p\!\!\!/' + p\!\!\!/ )
\right\rbrace , \;\;\;\;\;\;\;
\label{vertex_pomNN}
\end{eqnarray}
where $\beta_{I\!\!P NN} = 1.87$~GeV$^{-1}$ and $p\!\!\!/ = \gamma_{\mu} p^{\mu}$.
The explicit factor 3 above counts the number of valence quarks in each proton.
Following Donnachie and Landshoff \cite{DL} we use in (\ref{vertex_pomNN})
for describing the proton's extension
the proton's Dirac electromagnetic form factor $F_{1}(t)$.
A good representation of this form factor is given by the dipole formula
\begin{eqnarray}
F_{1}(t)= \frac{4 m_{p}^{2}-2.79\,t}{(4 m_{p}^{2}-t)(1-t/m_{D}^{2})^{2}} \,,
\label{Fpomproton}
\end{eqnarray}
where $m_{p}$ is the proton mass and $m_{D}^{2} = 0.71$~GeV$^{2}$
is the dipole mass squared.

The propagator of the tensor-pomeron exchange is given by
\begin{eqnarray}
i\Delta_{\mu \nu, \kappa \lambda}^{(I\!\!P)}(s,t)=
\frac{1}{4s} 
\left( g_{\mu \kappa} g_{\nu \lambda} + g_{\mu \lambda} g_{\nu \kappa}
-\frac{1}{2} g_{\mu \nu} g_{\kappa \lambda} \right)
\left(-i s \alpha'_{I\!\!P}\right)^{\alpha_{I\!\!P}(t)-1}\,;
\label{prop_pom}
\end{eqnarray}
see \cite{EMN13}.
Here the pomeron trajectory $\alpha_{I\!\!P}(t)$ is assumed to be 
of standard form, see for instance \cite{DDLN}, that is, linear in $t$
and with intercept slightly above 1:
\begin{eqnarray}
&&\alpha_{I\!\!P}(t) = \alpha_{I\!\!P}(0)+\alpha'_{I\!\!P}\,t \,,\nonumber\\
&&\alpha_{I\!\!P}(0) = 1.0808,\quad \alpha'_{I\!\!P} = 0.25 \; \mathrm{GeV}^{-2}\,.
\label{pomeron_trajectory}
\end{eqnarray}
%
The tensor-pomeron propagator fulfils the following relations
\begin{eqnarray}
&&\Delta_{\mu \nu, \kappa \lambda}^{(I\!\!P)}=
\Delta_{\nu \mu, \kappa \lambda}^{(I\!\!P)}=
\Delta_{\mu \nu, \lambda \kappa}^{(I\!\!P)}=
\Delta_{\kappa \lambda, \mu \nu}^{(I\!\!P)}\,,\nonumber\\
&&g^{\mu \nu} \Delta_{\mu \nu, \kappa \lambda}^{(I\!\!P)} = 0, \quad
g^{\kappa \lambda} \Delta_{\mu \nu, \kappa \lambda}^{(I\!\!P)} = 0\,.
\label{eqn2_delta}
\end{eqnarray}

Using now (\ref{vertex_pomNN}) - (\ref{pomeron_trajectory})
we can calculate the pomeron contribution to
the amplitude of $pp$ elastic scattering
\begin{eqnarray}
p(p_{a},\lambda_{a}) + p(p_{b},\lambda_{b}) \to
p(p_{1},\lambda_{1}) + p(p_{2},\lambda_{2}) \,.
\label{2to2}
\end{eqnarray}
With tensorial pomeron we get for the $\cal{T}$-matrix element
\begin{eqnarray}
&&
\Braket{p(p_{1},\lambda_{1}),p(p_{2},\lambda_{2})|{\cal T}|p(p_{a},\lambda_{a}),p(p_{b},\lambda_{b})}
\mid_{I\!\!P} \;\equiv \nonumber\\
&&{\cal M}^{2 \to 2}_{\lambda_{a}\lambda_{b} \to \lambda_{1}\lambda_{2}}\mid_{I\!\!P} \;=
(-i) 
\bar{u}(p_{1},\lambda_{1}) i \Gamma_{\mu_{1} \nu_{1}}^{(I\!\!P pp)} (p_{1},p_{a}) u(p_{a},\lambda_{a})
\nonumber \\  
&&\qquad \qquad \qquad \qquad  \times 
i \Delta^{(I\!\!P) \, \mu_{1} \nu_{1}, \mu_{2} \nu_{2}}(s,t)
\nonumber \\ 
&&\qquad \qquad \qquad \qquad  \times 
\bar{u}(p_{2},\lambda_{2}) i \Gamma_{\mu_{2} \nu_{2}}^{(I\!\!P pp)} (p_{2},p_{b}) u(p_{b},\lambda_{b})\,,
\label{2to2_matrix}
\end{eqnarray}
where
\begin{eqnarray}
&& s = (p_{a} + p_{b})^{2} = (p_{1} + p_{2})^{2} \,,\nonumber\\
&& t = (p_{1} - p_{a})^{2} = (p_{2} - p_{b})^{2} \,.
\label{2to2_kinematics}
\end{eqnarray}
Inserting in (\ref{2to2_matrix}) the expressions for the $I\!\!P pp$ vertex (\ref{vertex_pomNN})
and the $I\!\!P$ propagator (\ref{prop_pom}) we get at high energies, $s \gg m_{p}^{2}$,
\begin{eqnarray}
{\cal M}^{2 \to 2}_{\lambda_{a}\lambda_{b} \to \lambda_{1}\lambda_{2}}\mid_{I\!\!P} &\cong&
i \, 2s \, \left[ 3 \beta_{I\!\!P NN} \,  F_{1}(t) \right]^{2} \,
\left(-i s \alpha'_{I\!\!P}\right)^{\alpha_{I\!\!P}(t)-1} \,
\delta_{\lambda_{1}\lambda_{a}} \, \delta_{\lambda_{2}\lambda_{b}} \,.
\label{2to2_limit}
\end{eqnarray}
This is exactly the same expression as obtained
with the famous Donnachie-Landshoff-pomeron approach;
see \cite{DL, DDLN}, and Appendix~\ref{section:Vectorial_Pomeron} below.
One advantage of the tensorial-pomeron ansatz
is that it gives automatically, just using the rules of QFT,
the same $I\!\!P$ contributions to the amplitudes
of proton-proton and proton-antiproton scattering; see \cite{EMN13}.

\begin{figure}[!ht]
\includegraphics[width=0.45\textwidth]{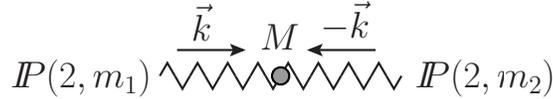}
  \caption{\label{pomMpom}
  \small
The fictitious reaction of two ``real spin 2 pomerons'' 
of momenta $\vec{k}$ and $-\vec{k}$ annihilating to a meson $M$.}
\end{figure}

We turn now to the $I\!\!P I\!\!P M$ vertices which
we want to construct in a field-theoretic manner, that is,
using a meson field operator and two effective pomeron field operators $I\!\!P_{\mu \nu}(x)$.
To get an overview of the possible couplings
of this type we shall first consider a fictitious reaction:
two ``real pomeron particles'' of spin 2 giving a meson $M$; see Fig.~\ref{pomMpom}.
From this exercise we can then easily learn how to classify and
write down covariant expressions for the $I\!\!P I\!\!P M$ vertices.

We consider, thus, the annihilation of two ``pomeron particles'' of spin 2
and $z$-components of spin $m_{1}$ and $m_{2}$ giving a meson of spin $J$
and $z$-component $J_{z}$ in the c.m.~system, that is, the rest system of $M$:
\begin{eqnarray}
&& I\!\!P (\vec{k},2,m_{1}) + I\!\!P (-\vec{k},2,m_{2}) \to M(J,J_{z}) \,, \nonumber\\
&& m_{1,2} \in \{-2, \ldots, 2 \} \,, 
\quad J_{z} \in \{ -J, \ldots, J \} \,.
\label{pom_to_pom}
\end{eqnarray}
Note that we use here the Wigner basis for all particles;
see \cite{Wigner}, and for instance, chapter 16.2 of \cite{Nachtmann90}, 
and Appendix~\ref{section:Covariant_Couplings}.
Clearly, in (\ref{pom_to_pom}) $M$ must have isospin and $G$ parity $I^{G} = 0^{+}$
and charge conjugation $C = +1$.
The question is: what are the possible values of spin $J$ and parity $P$
for meson $M$?

Let $a_{2,m}^{\dagger}(\vec{k})$, $a_{2,m}^{\dagger}(-\vec{k})$ be the creation operators
for the ``pomeron particles''.
We can first construct the states of the two ``pomerons''
with definite orbital angular momentum $l$, $l_{z}$ and then those with
given $l$, $l_{z}$ and total spin $S$, $S_{z}$.
We get with $\hat{k} = \vec{k}/|\vec{k}|$, $Y_{l}^{l_{z}}(\hat{k})$
the spherical harmonics, and the usual Clebsch-Gordan coefficients
\begin{eqnarray}
\left\vert 2,m_{1};2,m_{2};l,l_{z} \right\rangle =
\int d\Omega_{k} \,Y_{l}^{l_{z}}(\hat{k}) \, a_{2,m_{1}}^{\dagger}(\vec{k}) \,a_{2,m_{2}}^{\dagger}(-\vec{k}) 
\left\vert 0 \right\rangle \,,
\label{formula_A1_11}
\end{eqnarray}
\begin{eqnarray}
\left\vert S,S_{z};l,l_{z} \right\rangle =
\sum_{m_{1},m_{2}} 
\left\langle 2,m_{1};2,m_{2} \vert S,S_{z} \right\rangle
\left\vert 2,m_{1};2,m_{2};l,l_{z} \right\rangle \,.
\label{formula_A1_12}
\end{eqnarray}
Here we have
\begin{eqnarray}
 l &=& 0,1,2,\ldots, \nonumber \\ 
-l &\leqslant& l_{z} \leqslant l\,, \nonumber \\
 S &=& 0,1,2,3,4\,,\nonumber \\
-S &\leqslant& S_{z} \leqslant S\,.
\label{formula_A1_13}
\end{eqnarray}
From Bose symmetry of our ``pomeron particles'' we find that
\begin{eqnarray}
\left\vert S,S_{z};l,l_{z} \right\rangle = 0 \quad \mathrm{for} \; l-S \; \mathrm{odd} \,.
\label{formula_A1_14}
\end{eqnarray}
The parity transformation $U(P)$ gives
\begin{eqnarray}
U(P) \left\vert S,S_{z};l,l_{z} \right\rangle = (-1)^{l} \left\vert S,S_{z};l,l_{z} \right\rangle \,.
\label{formula_A1_15}
\end{eqnarray}
It is straightforward to construct the two-pomeron states of definite total
angular momentum $J$, $J_{z}$:
\begin{eqnarray}
\left\vert l,S;J,J_{z} \right\rangle = 
\sum_{S_{z},l_{z}} 
\left\langle S,S_{z};l,l_{z} \vert J,J_{z} \right\rangle
\left\vert S,S_{z};l,l_{z} \right\rangle \,.
\label{formula_A1_16}
\end{eqnarray}
Clearly, $J$ is then the spin of the produced meson in (\ref{pom_to_pom})
and $P = (-1)^{l}$ its parity.
In Table~\ref{tab:table1_A1} we list the values of $J$ and $P$
of mesons which can be produced in our fictitious reaction (\ref{pom_to_pom})
where we restrict ourselves to $l \leqslant 4$.
\begin{table}
\caption{The values, for orbital angular momentum $l$,
of total spin $S$, total angular momentum $J$, and parity $P$,
possible in the annihilation reaction (\ref{pom_to_pom}).
The continuation of the table for $l > 4$ is straightforward.
}
\label{tab:table1_A1}
\begin{center}
\begin{tabular}{|c|c|l|c|}
\hline
$l$ & $S$ & $J$ & $P$ \\
\hline 
0 & 0 & 0 & $+$ \\ 
  & 2 & 2 &   \\ 
  & 4 & 4 &   \\ 
\hline 
1 & 1 & 0, 1, 2  & $-$ \\ 
  & 3 & 2, 3, 4  &   \\ 
\hline 
2 & 0 & 2          & $+$ \\ 
  & 2 & 0,1,2,3,4  &   \\ 
  & 4 & 2,3,4,5,6  &   \\ 
\hline 
3 & 1 & 2,3,4          & $-$ \\ 
  & 3 & 0,1,2,3,4,5,6  &   \\ 
\hline 
4 & 0 & 4                  & $+$ \\ 
  & 2 & 2,3,4,5,6          &   \\ 
  & 4 & 0,1,2,3,4,5,6,7,8  &   \\ 
\hline
\end{tabular}
\end{center}
\end{table}

It is clear that for each value of $l$, $S$, $J$, and $P$ listed in Table~\ref{tab:table1_A1}
we can construct a covariant Lagrangian density ${\cal L}'$ coupling
the field operator for the meson $M$ to the pomeron fields $I\!\!P_{\mu \nu}$.
There, $l$ is related to the number of derivatives in ${\cal L}'$,
thus giving an indication of the angular momentum barrier in the production of $M$
in (\ref{pom_to_pom}).
In Table~\ref{tab:table2_A1} we list interesting candidates for mesons $M$
in central production and the corresponding minimal values of $l$ and $S$
which can lead to the meson states according to Table~\ref{tab:table1_A1}.
\begin{table}
\caption{Candidates for mesons producible in pomeron-pomeron annihilation.
The values of the minimal orbital angular momentum $l$ and
of the corresponding total spin $S$ for the reactions (\ref{pom_to_pom}) and (\ref{pomV_to_pomV})
with tensorial ($I\!\!P_{T}$) and vectorial ($I\!\!P_{V}$) ``pomeron particles'', respectively, 
are also indicated.
}
\label{tab:table2_A1}
\begin{center}
\begin{tabular}{cl|c|c|c|c|}
\cline{3-6}
& & \multicolumn{2}{c|}{$I\!\!P_{T}$} & \multicolumn{2}{c|}{$I\!\!P_{V}$}\\ 
\cline{1-6} 
\multicolumn{1}{ |c| }{\multirow{1}{*}{$J^{PC}$} } &
\multicolumn{1}{ c| }{meson $M$} & $l$ & $S$ & $l$ & $S$  \\ 
\cline{1-6}
\multicolumn{1}{ |c| }{\multirow{2}{*}{$0^{-+}$} } &
\multicolumn{1}{ c| }{$\eta$} & 1 & 1 & 1 & 1  \\ 
\multicolumn{1}{ |c  }{}                       &
\multicolumn{1}{ |c| }{$\eta'(958)$} &  &  &  &   \\ 
\cline{1-6}
\multicolumn{1}{ |c| }{\multirow{3}{*}{$0^{++}$} } &
\multicolumn{1}{ c| }{$f_{0}(980)$} & 0 & 0 & 0 & 0  \\ 
\multicolumn{1}{ |c  }{}                       &
\multicolumn{1}{ |c| }{$f_{0}(1370)$} &  &  &  &   \\ 
\multicolumn{1}{ |c  }{}                       &
\multicolumn{1}{ |c| }{$f_{0}(1500)$} &  &  &  &   \\ 
\cline{1-6}
\multicolumn{1}{ |c| }{\multirow{2}{*}{$1^{++}$} } &
\multicolumn{1}{ |c| }{$f_{1}(1285)$} & 2 & 2 & 2 & 2  \\ 
\multicolumn{1}{ |c  }{}                       &
\multicolumn{1}{ |c| }{$f_{1}(1420)$} &  &  &  &   \\ 
\cline{1-6}
\multicolumn{1}{ |c| }{\multirow{2}{*}{$2^{++}$} } &
\multicolumn{1}{ |c| }{$f_{2}(1270)$} & 0 & 2 & 0 & 2  \\ 
\multicolumn{1}{ |c  }{}                       &
\multicolumn{1}{ |c| }{$f'_{2}(1525)$} &  &  &  &   \\ 
\cline{1-6}
\multicolumn{1}{ |c| }{\multirow{1}{*}{$4^{++}$} } &
\multicolumn{1}{ |c| }{$f_{4}(2050)$} & 0 & 4 & 2 & 2  \\ 
\cline{1-6}
\end{tabular}
\end{center}
\end{table}

The strategy is now to construct for a given meson $M$ of Table~\ref{tab:table2_A1}
a coupling Lagrangian ${\cal L}_{I\!\!P I\!\!P M}'$
corresponding to the $l$ and $S$ values listed there.
We illustrate this here for the case of a $J^{PC} = 0^{++}$ meson $M$.
The case of a pseudoscalar meson $\tilde{M}$ is treated
in Section~\ref{subsection:Scalar_and_pseudoscalar_meson_production}.

\begin{figure}[!ht]
(a)\includegraphics[width=0.3\textwidth]{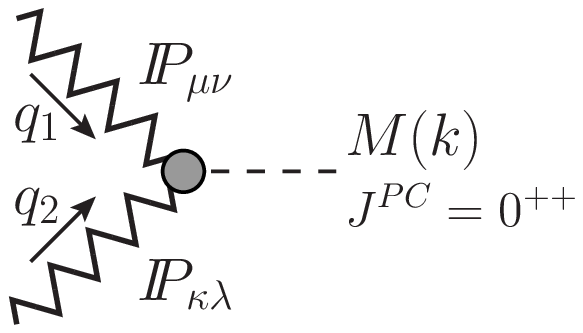}
(b)\includegraphics[width=0.3\textwidth]{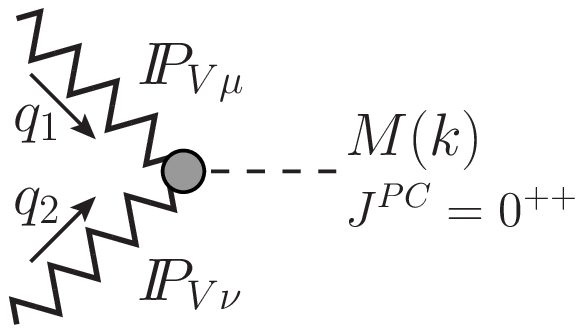}
  \caption{\label{fig:pMp_pom}
  \small
A sketch of the pomeron-pomeron-scalar meson vertex for the tensorial (a) 
and vectorial (b) pomeron fusion.}
\end{figure}

The Lagrangian ${\cal L}_{I\!\!P I\!\!P M}'$ for a scalar meson ($J^{PC} = 0^{++}$)
corresponding to $l = S = 0$ reads
\begin{eqnarray}
{\cal L}_{I\!\!P I\!\!P M}'(x) = 
M_{0} \, g_{I\!\!P I\!\!P M}' \, I\!\!P_{\mu \nu}(x) \, I\!\!P^{\mu \nu}(x) \, \chi(x) \,,
\label{formula_A1_17}
\end{eqnarray}
where $\chi(x)$ is the meson field operator, $M_{0} \equiv 1$~GeV,
and $g_{I\!\!P I\!\!P M}'$ is the dimensionless coupling constant.
The ``bare'' vertex obtained from (\ref{formula_A1_17}), see Fig.~\ref{fig:pMp_pom}~(a),
%
%
reads
\begin{eqnarray}
i\Gamma_{\mu \nu,\kappa \lambda}'^{(I\!\!P I\!\!P \to M)} \mid_{bare}=
i \, g_{I\!\!P I\!\!P M}' \, M_{0} \,  
\left( g_{\mu \kappa} g_{\nu \lambda} + g_{\mu \lambda} g_{\nu \kappa}
-\frac{1}{2} g_{\mu \nu} g_{\kappa \lambda} \right)\,.
\label{bare_vertex_pompomS}
\end{eqnarray}
Here we have made the vertex traceless since
the $I\!\!P_{\mu \nu}$ are supposed to have trace zero.

In Appendix~\ref{section:Covariant_Couplings} we use (\ref{bare_vertex_pompomS}) to calculate
the $T$-matrix element for the fictitious reaction (\ref{pom_to_pom})
with a scalar meson.
We show there that in the Wigner basis we get from (\ref{formula_A1_17})
an amplitude containing values of $(l,S)$ = $(0,0)$, $(2,2)$, and $(4,4)$.
But the higher terms are completely fixed by the lowest term $(l,S) = (0,0)$.
This justifies to call the coupling (\ref{formula_A1_17})
the one corresponding to $(l,S) = (0,0)$.

The coupling Lagrangian ${\cal L}_{I\!\!P I\!\!P M}''$ 
and vertex $\Gamma''^{(I\!\!P I\!\!P \to M)}$ 
corresponding to $l = S = 2$ read as follows:
\begin{eqnarray}
&& {\cal L}_{I\!\!P I\!\!P M}''(x) = 
\dfrac{1}{2 M_{0}} \, g_{I\!\!P I\!\!P M}'' \,
\left[ \partial^{\mu} I\!\!P^{\nu \rho}(x) - \partial^{\nu} I\!\!P^{\mu \rho}(x) \right] \, 
\left[ \partial_{\mu} I\!\!P_{\nu \rho}(x) - \partial_{\nu} I\!\!P_{\mu \rho}(x) \right] \, 
\chi(x)\,,
\label{formula_A1_17_bis} \\
&& i\Gamma_{\mu \nu,\kappa \lambda}''^{(I\!\!P I\!\!P \to M)} (q_{1}, q_{2}) \mid_{bare} = \nonumber
\dfrac{i \, g_{I\!\!P I\!\!P M}''}{2 M_{0}}\\   
&& \times \left[ q_{1 \kappa} q_{2 \mu} g_{\nu \lambda} + q_{1 \kappa} q_{2 \nu} g_{\mu \lambda}
         +q_{1 \lambda} q_{2 \mu} g_{\nu \kappa} + q_{1 \lambda} q_{2 \nu} g_{\mu \kappa} 
 -2 (q_{1}q_{2}) (g_{\mu \kappa} g_{\nu \lambda} + g_{\nu \kappa} g_{\mu \lambda}) \right]\,,
\label{bare_vertex_pompomS_bis}
\end{eqnarray}
where $g_{I\!\!P I\!\!P M}''$ is the dimensionless coupling constant.
The vertex (\ref{bare_vertex_pompomS_bis}) must be added coherently to the vertex 
(\ref{bare_vertex_pompomS}).

In the production reaction (\ref{2to3}) we cannot take the ``bare'' vertices 
((\ref{bare_vertex_pompomS}) and (\ref{bare_vertex_pompomS_bis})) directly.
We have to take into account that hadrons are extended objects, that is,
we shall have to introduce form factors.
The actual vertex which is assumed in this paper reads then as follows
\begin{eqnarray}
i\Gamma_{\mu \nu,\kappa \lambda}^{(I\!\!P I\!\!P \to M)} (q_{1},q_{2}) =
\left( i\Gamma_{\mu \nu,\kappa \lambda}'^{(I\!\!P I\!\!P \to M)}\mid_{bare} +
       i\Gamma_{\mu \nu,\kappa \lambda}''^{(I\!\!P I\!\!P \to M)} (q_{1}, q_{2})\mid_{bare} \right)
F_{I\!\!P I\!\!P M}(q_{1}^{2},q_{2}^{2}) \,.
\label{vertex_pompomS}
\end{eqnarray}

Unfortunately, the pomeron-pomeron-meson form factor
is not well known as it is due to
nonperturbative effects related to the internal structure of the respective meson.
In practical calculations we take the factorized form
with the following two approaches. Either we use
\begin{eqnarray}
F^{M}_{I\!\!P I\!\!P M}(t_{1},t_{2}) = F_{M}(t_{1}) F_{M}(t_{2})\,,
\label{Fpompommeson_pion}
\end{eqnarray}
with $F_{M}(t)$ the pion electromagnetic form factor in its simplest parametrization,
valid for $t < 0$,
\begin{eqnarray}
F_{M}(t)=F_{\pi}(t)=
\frac{1}{1-t/\Lambda_{0}^{2}}\,,
\label{Fpion}
\end{eqnarray}
where $\Lambda_{0}^{2} = 0.5$~GeV$^{2}$; see e.g.~(3.22) of \cite{DDLN}.
Alternatively, we use the exponential form given as
\begin{eqnarray}
F^{E}_{I\!\!P I\!\!P M}(t_{1},t_{2})=\exp\left(\frac{t_{1}+t_{2}}{\Lambda_{E}^{2}}\right)\,,
\label{Fpompommeson_exp}
\end{eqnarray}
where $\Lambda_{E}^{2} \approx 1$~GeV$^{2}$.
This discussion of form factors applies also to the other 
pomeron-pomeron-meson vertices considered in this paper.

In the case of meson-exchange diagrams we use the monopole form factor
which is normalized to unity at the on-shell point $t=m_{M}^{2}$
\begin{eqnarray}
F(t) = \frac{\Lambda_{M}^2 - m_{M}^2}{\Lambda_{M}^2 - t} \,,
\label{F_monopol_formfactor}
\end{eqnarray}
where $\Lambda_{M} > m_{M}$ and $t < 0$.
Alternatively, we use the exponential form
\begin{eqnarray}
F(t) = \exp\left(\frac{t - m_{M}^2}{\Lambda_{E}^2} \right) \,.
\label{F_exp_formfactor}
\end{eqnarray}
The influence of the choice of the form-factor parameters 
is discussed in the results section.

\section{Vectorial pomeron}
\label{section:Vectorial_Pomeron}
In this section we perform the same analysis for the vectorial pomeron ansatz
as is done for the tensorial pomeron in Appendix~\ref{section:Tensorial_Pomeron}.

In the vectorial approach, see \cite{DDLN}, \cite{DL},
the pomeron is treated as a ``$C = +1$ photon''.
Its coupling to the proton reads
\begin{eqnarray}
i\Gamma_{\mu}^{(I\!\!P_{V} pp)}(p',p)=
-i \,3 \beta_{I\!\!P NN} \,  F_{1}\bigl((p'-p)^2\bigr) \, M_{0} \, \gamma_{\mu}\,,
\label{vertex_pomVNN}
\end{eqnarray}
where $\beta_{I\!\!P NN} = 1.87$~GeV$^{-1}$, $M_{0} \equiv 1$~GeV; 
compare to (\ref{vertex_pomNN}).
The effective $I\!\!P_{V}$ propagator is given by
\begin{eqnarray}
i\Delta_{\mu \nu}^{(I\!\!P_{V})}(s,t)=
\frac{1}{M_{0}^{2}} \,
g_{\mu \nu}
\left(-i s \alpha'_{I\!\!P}\right)^{\alpha_{I\!\!P}(t)-1}\,,
\label{prop_pomV}
\end{eqnarray}
with $\alpha_{I\!\!P}(t)$ and $\alpha'_{I\!\!P}$ as in (\ref{pomeron_trajectory}).

From (\ref{vertex_pomVNN}) and (\ref{prop_pomV}) 
we get for proton-proton elastic scattering
\begin{eqnarray}
&&
\Braket{p(p_{1},\lambda_{1}),p(p_{2},\lambda_{2})|{\cal T}|p(p_{a},\lambda_{a}),p(p_{b},\lambda_{b})}
\mid_{I\!\!P_{V}}
\;\equiv \nonumber\\
&&{\cal M}^{2 \to 2}_{\lambda_{a}\lambda_{b} \to \lambda_{1}\lambda_{2}} \mid_{I\!\!P_{V}} =
(-i) \bar{u}(p_{1},\lambda_{1}) i \Gamma_{\mu}^{(I\!\!P_{V} pp)} (p_{1},p_{a}) u(p_{a},\lambda_{a})
\nonumber \\ 
&&\qquad \qquad \qquad \qquad \times 
i \Delta^{(I\!\!P_{V}) \, \mu \nu}(s,t)
\nonumber \\ 
&&\qquad \qquad \qquad \qquad \times 
\bar{u}(p_{2},\lambda_{2}) i \Gamma_{\nu}^{(I\!\!P_{V} pp)} (p_{2},p_{b}) u(p_{b},\lambda_{b})
\nonumber \\
&&\qquad \qquad \qquad
\xrightarrow{s \, \gg\, m_{p}^{2}}
i \, 2s \, \left[ 3 \beta_{I\!\!P NN} \,  F_{1}(t) \right]^{2} \,
\left(-i s \alpha'_{I\!\!P}\right)^{\alpha_{I\!\!P}(t)-1} \,
\delta_{\lambda_{1}\lambda_{a}} \, \delta_{\lambda_{2}\lambda_{b}} \,.
\label{2to2_pomV}
\end{eqnarray}
Comparing with (\ref{2to2_limit}) we see that for $s \gg m_{p}^{2}$,
both, the tensorial and the vectorial pomeron give the same
amplitude for $pp$ elastic scattering.

In the next step we consider the annihilation 
of two ``vector-pomeron particles'' into a meson $M$
\begin{eqnarray}
&& I\!\!P_{V} (\vec{k},1,m_{1}) + I\!\!P_{V} (-\vec{k},1,m_{2}) \to M(J,J_{z}) \,, \nonumber\\
&& m_{1,2} \in \{-1, 0, 1 \} \,, 
\quad J_{z} \in \{ -J, \ldots, J \} \,;
\label{pomV_to_pomV}
\end{eqnarray}
compare to (\ref{pom_to_pom}). Here, again, we use the Wigner basis.
The same analysis as done after (\ref{pom_to_pom}) for 
the tensorial pomeron can now be performed for the vectorial one.
The result is given in Table~\ref{tab:table1_A2}
which is the analogue of Table~\ref{tab:table1_A1}
for the tensorial pomeron.
\begin{table}
\caption{The values of $l$, $S$, $J$, and $P$,
of orbital angular momentum, total spin of the two ``vector-pomeron particles'',
total angular momentum, and parity of the state, respectively,
possible in the vectorial pomeron annihilation reaction (\ref{pomV_to_pomV}).
We have $S \in \{0,1,2 \}$, $P = (-1)^{l}$,
$|l-S| \leqslant J \leqslant l+S$, and Bose symmetry requires $l-S$ to be even.
The continuation of the table for $l > 4$ is straightforward.
}
\label{tab:table1_A2}
\begin{center}
\begin{tabular}{|c|c|l|c|}
\hline
$l$ & $S$ & $J$ & $P$ \\
\hline 
0 & 0 & 0 & $+$ \\ 
  & 2 & 2 &   \\ 
\hline 
1 & 1 & 0, 1, 2  & $-$ \\ 
\hline 
2 & 0 & 2          & $+$ \\ 
  & 2 & 0,1,2,3,4  &   \\ 
\hline 
3 & 1 & 2,3,4          & $-$ \\ 
\hline 
4 & 0 & 4                  & $+$ \\ 
  & 2 & 2,3,4,5,6          &   \\ 
\hline
\end{tabular}
\end{center}
\end{table}

As in Appendix~\ref{section:Tensorial_Pomeron} we illustrate the use of Table~\ref{tab:table1_A2}
by discussing the coupling of two vectorial pomerons to a $J^{PC} = 0^{++}$ meson $M$.
Let $\chi$ be the meson field, $I\!\!P_{V}^{\mu}$ the effective vector-pomeron field.
The coupling corresponding to $(l,S) = (0,0)$ reads
\begin{eqnarray}
{\cal L'}_{I\!\!P_{V} I\!\!P_{V} M}(x) = 
M_{0} \, g_{I\!\!P_{V} I\!\!P_{V} M}' \, I\!\!P_{V \mu}(x) \, I\!\!P_{V}^{\mu}(x) \, \chi(x) \,
\label{formula_A2_31}
\end{eqnarray}
with $M_{0} \equiv 1$~GeV,
and $g_{I\!\!P_{V} I\!\!P_{V} M}'$ the dimensionless coupling constant.
From (\ref{formula_A2_31}) we get the ``bare'' vertex, see Fig.~\ref{fig:pMp_pom}~(b),
\begin{eqnarray}
i\Gamma_{\mu \nu}'^{(I\!\!P_{V} I\!\!P_{V} \to M)} \mid_{bare}=
i \, g_{I\!\!P_{V} I\!\!P_{V} M}' \, M_{0} \, 2 g_{\mu \nu}  
\,.
\label{bare_vertex_pomVpomVS}
\end{eqnarray}
Using this vertex to calculate the amplitude for
the fictitious reaction (\ref{pomV_to_pomV}) we find,
in the Wigner basis, contributions with $(l,S) = (0,0)$
and $(2,2)$ with the $(2,2)$ part completely fixed by the $(0,0)$ part;
see Appendix~\ref{section:Covariant_Couplings}.
Thus, we shall refer to the coupling (\ref{bare_vertex_pomVpomVS})
as the one corresponding to $(l,S) = (0,0)$.

For $l = S = 2$ the coupling Lagrangian and vertex read as follows:
\begin{eqnarray}
&& {\cal L}_{I\!\!P_{V} I\!\!P_{V} M}''(x) = 
\dfrac{1}{2 M_{0}} \, g_{I\!\!P_{V} I\!\!P_{V} M}'' \,
\left[ \partial^{\mu} I\!\!P_{V}^{\nu}(x) - \partial^{\nu} I\!\!P_{V}^{\mu}(x) \right] \, 
\left[ \partial_{\mu} I\!\!P_{V \nu}(x) - \partial_{\nu} I\!\!P_{V \mu}(x) \right] \, 
\chi(x)\,,
\label{formula_A2_31_bis} \\
&& i\Gamma_{\mu \nu}''^{(I\!\!P_{V} I\!\!P_{V} \to M)} (q_{1}, q_{2}) \mid_{bare} =
\dfrac{2i \, g_{I\!\!P_{V} I\!\!P_{V} M}''}{M_{0}}   
\left[ q_{2 \mu} q_{1 \nu} - (q_{1}q_{2}) g_{\mu \nu} \right]\,,
\label{bare_vertex_pomVpomVS_bis}
\end{eqnarray}
where $g_{I\!\!P_{V} I\!\!P_{V} M}''$ is the dimensionless coupling constant.

The discussion of form factors for these vertices
is identical to the one for the tensorial pomeron in Appendix~\ref{section:Tensorial_Pomeron}.
Thus, for the full vertex for two vectorial pomerons giving a $0^{++}$ meson
we add (\ref{bare_vertex_pomVpomVS}) and (\ref{bare_vertex_pomVpomVS_bis}) 
and multiply the sum by a form factor
\begin{eqnarray}
i\Gamma_{\mu \nu}^{(I\!\!P_{V} I\!\!P_{V} \to M)} (q_{1},q_{2}) =
\left( 
i\Gamma_{\mu \nu}'^{(I\!\!P_{V} I\!\!P_{V} \to M)} \mid_{bare} +
i\Gamma_{\mu \nu}''^{(I\!\!P_{V} I\!\!P_{V} \to M)} (q_{1}, q_{2}) \mid_{bare} 
\right)
F_{I\!\!P I\!\!P M}(q_{1}^{2},q_{2}^{2}).
\label{vertex_pomVpomVS}
\end{eqnarray}
The coupling of two vectorial pomerons to a pseudoscalar mesons $\tilde{M}$
is given in Section~\ref{subsection:Scalar_and_pseudoscalar_meson_production}; 
cf. (\ref{Lagrangian_pseudoscalar_pomV}) and (\ref{vertex_pomVpomVPS}).

\section{Covariant $I\!\!P I\!\!P M$ couplings and the Wigner basis}
\label{section:Covariant_Couplings}
In this appendix we discuss the relation of the covariant $I\!\!P I\!\!P M$
couplings to the classification of partial wave amplitudes
in the Wigner basis as given in Table~\ref{tab:table1_A1}
for the tensorial and in Table~\ref{tab:table1_A2}
for the vectorial pomeron.

Let us consider as an example of the reaction (\ref{pomV_to_pomV})
the annihilation of two fictitious ``vectorial pomeron particles''
of mass $m$ giving a $J^{PC} = 0^{++}$ meson $M$:
\begin{eqnarray}
I\!\!P_{V} ( \vec{k},\vec{\varepsilon}_{1}^{\,W}) + 
I\!\!P_{V} (-\vec{k},\vec{\varepsilon}_{2}^{\,W}) \to M \,.
\label{B_1}
\end{eqnarray}
Here $\vec{\varepsilon}_{1,2}^{\,W}$ are the polarization vectors
in the Wigner basis with
\begin{eqnarray}
|\vec{\varepsilon}_{1}^{\,W}| = |\vec{\varepsilon}_{2}^{\,W}| = 1 \,.
\label{B_2}
\end{eqnarray}
To transform to the covariant polarization vectors
$\varepsilon_{i}\,^{\mu}$ ($i = 1,2$) we need the boost transformation $\Lambda_{\vec{k}}$:
\begin{eqnarray}
&& (\Lambda_{\vec{k}} \,^{\mu}_{\;\;\nu}) = 
\left( \begin{array}{cl}
\dfrac{k^{0}}{m} & \;\, \dfrac{k^{j}}{m}
\\ \\
\dfrac{k^{i}}{m} & \;\, \delta^{ij}+\hat{k}^{i}\hat{k}^{j} \Bigl( \dfrac{k^{0}}{m} -1 \Bigr)  \\
\end{array} \right) \,, \nonumber\\
&& i,j \in \left\lbrace 1,2,3 \right\rbrace , \quad \hat{k} = \vec{k}/|\vec{k}|\,.
\label{B_3}
\end{eqnarray}
We have
\begin{eqnarray}
&& (\varepsilon_{1}\,^{\mu}) = \Lambda_{\vec{k}}
\left( \begin{array}{c}
0 \\
\vec{\varepsilon}_{1}^{\,W} \\
\end{array} \right)\,, \nonumber\\
&& (\varepsilon_{2}\,^{\mu}) = \Lambda_{-\vec{k}}
\left( \begin{array}{c}
0 \\
\vec{\varepsilon}_{2}^{\,W} \\
\end{array} \right) \,.
\label{B_4}
\end{eqnarray}
From the vertex (\ref{bare_vertex_pomVpomVS}) we get
the amplitude for reaction (\ref{B_1}) as follows
\begin{eqnarray}
&& \Braket{M|{\cal T}|I\!\!P_{V} ( \vec{k},\vec{\varepsilon}_{1}^{\,W}),
                   I\!\!P_{V} (-\vec{k},\vec{\varepsilon}_{2}^{\,W})}
= \Gamma_{\mu \nu}'^{ ( I\!\!P_{V} I\!\!P_{V} \to M) }
\varepsilon_{1}\,^{\mu}
\varepsilon_{2}\,^{\nu} 
= -2 \,M_{0} \,g_{I\!\!P_{V} I\!\!P_{V} M}' \nonumber\\
&&\times
\Bigl[
\Bigl( 
1 + \frac{2}{3} \frac{\vec{k}^{2} }{m^{2}} 
\Bigr)
\vec{\varepsilon}_{1}^{\,W} \cdot \vec{\varepsilon}_{2}^{\,W}
+ \frac{1}{m^{2}} 
\Bigl(  
k^{i}k^{j}-\frac{1}{3}\delta^{ij} \vec{k}^{2} 
\Bigr)
\Bigl( 
       \varepsilon_{1}^{W \,i} \varepsilon_{2}^{W \,j} +
       \varepsilon_{2}^{W \,i} \varepsilon_{1}^{W \,j} -
       \frac{2}{3} \delta^{ij} \vec{\varepsilon}_{1}^{\,W} \cdot \vec{\varepsilon}_{2}^{\,W}
\Bigr)
\Bigr]. \, \, \, \, \, \, \, \, 
\label{B_5}
\end{eqnarray}
%
From the vertex (\ref{bare_vertex_pomVpomVS_bis}) we get
\begin{eqnarray}
&&\Braket{M|{\cal T}|I\!\!P_{V} ( \vec{k},\vec{\varepsilon}_{1}^{\,W}),
                   I\!\!P_{V} (-\vec{k},\vec{\varepsilon}_{2}^{\,W})}
= \frac{2 g_{I\!\!P_{V} I\!\!P_{V} M}''}{M_{0}}
\Bigl[ 
\Bigl( k_{2} \varepsilon_{1} \Bigr) 
\Bigl( k_{1} \varepsilon_{2} \Bigr)-
\Bigl( k_{1}k_{2}\Bigr)
\Bigl(\varepsilon_{1}\varepsilon_{2}\Bigr)
\Bigr] 
= \frac{2 g_{I\!\!P_{V} I\!\!P_{V} M}''}{M_{0}} \nonumber\\
&&\times
\Bigl[ 
\Bigl( 
\frac{4}{3} \vec{k}^{2} + m^{2}
\Bigr)
\vec{\varepsilon}_{1}^{\,W} \cdot \vec{\varepsilon}_{2}^{\,W}
-
\Bigl( 
k^{i}k^{j}-\frac{1}{3}\delta^{ij} \vec{k}^{2} 
\Bigr) 
\Bigl( 
       \varepsilon_{1}^{W \,i} \varepsilon_{2}^{W \,j} +
       \varepsilon_{2}^{W \,i} \varepsilon_{1}^{W \,j} -
       \frac{2}{3} \delta^{ij} \vec{\varepsilon}_{1}^{\,W} \cdot \vec{\varepsilon}_{2}^{\,W}
\Bigr)
\Bigr]. 
\label{B_5a}
\end{eqnarray}
Thus, in the Wigner basis we get from both vertices,
(\ref{bare_vertex_pomVpomVS}) and (\ref{bare_vertex_pomVpomVS_bis}),
partial wave amplitudes with $(l,S) = (0,0)$ and $(2,2)$.
Multiplying the vertices (\ref{bare_vertex_pomVpomVS}) and (\ref{bare_vertex_pomVpomVS_bis})
with suitable form factors and forming linear combinations
of them it would be possible to construct vertices giving
only $(l,S) = (0,0)$ or $(2,2)$ in the Wigner basis.
But this would be a very cumbersome procedure.
Therefore, we shall in this paper stick to the simple vertices as given
above and label (\ref{bare_vertex_pomVpomVS}) with $(l,S) = (0,0)$
and (\ref{bare_vertex_pomVpomVS_bis}) with $(l,S) = (2,2)$
since (\ref{bare_vertex_pomVpomVS}) has no momenta and
(\ref{bare_vertex_pomVpomVS_bis}) two momenta.
But we have keep in mind that the translation of the power of momenta
in the covariant vertices to the angular momentum $l$
in the Wigner basis is not one to one.

For the tensorial pomeron the situation is similar.
We discuss the reaction (\ref{pom_to_pom}) for a scalar meson $M$
\begin{eqnarray}
&&
I\!\!P ( \vec{k},\varepsilon_{1}^{W \,ij}) + 
I\!\!P (-\vec{k},\varepsilon_{2}^{W \,hl}) \to M \,, \nonumber\\
&& i,j,h,l \in \left\lbrace 1,2,3 \right\rbrace \,.
\label{B_6}
\end{eqnarray}
Here $\varepsilon_{1,2}^{W \,ij}$ are the polarization tensors
of the fictitious ``tensor-pomeron particle'' of mass $m$ in the Wigner basis.
We have:
\begin{eqnarray}
&& \varepsilon_{1}^{W \,ij} = \varepsilon_{1}^{W \,ji}, \quad
   \varepsilon_{2}^{W \,ij} = \varepsilon_{2}^{W \,ji}, \nonumber\\
&& \varepsilon_{1}^{W \,ij} \delta_{ij} = \varepsilon_{2}^{W \,ij} \delta_{ij} = 0 \,, \nonumber\\
&& (\varepsilon_{1}^{W \,ij})^{*} (\varepsilon_{1}^{W \,ji}) =1, \quad
   (\varepsilon_{2}^{W \,ij})^{*} (\varepsilon_{2}^{W \,ji}) =1\,.
\label{B_7}
\end{eqnarray}
The covariant polarization tensors are
\begin{eqnarray}
&& \varepsilon_{1}\,^{\mu \nu} = \Lambda_{ \vec{k}}\,^{\mu}_{\;\;i}\,
                                 \Lambda_{ \vec{k}}\,^{\nu}_{\;\;j}\,
                                 \varepsilon_{1}^{W \,ij} \,, \nonumber\\
&& \varepsilon_{2}\,^{\mu \nu} = \Lambda_{-\vec{k}}\,^{\mu}_{\;\;i}\,
                                 \Lambda_{-\vec{k}}\,^{\nu}_{\;\;j}\,
                                 \varepsilon_{2}^{W \,ij} \,.
\label{B_8}
\end{eqnarray}
With (\ref{B_8}) we obtain the amplitude for (\ref{B_6})
from the vertex (\ref{bare_vertex_pompomS}) as follows:
\begin{eqnarray}
&&\Braket{M|{\cal T}|I\!\!P ( \vec{k},\varepsilon_{1}^{W \,ij}),
                     I\!\!P (-\vec{k},\varepsilon_{2}^{W \,hl})}
= 2 \,M_{0} \,g_{I\!\!P I\!\!P M}' \,
\varepsilon_{1}\,^{ \mu \nu} \varepsilon_{2}\,_{\mu \nu} \,.
\label{B_9}
\end{eqnarray}
Inserting here the explicit expressions from (\ref{B_8})
we see easily that the amplitude (\ref{B_9}) has,
in the Wigner basis, partial wave parts with $(l,S) = (0,0), (2,2)$, and $(4,4)$.
Similarly, also the vertex (\ref{bare_vertex_pompomS_bis}) gives contributions
with $(l,S) = (0,0), (2,2)$, and $(4,4)$.
We label the vertex (\ref{bare_vertex_pompomS}) with $(l,S) = (0,0)$
since it has no momenta, and (\ref{bare_vertex_pompomS_bis}) with $(l,S) = (2,2)$
since it is quadratic in the momenta.

The discussion of other pomeron-pomeron-meson couplings when
going from the covariant forms to the partial wave amplitudes
in the Wigner basis can be done in a completely analogous way.

\section{Kinematic relations and the high-energy small-angle limit}
\label{section:Kinematic_Relations}
The following relations hold, cf. (\ref{2to3_kinematics}),
\begin{eqnarray}
s_{13} &=& (p_{a} + q_{2})^{2}\nonumber \\
       &=& (s- 2 m_{p}^{2}) \, \xi_{2} + m_{p}^{2} + t_{2}\,, \nonumber \\
s_{23} &=& (p_{b} + q_{1})^{2}\nonumber \\
       &=& (s- 2 m_{p}^{2}) \, \xi_{1} + m_{p}^{2} + t_{1}\,,
\label{D_1}
\end{eqnarray}
where $\xi_{1} = \dfrac{p_{b} \cdot q_{1}}{p_{b} \cdot p_{a}}$ and
$\xi_{2} = \dfrac{p_{a} \cdot q_{2}}{p_{a} \cdot p_{b}}$
are the fractional energy losses
of the protons with momenta $p_{a}$ and $p_{b}$, respectively.
We consider now the reaction (\ref{2to3}) in the overall c.m.~system
with the $z$ axis along $\vec{p}_{a}$.
We have then
\begin{eqnarray}
&& p_{a} = 
\left( \begin{array}{c}
p_{a}^{0}  \\
0 \\
0 \\
|\vec{p}_{a}|  \\
\end{array} \right), \quad
p_{b} = 
\left( \begin{array}{c}
p_{b}^{0}  \\
0 \\
0 \\
-|\vec{p}_{b}|  \\
\end{array} \right), 
\nonumber\\
&& p_{a}^{0} = p_{b}^{0} = \frac{\sqrt{s}}{2}, \quad
|\vec{p}_{a}| = |\vec{p}_{b}| = \frac{1}{2}\sqrt{s - 4 m_{p}^{2}} \,.
\label{D_2}
\end{eqnarray}
With $i =1,2$ we get
\begin{eqnarray}
&& p_{i} = 
\left( \begin{array}{c}
p_{i}^{0}  \\
\vec{p}_{i \perp} \\
p_{i z}  \\
\end{array} \right), \quad
q_{i} = 
\left( \begin{array}{c}
q_{i}^{0}  \\
\vec{q}_{i \perp} \\
q_{i z}  \\
\end{array} \right), 
\nonumber\\
&& \vec{p}_{i \perp} = 
|\vec{p}_{i \perp}| 
\left( \begin{array}{c}
\cos\phi_{i}  \\
\sin\phi_{i}  \\
\end{array} \right), \quad  \vec{q}_{i \perp} =  -\vec{p}_{i \perp} \,.
\label{D_3}
\end{eqnarray}
The azimuthal angle $\phi_{pp}$ between the two outgoing protons in (\ref{2to3}) is given by
\begin{eqnarray}
\phi_{pp} = \phi_{1} - \phi_{2} \,.
\label{D_4}
\end{eqnarray}
The ``glueball variable'' \cite{CK97} $dP_{\perp} = |d\vec{P}_{\perp}|$
is defined by the difference of the transverse momentum vectors
\begin{eqnarray}
d\vec{P}_{\perp} = \vec{q}_{1\perp} - \vec{q}_{2\perp} = \vec{p}_{2\perp} - \vec{p}_{1\perp} \,.
\label{D_4a}
\end{eqnarray}
Further relations are as follows
(no summation over $i$ in (\ref{D_6}) for $\xi_{i}t_{i}$)
\begin{eqnarray}
&&m_{M}^{2} = k^{2} = 2 q_{1} q_{2} + t_{1} + t_{2} \nonumber \\
&& \quad  = \frac{(s-2 m_{p}^{2})^{3}}{s(s-4 m_{p}^{2})} \xi_{1} \xi_{2}
          + t_{1} + t_{2} - 2 \vec{q}_{1\perp} \cdot \vec{q}_{2\perp}
          + \frac{(s-2 m_{p}^{2})}{s(s-4 m_{p}^{2})} 
          \left[ t_{1}t_{2} - 2m_{p}^{2}(t_{1} \xi_{2} + t_{2} \xi_{1}) \right],
\label{D_5}\\
&&t_{i} = -\vec{q}_{i\perp}^{\,2} + \frac{1}{s(s-4 m_{p}^{2})} 
\left[ (s-2 m_{p}^{2})^{2} (\xi_{i}t_{i} - \xi_{i}^{2}m_{p}^{2}) - t_{i}^{2}m_{p}^{2} \right],
\label{D_6}\\
&&p_{i}^{0} = \frac{\sqrt{s}}{2} (1-\xi_{i}) + \frac{1}{2\sqrt{s}} (2 m_{p}^{2} \xi_{i} - t_{i})\,,
\label{D_7}
\end{eqnarray}
\begin{eqnarray}
&&\varepsilon_{\mu_{1} \mu_{2} \rho \sigma} 
(p_{1}+p_{a})^{\mu_{1}} 
(p_{2}+p_{b})^{\mu_{2}} 
(q_{1}-q_{2})^{\rho} 
(q_{1}+q_{2})^{\sigma}
=-8 \sqrt{s} \, \vec{p}_{a} \cdot (\vec{p}_{1\perp} \times \vec{p}_{2\perp}) \nonumber\\
&& \qquad \qquad \qquad \qquad \qquad \qquad \qquad \qquad \qquad \quad \quad \; \; \,
= 8 \sqrt{s} \, |\vec{p}_{a}| |\vec{p}_{1\perp}| |\vec{p}_{2\perp}| \sin\phi_{pp}\,.
\label{D_8a}
\end{eqnarray}

In Figs.~\ref{fig:dsig_dy} and \ref{fig:5} we have shown distributions
in rapidity, $\mathrm{y}_{M}$, and pseudorapidity, $\eta_{M}$,
of the produced meson $M$ in the overall c.m. system.
We discuss here their kinematic relation.
We have with
\begin{eqnarray}
k = 
\left( \begin{array}{c}
k^{0}  \\
\vec{k}_{\perp} \\
k_{z}  \\
\end{array} \right) \,,
\label{D_13}
\end{eqnarray}
the four-momentum of meson $M$ ($k^{2}=m_{M}^{2}$),
\begin{eqnarray}
\mathrm{y}_{M} &=&\frac{1}{2}\,\ln \frac{k^{0}+k_{z}}
                             {k^{0}-k_{z}} 
       = \ln \frac{\sqrt{k_{z}^{2} + \vec{k}^{2}_{\perp} + m_{M}^{2}} + k_{z}}
                {\sqrt{\vec{k}^{2}_{\perp} + m_{M}^{2}}}\,,\\
\eta_{M} &=&\frac{1}{2}\,\ln \frac{|\vec{k}|+k_{z}}
                                {|\vec{k}|-k_{z}} 
          = \ln \frac{\sqrt{k_{z}^{2} + \vec{k}^{2}_{\perp}} + k_{z}}
                   {\sqrt{\vec{k}^{2}_{\perp}}}\,.
\label{D_14}
\end{eqnarray}
Consider now the distributions of meson $M$ in ($\mathrm{y}_{M}, \vec{k}^{2}_{\perp}$)
and ($\eta_{M}, \vec{k}^{2}_{\perp}$).
We have
\begin{eqnarray}
&& f(\mathrm{y}_{M}, \vec{k}^{2}_{\perp}) d\mathrm{y}_{M} d(\vec{k}^{2}_{\perp}) =
   \tilde{f}(\eta_{M}, \vec{k}^{2}_{\perp}) d\eta_{M} d(\vec{k}^{2}_{\perp})\,,
\label{D_15}\\
&& \tilde{f}(\eta_{M}, \vec{k}^{2}_{\perp}) = 
   f(\mathrm{y}_{M}, \vec{k}^{2}_{\perp}) 
   \frac{\partial \mathrm{y}_{M}/\partial k_{z}}{\partial \eta_{M}/\partial k_{z}} 
   \mid_{\vec{k}^{2}_{\perp} \; \mathrm{fixed}}\,,
\label{D_16}
\end{eqnarray}
where
\begin{eqnarray}
\frac{\partial \mathrm{y}_{M}/\partial k_{z}}{\partial \eta_{M}/\partial k_{z}} 
   \mid_{\vec{k}^{2}_{\perp} \; \mathrm{fixed}} =
   \frac{\sqrt{k_{z}^{2} + \vec{k}^{2}_{\perp}}}
        {\sqrt{k_{z}^{2} + \vec{k}^{2}_{\perp} + m_{M}^{2}}} \equiv
   w(k_{z},\vec{k}^{2}_{\perp})\,.
\label{D_17}
\end{eqnarray}
Clearly, for large $|\mathrm{y}_{M}|$ and correspondingly large $|\eta_{M}|$
we have $|k_{z}| \gg m_{M}$ and the transformation factor $w(k_{z},\vec{k}^{2}_{\perp}) \to 1$.
On the other hand, for $|\mathrm{y}_{M}| \to 0$ corresponding to $|\eta_{M}| \to 0$
and $k_{z} \to 0$ we have $w(0,\vec{k}^{2}_{\perp}) < 1$.
Thus, we conclude that, for fixed $\vec{k}^{2}_{\perp} \neq 0$ 
a $\mathrm{y}_{M}$ distribution which is roughly constant for $|\mathrm{y}_{M}| \to 0$
will give a dip in the $\eta_{M}$ distribution for $|\eta_{M}| \to 0$.
A dip in the $\mathrm{y}_{M}$ distribution for $|\mathrm{y}_{M}| \to 0$ 
will be deepened in the $\eta_{M}$ distribution.
To get the $\mathrm{y}_{M}$ and $\eta_{M}$ distributions of Figs.~\ref{fig:dsig_dy} and \ref{fig:5} 
we still have to integrate in (\ref{D_15}) over $\vec{k}^{2}_{\perp}$.
We note, however, that integration over $\vec{k}^{2}_{\perp}$ at fixed $\mathrm{y}_{M}$ is,
in general, not the same as integration at fixed $\eta_{M}$.
Nevertheless, if the unintegrated distributions of (\ref{D_15}) in $(\mathrm{y}_{M}, \vec{k}^{2}_{\perp})$,
respectively $(\eta_{M}, \vec{k}^{2}_{\perp})$, behave ``reasonably''
we should be able to replace in the above arguments fixed $\vec{k}^{2}_{\perp}$
by some mean value $\langle \vec{k}^{2}_{\perp} \rangle$.
Then the above features will survive.
That is, a $\mathrm{y}_{M}$ distribution being roughly constant
for $|\mathrm{y}_{M}| \to 0$ will give a dip for $|\eta_{M}| \to 0$,
as observed in Fig.~\ref{fig:dsig_dy}.
A dip in the $\mathrm{y}_{M}$ distribution for $|\mathrm{y}_{M}| \to 0$ will be deepened
in the $\eta_{M}$ distribution, as observed in Fig.~\ref{fig:5}.

We consider now the high-energy small-angle limit where we require in reaction (\ref{2to3})
\begin{eqnarray}
|t_{1}|, |t_{2}| \ll m_{p}^{2}\,, \quad
m_{M}^{2} \ll s\,, \quad
\xi_{1}, \xi_{2} = \mathcal{O}(m_{M}/\sqrt{s}) \,.
\label{D_8}
\end{eqnarray}
In this limit 
we have the simple relations
\begin{eqnarray}
&&\xi_{1} \cong \frac{s_{23}}{s}\,,\quad \xi_{2} \cong \frac{s_{13}}{s}\,,\quad m_{M}^{2} \cong s \xi_{1} \xi_{2} = \frac{s_{13} s_{23}}{s}\,, \quad t_{1} \cong -\vec{q}_{1\perp}^{\,2}\,,\quad t_{2} \cong -\vec{q}_{2\perp}^{\,2}\,;
\label{D_9}\\
&& \bar{u}(p_{1},\lambda_{1}) \gamma^{\mu} u(p_{a},\lambda_{a}) \cong (p_{1} + p_{a})^{\mu}
   \delta_{\lambda_{1}\lambda_{a}}\,, \nonumber \\
&& \bar{u}(p_{2},\lambda_{2}) \gamma^{\mu} u(p_{b},\lambda_{b}) \cong (p_{2} + p_{b})^{\mu}
   \delta_{\lambda_{2}\lambda_{b}}\,;
\label{D_10}\\
&&(p_{1} + p_{a}, p_{2} + p_{b}) \cong 2s\,;
\label{D_11}\\
&&(q_{1},p_{2}+p_{b})(q_{2},p_{1}+p_{a})-
(q_{1},q_{2})(p_{1}+p_{a},p_{2}+p_{b}) \cong 2s \, \vec{p}_{1\perp} \cdot \vec{p}_{2\perp} 
   = 2s \, |\vec{p}_{1\perp}| |\vec{p}_{2\perp}| \cos\phi_{pp}\,.\nonumber\\
\label{D_11a}
\end{eqnarray}
We see from (\ref{D_8}) and (\ref{D_9}) that in this limit
both subenergies squared become large
\begin{eqnarray}
s_{13}, s_{23} = \mathcal{O}(m_{M}\sqrt{s})\,.
\label{D_12}
\end{eqnarray}
%


\end{document}